\documentclass[11pt,a4paper]{article}
\usepackage{jheppub}

\title{Study of Monte Carlo approach to experimental uncertainty propagation with MSTW 2008 PDFs}

\author[a]{G.~Watt}
\author[b]{and R.~S.~Thorne}

\affiliation[a]{Theory Group, Physics Department, CERN, CH-1211 Geneva 23, Switzerland}
\affiliation[b]{Department of Physics and Astronomy, University College London, WC1E 6BT, UK}

\emailAdd{Graeme.Watt@cern.ch}
\emailAdd{thorne@hep.ucl.ac.uk}

\abstract{We investigate the Monte Carlo approach to propagation of experimental uncertainties within the context of the established ``MSTW 2008'' global analysis of parton distribution functions (PDFs) of the proton at next-to-leading order in the strong coupling.  We show that the Monte Carlo approach using replicas of the original data gives PDF uncertainties in good agreement with the usual Hessian approach using the standard $\Delta\chi^2=1$ criterion, then we explore potential parameterisation bias by increasing the number of free parameters, concluding that any parameterisation bias is likely to be small, with the exception of the valence-quark distributions at low momentum fractions $x$.  We motivate the need for a larger tolerance, $\Delta\chi^2>1$, by making fits to restricted data sets and idealised consistent or inconsistent pseudodata.  Instead of using data replicas, we alternatively produce PDF sets randomly distributed according to the covariance matrix of fit parameters including appropriate tolerance values, then we demonstrate a simpler method to produce an arbitrary number of random predictions on-the-fly from the existing eigenvector PDF sets.  Finally, as a simple example application, we use Bayesian reweighting to study the effect of recent LHC data on the lepton charge asymmetry from $W$ boson decays.}


\arxivnumber{1205.4024}

\begin{document}

\begin{flushright}
  CERN-PH-TH/2012-132 \\
  LCTS/2012-11 \\
  17th July 2012
\end{flushright}

\maketitle

\section{Introduction} \label{sec:introduction}
The parton distribution functions (PDFs) of the proton are best determined from global analysis of a wide variety of deep-inelastic scattering (DIS) and related hard-scattering data taken from both fixed-target experiments and colliders (HERA, the Tevatron, and most recently the LHC).  Propagation of the experimental errors on the fitted data points to the uncertainties on the PDFs is a non-trivial task.  The traditional Hessian method requires effective error inflation by a \emph{tolerance} parameter to accommodate minor inconsistencies between the fitted data sets.  This means that the PDF uncertainties cannot be considered to be statistically rigorous, despite the r\^ole of PDF uncertainties as an important (and sometimes dominant) source of theoretical uncertainty on predicted quantities, such as the cross sections for Drell--Yan processes or Higgs boson production at the Tevatron and LHC~\cite{Watt:2011kp,Thorne:2011kq}.  Moreover, the number of fitted parameters for error propagation in the Hessian method must be kept sufficiently small to avoid large correlations, often requiring several parameters to be held fixed and thereby introducing a potential parameterisation bias.  Some insight into these problems may be gained using Monte Carlo techniques~\cite{Giele:1998gw,Giele:2001mr}, recently used in conjunction with a neural-network parameterisation by the NNPDF Collaboration (\cite{Ball:2011eq}, and references therein), where a large number $N_{\rm rep}\sim \mathcal{O}(10$--$1000)$ of fits are performed, each to a sample of replica pseudodata generated by shifting the original data points by random amounts dependent on the data errors.  Then the PDF uncertainties can be calculated by simply taking the standard deviation of the resulting $N_{\rm rep}$ PDF sets.

In this paper we make a first study of the Monte Carlo approach to experimental error propagation within the context of the established ``MSTW 2008'' PDF determination~\cite{Martin:2009iq}.  We retain the usual functional-form parameterisation and least-squares $\chi^2$-minimisation (using the Levenberg--Marquardt algorithm) rather than moving to the neural-network parameterisation and genetic-algorithm $\chi^2$-minimisation of the NNPDF approach~\cite{Ball:2011eq}.  We focus on the most widely-used PDF determination at next-to-leading order (NLO) in the strong coupling $\alpha_S$, although the results would be expected to be similar at leading-order (LO) and at next-to-next-to-leading order (NNLO).  Moreover, to avoid complications associated with simultaneously fitting $\alpha_S$ with the PDFs, throughout this paper we keep the value of $\alpha_S(M_Z^2)$ held fixed at the MSTW 2008 NLO best-fit value.  First in section~\ref{sec:generation} we describe the Monte Carlo approach using data replicas and compare results to the usual Hessian method, then in section~\ref{sec:parambias} we explore potential parameterisation bias by increasing the number of free parameters.  We then motivate the need for a tolerance parameter by fitting restricted data sets in section~\ref{sec:restricted} and by fitting idealised pseudodata in section~\ref{sec:theory}.  In section~\ref{sec:random} we explain how to produce PDF sets randomly distributed in the space of parameters rather than in the space of data, which allows the inclusion of a suitable tolerance.  As an example application of these random PDFs, in section~\ref{sec:reweighting} we demonstrate the use of Bayesian reweighting to study the effect of recent LHC data on the $W\to\ell\nu$ charge asymmetry~\cite{Chatrchyan:2011jz,Aad:2011dm}.  Finally, we conclude in section~\ref{sec:conclusions}.

\section{Comparison of Hessian and Monte Carlo uncertainties} \label{sec:generation}

\subsection{Recap of the Hessian method}
The basic procedure for propagating experimental uncertainties in global PDF analyses using the Hessian method is discussed in detail in refs.~\cite{Pumplin:2001ct,Pumplin:2002vw,Martin:2002aw,Martin:2009iq}.  Here, we briefly review the salient points.  We assume that the global goodness-of-fit quantity, $\chi^2_{\rm global}$, is quadratic about the global minimum, which has $n$ best-fit parameters $\{a_1^0,\ldots,a_n^0\}$.  In this case we can write
\begin{equation} \label{eq:hessian}
  \Delta\chi^2_{\rm global} \equiv \chi^2_{\rm global} - \chi_{\rm min}^2 = \sum_{i,j=1}^n H_{ij}(a_i-a_i^0)(a_j-a_j^0),
\end{equation}
where the Hessian matrix $H$ has components
\begin{equation}
  H_{ij} = \left.\frac{1}{2}\frac{\partial^2\chi^2_{\rm global}}{\partial a_i\partial a_j}\right|_{\rm min}.
\end{equation}
It is convenient to diagonalise the covariance (inverse Hessian) matrix, $C\equiv H^{-1}$, also known as the error matrix, and work in terms of the eigenvectors and eigenvalues.  Since the covariance matrix is symmetric it has a set of orthonormal eigenvectors $\vec{v}_k$ defined by
\begin{equation} \label{eq:eigeq}
  \sum_{j=1}^n C_{ij} v_{jk} = \lambda_k v_{ik},
\end{equation}
where $\lambda_k$ is the $k$th eigenvalue and $v_{ik}$ is the $i$th component of the $k$th orthonormal eigenvector ($k = 1,\ldots,n$).  The parameter displacements from the global minimum can be expanded in a basis of rescaled eigenvectors $e_{ik}\equiv \sqrt{\lambda_k}v_{ik}$, that is,
\begin{equation} \label{eq:eigbasis}
  a_i - a_i^0 = \sum_{k=1}^n e_{ik} z_k.
\end{equation}
Then it can be shown, using the orthonormality of $\vec{v}_k$, that eq.~\eqref{eq:hessian} reduces to
\begin{equation} \label{eq:hessiandiag}
  \chi^2_{\rm global} = \chi^2_{\rm min} + \sum_{k=1}^n z_k^2,
\end{equation}
that is, $\sum_{k=1}^n z_k^2\le T^2$ is the interior of a hypersphere of radius $T$.  Pairs of eigenvector PDF sets $S_k^\pm$ can then be produced to span this hypersphere, with parameters given by
\begin{equation} \label{eq:eigenstept}
  a_i(S_k^\pm) = a_i^0 \pm t\,e_{ik}.
\end{equation}
In the quadratic approximation, $t=T\equiv(\Delta\chi^2_{\rm global})^{1/2}$, but particularly for the larger eigenvalues $\lambda_k$ there are significant deviations from the ideal quadratic behaviour, so in general $t$ is adjusted iteratively to give the desired value of $T$.  Then asymmetric PDF uncertainties on a quantity $F$, which may be an individual PDF at particular values of $x$ and $Q^2$, or a derived quantity such as a cross section, can be calculated with the following ``master equations'':
\begin{align}
  (\Delta F)_+ &= \sqrt{\sum_{k=1}^n \left\{{\rm max}\left[\;F(S_k^+)-F(S_0),\;F(S_k^-)-F(S_0),\;0\right]\right\}^2}, \label{eq:Fp} \\
  (\Delta F)_- &= \sqrt{\sum_{k=1}^n \left\{{\rm max}\left[\;F(S_0)-F(S_k^+),\;F(S_0)-F(S_k^-),\;0\right]\right\}^2}, \label{eq:Fm}
\end{align}
where $S_0$ is the central PDF set.  Symmetric PDF uncertainties can be calculated with
\begin{equation} \label{eq:symmunc}
  \Delta F = \frac{1}{2}\sqrt{\sum_{k=1}^n \left[F(S_k^+)-F(S_k^-)\right]^2}.
\end{equation}
Ideally, with the standard ``parameter-fitting'' criterion~\cite{Collins:2001es}, we would expect the errors to be given by the choice of tolerance $T=1$ for the $68\%$ (one-sigma) confidence-level (C.L.) limit or $T=1.64$ for the $90\%$ C.L.~limit~\cite{Nakamura:2010zzi}.  This criterion is appropriate if fitting consistent data sets with ideal Gaussian errors to a well-defined theory.  However, in practice, there are some inconsistencies between the independent fitted data sets, and unknown experimental and theoretical uncertainties, so the parameter-fitting criterion is not appropriate for global PDF analyses.  Historically, the CTEQ~\cite{Pumplin:2002vw} and MRST~\cite{Martin:2002aw} groups defined $90\%$ C.L.~uncertainties using $T = \sqrt{100}$ and $T = \sqrt{50}$, respectively.  Instead, the ``MSTW 2008'' analysis~\cite{Martin:2009iq} introduced a new ``dynamic'' determination of the tolerance, chosen separately for each eigenvector direction according to a ``hypothesis-testing'' criterion~\cite{Collins:2001es} to maintain an adequate description of each individual data set in the global fit.  Therefore, the distance $t$ in eq.~\eqref{eq:eigenstept} was replaced by $t_k^\pm$, adjusted to give the desired $T_k^\pm$, with an average value of $\langle t_k^\pm\rangle\approx\langle T_k^\pm\rangle\approx 3$ for 68\% C.L.~uncertainties, and $\langle t_k^\pm\rangle\approx\langle T_k^\pm\rangle\approx 6$ for 90\% C.L.~uncertainties; see figure 10 of ref.~\cite{Martin:2009iq} for the individual $T_k^\pm$ values in the MSTW 2008 NLO fit.

\subsection{Generation of Monte Carlo replica sets} \label{sec:MCgen}
We generate replica data sets with the central values shifted according to
\begin{equation} \label{eq:MCgen}
  D_{m,i} \to \left(D_{m,i}+R_{m,i}^{\rm uncorr.}\,\sigma_{m,i}^{\rm uncorr.}+\sum_{k=1}^{N_{\rm corr.}}R_{m,k}^{\rm corr.}\,\sigma_{m,k,i}^{\rm corr.}\right)\cdot\left(1+R_m^{\mathcal{N}}\,\sigma_m^{\mathcal{N}}\right).
\end{equation}
Here, ``$m$'' labels a particular data set, or a combination of data sets, with a common (fitted) normalisation $\mathcal{N}_m$, ``$i$'' labels the individual data points in that data set, and ``$k$'' labels the individual correlated systematic errors for a particular data set.  The individual data points $D_{m,i}$ have uncorrelated (statistical and systematic) errors $\sigma_{m,i}^{\rm uncorr.}$ and correlated systematic errors $\sigma_{m,k,i}^{\rm corr.}$.  Treating the correlated errors as uncorrelated leads to the alternative form used for most of the data sets in the MSTW 2008 fit:
\begin{equation} \label{eq:MCgenQuad}
  D_{m,i} \to \left(D_{m,i}+R_{m,i}^{\rm uncorr.}\,\sigma_{m,i}^{\rm tot.}\right)\cdot\left(1+R_m^{\mathcal{N}}\,\sigma_m^{\mathcal{N}}\right),
\end{equation}
where the total error is simply obtained by adding all errors (except normalisation) in quadrature,
\begin{equation}
  \left(\sigma_{m,i}^{\rm tot.}\right)^2 = \left(\sigma_{m,i}^{\rm uncorr.}\right)^2 + \sum_{k=1}^{N_{\rm corr.}}\left(\sigma_{m,k,i}^{\rm corr.}\right)^2.
\end{equation}
We shift the data points in a way to be as consistent as possible with the $\chi^2$ definition used in the MSTW 2008 fit~\cite{Martin:2009iq}.  The random numbers $R_{m,i}^{\rm uncorr.}$ or $R_{m,k}^{\rm corr.}$ are obtained from a Gaussian distribution of mean zero and variance one.  A complication arises with the treatment of normalisation uncertainties in the MSTW 2008 analysis, where a \emph{quartic} penalty term was used in the $\chi^2$ definition instead of the usual quadratic penalty term, cf.~eqs.~(35) and (37) of ref.~\cite{Martin:2009iq}.  This modification was made to discourage large normalisation shifts in the fit.  It was partly motivated by claims (see section 6.7.4 on ``Normalizations'', pg.~170 in ref.~\cite{Devenish:2004pb}) that, for many experiments, quoted normalisation uncertainties represent the limits of a box-shaped distribution rather than the standard deviation of a Gaussian distribution; see further discussion in section 5.2.1 of ref.~\cite{Martin:2009iq}.  The quartic $\chi^2$ penalty term is small if the fitted normalisation $\mathcal{N}_m\in[1-\sigma_m^{\mathcal{N}},1+\sigma_m^{\mathcal{N}}]$, then it rises rapidly outside this range, with the effect that the normalisation uncertainty is perhaps closer to being described by a box-shaped distribution than by a Gaussian distribution (which would correspond to a quadratic $\chi^2$ penalty term).  Therefore, by default we take $R_m^{\mathcal{N}}$ in eqs.~\eqref{eq:MCgen} and \eqref{eq:MCgenQuad} to be uniformly distributed in the interval $(-1,1)$, so that the normalisation $\mathcal{N}_m$ is uniformly distributed in the interval $(1-\sigma_m^{\mathcal{N}},1+\sigma_m^{\mathcal{N}})$.  However, we have also looked at the effect of obtaining $R_m^{\mathcal{N}}$ from a Gaussian distribution or alternatively simply fixing $R_m^{\mathcal{N}}=0$, i.e.~the case of fixed data set normalisations.  As expected, fixing normalisations in the data replicas generally gives slightly smaller PDF uncertainties, while assuming normalisation uncertainties to be Gaussian gives larger PDF uncertainties, particularly for the up-valence distribution.  However, it is perhaps inconsistent to assume Gaussian uncertainties in the replica generation with a quartic penalty term in the $\chi^2$: changing to a quadratic penalty term would allow more freedom in the fitted normalisations and so the PDF parameters would move less, likely reducing the PDF uncertainty compared to the case of a quartic penalty term.  The default treatment of uniform $R_m^{\mathcal{N}}\in(-1,1)$ is probably reasonable and is closer to the treatment of normalisation uncertainties in the $\chi^2$ definition than a Gaussian $R_m^{\mathcal{N}}$.  The Hessian error propagation via eigenvector PDF sets includes theoretical uncertainties on the hadronisation corrections for the CDF jet data (treated as a correlated systematic) and the small modification for the nuclear corrections ($r_1$, $r_2$, $r_3$)~\cite{Martin:2009iq}.  It is currently not obvious how to treat these theoretical uncertainties in the replica generation, so the effect on PDF uncertainties will be assumed to be small.

\begin{figure}
  \centering
  \begin{minipage}{0.5\textwidth}
    (a)\\
    \includegraphics[width=\textwidth]{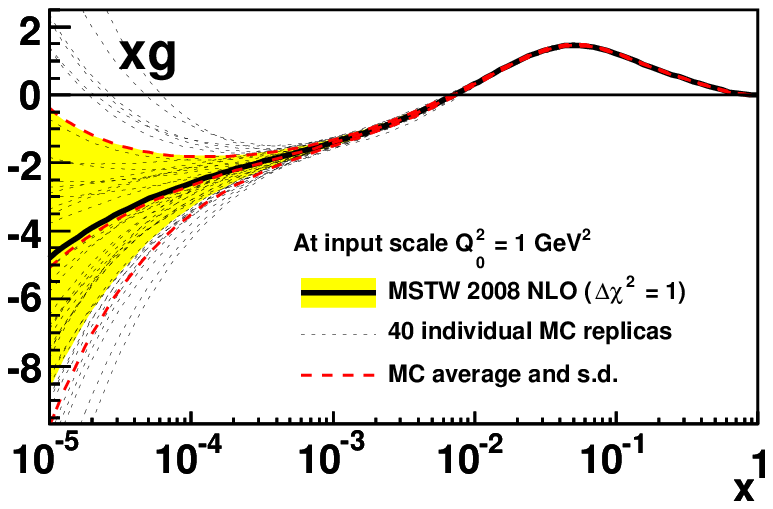}
  \end{minipage}%
  \begin{minipage}{0.5\textwidth}
    (b)\\
    \includegraphics[width=\textwidth]{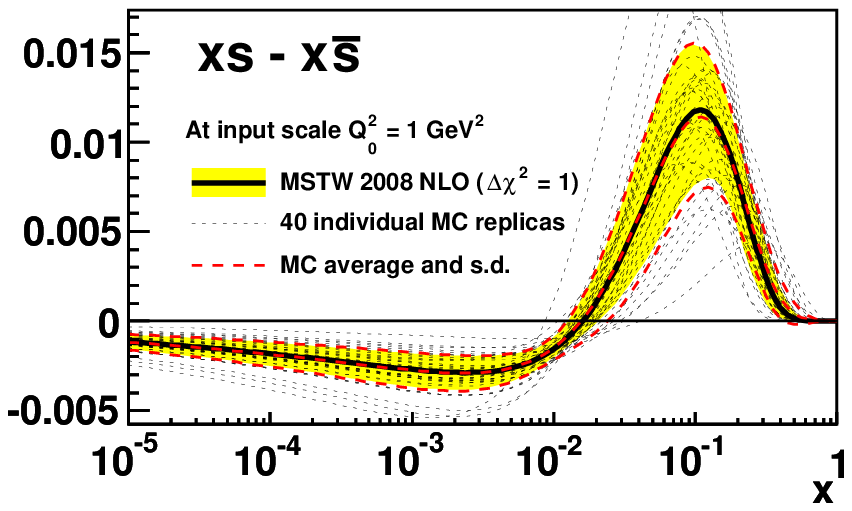}
  \end{minipage}
  \caption{Comparison of Hessian and Monte Carlo results at the input scale of $Q_0^2=1$~GeV$^2$ for the (a)~gluon distribution and (b)~strange asymmetry.  Both results allow $n=20$ free PDF parameters and do not apply a tolerance (i.e.~$T=1$ in the Hessian case).  The best-fit (solid curves) and Hessian uncertainty (shaded region) are in good agreement with the average and standard deviation (thick dashed curves) of the $N_{\rm rep} = 40$ Monte Carlo replica PDF sets (thin dotted curves).}
  \label{fig:inputpdfs}
\end{figure}
We perform a separate PDF fit to each replica data set, then we can take the average $\langle F\rangle$ and standard deviation $\Delta F$ of an observable $F$ calculated with each PDF replica set, $\mathcal{S}_k$ ($k=1,\ldots,N_{\rm rep}$), that is,
\begin{align}
  \langle F\rangle = \frac{1}{N_{\rm rep}}\sum_{k=1}^{N_{\rm rep}}F(\mathcal{S}_k), \label{eq:MCav} \\
  \Delta F = \sqrt{\frac{N_{\rm rep}}{N_{\rm rep}-1}\left(\langle F^2\rangle-\langle F\rangle^2\right)}. \label{eq:MCsd}
\end{align}
The number of replicas $N_{\rm rep}$ is arbitrary, but in all cases we choose to generate $N_{\rm rep} = 40$ replica PDF sets, where this number is chosen to be equal to the number of eigenvector PDF sets mostly for practical reasons, i.e.~to demonstrate that the implementation of the Monte Carlo (MC) method does not necessarily require more computer resources than the Hessian method.  It could easily be increased in further studies, but first indications are that $N_{\rm rep} = 40$ is sufficiently large to avoid significant fluctuations.  To allow a fair comparison with the existing Hessian eigenvector PDF sets, we take $n=20$ free PDF parameters, i.e.~8 PDF parameters are held fixed at their global best-fit values, and we do not apply a tolerance, i.e.~we use the Hessian eigenvector PDF sets corresponding to $T=1$ (see section~6.6 of ref.~\cite{Martin:2009iq}).  In figure~\ref{fig:inputpdfs} we show the input gluon distribution and strange asymmetry for the $N_{\rm rep} = 40$ MC replica PDF sets (thin dotted curves), and their average and standard deviation (thick dashed curves), and we compare to the best-fit and Hessian uncertainty (solid curves and shaded region).  We find good agreement of the Hessian and MC results at all $x$ and $Q^2$ values, and for all parton flavours, as will be demonstrated more explicitly in the next section.

Similar comparisons between Hessian and MC results were performed in a fit only to the H1 data from HERA I on neutral- and charged-current $e^\pm p$ cross sections~\cite{Dittmar:2009ii}, but it is still reassuring that we find a similar good agreement in the context of a more complicated global fit.  On the other hand, in section~6.6 of ref.~\cite{Martin:2009iq} we also performed a fit to a reduced data set consisting of a limited number of inclusive DIS data sets (BCDMS, NMC, H1, ZEUS) with fairly conservative cuts of $Q^2\ge9$~GeV$^2$ and $W^2\ge 15$~GeV$^2$, where eigenvector PDF sets were produced with $n=16$ free PDF parameters for both a dynamic tolerance and with $T=1$.  We find that there are some differences between the MC results with $n=16$ free PDF parameters and the Hessian results with $T=1$.  The approximate equivalence between the Hessian and MC methods may break down, therefore, when fitting a limited selection of discrepant data sets that are insufficient to unambiguously constrain all fitted parameters.

\section{Investigation of potential parameterisation bias} \label{sec:parambias}
Recall the MSTW 2008 NLO PDF parameterisation at the input scale $Q_0^2 = 1$ GeV$^2$~\cite{Martin:2009iq}:
\begin{align}
  xu_v & \equiv xu-x\bar{u} = A_u\,x^{\mbox{\large $\boldsymbol{\color{red} \eta_1}$}} (1-x)^{\mbox{\large $\boldsymbol{\color{red} \eta_2}$}} (1 + \mbox{\Large $\boldsymbol{\color{red} \epsilon_u}$}\,\sqrt{x} + {\color{blue}\gamma_u}\,x), \\
  xd_v & \equiv xd-x\bar{d} = A_d\,x^{\mbox{\large $\boldsymbol{\color{red} \eta_3}$}} (1-x)^{\mbox{\large $\boldsymbol{\color{red} \eta_4}$}} (1 + \mbox{\Large $\boldsymbol{\color{red} \epsilon_d}$}\,\sqrt{x} + {\color{blue}\gamma_d}\,x), \\
  xS & \equiv 2x\bar{u}+2x\bar{d}+xs+x\bar{s} = \mbox{\Large $\boldsymbol{\color{red} A_S}$}\,x^{{\color{blue}\delta_S}} (1-x)^{\mbox{\large $\boldsymbol{\color{red} \eta_S}$}} (1 + \mbox{\Large $\boldsymbol{\color{red} \epsilon_S}$}\,\sqrt{x} + {\color{blue}\gamma_S}\,x), \\
  x\Delta & \equiv x\bar{d} - x \bar{u} = \mbox{\Large $\boldsymbol{\color{red} A_\Delta}$}\,x^{\mbox{\Large $\boldsymbol{\color{red} \eta_\Delta}$}} (1-x)^{\eta_S+2} (1 + \mbox{\Large $\boldsymbol{\color{red} \gamma_\Delta}$}\,x + {\color{blue}\delta_\Delta}\,x^2), \\
  xg &= A_g\,x^{\mbox{\large $\boldsymbol{\color{red} \delta_g}$}} (1-x)^{\mbox{\large $\boldsymbol{\color{red} \eta_g}$}} (1 + {\color{blue}\epsilon_g}\,\sqrt{x} + {\color{blue}\gamma_g}\,x) + {\color{blue}A_{g^\prime}}\,x^{\mbox{\large $\boldsymbol{\color{red} \delta_{g^\prime}}$}} (1-x)^{\mbox{\large $\boldsymbol{\color{red} \eta_{g^\prime}}$}}, \label{eq:gluon} \\
  xs + x\bar{s} &= \mbox{\Large $\boldsymbol{\color{red} A_{+}}$}\,x^{\delta_S}\,(1-x)^{\mbox{\large $\boldsymbol{\color{red} \eta_{+}}$}} (1 + \epsilon_S\,\sqrt{x} + \gamma_S\,x), \label{eq:splus} \\
 xs - x\bar{s} &= \mbox{\Large $\boldsymbol{\color{red} A_{-}}$}\,x^{0.2} (1-x)^{\mbox{\large $\boldsymbol{\color{red} \eta_{-}}$}} (1-x/x_0). \label{eq:sminus}
\end{align}
The parameters $A_u$, $A_d$, $A_g$ and $x_0$ were fixed by enforcing number- and momentum-sum rule constraints, while the other parameters were allowed to go free to determine the overall best fit.  The 20 highlighted (red) parameters were those allowed to go free when producing the eigenvector PDF sets, where the other 8 (blue) parameters were held fixed, as for the MC results in the previous section.  However, this is not in fact necessary in the MC approach where it is only needed to find the best fit for each replica data set, and the Hessian matrix is not used for error propagation.  Therefore, we can perform MC replica fits with all 28 free parameters to examine the effect on PDF uncertainties of the greater freedom in parameterisation, and to explore the extent that the Hessian uncertainties are limited by the restricted parameterisation.

\begin{figure}
  \centering
  \begin{minipage}{0.5\textwidth}
    (a)\\
    \includegraphics[width=\textwidth]{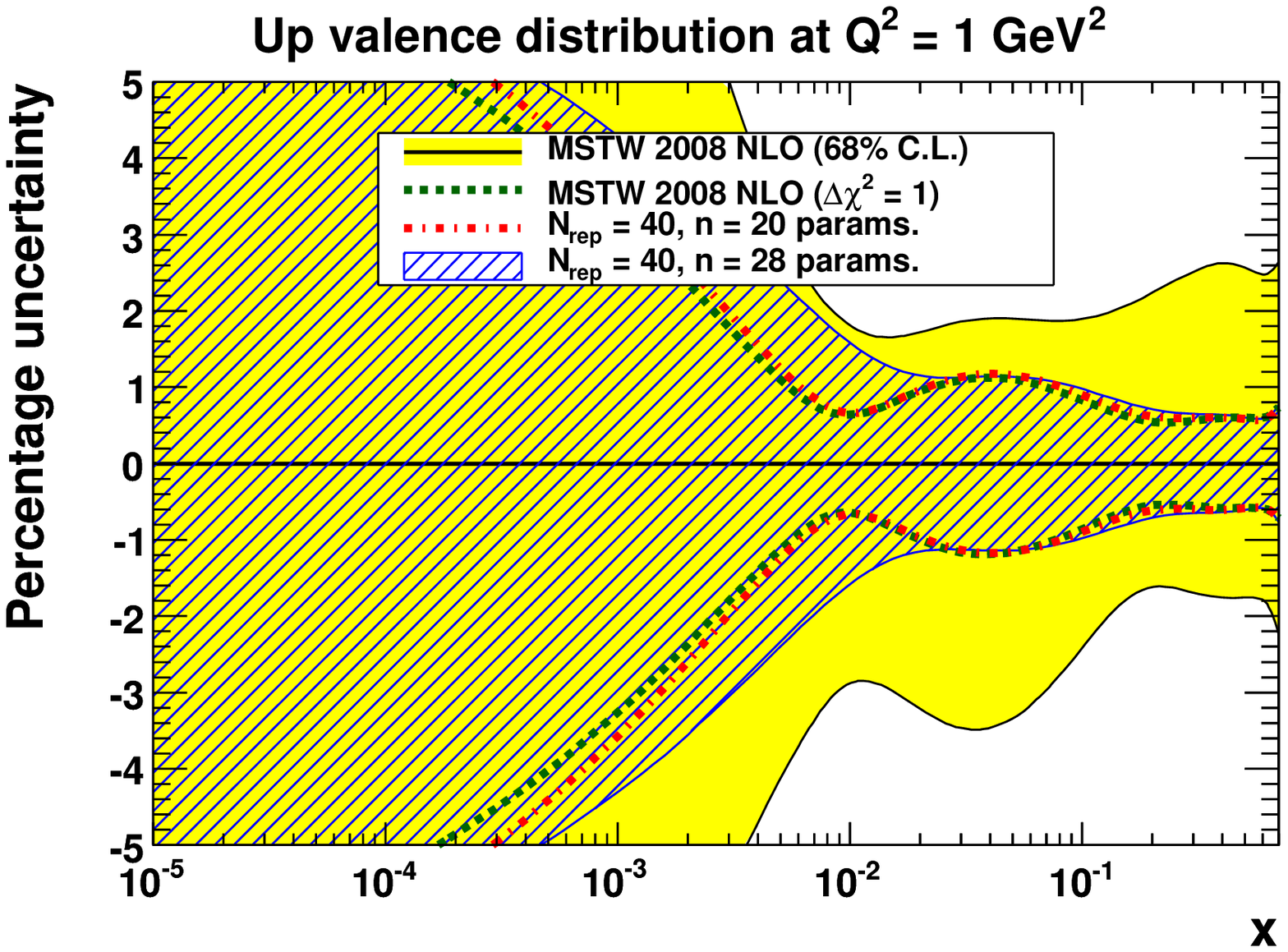}
  \end{minipage}%
  \begin{minipage}{0.5\textwidth}
    (b)\\
    \includegraphics[width=\textwidth]{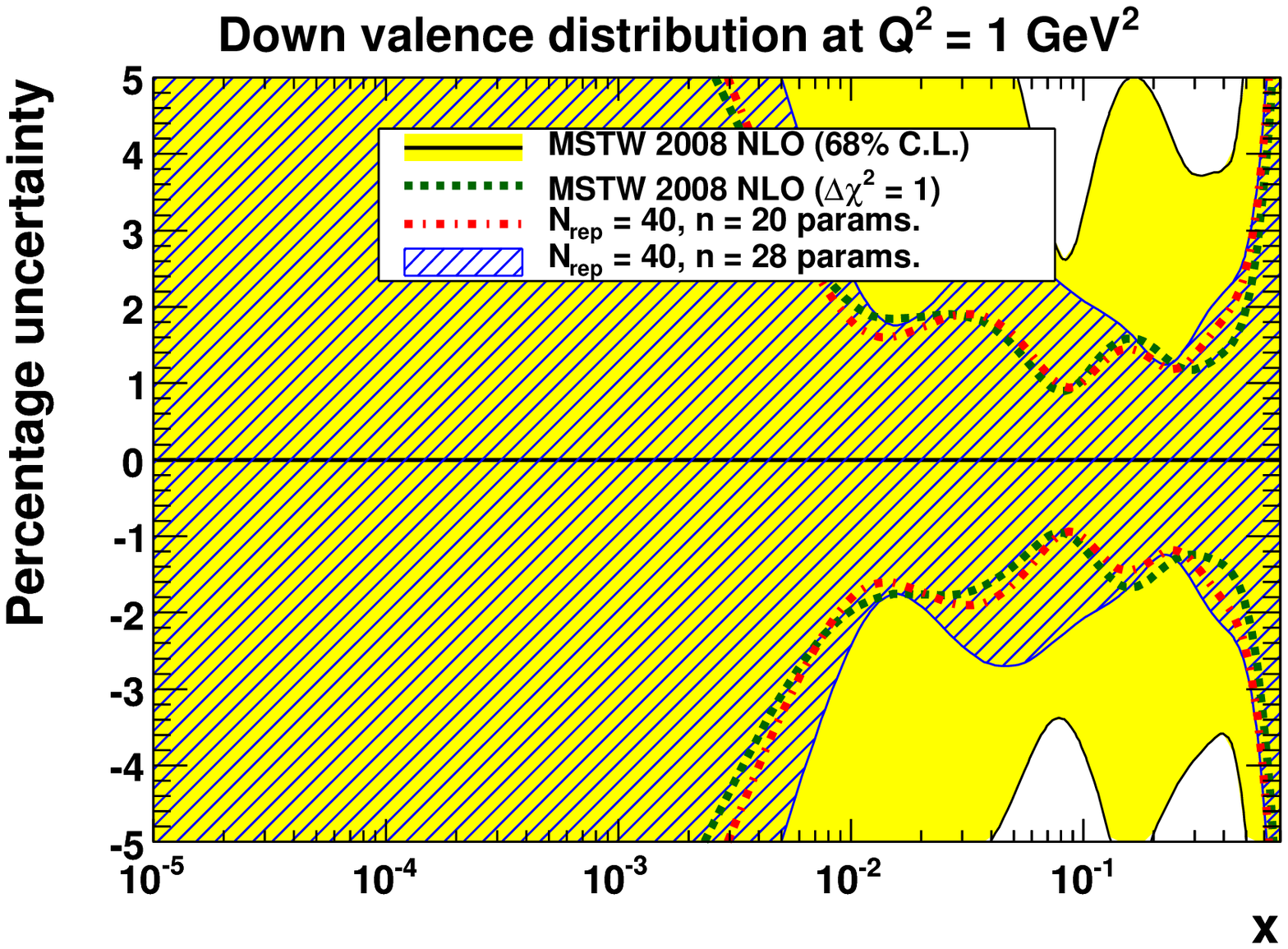}
  \end{minipage}
  \begin{minipage}{0.5\textwidth}
    (c)\\
    \includegraphics[width=\textwidth]{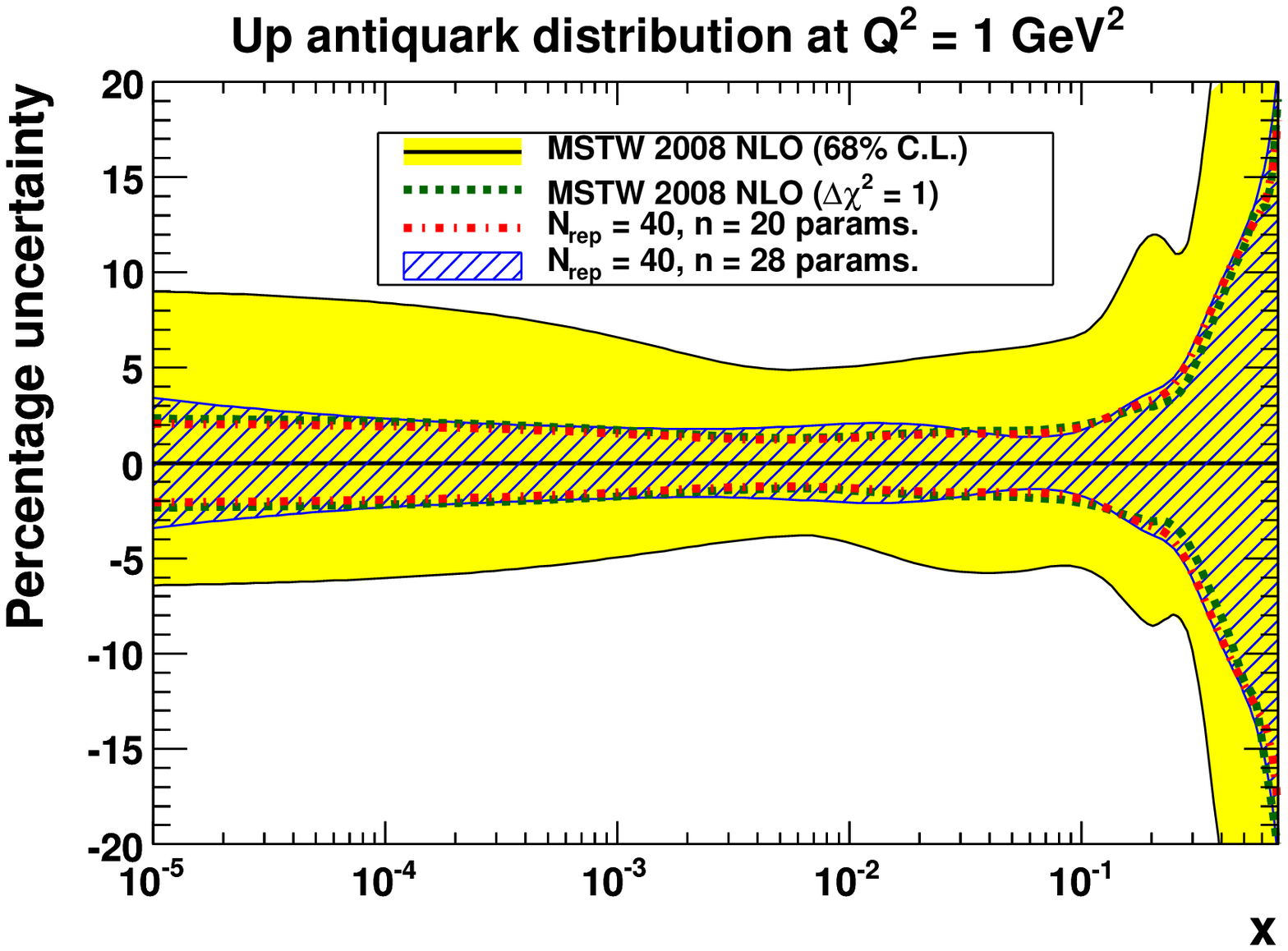}
  \end{minipage}%
  \begin{minipage}{0.5\textwidth}
    (d)\\
    \includegraphics[width=\textwidth]{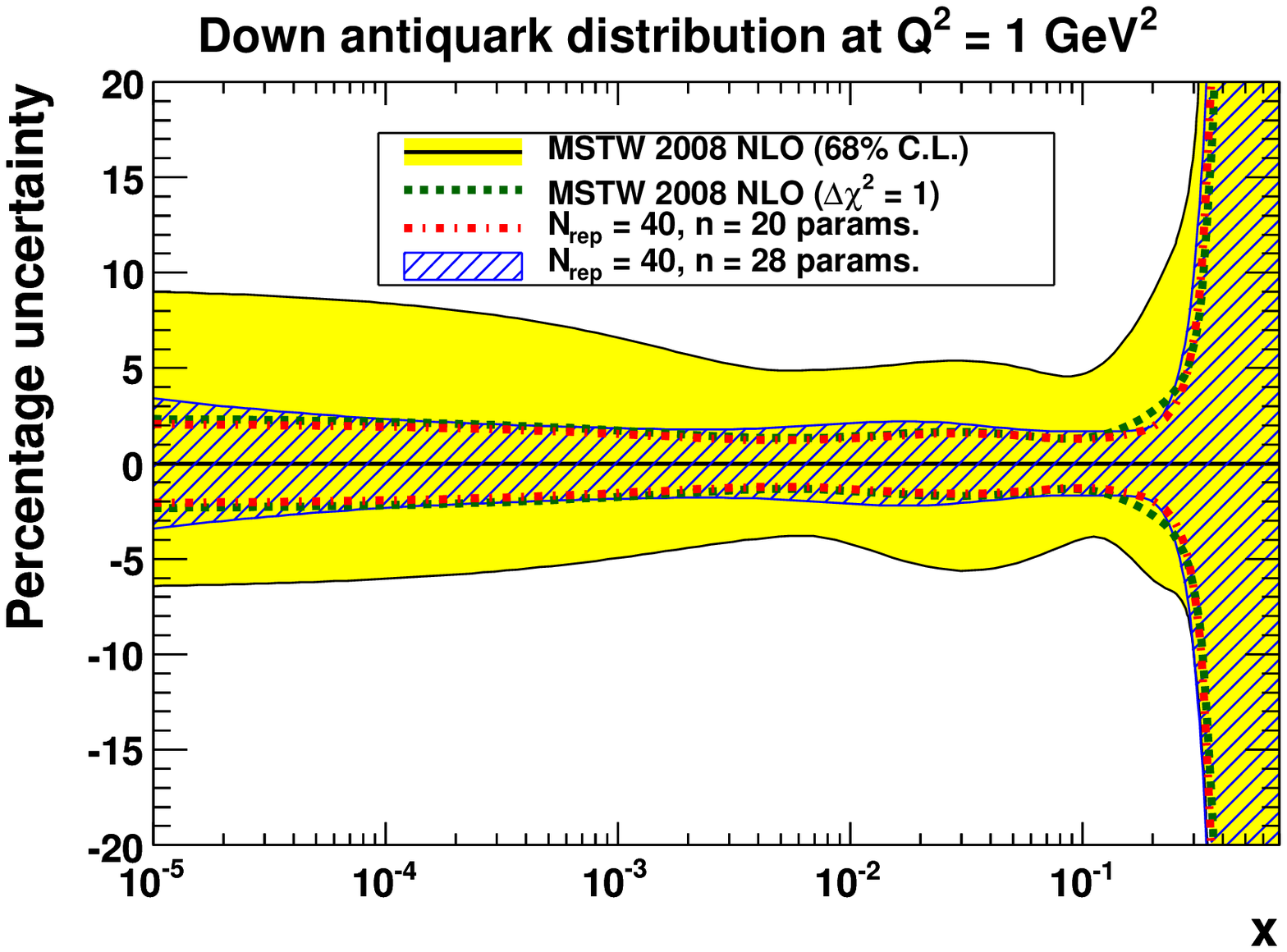}
  \end{minipage}
  \begin{minipage}{0.5\textwidth}
    (e)\\
    \includegraphics[width=\textwidth]{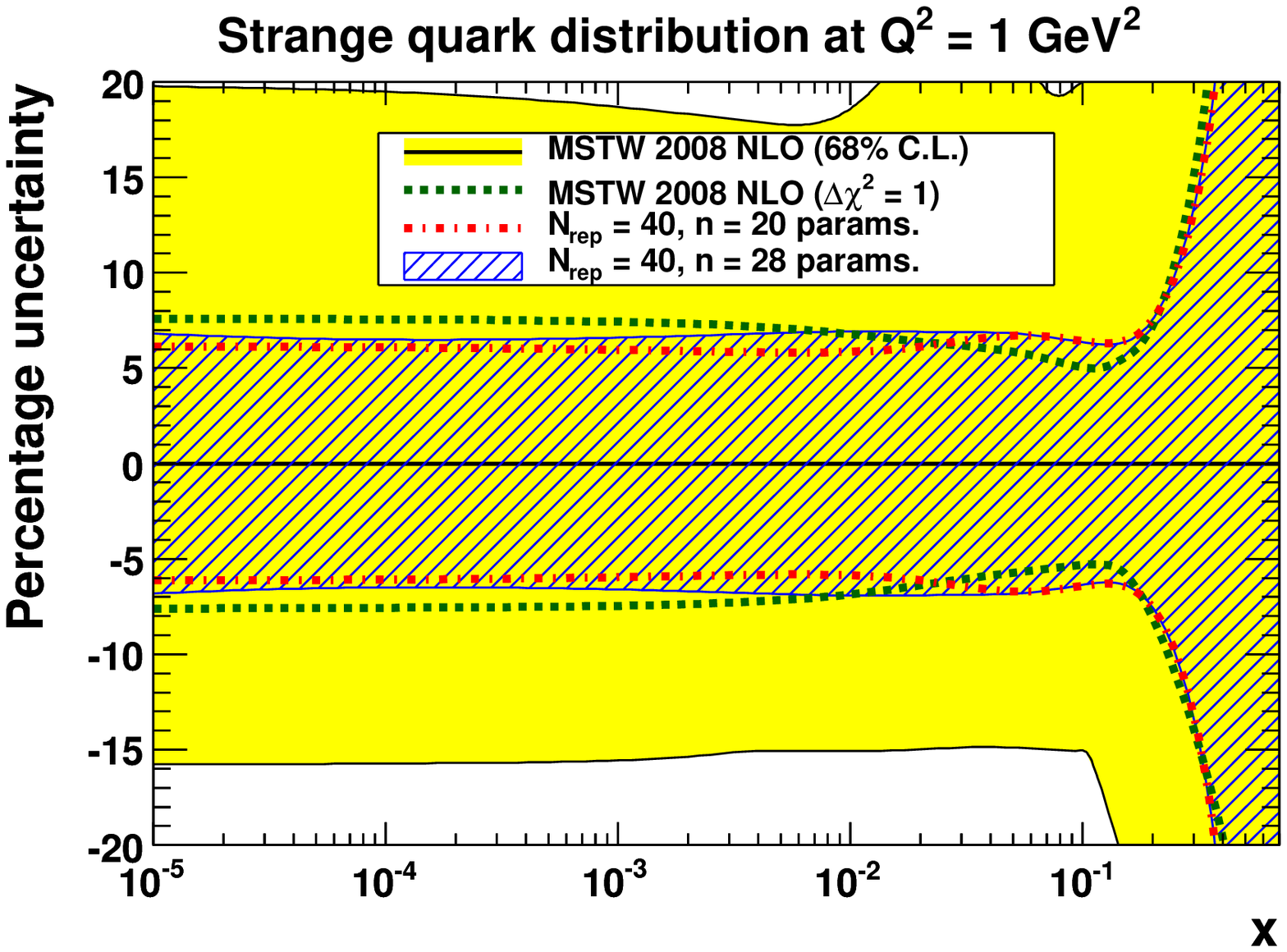}
  \end{minipage}%
  \begin{minipage}{0.5\textwidth}
    (f)\\
    \includegraphics[width=\textwidth]{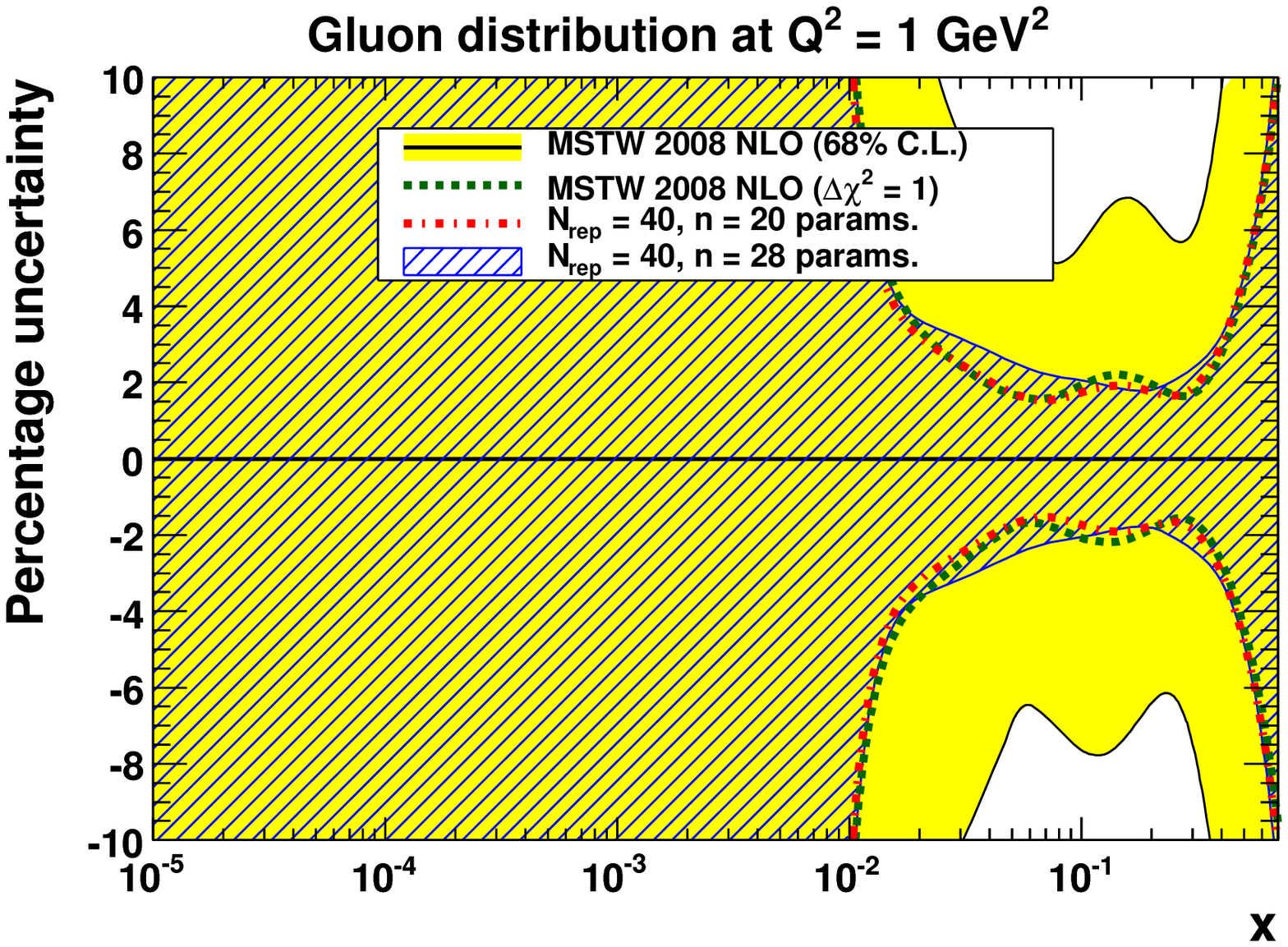}
  \end{minipage}
  \caption{Effect of $n=20\to28$ parameters on percentage PDF uncertainties at $Q^2=1~{\rm GeV}^2$.}
  \label{fig:frac_1GeV}
\end{figure}
\begin{figure}
  \centering
  \begin{minipage}{0.5\textwidth}
    (a)\\
    \includegraphics[width=\textwidth]{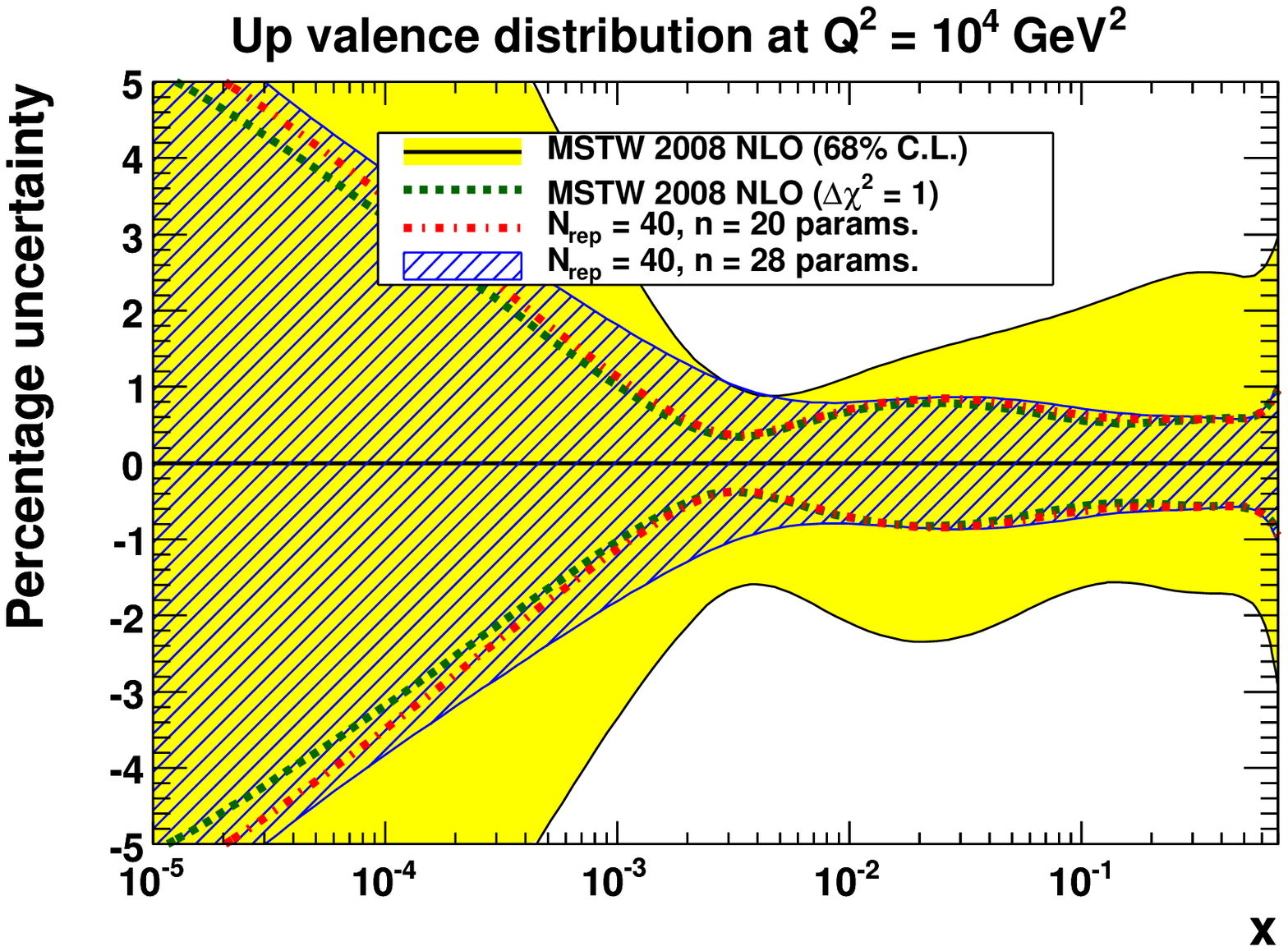}
  \end{minipage}%
  \begin{minipage}{0.5\textwidth}
    (b)\\
    \includegraphics[width=\textwidth]{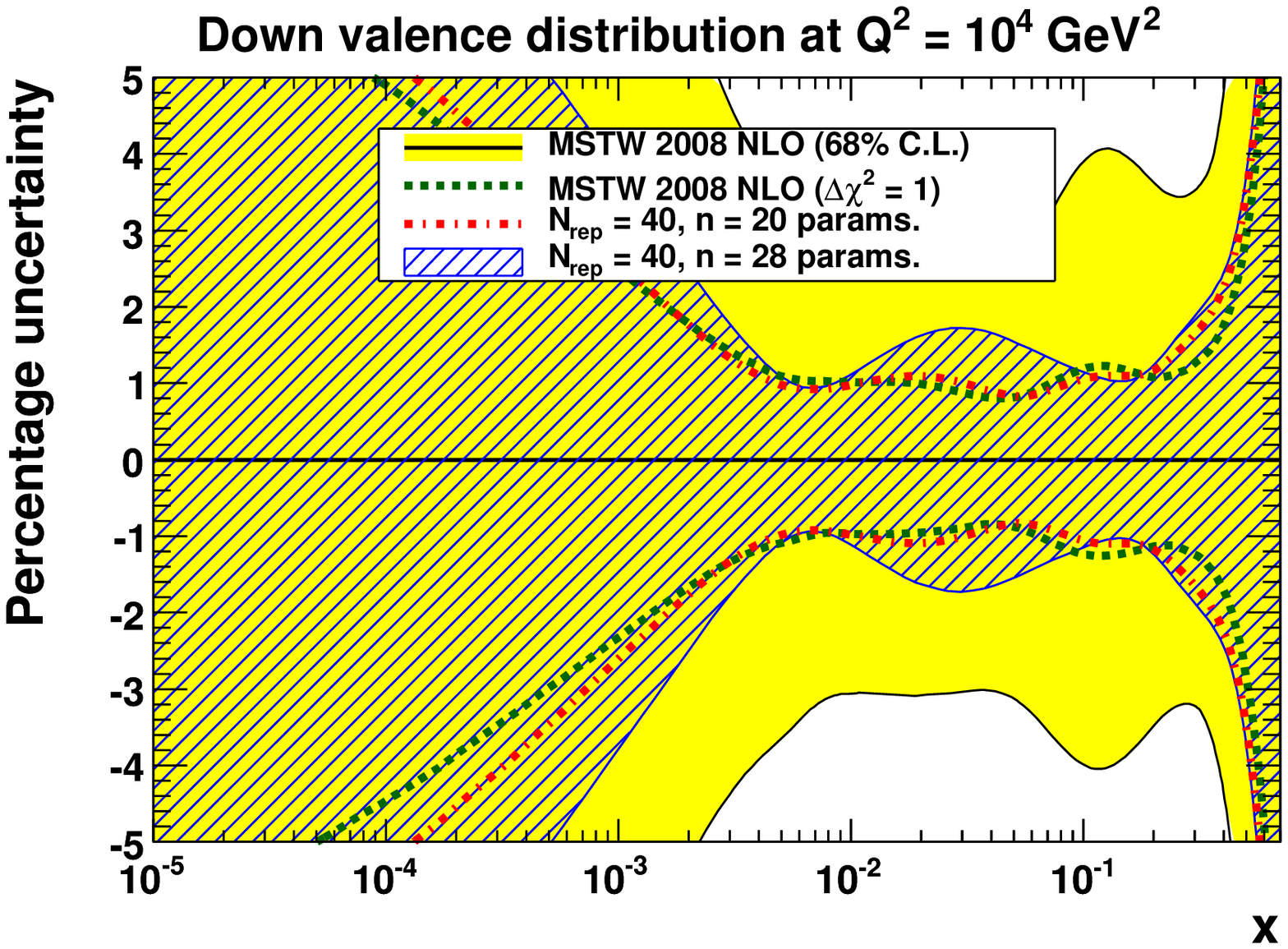}
  \end{minipage}
  \begin{minipage}{0.5\textwidth}
    (c)\\
    \includegraphics[width=\textwidth]{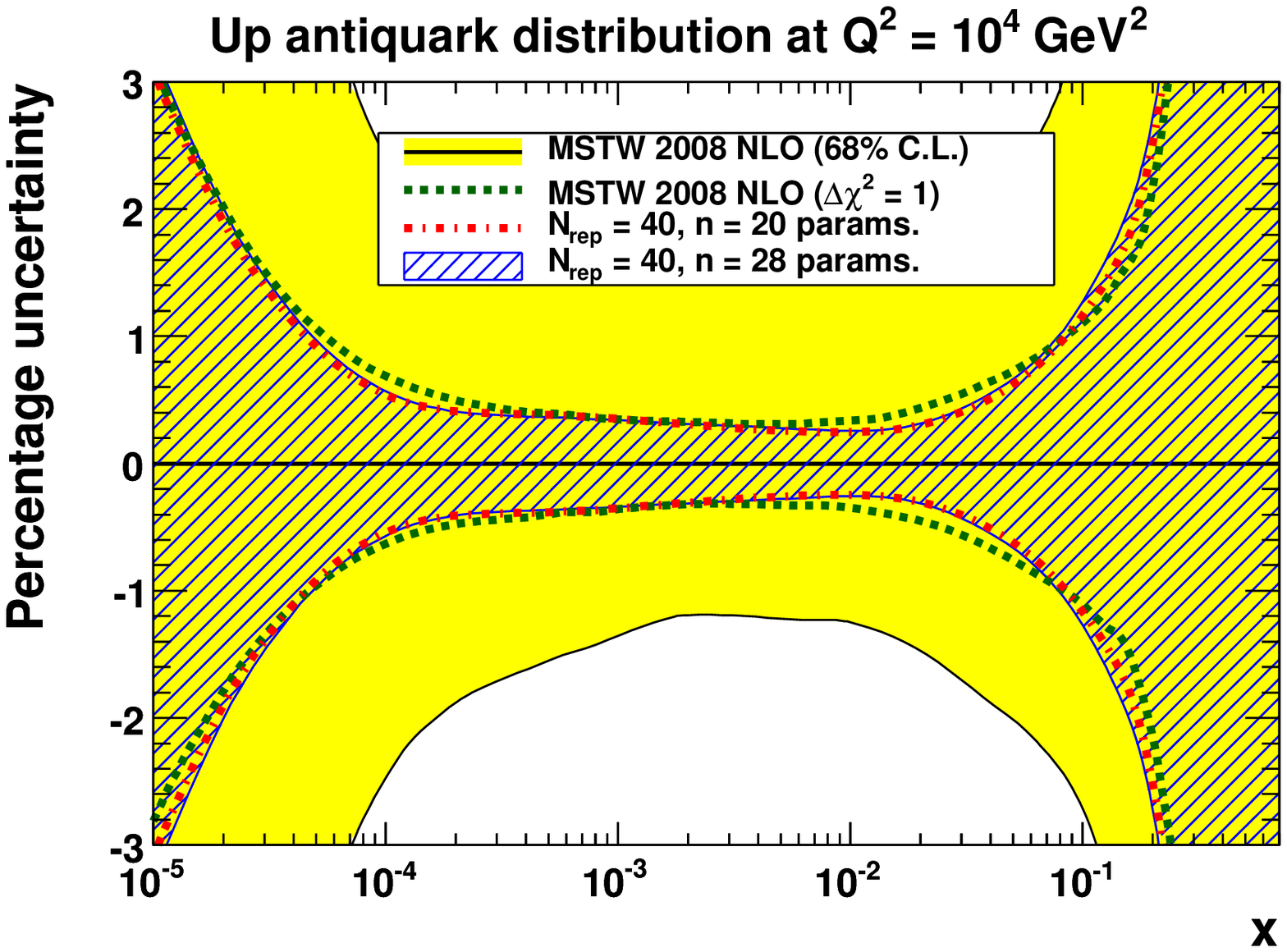}
  \end{minipage}%
  \begin{minipage}{0.5\textwidth}
    (d)\\
    \includegraphics[width=\textwidth]{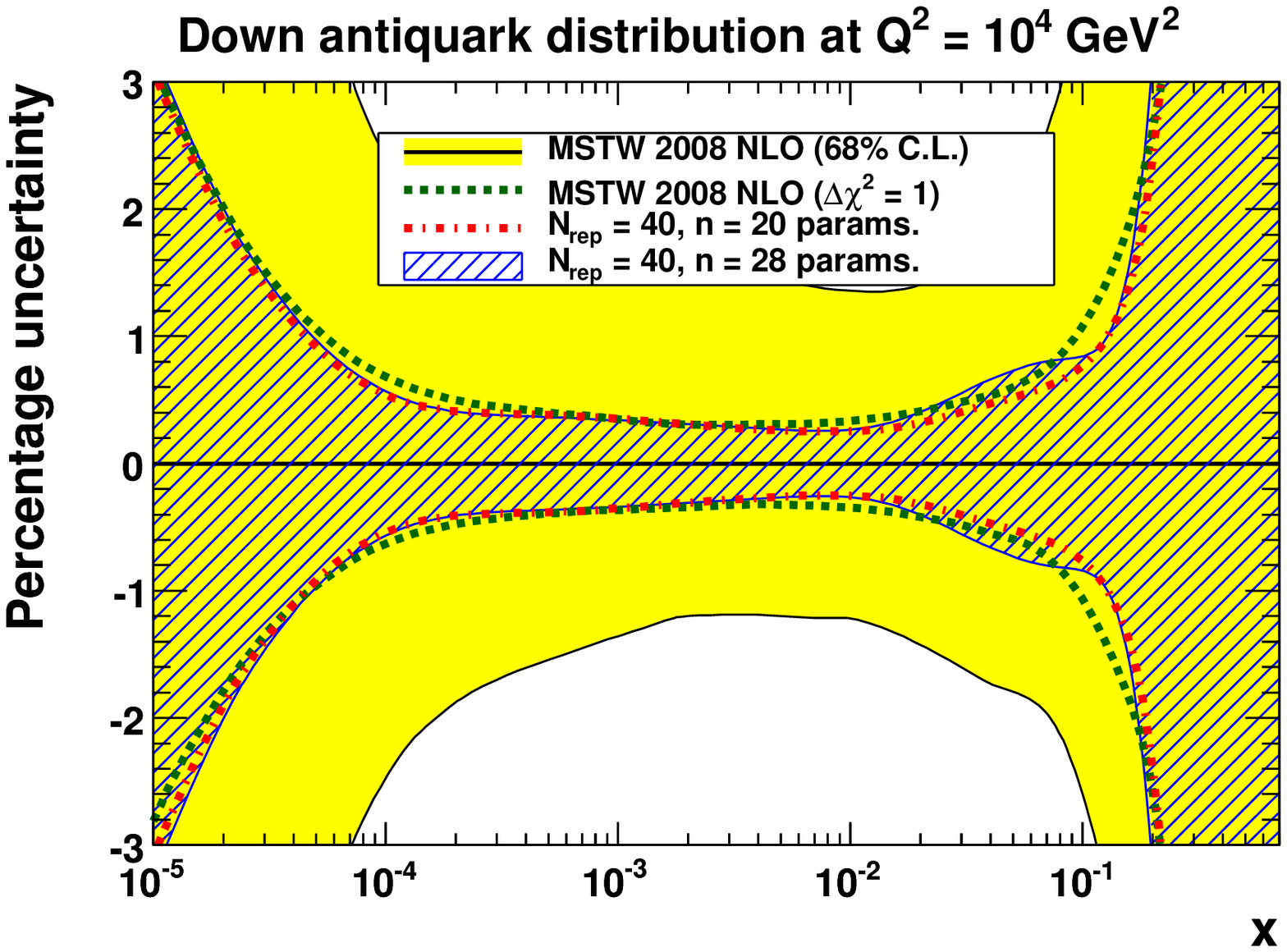}
  \end{minipage}
  \begin{minipage}{0.5\textwidth}
    (e)\\
    \includegraphics[width=\textwidth]{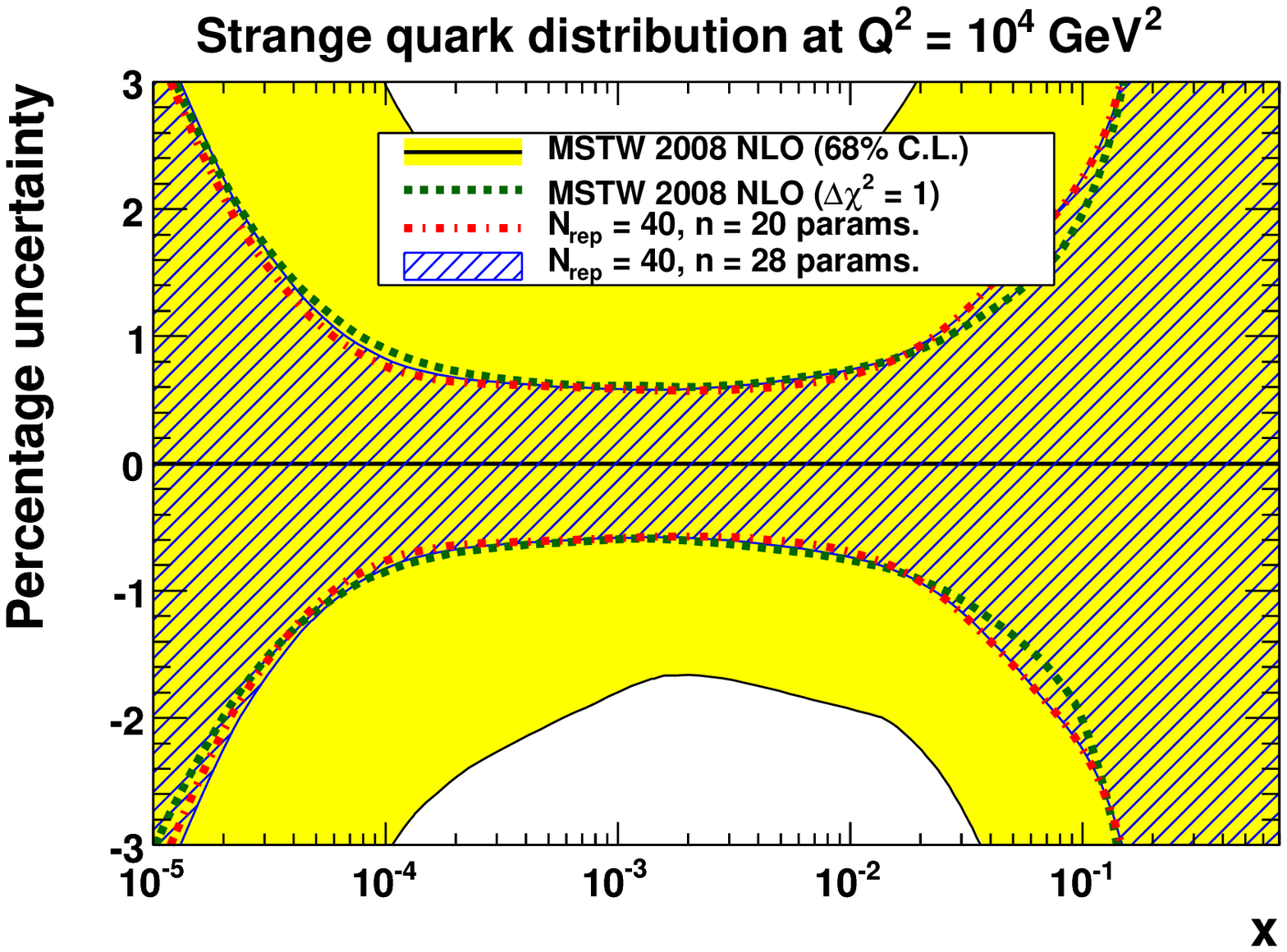}
  \end{minipage}%
  \begin{minipage}{0.5\textwidth}
    (f)\\
    \includegraphics[width=\textwidth]{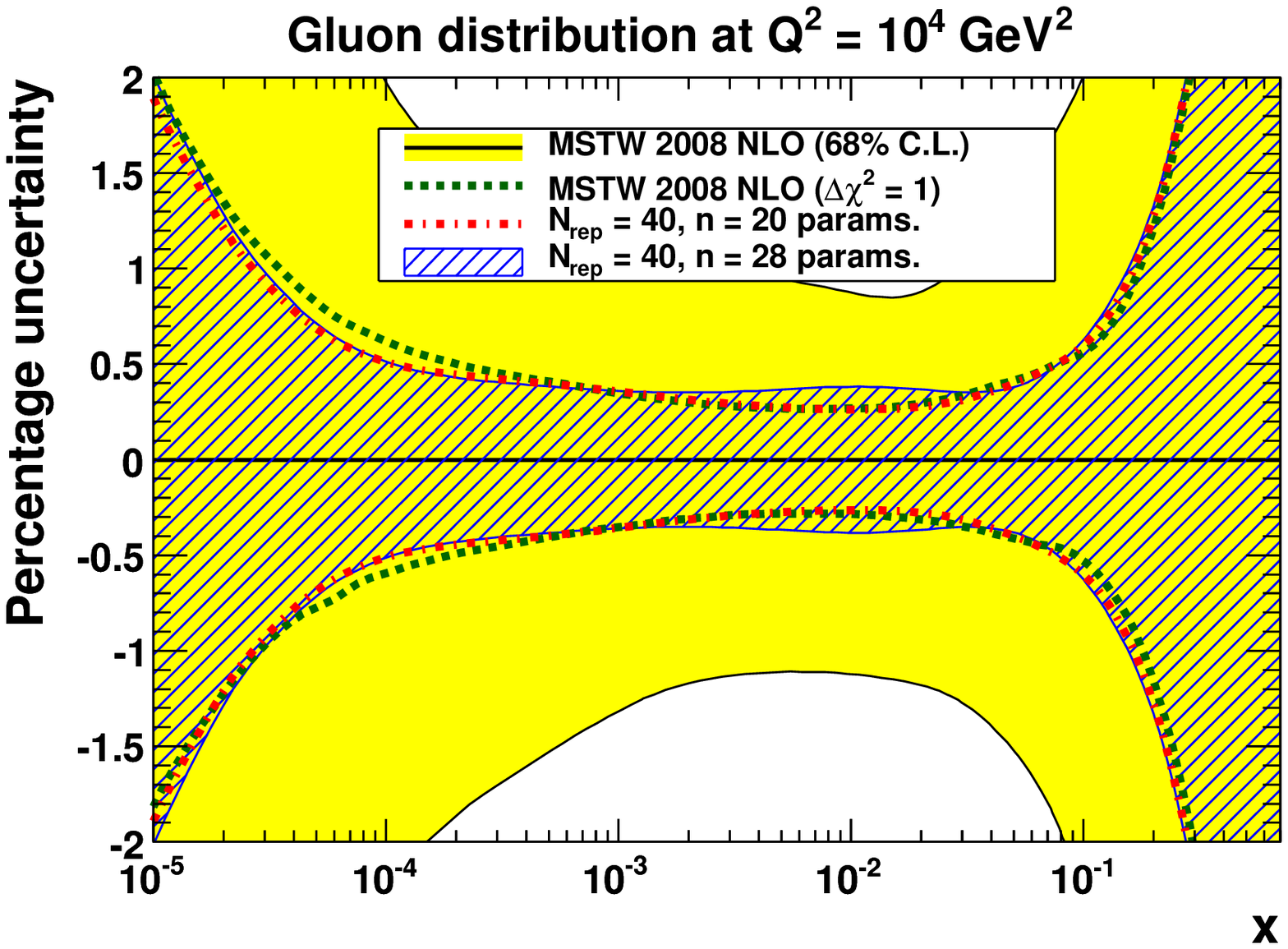}
  \end{minipage}
  \caption{Effect of $n=20\to28$ parameters on percentage PDF uncertainties at $Q^2=(100~{\rm GeV})^2$.}
  \label{fig:frac}
\end{figure}
Recall~\cite{Martin:2009iq} that the reason to freeze several parameters before applying the Hessian method was to reduce the large correlations between some parameters, which would lead to severe breaking of the quadratic behaviour of $\Delta\chi^2$, meaning that linear error propagation would not be applicable.  (A similar procedure was used in the CTEQ global fits; see, for example, section~5 of ref.~\cite{Tung:2006tb}.)  We observed some departure from the ideal quadratic behaviour of $\Delta\chi^2$ even with only 20 parameters; see figures~5 and 6 of ref.~\cite{Martin:2009iq}.  However, even with all 28 parameters free, the Hessian matrix is generally still positive-definite (has positive eigenvalues) and therefore we can still be relatively confident that the best fit is correctly determined.  Note that we use the Levenberg--Marquardt algorithm for $\chi^2$-minimisation, which combines the advantages of the inverse-Hessian method and the steepest-descent method, and therefore simply finding the best fit is less reliant on accurate knowledge of the Hessian matrix compared to subsequent error propagation using the Hessian method.

In figure~\ref{fig:frac_1GeV} we show percentage uncertainties at the input scale $Q_0^2=1$~GeV$^2$, and in figure~\ref{fig:frac} we show percentage uncertainties after evolving to $Q^2=(100~{\rm GeV})^2$.  We show only the uncertainties since the MC average is very close to the Hessian best-fit, with residual differences likely explained by statistical fluctuations.  Again the MC uncertainties with $n=20$ input PDF parameters are in good agreement with the Hessian uncertainties with $\Delta\chi^2=1$, and both are much smaller than the 68\% C.L.~uncertainties including the dynamic tolerance.  We show the effect of moving to $n=28$ input PDF parameters, which gives significantly larger $u_v$ and $d_v$ uncertainties mainly at low $x$ values (removing some unusual shapes in the $x$ dependence) and slightly larger gluon uncertainties around $x\sim 0.05$ in figure~\ref{fig:frac_1GeV}(f) and around $x\sim 0.01$ in figure~\ref{fig:frac}(f), but in all cases the MC uncertainties are still much smaller than the Hessian uncertainties at 68\% C.L.
One can see from the equations above that in going from a total of $20\to28$ input PDF parameters, the number of free parameters for both $xu_v$ and $xd_v$ goes from $3\to4$, for $xS$ ($\equiv 2x\bar{u}+2x\bar{d}+xs+x\bar{s}$) goes from $3\to5$, and for $xg$ goes from $4\to 7$.  While there is perhaps some degree of parameterisation bias in the valence-quark distributions, the insensitivity of the sea-quark and gluon distributions to the relatively large increase in the number of free parameters suggests that parameterisation bias is likely to be small in those cases.  Of course, an exception is the strange-quark and -antiquark distributions which are certainly constrained by the choice of parameterisation outside the limited data region ($0.01\lesssim x\lesssim 0.2$) of the CCFR/NuTeV dimuon cross sections.  For example, the low-$x$ behaviour of $s$ and $\bar{s}$ is assumed to be the same as for $\bar{u}$ and $\bar{d}$, as suggested by arguments based on both Regge theory and perturbative QCD (see discussion in section~6.5.5 of ref.~\cite{Martin:2009iq}).

The study of potential parameterisation bias presented here is indicative rather than exhaustive.  It could be followed up by a more involved study, for example, using Chebyshev polynomials along the lines of refs.~\cite{Pumplin:2009bb,Glazov:2010bw}.  However, switching to an extremely flexible parameterisation brings the danger of fitting the statistical fluctuations of the data unless some method is used to enforce smoothness.  We note that the limiting power-law behaviour as $x\to 0$ and $x\to 1$ is well-motivated by Regge theory and counting rules, respectively, and it is difficult to perceive a sensible alternative.  More discussion and justification for the MSTW 2008 input parameterisation was given in section~6.5 of ref.~\cite{Martin:2009iq}.

\section{Fits to restricted data sets using data replicas} \label{sec:restricted}
Although we see little evidence for significant parameterisation bias in the MSTW 2008 \emph{global} fit, this might not be true for some ``non-global'' fits which tend to be constrained by parameterisation choices in the absence of relevant data.  For example, the input parameterisation at $Q_0^2 = 1.9$~GeV$^2$ in the HERAPDF1.0~\cite{HERA:2009wt} or HERAPDF1.5 NLO~\cite{HERA:2010} analyses takes the form:
\begin{align*}
  xu_v &= A_{u_v}\,x^{{\color{red} B_{q_v}}} (1-x)^{{\color{red} C_{u_v}}} (1 + {\color{red} E_{u_v}}\,x^2), \\
  xd_v &= A_{d_v}\,x^{B_{q_v}} (1-x)^{{\color{red} C_{d_v}}}, \\
  x\bar{u} &= {\color{red} A_{\bar{q}}}\,x^{{\color{red} B_{\bar{q}}}} (1-x)^{{\color{red} C_{\bar{u}}}}, \\
  x\bar{d} &= A_{\bar{q}}\,x^{B_{\bar{q}}} (1-x)^{{\color{red} C_{\bar{d}}}}, \\
  x\bar{s} &= 0.45\,x\bar{d}, \\
  xs &= x\bar{s}, \\
  xg &= A_g\,x^{{\color{red} B_g}} (1-x)^{{\color{red} C_g}}.
\end{align*}
\begin{figure}
  \centering
  \begin{minipage}{0.5\textwidth}
    (a)\\
    \includegraphics[width=\textwidth]{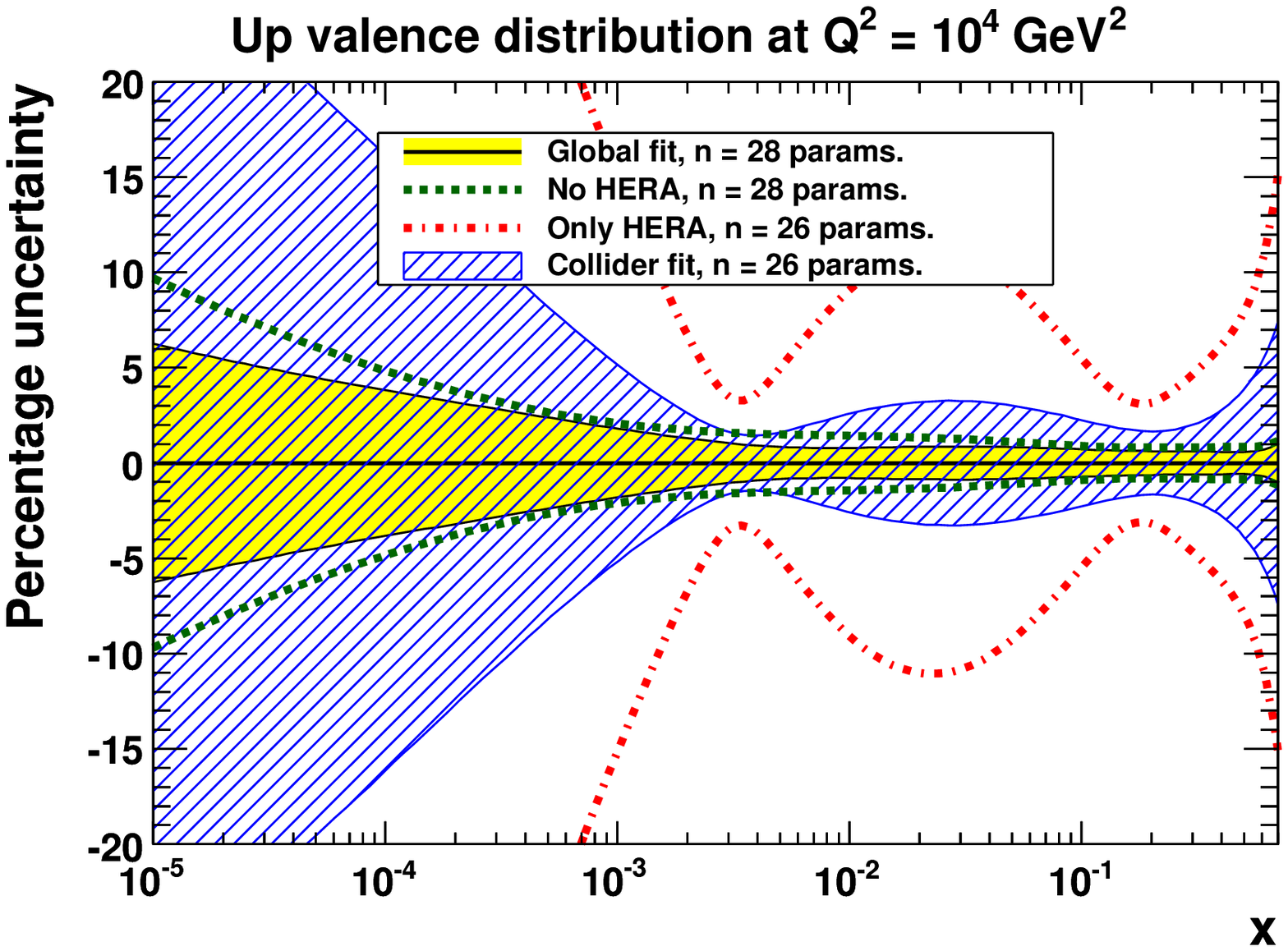}
  \end{minipage}%
  \begin{minipage}{0.5\textwidth}
    (b)\\
    \includegraphics[width=\textwidth]{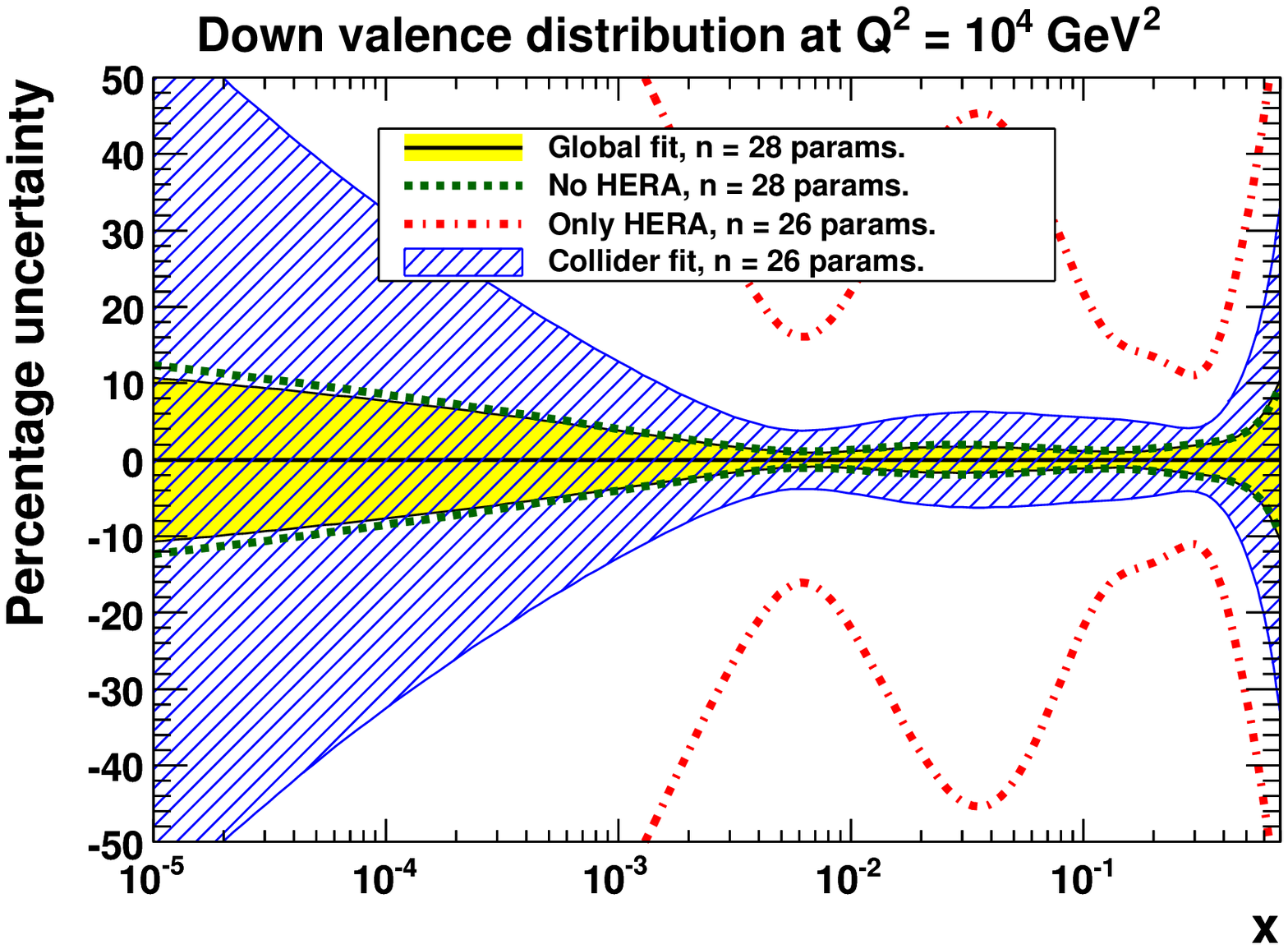}
  \end{minipage}
  \begin{minipage}{0.5\textwidth}
    (c)\\
    \includegraphics[width=\textwidth]{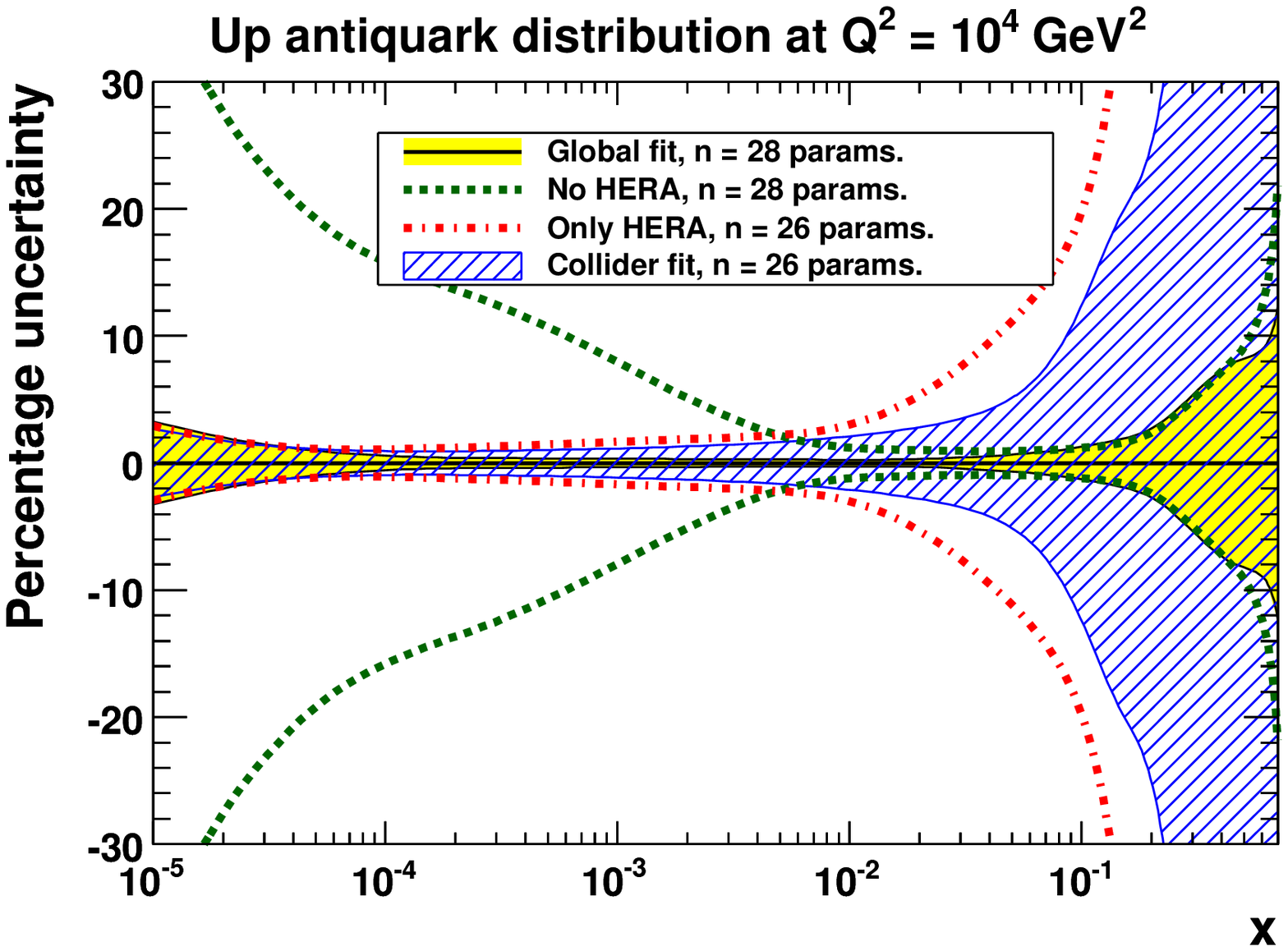}
  \end{minipage}%
  \begin{minipage}{0.5\textwidth}
    (d)\\
    \includegraphics[width=\textwidth]{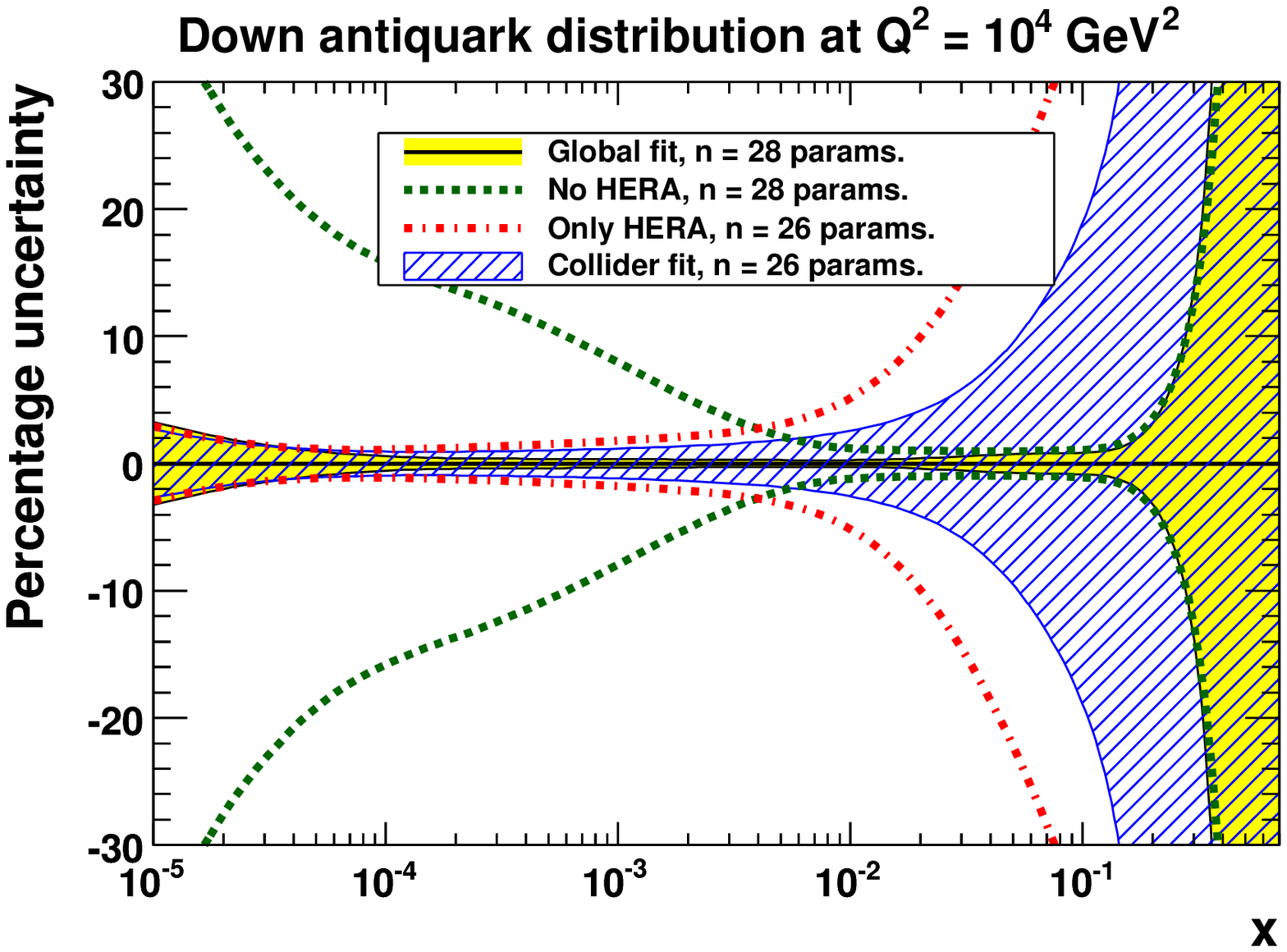}
  \end{minipage}
  \begin{minipage}{0.5\textwidth}
    (e)\\
    \includegraphics[width=\textwidth]{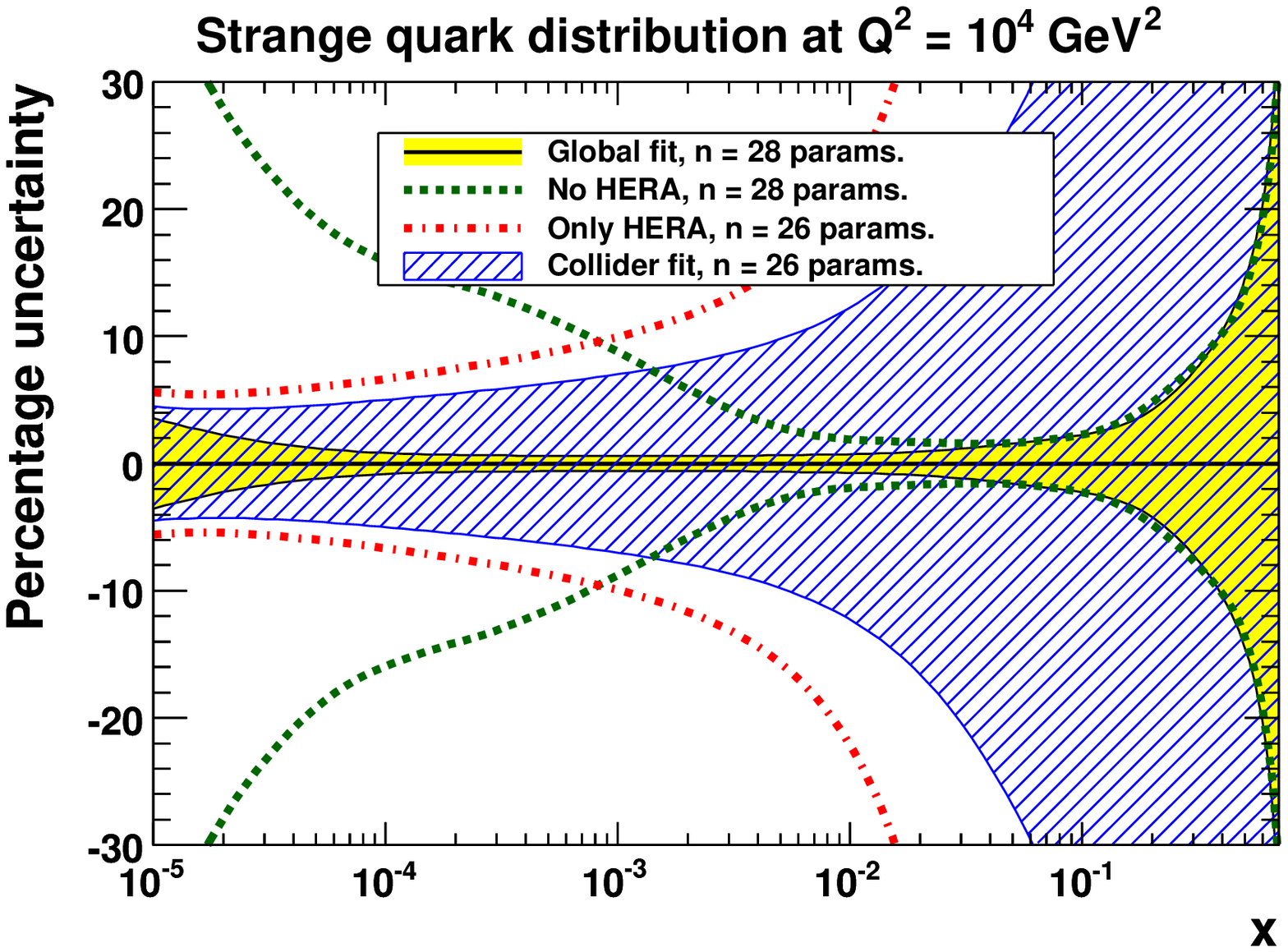}
  \end{minipage}%
  \begin{minipage}{0.5\textwidth}
    (f)\\
    \includegraphics[width=\textwidth]{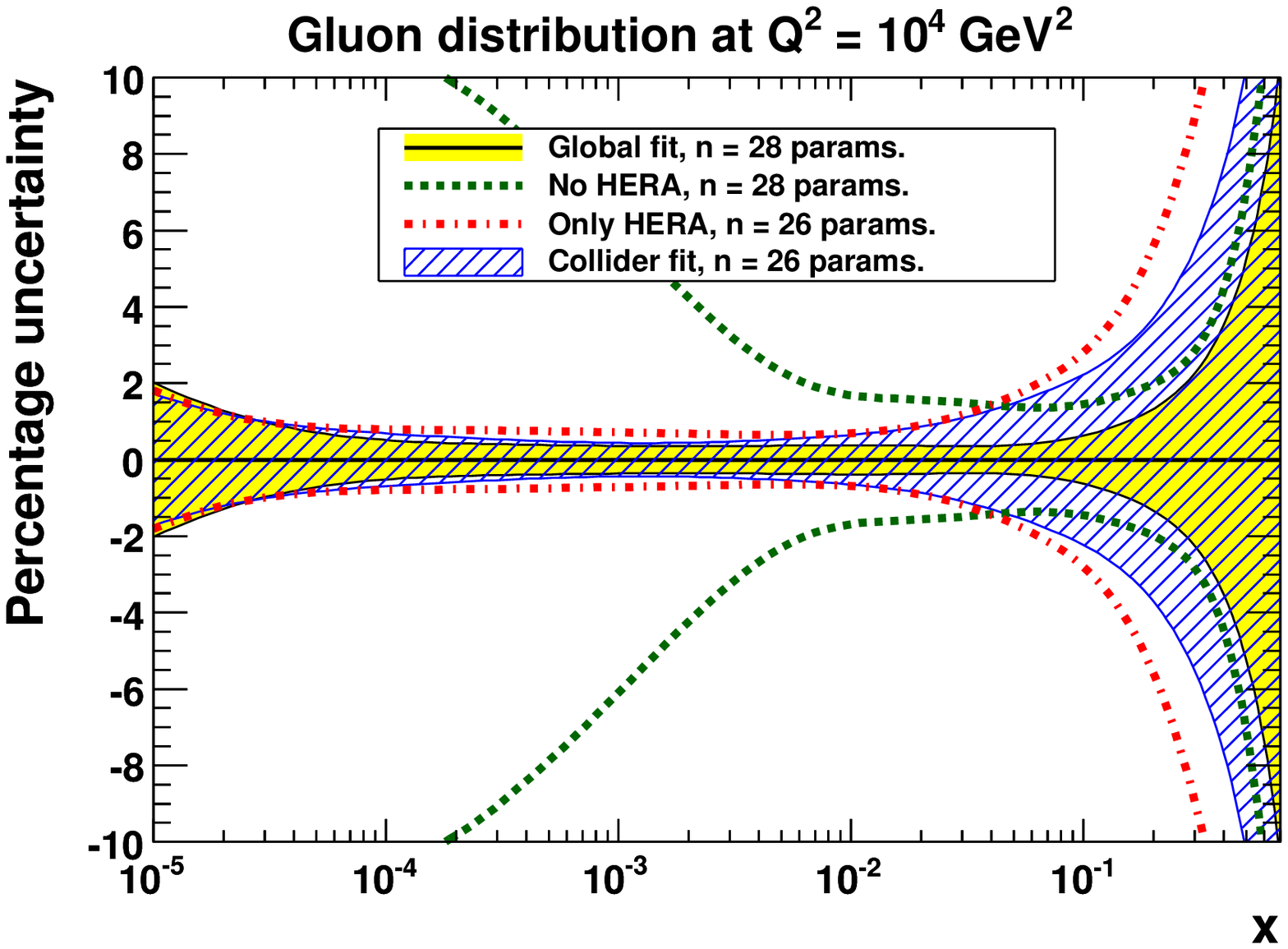}
  \end{minipage}
  \caption{Effect on percentage PDF uncertainties of fitting subsets of MSTW 2008 global data.}
  \label{fig:frac_varyhera}
\end{figure}
\begin{figure}
  \centering
  \begin{minipage}{0.5\textwidth}
    (a)\\
    \includegraphics[width=\textwidth]{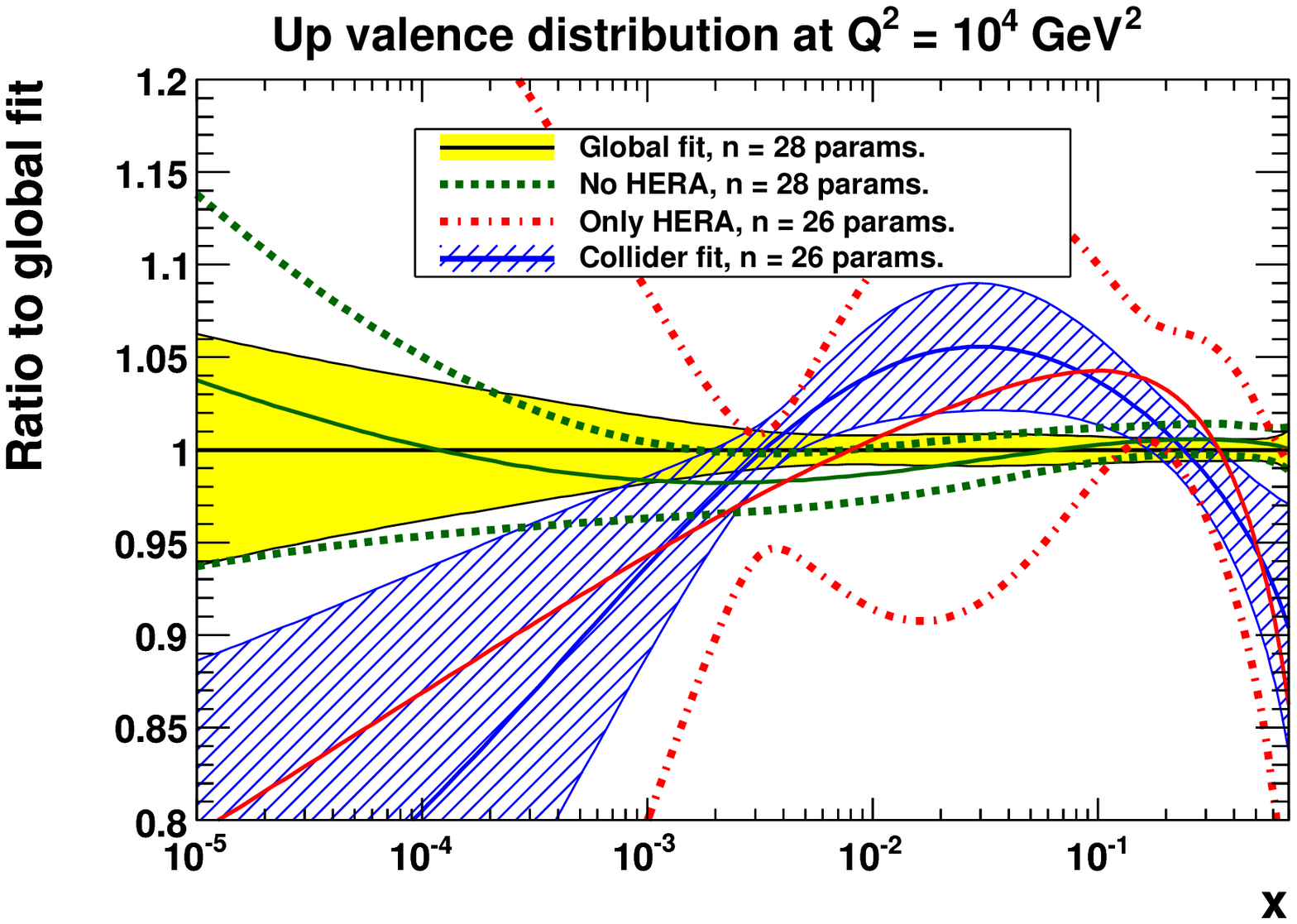}
  \end{minipage}%
  \begin{minipage}{0.5\textwidth}
    (b)\\
    \includegraphics[width=\textwidth]{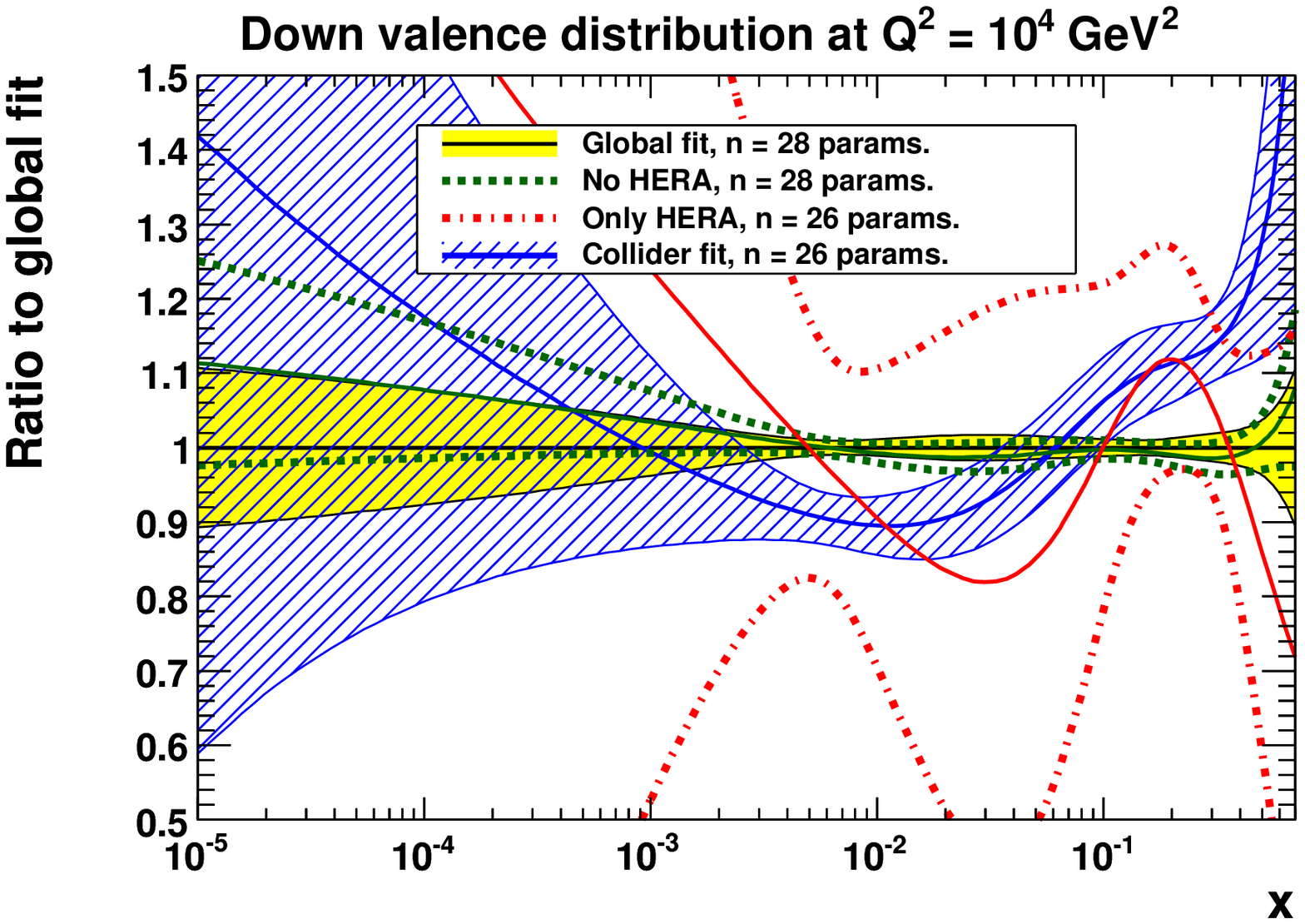}
  \end{minipage}
  \begin{minipage}{0.5\textwidth}
    (c)\\
    \includegraphics[width=\textwidth]{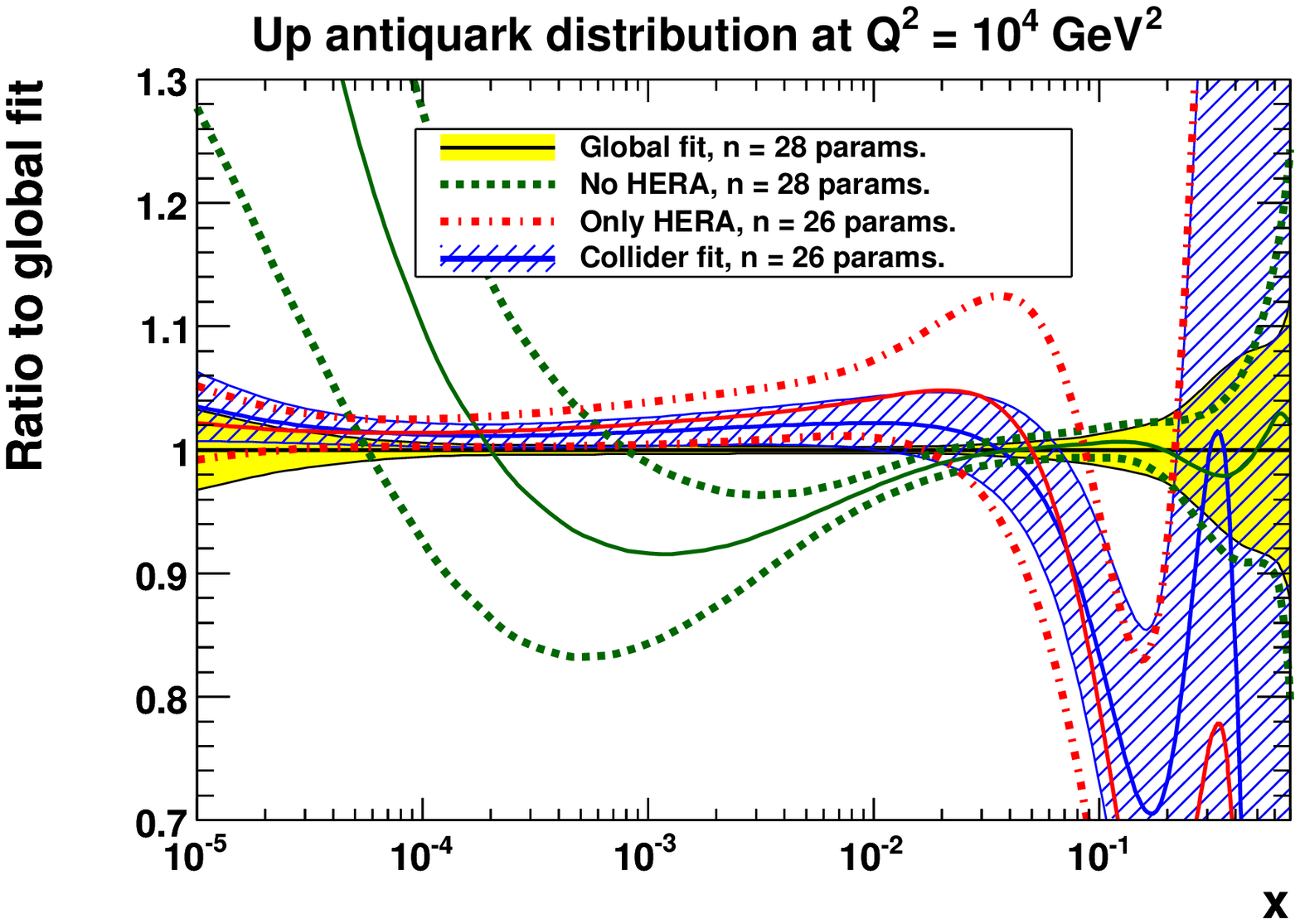}
  \end{minipage}%
  \begin{minipage}{0.5\textwidth}
    (d)\\
    \includegraphics[width=\textwidth]{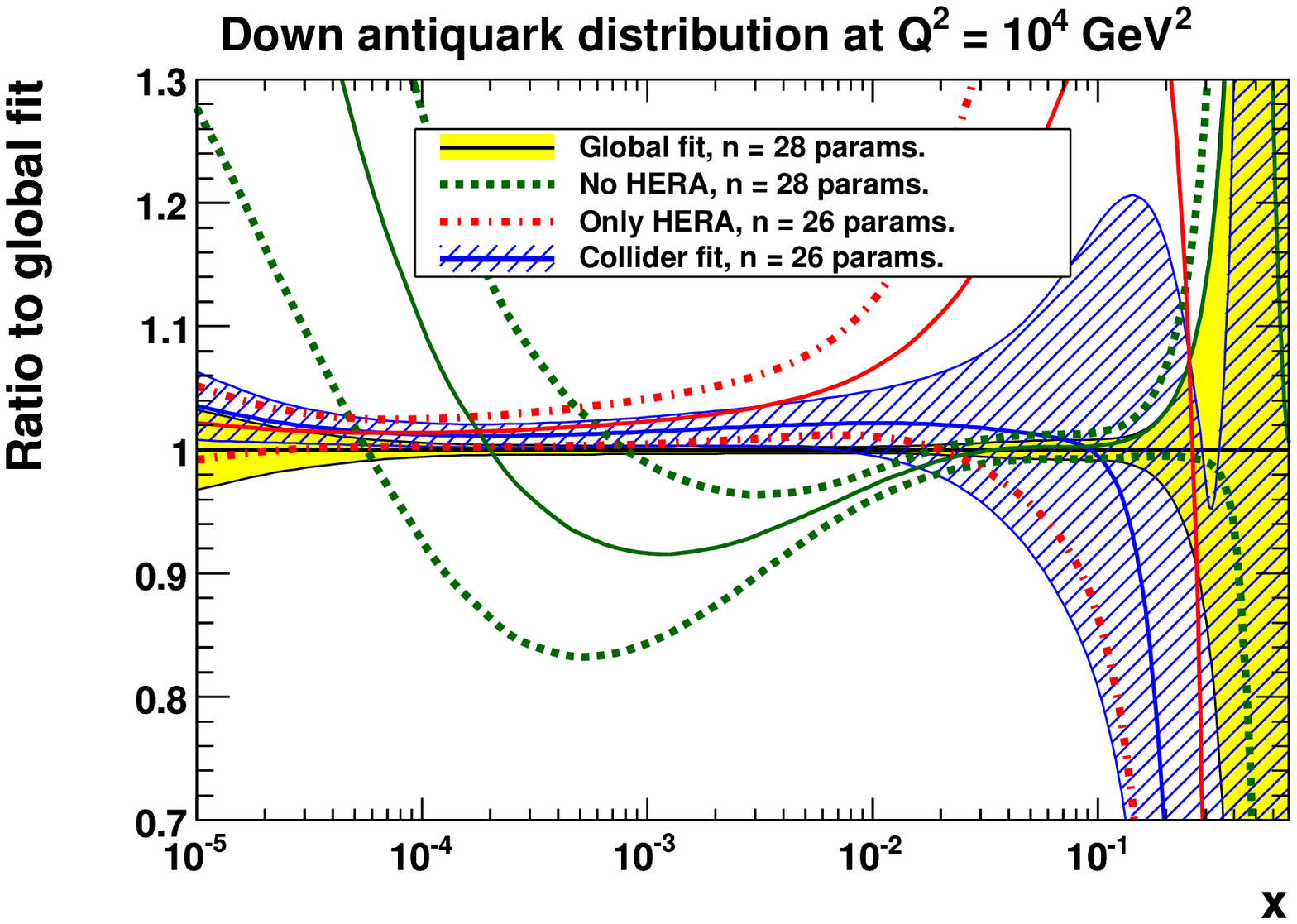}
  \end{minipage}
  \begin{minipage}{0.5\textwidth}
    (e)\\
    \includegraphics[width=\textwidth]{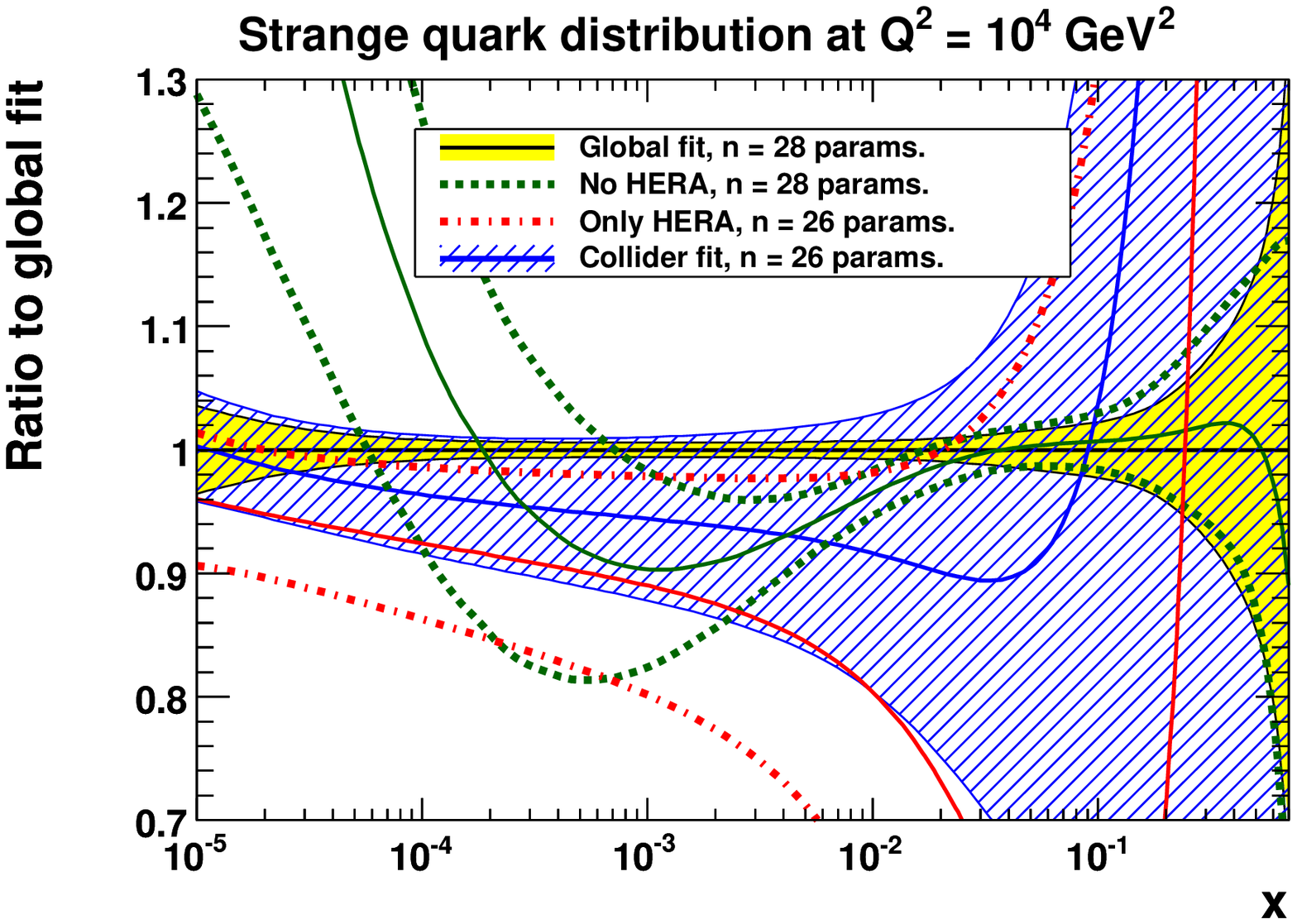}
  \end{minipage}%
  \begin{minipage}{0.5\textwidth}
    (f)\\
    \includegraphics[width=\textwidth]{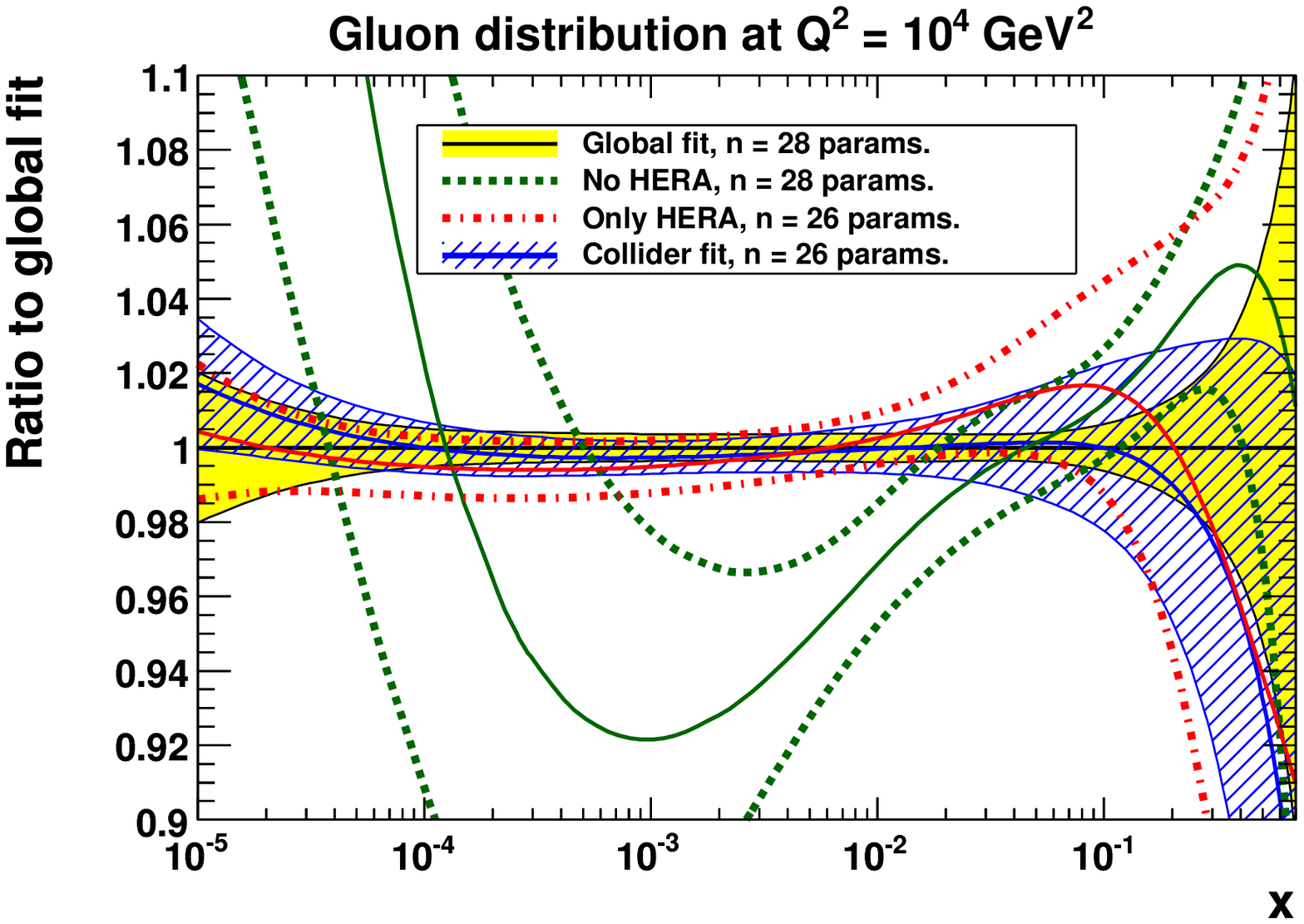}
  \end{minipage}
  \caption{Effect on PDFs of fitting subsets of MSTW 2008 global data.}
  \label{fig:ratio_varyhera}
\end{figure}
There are only 10 parameters used to obtain the central fit and ``experimental'' uncertainties, although the more recent HERAPDF1.5 NNLO~\cite{HERA:2011} analysis introduces 4 more parameters (2 for $g$ and 1 each for $u_v,d_v$).  The HERAPDF analyses additionally include ``model'' and ``parameterisation'' uncertainties that can be much larger than the ``experimental'' uncertainties.  For example, quantities sensitive to the high-$x$ gluon distribution have a very large ``model'' uncertainty in the HERAPDF1.5 NNLO analysis due to variation of the minimum $Q^2$ cut~\cite{Watt:2012fj}.  Nevertheless, it is interesting to investigate the potentially more realistic constraint arising only from HERA data with a flexible parameterisation; see also similar studies by the NNPDF Collaboration~\cite{Ball:2011uy}.  This would be difficult to achieve in the Hessian method where several parameters would need to be held fixed to use the covariance matrix for error propagation, but it is straightforward using the MC method.  We fit subsets of the global data included in the MSTW 2008 NLO analysis~\cite{Martin:2009iq}, specifically (i)~\emph{excluding} all HERA data (neutral-current $e^\pm p$ and charged-current $e^+ p$ cross sections, $F_2^{\rm charm}$, and inclusive jet production in DIS), (ii)~including \emph{only} HERA data, (iii)~performing a ``collider'' fit meaning data from HERA and the Tevatron (inclusive jet production, the $W\to\ell\nu$ charge asymmetry, and the $Z$ rapidity distribution) with no fixed-target data.  The HERA-only fit uses the older separate H1 and ZEUS inclusive cross sections compared to the more precise combined HERA I data~\cite{HERA:2009wt} used in the HERAPDF fits.  On the other hand, the public HERAPDF fits~\cite{HERA:2009wt,HERA:2010,HERA:2011} do not use data on $F_2^{\rm charm}$ or jet production.  In all cases we use the MC method with $n=28$ free parameters wherever possible.  However, for the HERA-only and HERA+Tevatron fits, there is no constraint at all on the strange asymmetry since the CCFR/NuTeV dimuon cross sections are missing, so we fix $s-\bar{s}$ at the global best-fit value, leaving $n=26$ free parameters.  The percentage uncertainties on the PDFs at $Q^2=(100~{\rm GeV})^2$ from the various fits are shown in figure~\ref{fig:frac_varyhera}.  The results reflect what might na\"ively be expected.  For example, removing HERA data gives a huge increase in the small-$x$ uncertainties for the sea-quarks and gluon, but the valence-quark uncertainties are almost unchanged.  With only HERA data, the gluon and antiquarks are still well-constrained at small $x$, but not at large $x$, and there are huge uncertainties in the valence- and strange-quark distributions.  Adding the Tevatron data helps, but the collider-only uncertainty is still much larger than in the global fit, so really we need data from HERA, the Tevatron \emph{and} the fixed-target experiments to get a meaningful result.  The corresponding ratios to the global fit are shown in figure~\ref{fig:ratio_varyhera}.  Here, we see that the uncertainty bands from fits to subsets of the global data do not always overlap with those from the global fit, implying some tension between the different data sets, and suggesting that some kind of error inflation (or \emph{tolerance}) is necessary.  A similar exercise was performed in the MSTW 2008 paper~\cite{Martin:2009iq} to a ``reduced'' data set, with a slightly more constrained parameterisation, and we find similar results if fitting the same ``reduced'' data set using the MC method.

\section{Fits to idealised consistent and inconsistent pseudodata} \label{sec:theory}
As a further exercise to examine potential data set inconsistency within the global fit, we generate idealised pseudodata from the best-fit theory predictions, i.e.~we replace $D_{m,i}$ by $T_{m,i}$ on the right-hand side of eqs.~\eqref{eq:MCgen} and \eqref{eq:MCgenQuad}, where $T_{m,i}$ are the theory predictions evaluated using the global best-fit parameters.  The pseudodata are then simply given by the best-fit theory predictions with appropriate Gaussian noise added, and with uncertainties given by the genuine data uncertainties.  We can then introduce deliberate inconsistencies into this idealised pseudodata and investigate the effect on the fitted PDFs.  We choose the following deliberate inconsistencies, intended to simulate realistic, if somewhat large, incompatibilities that could potentially be present in the genuine data:
\begin{itemize}
\item We introduce a $Q^2$-dependent offset for the H1 and ZEUS inclusive neutral-current reduced cross sections, such that the pseudodata are multiplied by a factor of $\{1\pm0.005\log[Q^2/(10\,{\rm GeV}^2)]\}$, with the ``$+$'' sign for H1 and the ``$-$'' sign for ZEUS.
\item We generate the pseudodata for the CDF and D{\O} inclusive jet cross sections with a scale choice $\mu_R=\mu_F=p_T/2$, but fit it with $\mu_R=\mu_F=p_T$.
\item We normalise the CCFR/NuTeV dimuon cross sections downwards by 10\%.
\item We normalise the NuTeV/CHORUS $xF_3$ structure functions upwards by 5\%.
\item We introduce a rapidity-dependent offset for the CDF $Z$ rapidity distribution, such that the pseudodata are multiplied by a factor of $(1+0.03\,y_Z)$.
\item We introduce an $x$-dependent offset for the BCDMS/NMC/SLAC/E665 deuteron structure functions, intended to mimic a possible deuteron correction, such that the $F_2^d$ data are multiplied by a factor of
\[
f(x) = 
\begin{cases}
  (1+0.005)\left[1-0.003\log^2(x/x_1)\right] & :\quad x<x_1\\
  (1+0.005)\left[1-0.018\log^2(x/x_1)+3\cdot10^{-8}\log^{20}(x/x_1)\right] & :\quad x\ge x_1
\end{cases},
\]
where $x_1=\exp(-2.5)\simeq 0.0821$.
\item We introduce a $Q^2$-dependent offset for the BCDMS $F_2^p$ and $F_2^d$ structure functions, such that the pseudodata are multiplied by a factor of $\{1+0.01\log[Q^2/(1\,{\rm GeV}^2)]\}$.
\end{itemize}
\begin{figure}
  \centering
  \begin{minipage}{0.5\textwidth}
    (a)\\
    \includegraphics[width=\textwidth]{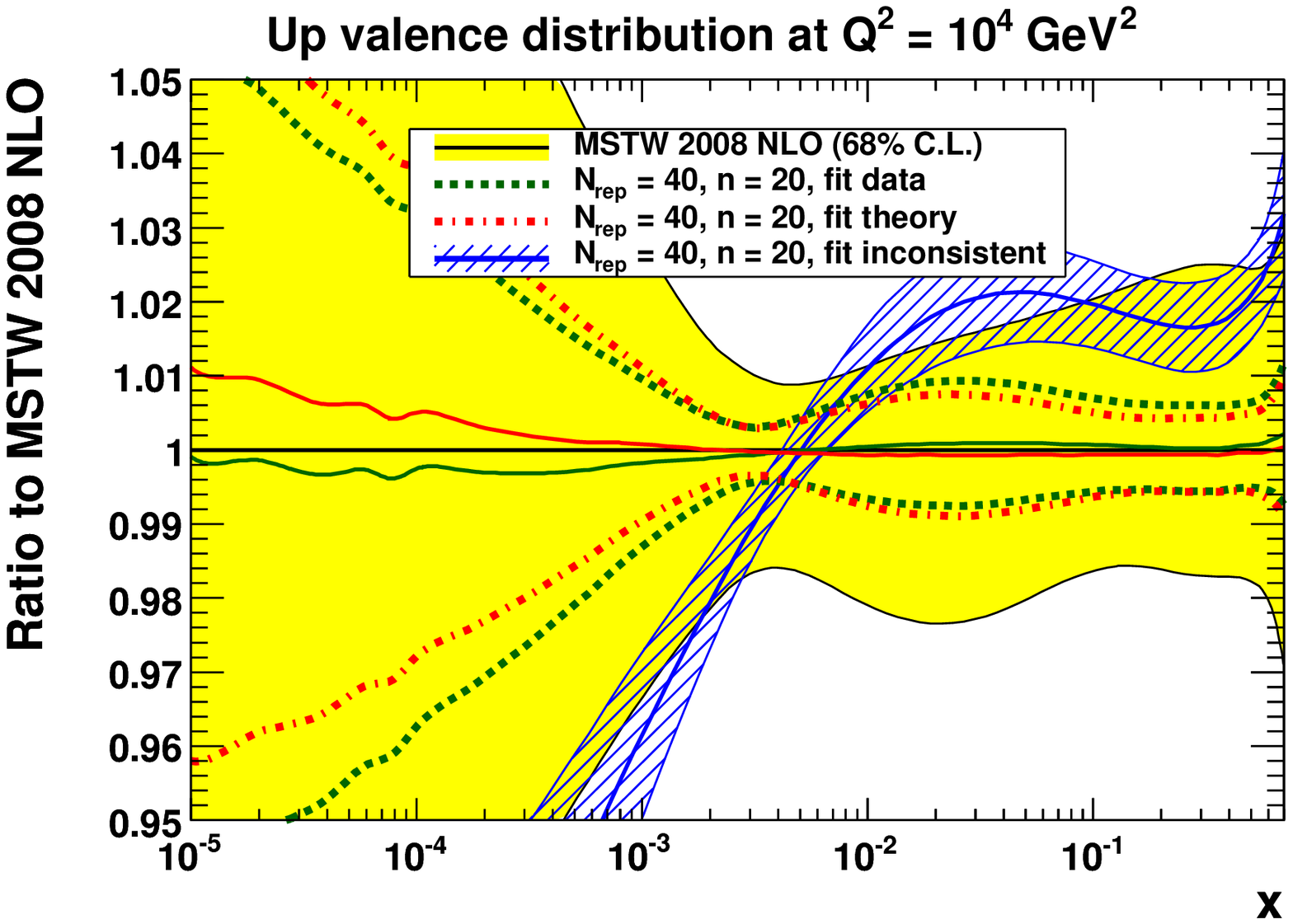}
  \end{minipage}%
  \begin{minipage}{0.5\textwidth}
    (b)\\
    \includegraphics[width=\textwidth]{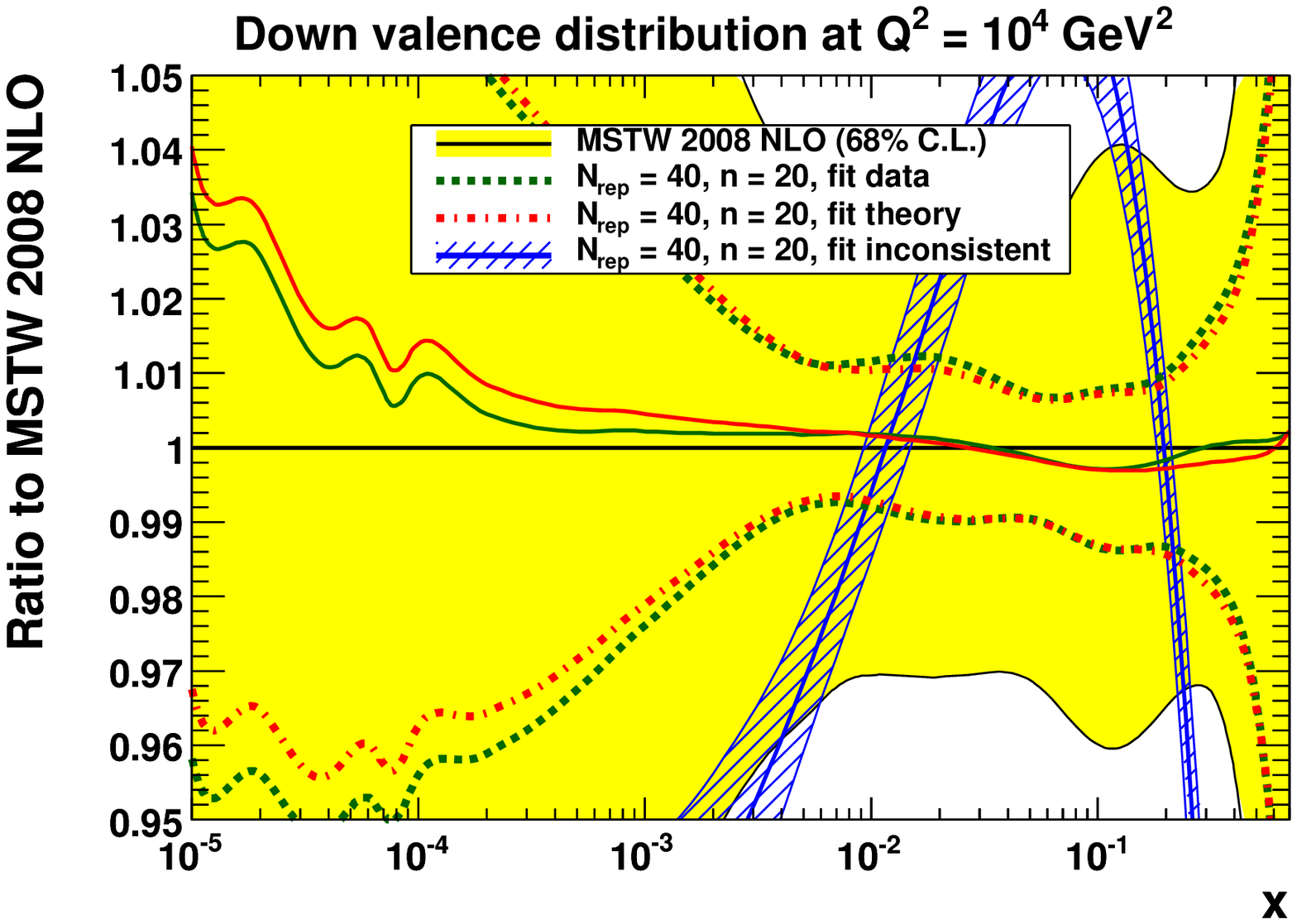}
  \end{minipage}
  \begin{minipage}{0.5\textwidth}
    (c)\\
    \includegraphics[width=\textwidth]{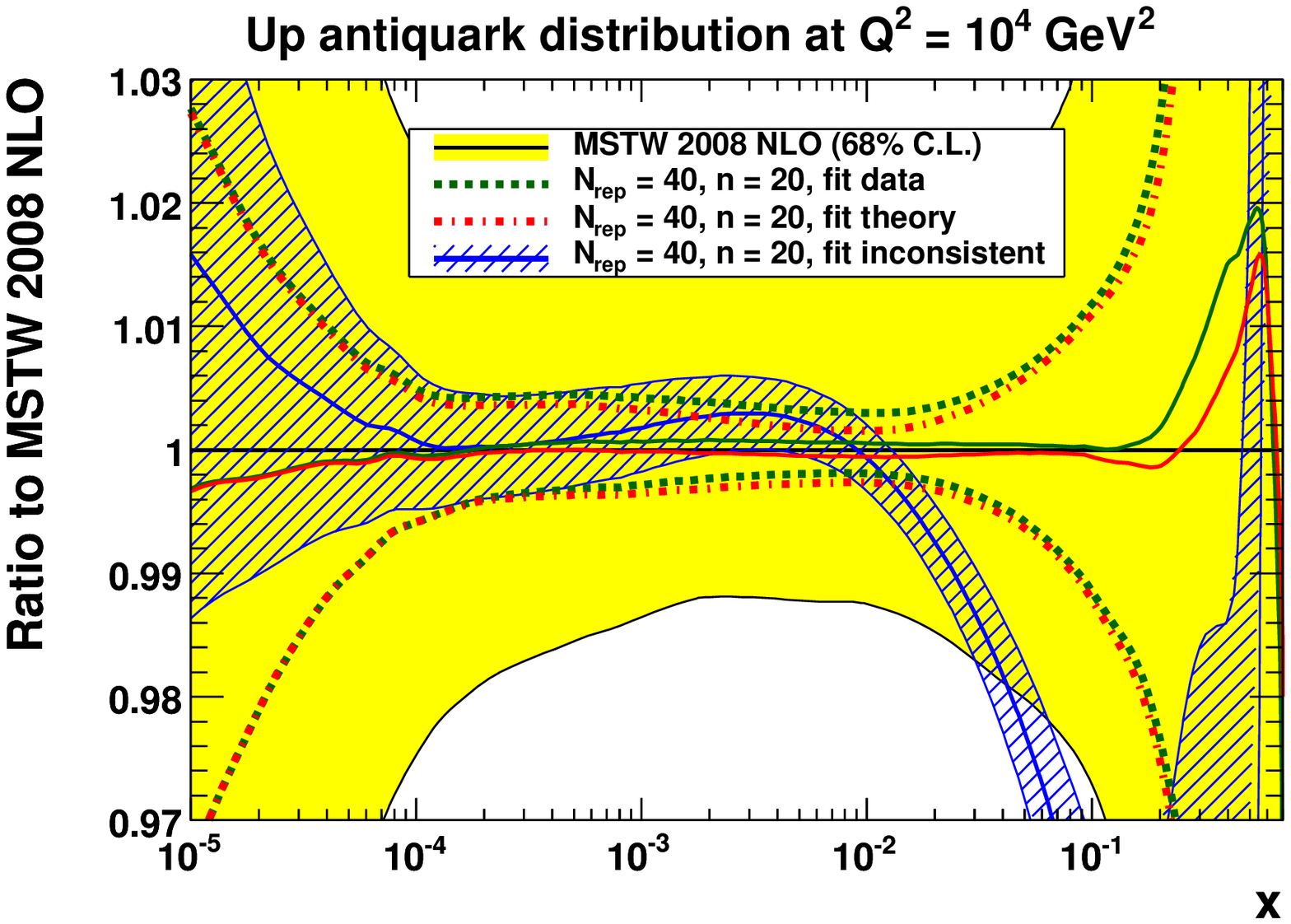}
  \end{minipage}%
  \begin{minipage}{0.5\textwidth}
    (d)\\
    \includegraphics[width=\textwidth]{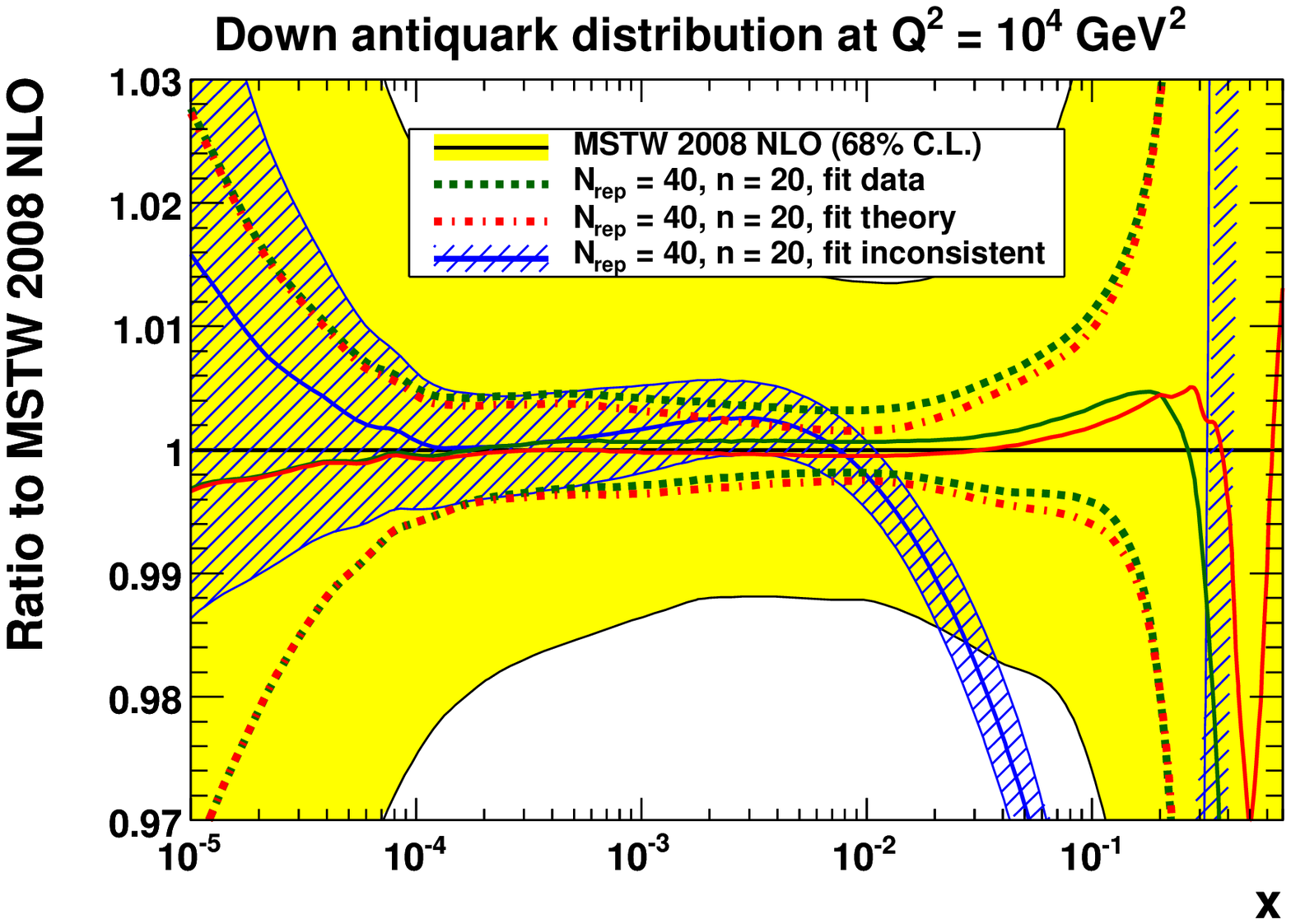}
  \end{minipage}
  \begin{minipage}{0.5\textwidth}
    (e)\\
    \includegraphics[width=\textwidth]{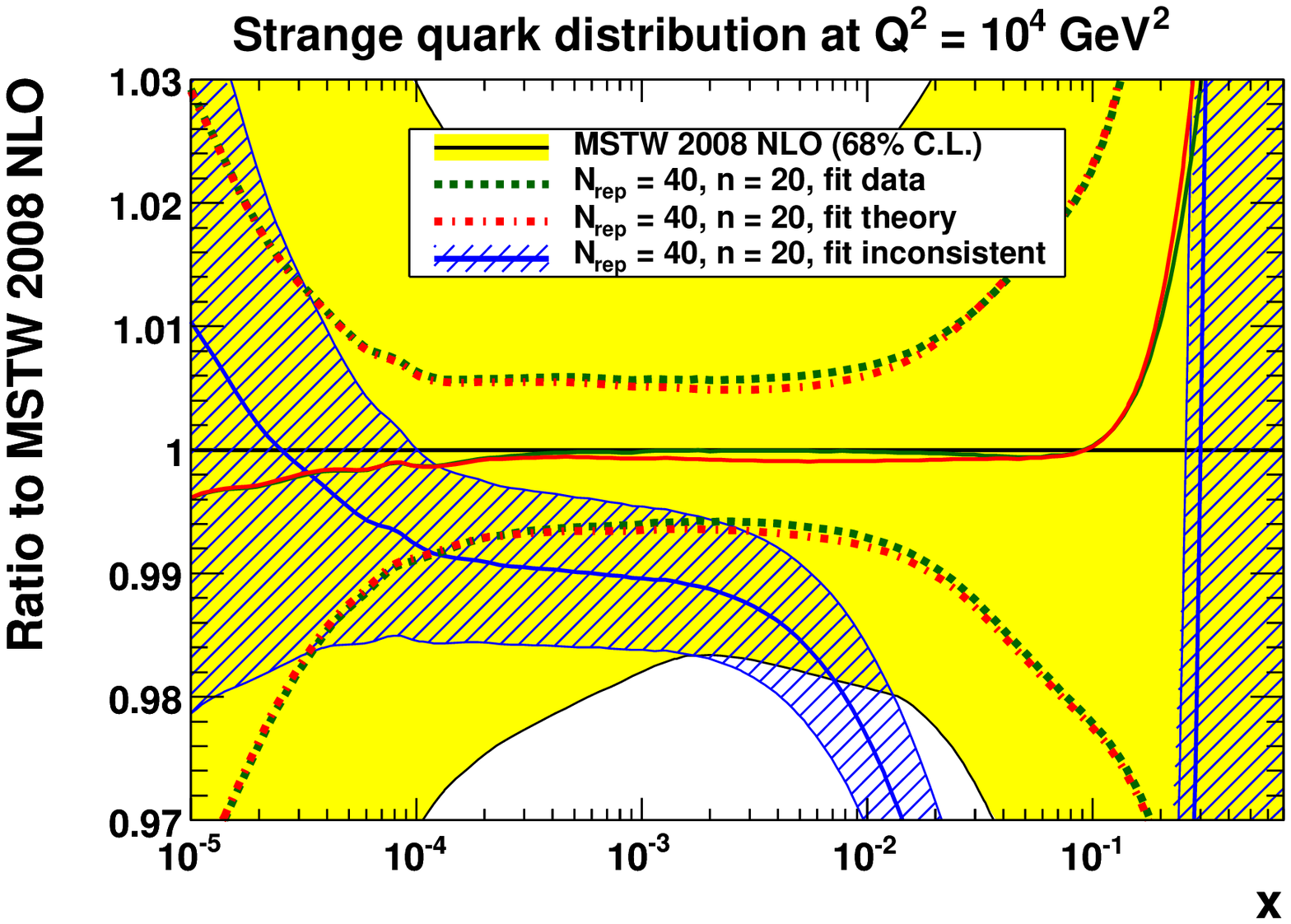}
  \end{minipage}%
  \begin{minipage}{0.5\textwidth}
    (f)\\
    \includegraphics[width=\textwidth]{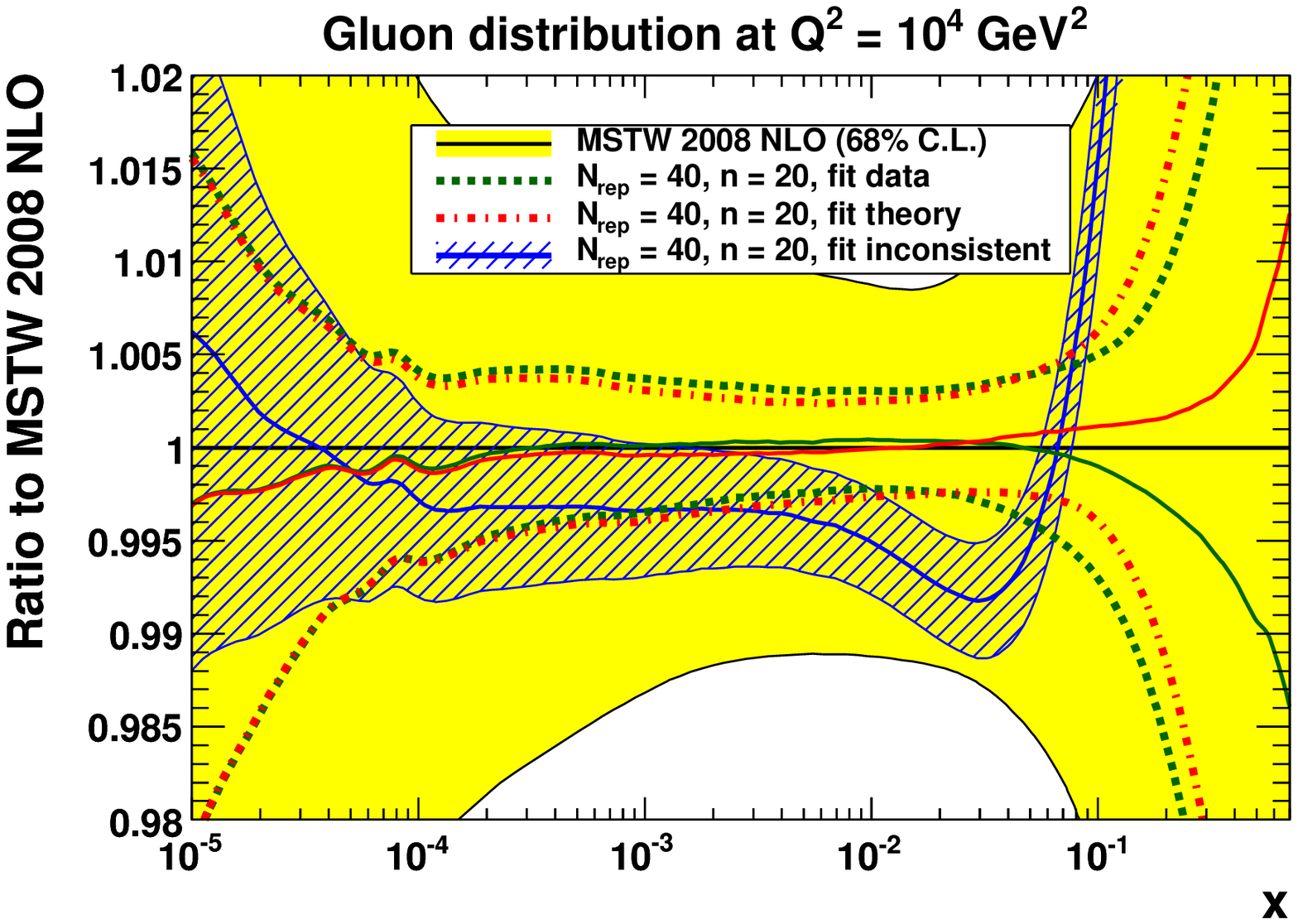}
  \end{minipage}
  \caption{Effect on PDFs of fitting consistent or inconsistent idealised pseudodata.}
  \label{fig:ratio_theory}
\end{figure}
In figures~\ref{fig:ratio_theory} and \ref{fig:frac_theory} we show the effect of fitting the genuine data, then the consistent or inconsistent idealised pseudodata, in each case using MC error propagation with $N_{\rm rep} = 40$ replica data sets and $n=20$ input PDF parameters, and we compare to the standard MSTW 2008 NLO fit with dynamic tolerance.  Despite the central values of the PDFs from the inconsistent fit shifting by significant amounts, the percentage uncertainties in figure~\ref{fig:frac_theory} are remarkably almost identical whether fitting either the genuine data, the consistent pseudodata or the inconsistent pseudodata.
\begin{figure}
  \centering
  \begin{minipage}{0.5\textwidth}
    (a)\\
    \includegraphics[width=\textwidth]{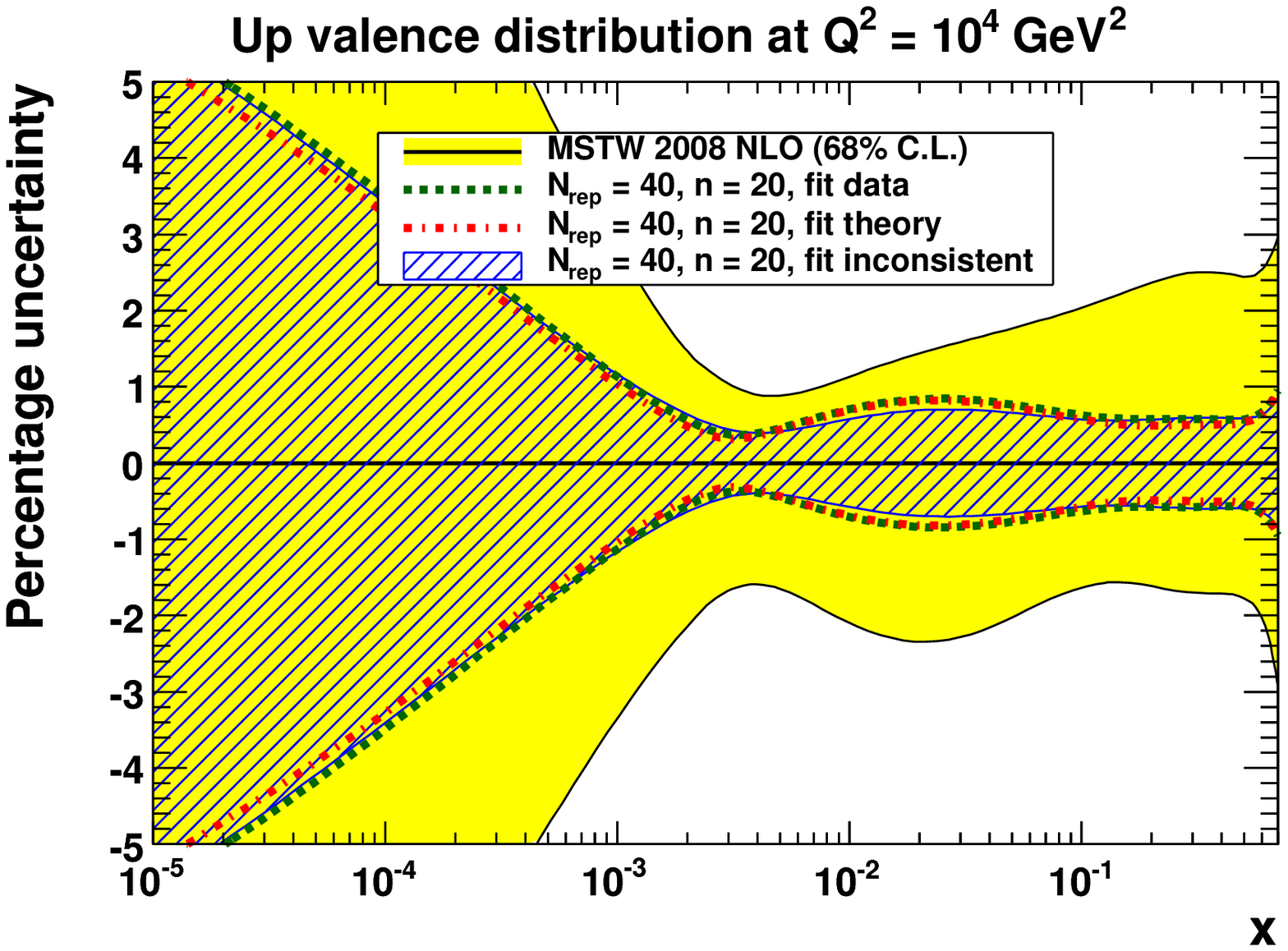}
  \end{minipage}%
  \begin{minipage}{0.5\textwidth}
    (b)\\
    \includegraphics[width=\textwidth]{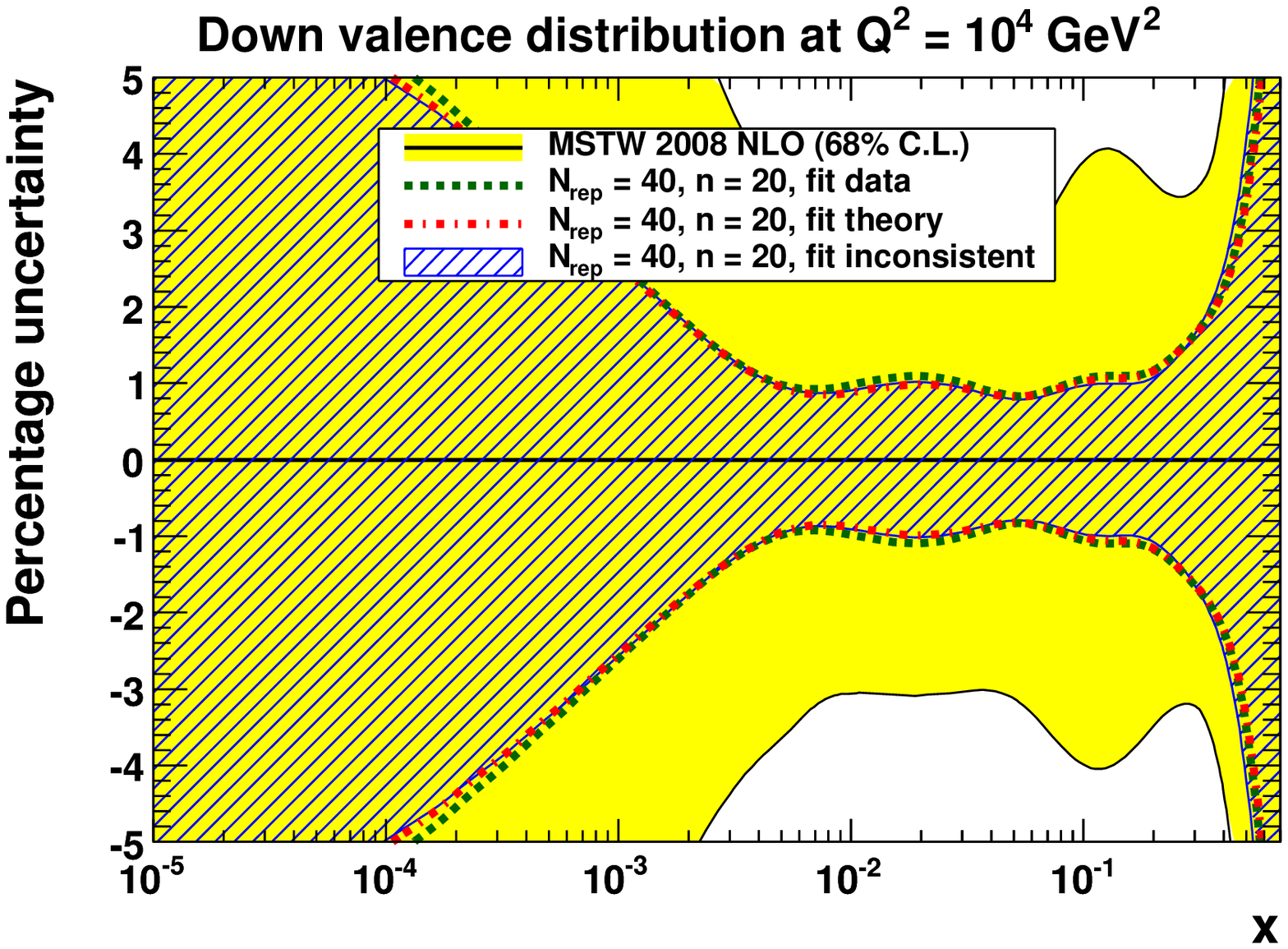}
  \end{minipage}
  \begin{minipage}{0.5\textwidth}
    (c)\\
    \includegraphics[width=\textwidth]{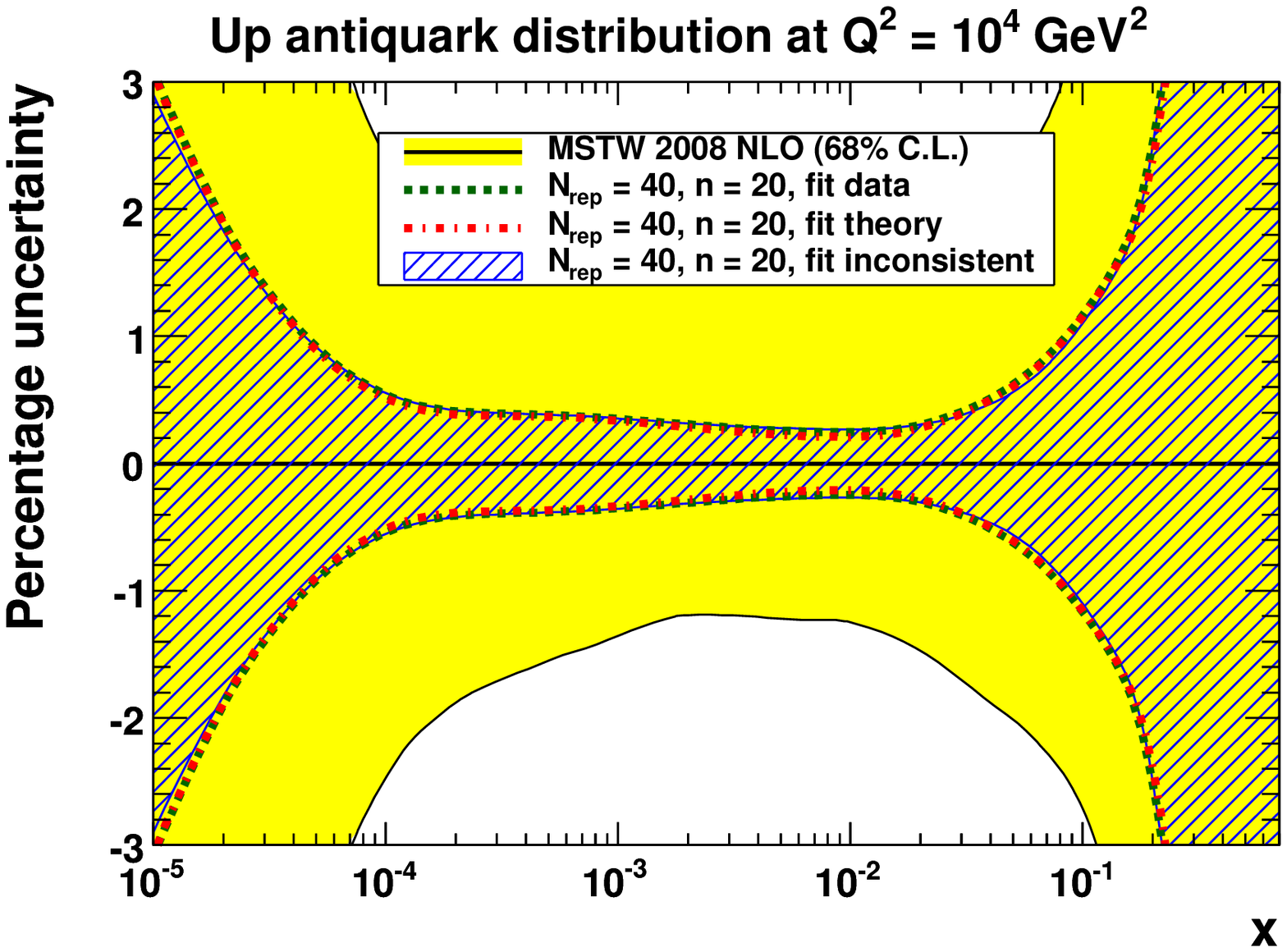}
  \end{minipage}%
  \begin{minipage}{0.5\textwidth}
    (d)\\
    \includegraphics[width=\textwidth]{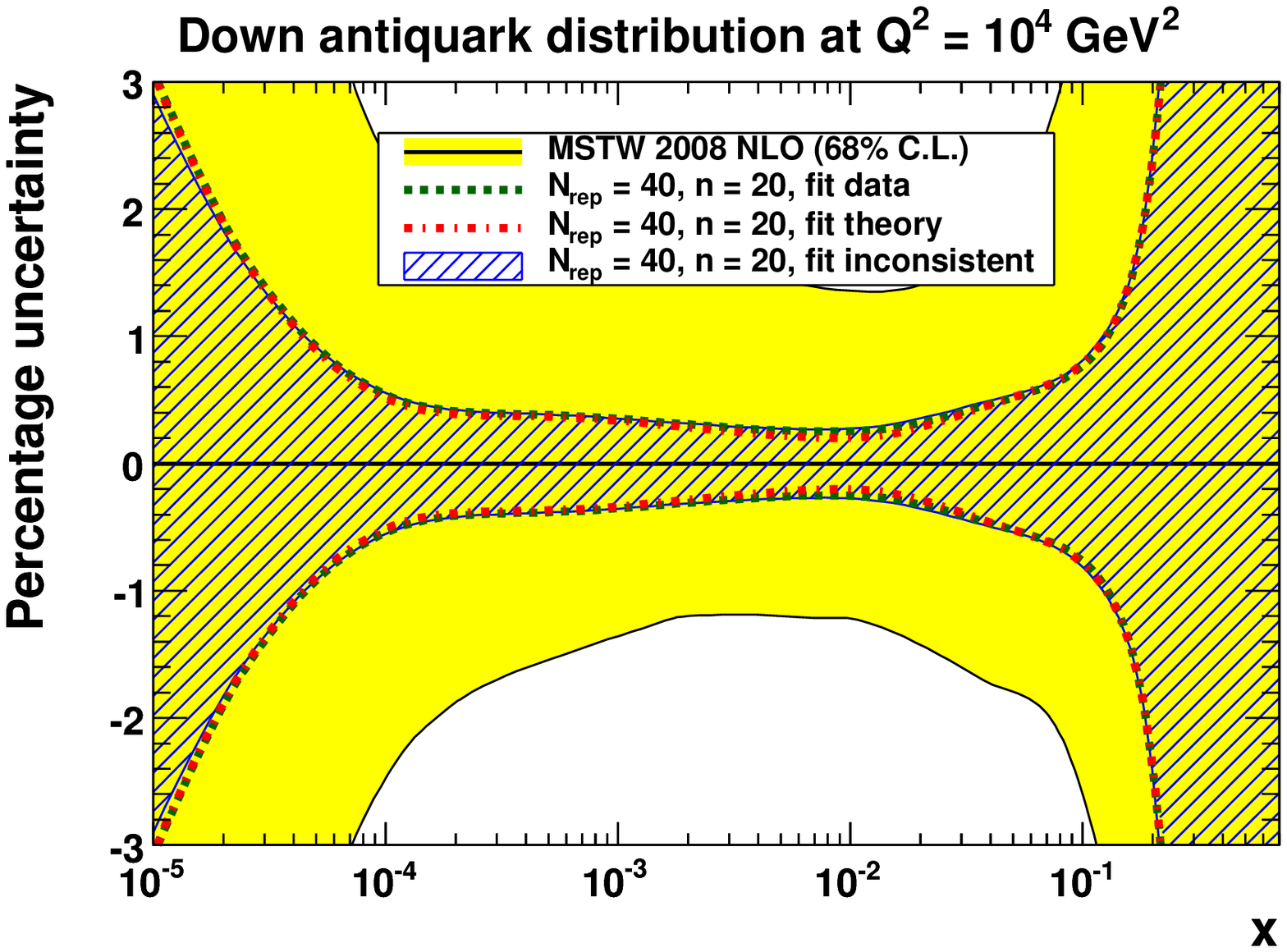}
  \end{minipage}
  \begin{minipage}{0.5\textwidth}
    (e)\\
    \includegraphics[width=\textwidth]{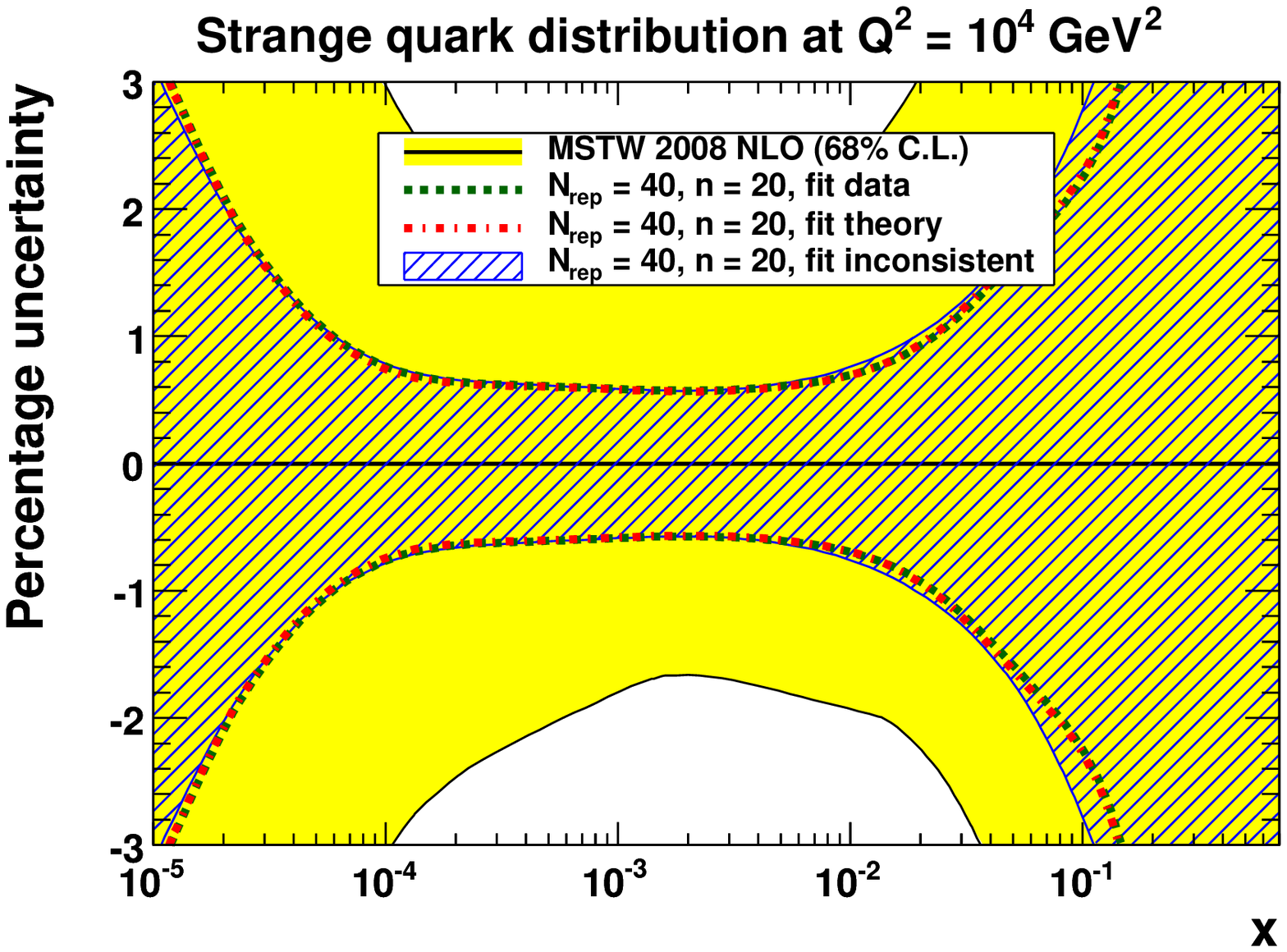}
  \end{minipage}%
  \begin{minipage}{0.5\textwidth}
    (f)\\
    \includegraphics[width=\textwidth]{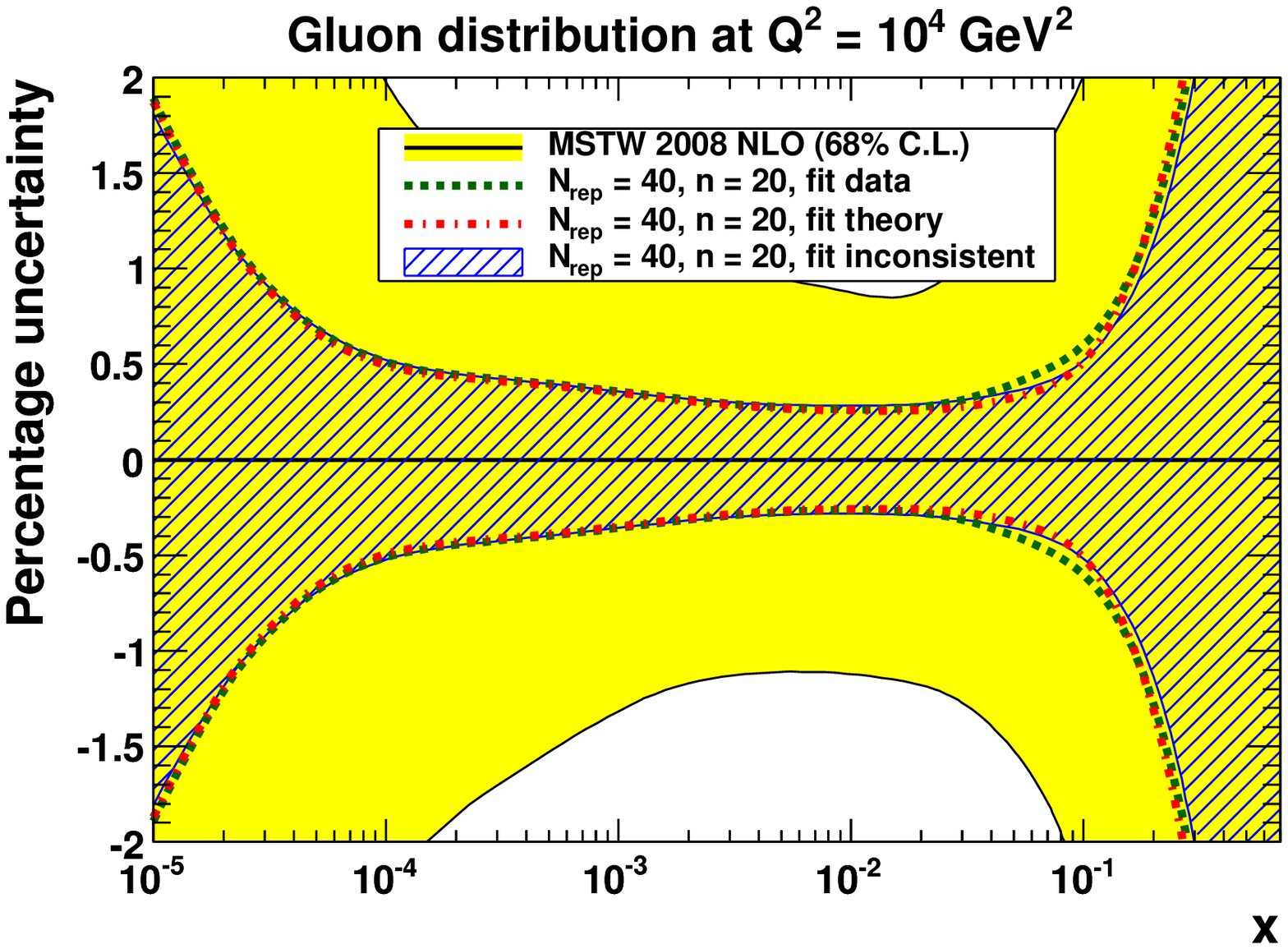}
  \end{minipage}
  \caption{Effect on percentage PDF uncertainties of fitting consistent or inconsistent pseudodata.}
  \label{fig:frac_theory}
\end{figure}
The MC fit to perfectly consistent pseudodata gives $\chi^2_{\rm global}/N_{\rm pts.}=0.98\pm0.03$, which by construction is exactly unity up to the statistical fluctuation, and similarly for the individual data sets included in the global fit; see table~\ref{tab:chisq}.  On the other hand, the MC fit to the inconsistent pseudodata gives $\chi^2_{\rm global}/N_{\rm pts.}=1.07\pm0.03$, so the fit quality has only deteriorated slightly, despite the central values of some PDFs shifting well outside their original uncertainty band; see figure~\ref{fig:ratio_theory}.  This result is in contradiction to what seems to be a widely held view that a fit to inconsistent data should lead to a $\chi^2/N_{\rm pts.}\gg 1$.  The values of the $\chi^2 / N_{\rm pts.}$ in table~\ref{tab:chisq} deviate further from unity for a few individual data sets such as BCDMS $F_2^d$, the NMC $F_2^d/F_2^p$ ratio, NuTeV $xF_3$ and the CDF $Z$ rapidity distribution, but not by such large amounts that the inconsistent fit would not be judged to be an ``acceptable'' fit.  Despite the fairly significant $Q^2$-dependent offset of the H1 and ZEUS inclusive cross sections, amounting to almost 4\% at $Q^2=500$~GeV$^2$, there is only a slight increase in the $\chi^2$ values in going from the consistent to the inconsistent fit.  Similarly, by looking at the MSTW08 fit to the genuine data in table~\ref{tab:chisq}, there are only a few individual data sets with values of $\chi^2 / N_{\rm pts.}$ significantly above unity, perhaps giving the false impression that there is not a large degree of incompatibility between individual data sets.
\begin{table}
  \centering
{\footnotesize
  \begin{tabular}{|l|c|c|c|}
    \hline
    Data set & MSTW08 & Fit consistent pseudodata & Fit inconsistent pseudodata \\ \hline
    BCDMS $\mu p$ $F_2$ & $1.12$ & $0.96\pm0.13$ & $1.10\pm0.15$ \\ 
    BCDMS $\mu d$ $F_2$ & $1.26$ & $0.99\pm0.13$ & $1.44\pm0.17$ \\ 
    NMC $\mu p$ $F_2$ & $0.98$ & $0.96\pm0.12$ & $0.97\pm0.12$ \\ 
    NMC $\mu d$ $F_2$ & $0.83$ & $1.00\pm0.12$ & $1.05\pm0.13$ \\ 
    NMC $\mu n/\mu p$ & $0.88$ & $1.02\pm0.12$ & $1.25\pm0.13$ \\ 
    E665 $\mu p$ $F_2$ & $1.08$ & $0.99\pm0.18$ & $0.99\pm0.18$ \\ 
    E665 $\mu d$ $F_2$ & $1.01$ & $1.00\pm0.18$ & $1.02\pm0.18$ \\ 
    SLAC $ep$ $F_2$ & $0.80$ & $0.97\pm0.22$ & $0.98\pm0.23$ \\ 
    SLAC $ed$ $F_2$ & $0.78$ & $0.98\pm0.16$ & $1.03\pm0.18$ \\ 
    NMC/BCDMS/SLAC $F_L$ & $1.22$ & $1.04\pm0.27$ & $1.04\pm0.27$ \\ \hline 
    E866/NuSea $pp$ DY & $1.24$ & $0.92\pm0.10$ & $0.98\pm0.10$ \\ 
    E866/NuSea $pd/pp$ DY & $0.93$ & $0.86\pm0.35$ & $0.96\pm0.35$ \\ \hline 
    NuTeV $\nu N$ $F_2$ & $0.92$ & $0.93\pm0.19$ & $1.07\pm0.19$ \\ 
    CHORUS $\nu N$ $F_2$ & $0.62$ & $1.01\pm0.24$ & $1.08\pm0.27$ \\ 
    NuTeV $\nu N$ $xF_3$ & $0.89$ & $0.99\pm0.19$ & $1.42\pm0.22$ \\ 
    CHORUS $\nu N$ $xF_3$ & $0.93$ & $0.89\pm0.21$ & $1.14\pm0.2$5 \\ 
    CCFR $\nu N\to \mu\mu X$ & $0.77$ & $0.98\pm0.14$ & $1.03\pm0.14$ \\ 
    NuTeV $\nu N\to \mu\mu X$ & $0.46$ & $0.96\pm0.16$ & $1.00\pm0.17$ \\ \hline 
    H1 MB 99 $e^+p$ NC & $1.15$ & $0.87\pm0.44$ & $0.92\pm0.44$ \\ 
    H1 MB 97 $e^+p$ NC & $0.66$ & $0.99\pm0.20$ & $1.01\pm0.20$ \\ 
    H1 low $Q^2$ 96--97 $e^+p$ NC & $0.56$ & $1.00\pm0.15$ & $1.03\pm0.15$ \\ 
    H1 high $Q^2$ 98--99 $e^-p$ NC & $0.97$ & $0.98\pm0.12$ & $1.00\pm0.12$ \\ 
    H1 high $Q^2$ 99--00 $e^+p$ NC & $0.89$ & $1.02\pm0.10$ & $1.05\pm0.10$ \\ 
    ZEUS SVX 95 $e^+p$ NC & $1.16$ & $0.94\pm0.25$ & $0.94\pm0.25$ \\ 
    ZEUS 96--97 $e^+p$ NC & $0.60$ & $1.01\pm0.11$ & $1.04\pm0.11$ \\ 
    ZEUS 98--99 $e^-p$ NC & $0.59$ & $0.98\pm0.14$ & $1.00\pm0.14$ \\ 
    ZEUS 99--00 $e^+p$ NC & $0.70$ & $1.02\pm0.16$ & $1.05\pm0.16$ \\ 
    H1 99--00 $e^+p$ CC & $1.04$ & $1.00\pm0.23$ & $1.03\pm0.24$ \\ 
    ZEUS 99--00 $e^+p$ CC & $1.27$ & $0.95\pm0.20$ & $1.02\pm0.21$ \\ 
    H1/ZEUS $ep$ $F_2^{\rm charm}$ & $1.29$ & $1.00\pm0.12$ & $1.00\pm0.12$ \\ 
    H1 99--00 $e^+p$ incl.~jets & $0.78$ & $1.00\pm0.30$ & $1.03\pm0.30$ \\ 
    ZEUS 96--97 $e^+p$ incl.~jets & $0.99$ & $1.07\pm0.26$ & $1.07\pm0.25$ \\ 
    ZEUS 98--00 $e^\pm p$ incl.~jets & $0.56$ & $0.95\pm0.25$ & $0.98\pm0.26$ \\ \hline 
    D{\O} II $p\bar{p}$ incl.~jets & $1.04$ & $0.96\pm0.14$ & $1.03\pm0.15$ \\ 
    CDF II $p\bar{p}$ incl.~jets & $0.73$ & $1.01\pm0.22$ & $1.08\pm0.23$ \\ 
    CDF II $W\to \ell\nu$ asym. & $1.32$ & $1.00\pm0.30$ & $1.03\pm0.3$3 \\ 
    D{\O} II $W\to \ell\nu$ asym. & $2.51$ & $0.94\pm0.40$ & $1.08\pm0.47$ \\ 
    D{\O} II $Z$ rap. & $0.68$ & $1.05\pm0.2$9 & $1.07\pm0.30$ \\ 
    CDF II $Z$ rap. & $1.70$ & $1.05\pm0.29$ & $1.62\pm0.4$3 \\ \hline 
    \textbf{All data sets} & $\mathbf{0.93}$ & $\mathbf{0.98\pm0.03}$ & $\mathbf{1.07\pm0.03}$ \\ 
    \hline
  \end{tabular}
}
\caption{Values of $\chi^2 / N_{\rm pts.}$ for the data sets in various NLO global fits.  The ``MSTW08'' column shows the best-fit numbers~\cite{Martin:2009iq}.  The pseudodata numbers in the other two columns are the average and standard deviation of the $\chi^2 / N_{\rm pts.}$ over $N_{\rm rep} = 40$ replica fits.  See ref.~\cite{Martin:2009iq} for data references.}
\label{tab:chisq}
\end{table}

\begin{figure}
  \centering
  \begin{minipage}{0.5\textwidth}
    (a)\\
    \includegraphics[width=\textwidth]{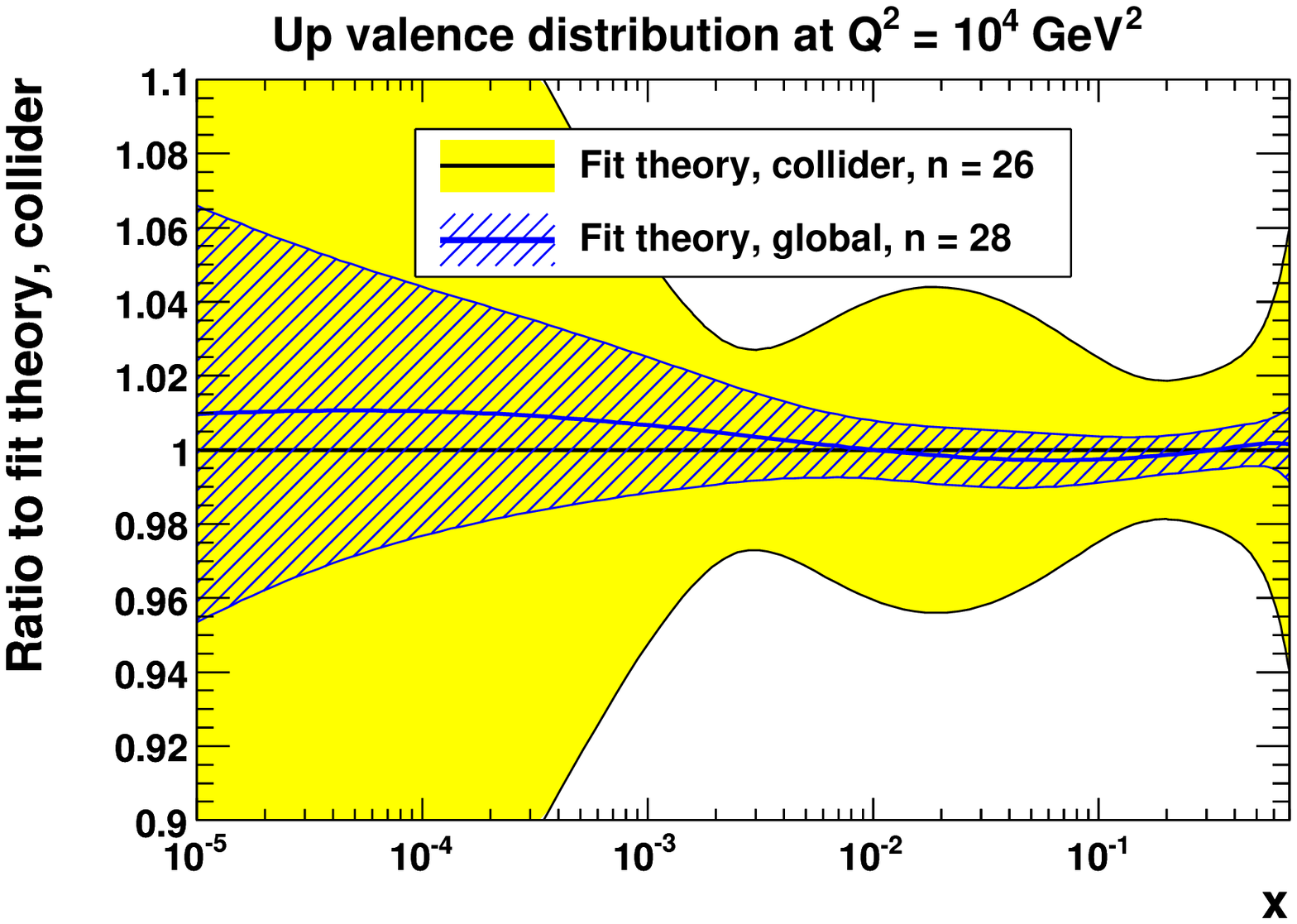}
  \end{minipage}%
  \begin{minipage}{0.5\textwidth}
    (b)\\
    \includegraphics[width=\textwidth]{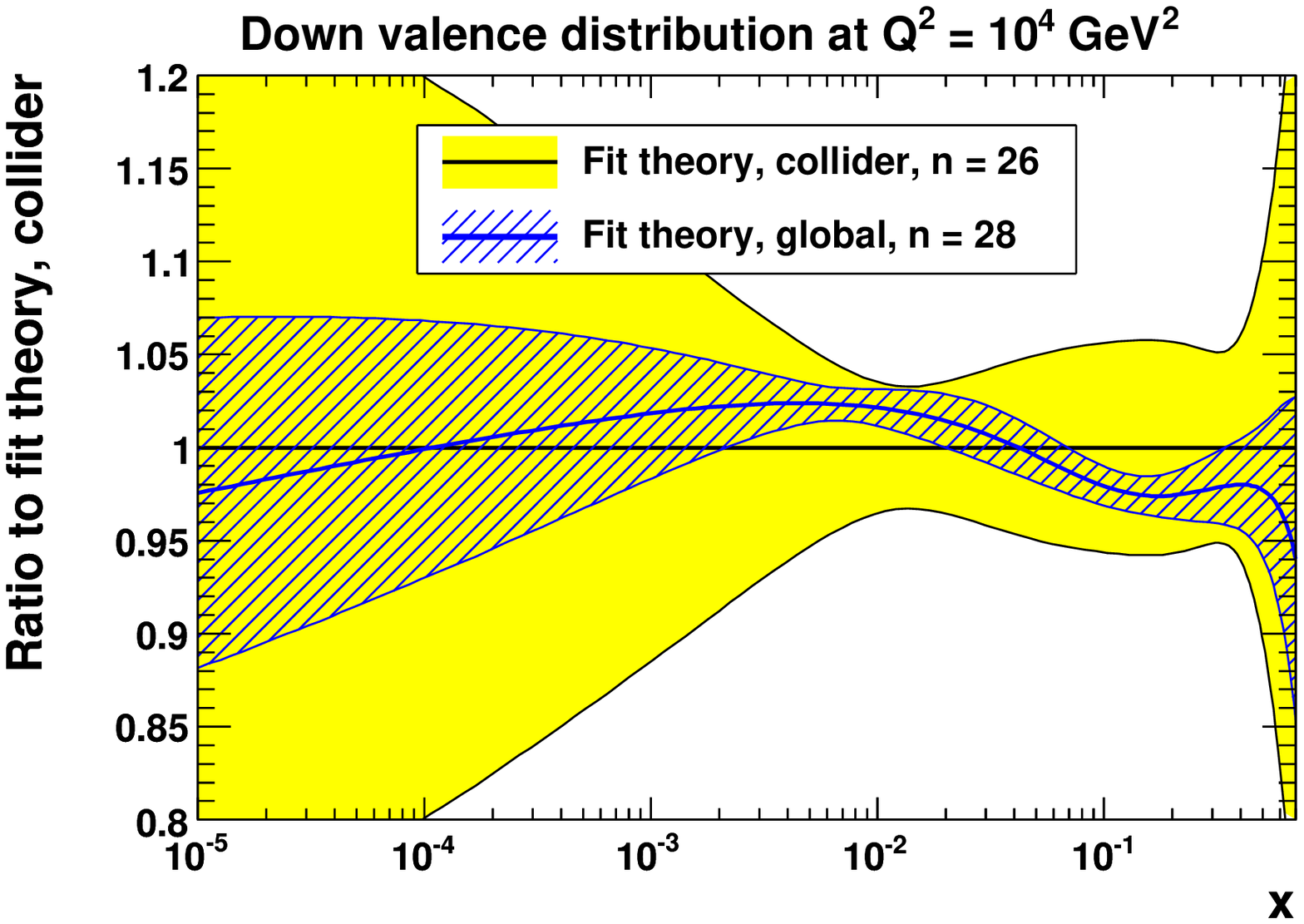}
  \end{minipage}
  \begin{minipage}{0.5\textwidth}
    (c)\\
    \includegraphics[width=\textwidth]{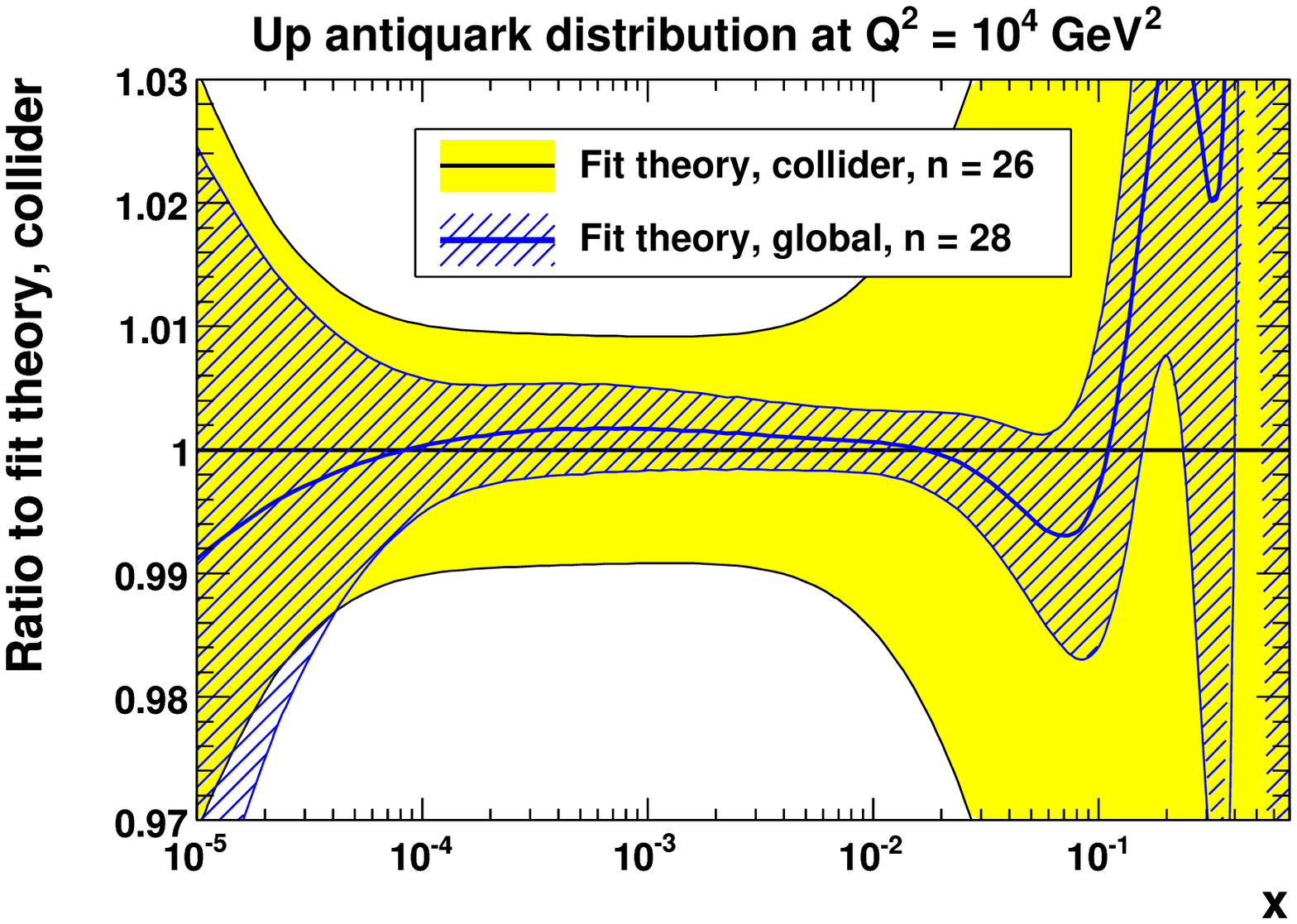}
  \end{minipage}%
  \begin{minipage}{0.5\textwidth}
    (d)\\
    \includegraphics[width=\textwidth]{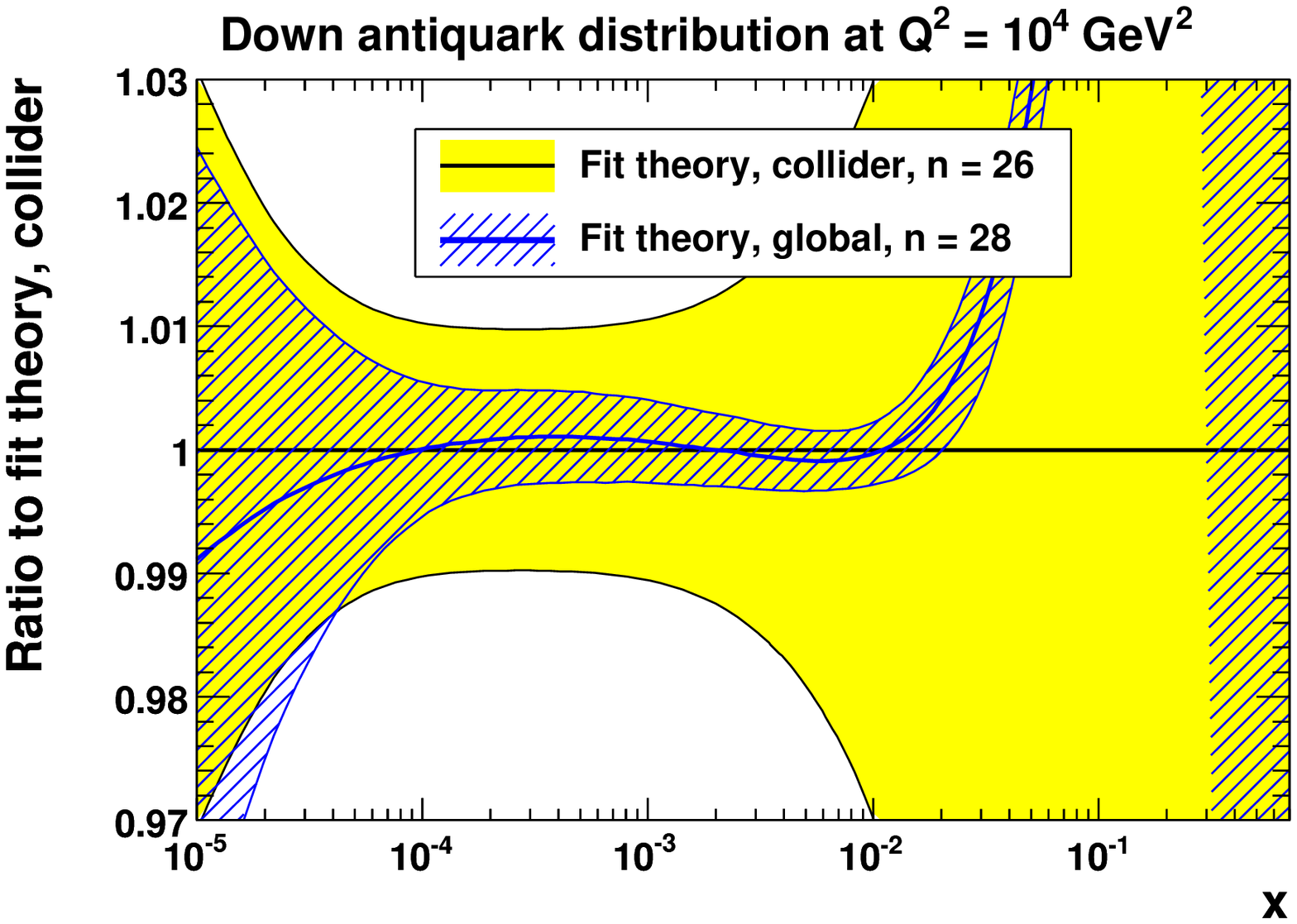}
  \end{minipage}
  \begin{minipage}{0.5\textwidth}
    (e)\\
    \includegraphics[width=\textwidth]{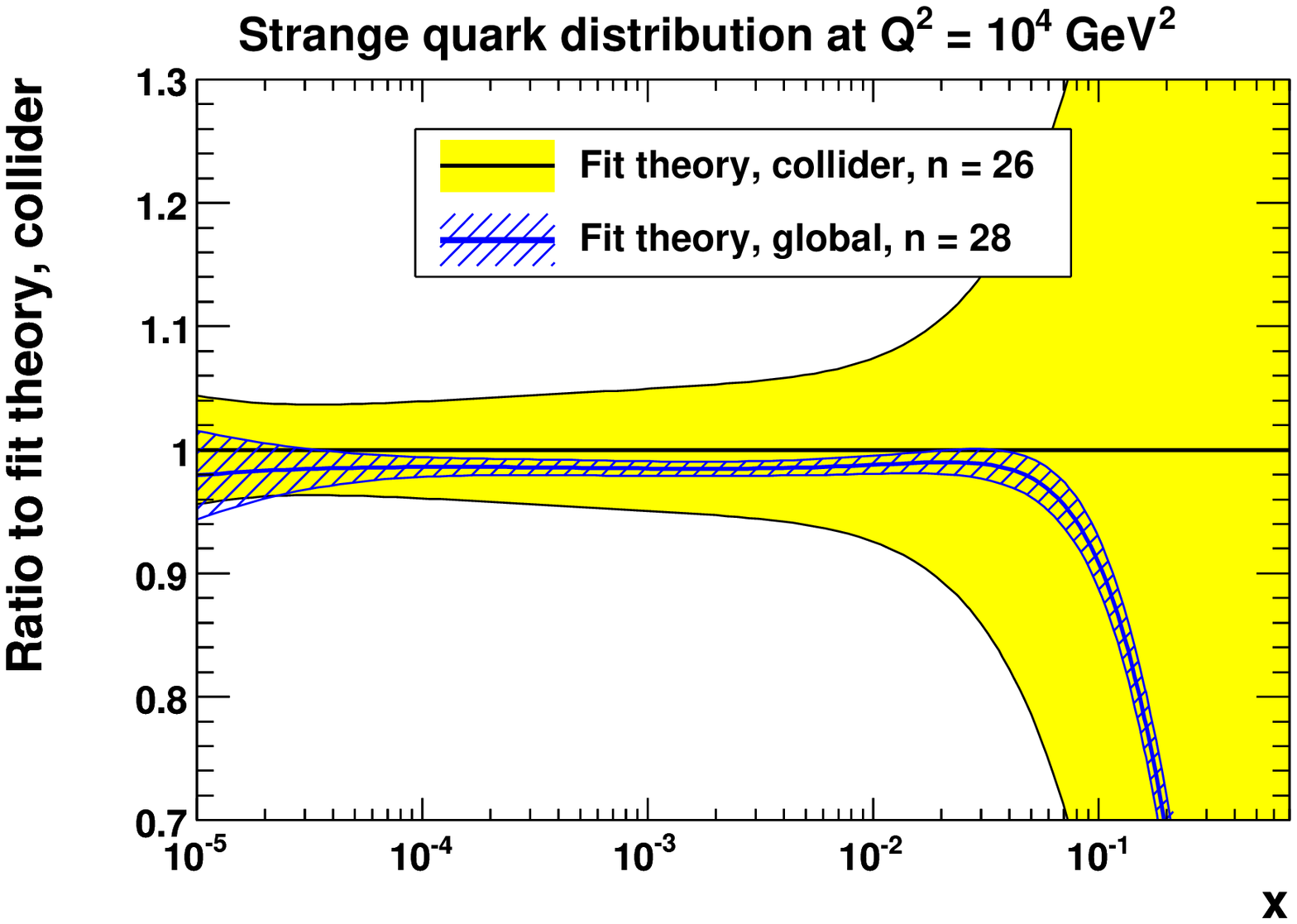}
  \end{minipage}%
  \begin{minipage}{0.5\textwidth}
    (f)\\
    \includegraphics[width=\textwidth]{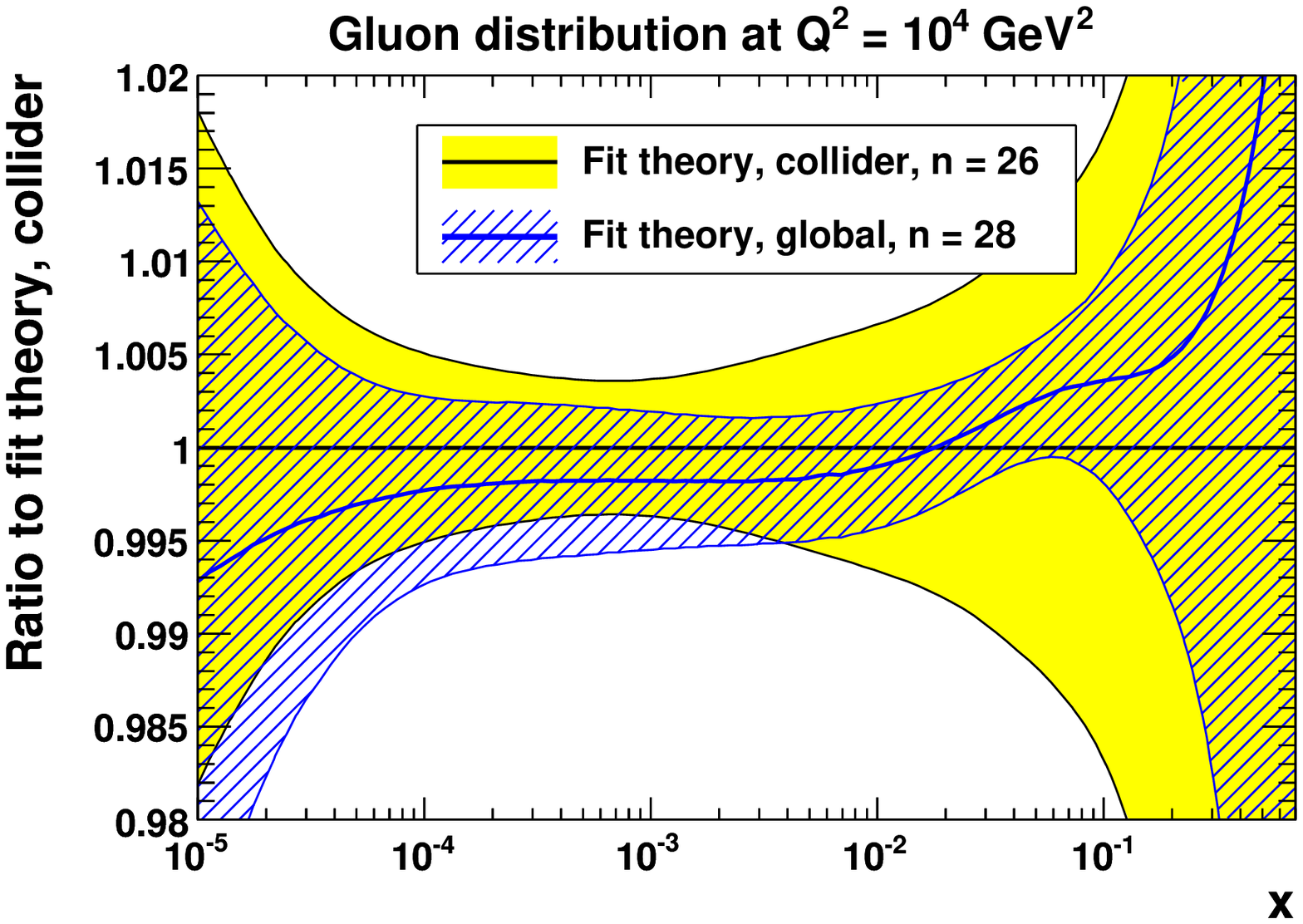}
  \end{minipage}
  \caption{Effect on PDFs of fitting consistent idealised pseudodata, either collider-only or global.}
  \label{fig:ratio_collider_theory}
\end{figure}

\begin{figure}
  \centering
  \begin{minipage}{0.5\textwidth}
    (a)\\
    \includegraphics[width=\textwidth]{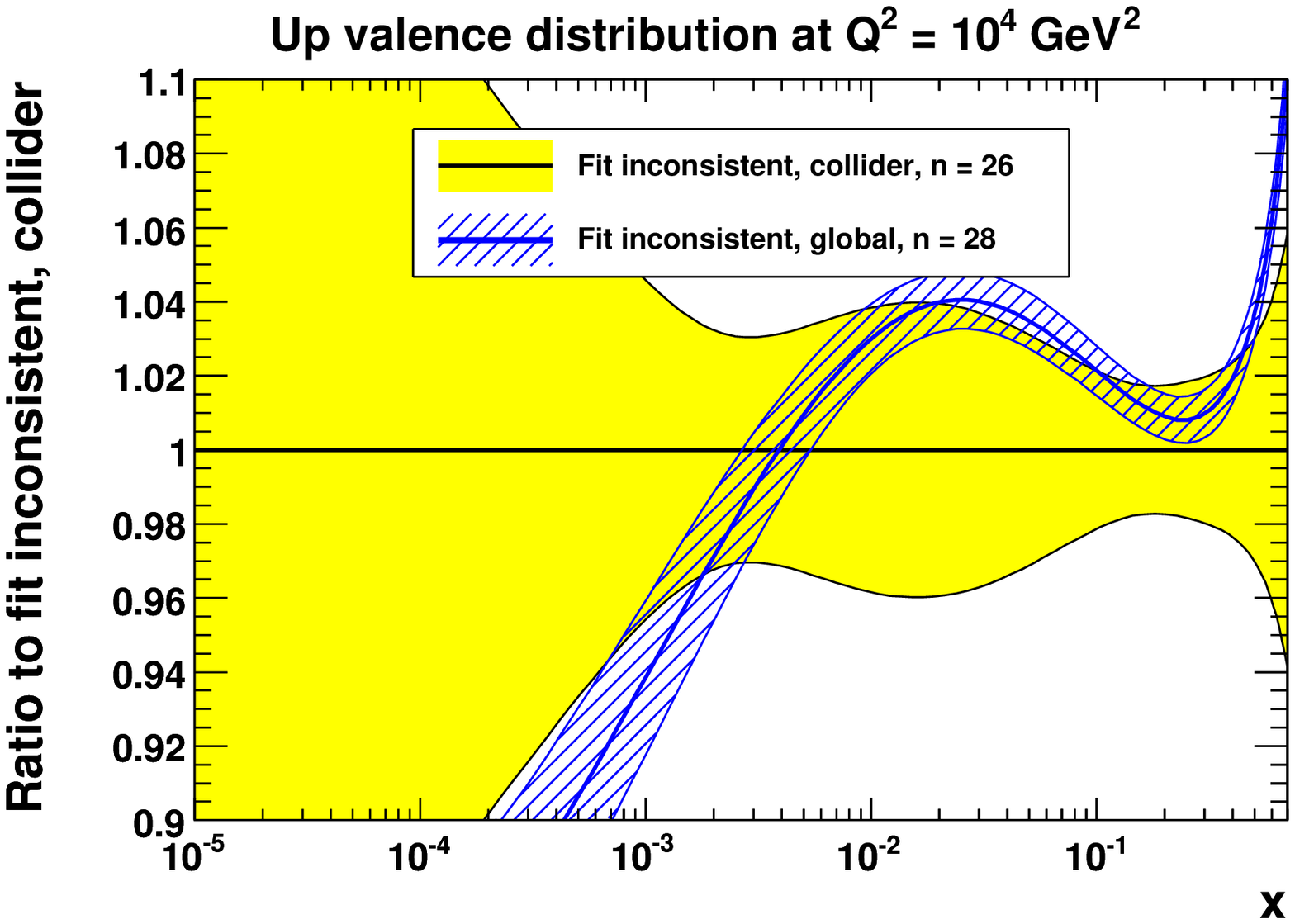}
  \end{minipage}%
  \begin{minipage}{0.5\textwidth}
    (b)\\
    \includegraphics[width=\textwidth]{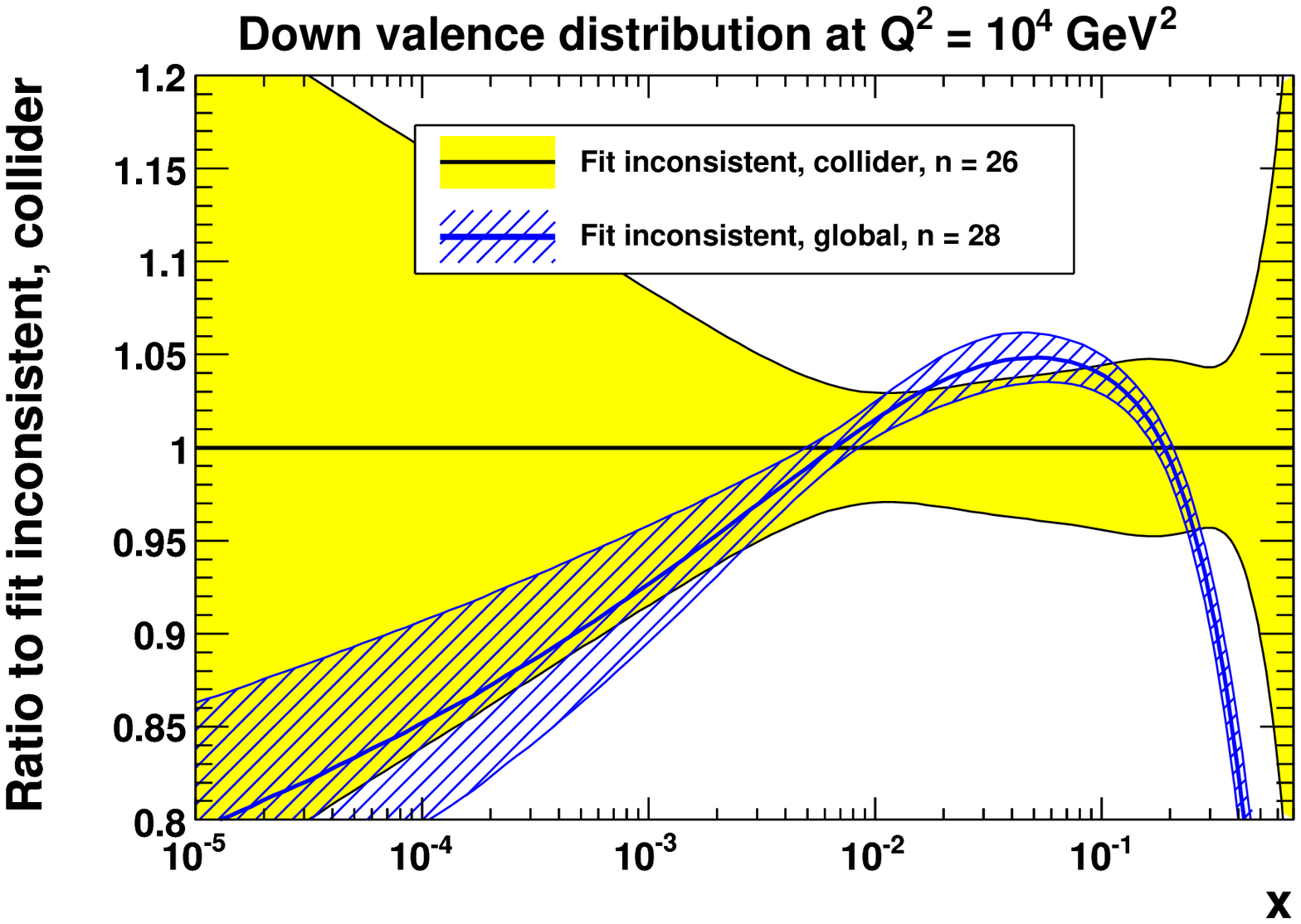}
  \end{minipage}
  \begin{minipage}{0.5\textwidth}
    (c)\\
    \includegraphics[width=\textwidth]{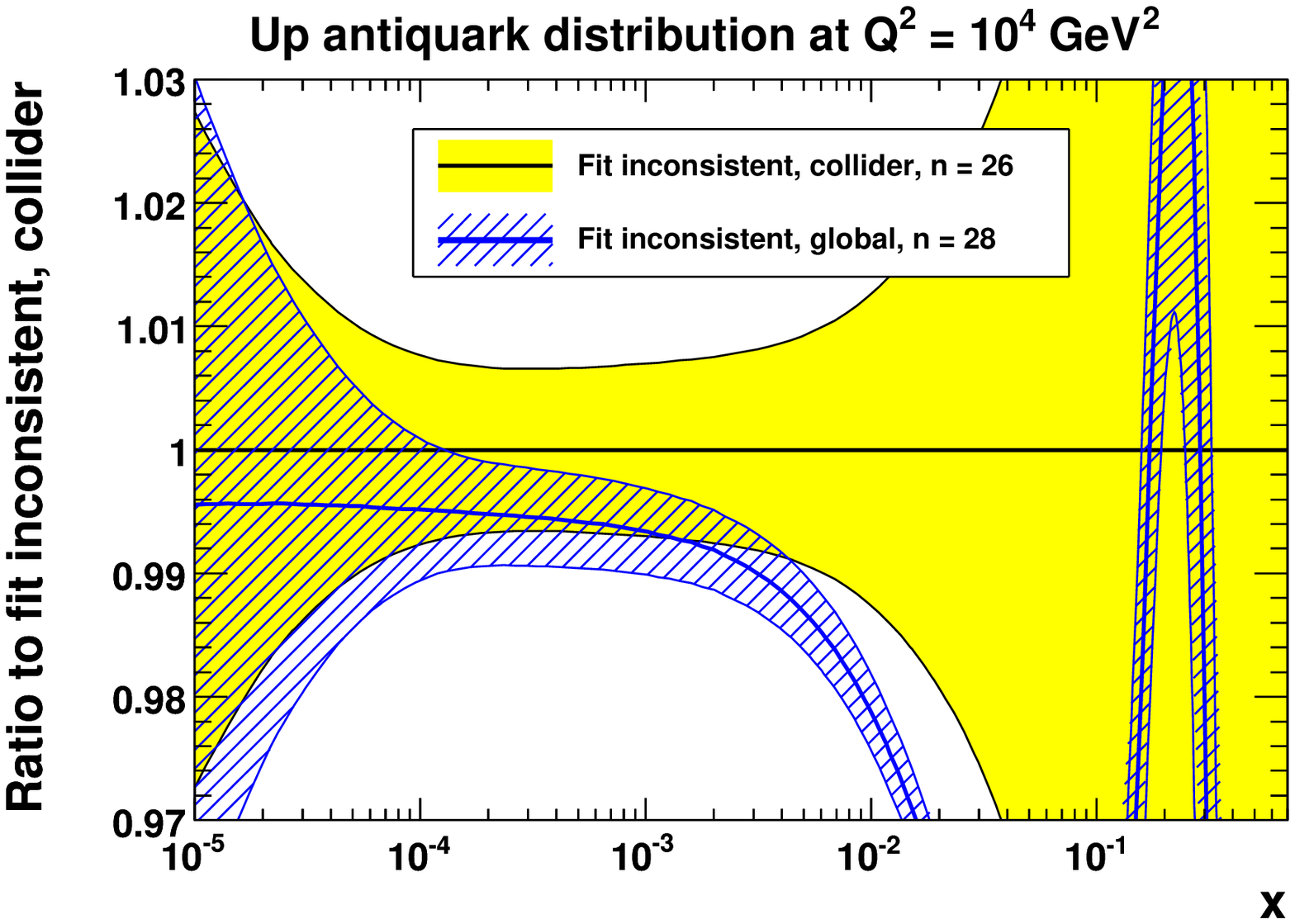}
  \end{minipage}%
  \begin{minipage}{0.5\textwidth}
    (d)\\
    \includegraphics[width=\textwidth]{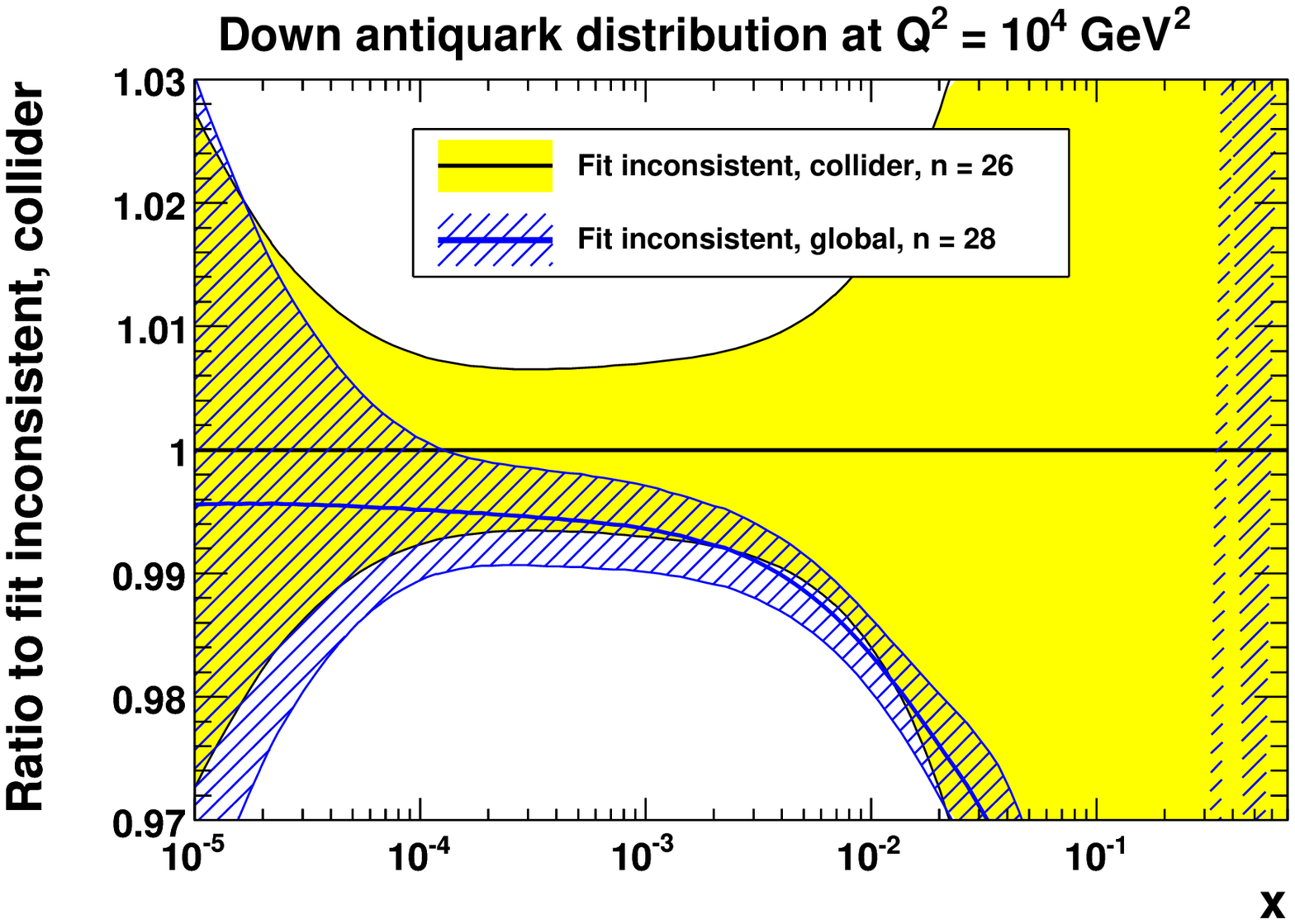}
  \end{minipage}
  \begin{minipage}{0.5\textwidth}
    (e)\\
    \includegraphics[width=\textwidth]{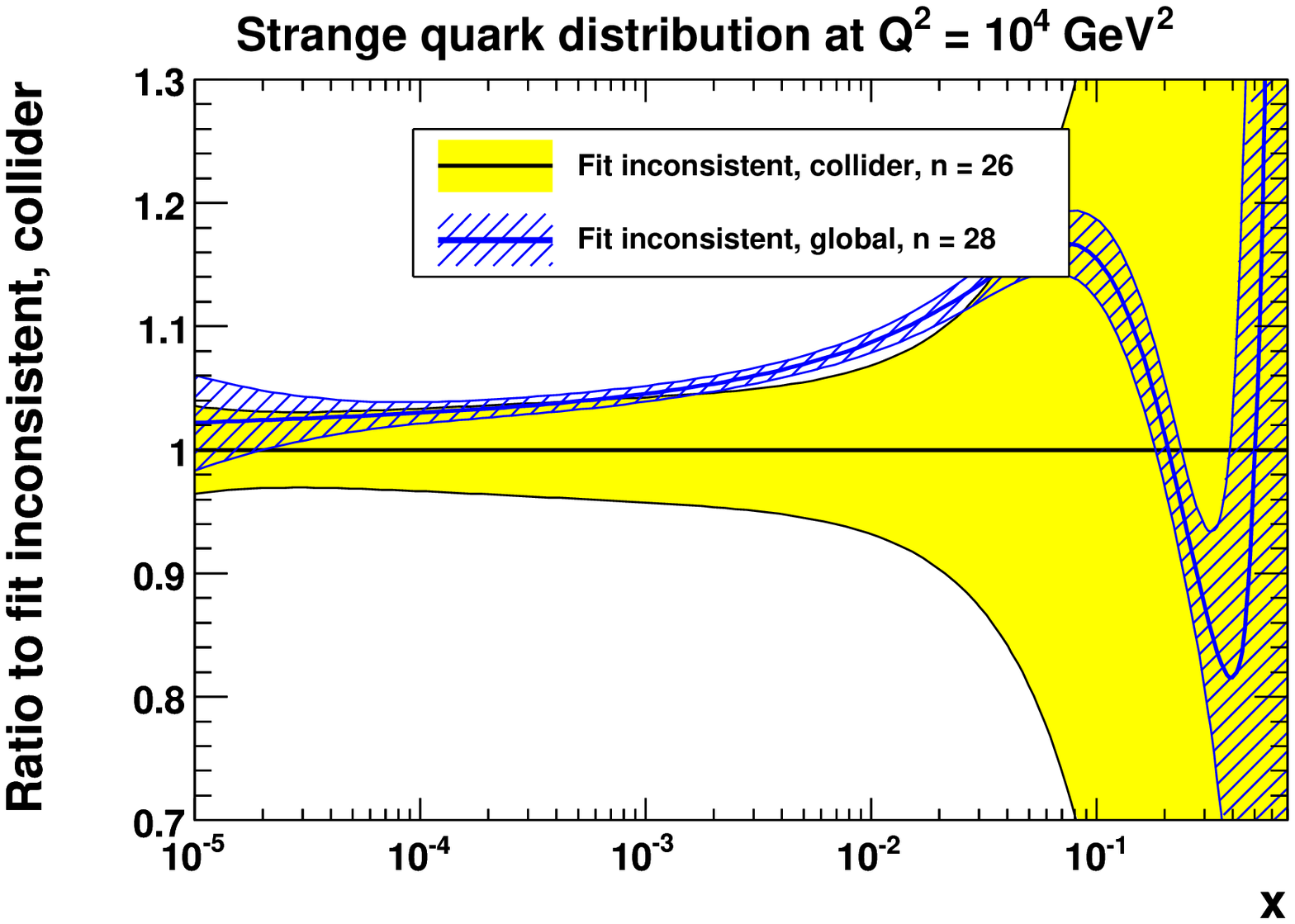}
  \end{minipage}%
  \begin{minipage}{0.5\textwidth}
    (f)\\
    \includegraphics[width=\textwidth]{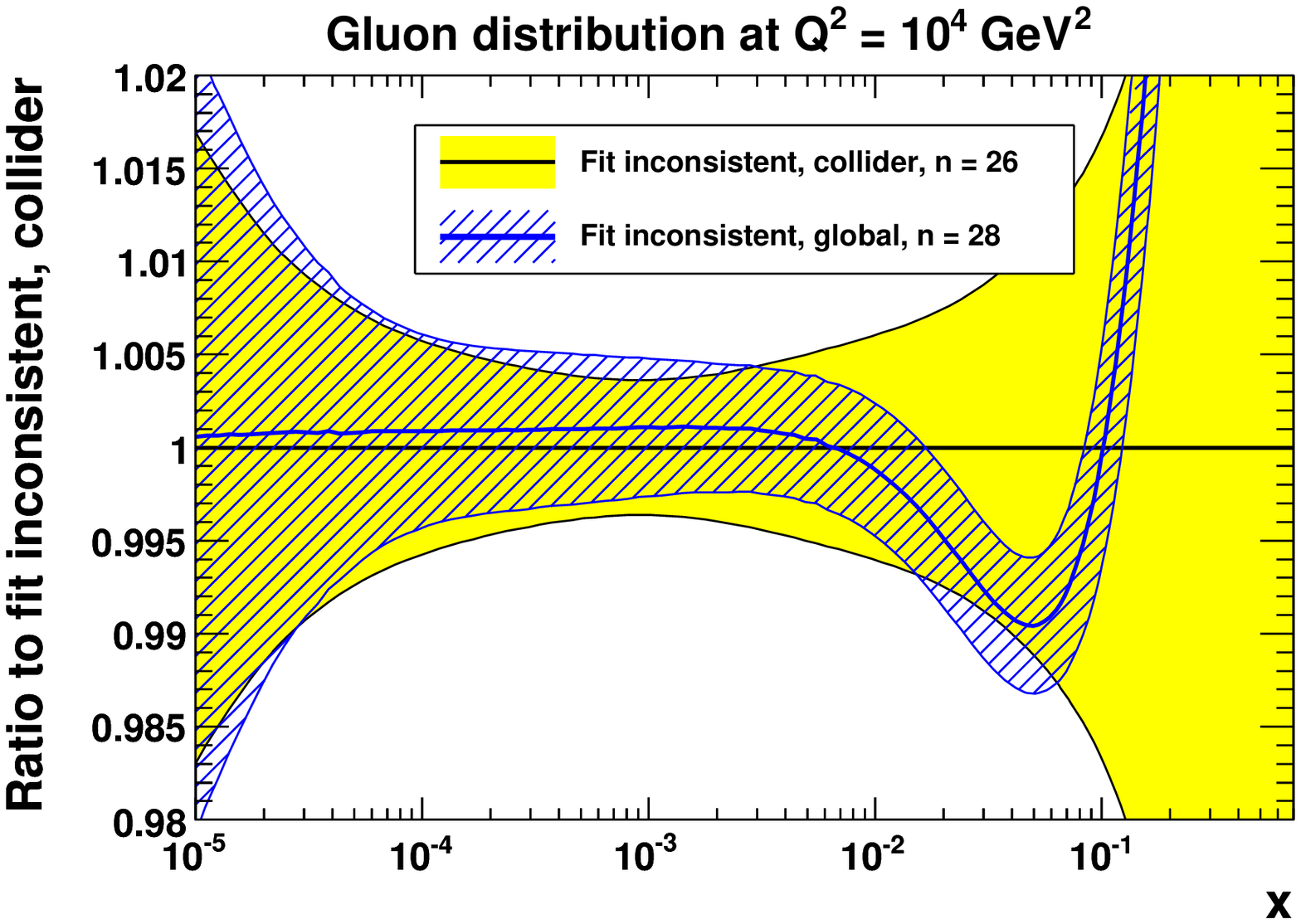}
  \end{minipage}
  \caption{Effect on PDFs of fitting inconsistent idealised pseudodata, either collider-only or global.}
  \label{fig:ratio_collider_inconsistent}
\end{figure}

In figures~\ref{fig:ratio_collider_theory} and \ref{fig:ratio_collider_inconsistent} we show the result of another study using the same consistent or inconsistent idealised pseudodata.  First we show the PDFs obtained from fitting only the collider (HERA+Tevatron) subset of the pseudodata, then we show the effect of adding the remaining fixed-target pseudodata.  In the ``theory'' case in figure~\ref{fig:ratio_collider_theory}, the fixed-target pseudodata are perfectly consistent with the collider pseudodata (by construction), so the global fit gives PDFs consistent with the collider fit, but with much smaller uncertainties.  This is not true for the ``inconsistent'' case in figure~\ref{fig:ratio_collider_inconsistent}, where the global fit gives PDFs often lying outside the uncertainty band for the collider fit.  The latter situation arises when fitting the genuine data in figure~\ref{fig:ratio_varyhera}, implying that the real collider data are inconsistent with the real fixed-target data.  Note that the peculiar behaviour at large $x$ in figures~\ref{fig:ratio_collider_theory}(c,d) and \ref{fig:ratio_collider_inconsistent}(c,d) is due to the antiquark distributions going negative in the collider fit at high $x$ where there is no data constraint.

The conclusion of these studies is that defining experimental uncertainties via $\Delta\chi^2_{\rm global}=1$ is overly optimistic for \emph{global} PDF analysis and that the more conservative ``dynamic'' tolerance~\cite{Martin:2009iq} based on a ``hypothesis-testing'' criterion~\cite{Collins:2001es} is much more appropriate.\footnote{A similar conclusion has been reached using very different arguments in ref.~\cite{Pumplin:2009sc}.}  As a final example of a situation where we believe it would make sense to introduce a tolerance to account for a potential discrepancy between data sets, consider the recent ATLAS determination~\cite{Aad:2012sb} of the ratio of the strange-to-down sea-quark distributions, $r_s(x,Q^2)\equiv 0.5(s+\bar{s})/\bar{d}$, from a fit to inclusive $W^\pm$ and $Z$ differential cross sections at the LHC, combined with inclusive DIS data from HERA.  This ratio took the surprising values of
\begin{equation*}
  r_s\left(x=0.023,Q_0^2=1.9~{\rm GeV}^2\right)=1.00^{+0.25}_{-0.28}\quad\text{and}\quad r_s\left(x=0.013,Q^2=M_Z^2\right)=1.00^{+0.09}_{-0.10},
\end{equation*}
where the $r_s$ uncertainty is dominated by the experimental PDF uncertainty, determined using $\Delta\chi^2=1$, of $\pm0.20$ and $\pm0.07$, respectively.  These values being consistent with unity indicate no strange suppression, contrary to previous determinations from CCFR/NuTeV dimuon cross sections ($\nu N\to \mu\mu X$), where the strange-quark distributions are suppressed to about half of the $\bar{d}$ and $\bar{u}$ distributions at the lower $Q^2$ value.  Even the HERA DIS data included in the ATLAS analysis~\cite{Aad:2012sb} shows some tension with the result of no strange suppression; the $\chi^2$ for the HERA DIS data increases by 2.9 units in going from fixed $r_s(x,Q_0^2)=0.5$ to free $r_s$ with two extra parameters.  The MSTW 2008 NNLO analysis~\cite{Martin:2009iq}, which included the CCFR/NuTeV dimuon cross sections, found central values and 68\% C.L.~PDF uncertainties (including the ``dynamic'' tolerance) of
\begin{equation*}
  r_s\left(x=0.023,Q^2=1.9~{\rm GeV}^2\right)=0.48\pm0.04\quad\text{and}\quad r_s\left(x=0.023,Q^2=M_Z^2\right)=0.79\pm0.02.
\end{equation*}
Rescaling the experimental PDF uncertainty of the ATLAS determination~\cite{Aad:2012sb} by a tolerance of $\approx 3$, corresponding to $\Delta\chi^2\approx 9$, would be enough to bring it into agreement with the MSTW08 result.  The conclusion that the uncertainty on $r_s$ in the ATLAS determination~\cite{Aad:2012sb} has been underestimated was also reached by the NNPDF Collaboration~\cite{Hartland:2012}.

\section{Random PDFs generated in space of fit parameters} \label{sec:random}
Given that we have now established that we need an appropriate tolerance, the question arises of how to include this into the MC method.  We can introduce a tolerance in the generation of the data replicas simply by rescaling all experimental errors in eqs.~\eqref{eq:MCgen} and \eqref{eq:MCgenQuad} by $\langle t\rangle\approx\langle T\rangle\approx 3$, corresponding to the average tolerance for 68\% C.L.~uncertainties.  We find that this simple approach, using $n=20$ input PDF parameters, reproduces the Hessian uncertainties with a dynamic tolerance surprisingly well for most parton flavours and kinematic regions.  However, it is not possible to implement exactly the ``dynamic'' tolerance (different for each eigenvector direction) in the MC method, since no reference is being made to the eigenvectors of the covariance matrix.

\begin{figure}
  \centering
  \begin{minipage}{0.5\textwidth}
    (a)\\
    \includegraphics[width=\textwidth]{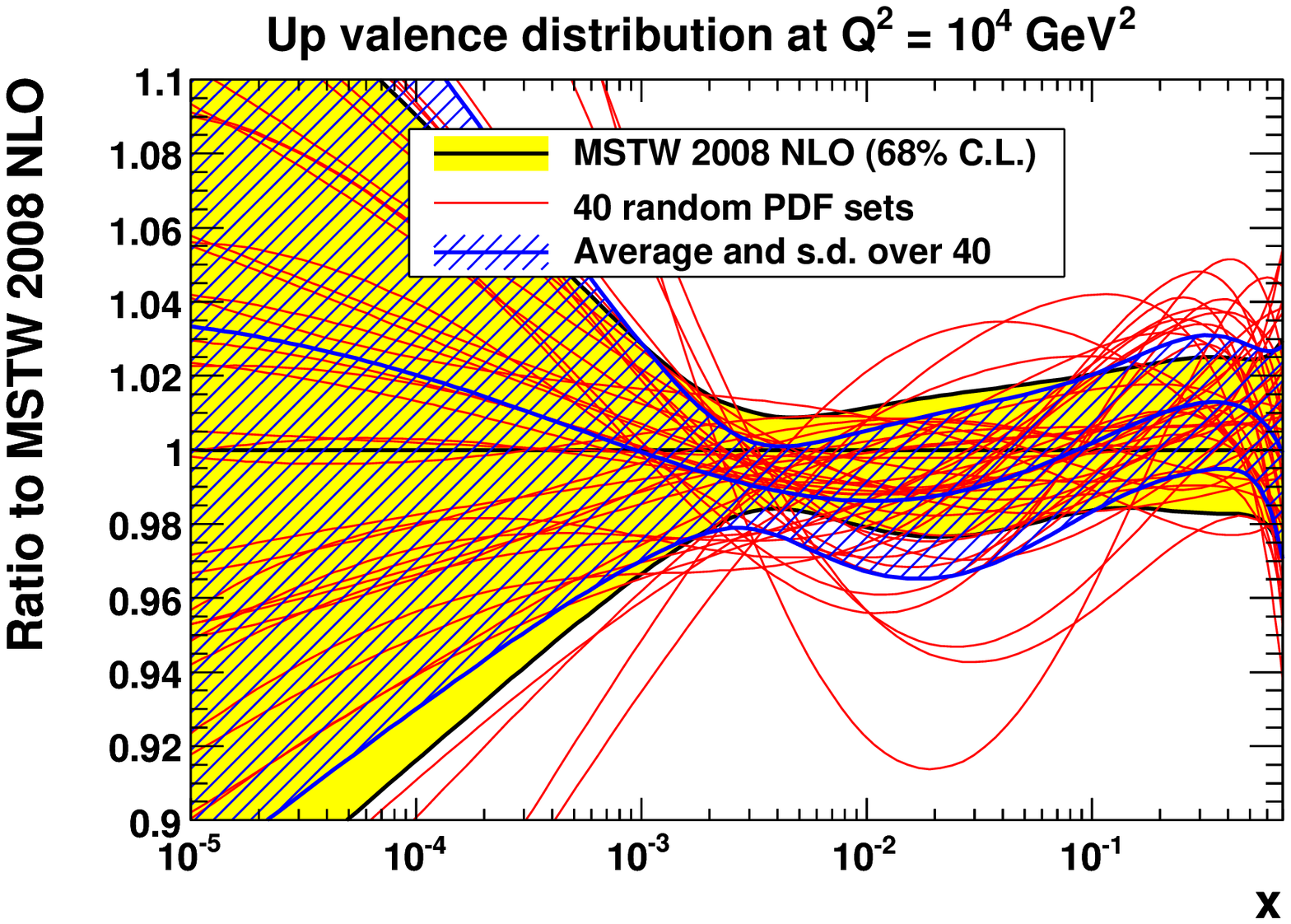}
  \end{minipage}%
  \begin{minipage}{0.5\textwidth}
    (b)\\
    \includegraphics[width=\textwidth]{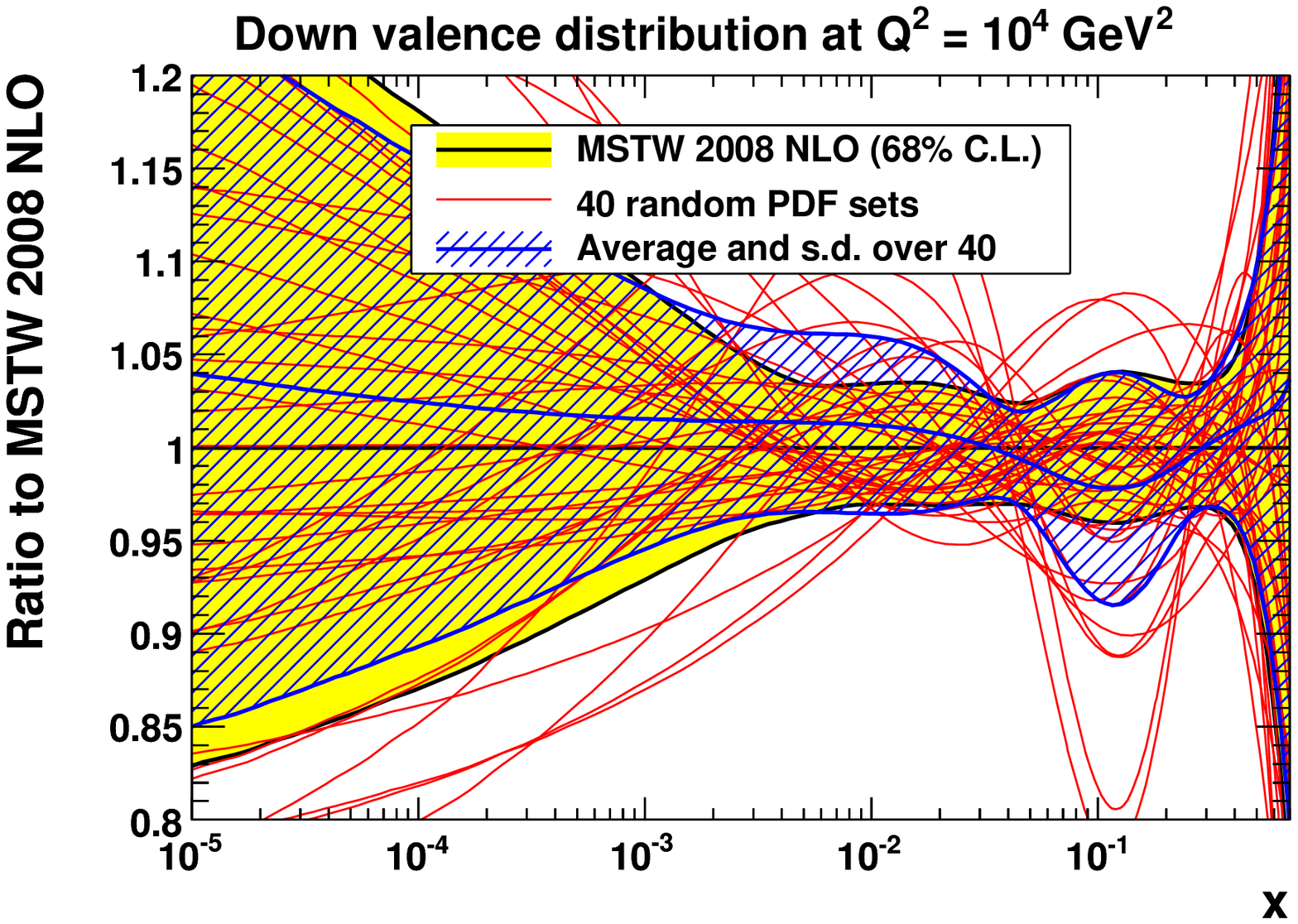}
  \end{minipage}
  \begin{minipage}{0.5\textwidth}
    (c)\\
    \includegraphics[width=\textwidth]{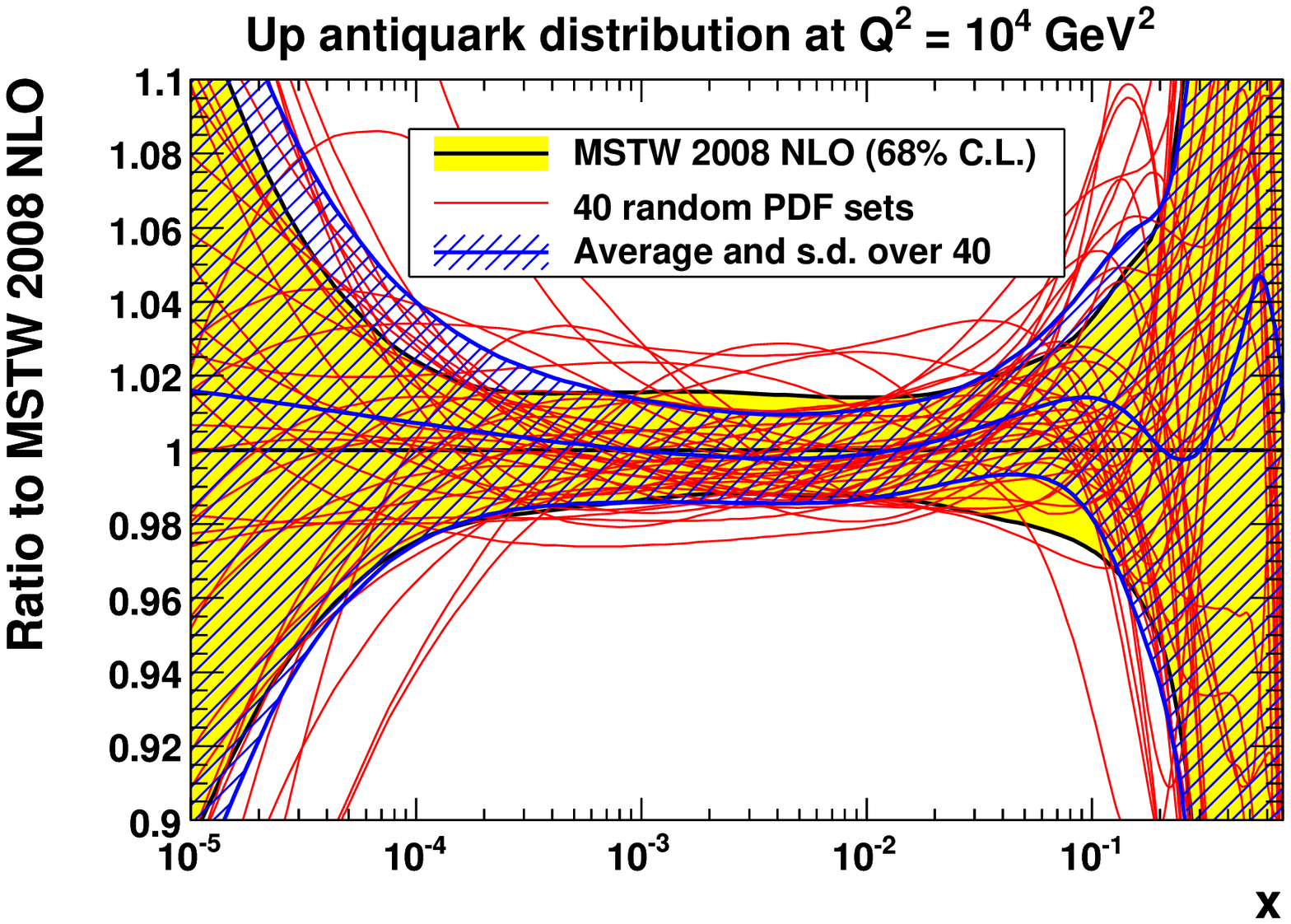}
  \end{minipage}%
  \begin{minipage}{0.5\textwidth}
    (d)\\
    \includegraphics[width=\textwidth]{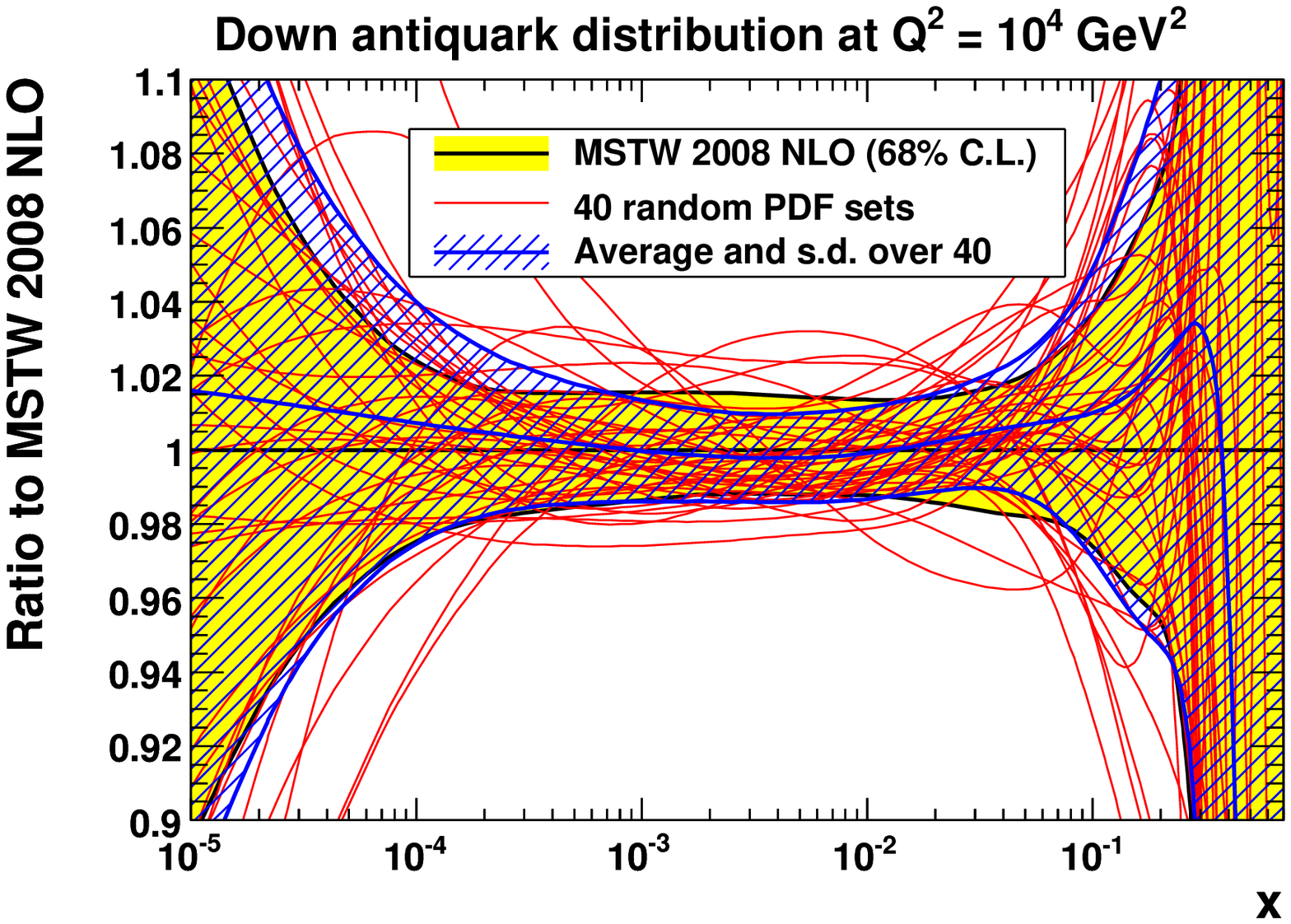}
  \end{minipage}
  \begin{minipage}{0.5\textwidth}
    (e)\\
    \includegraphics[width=\textwidth]{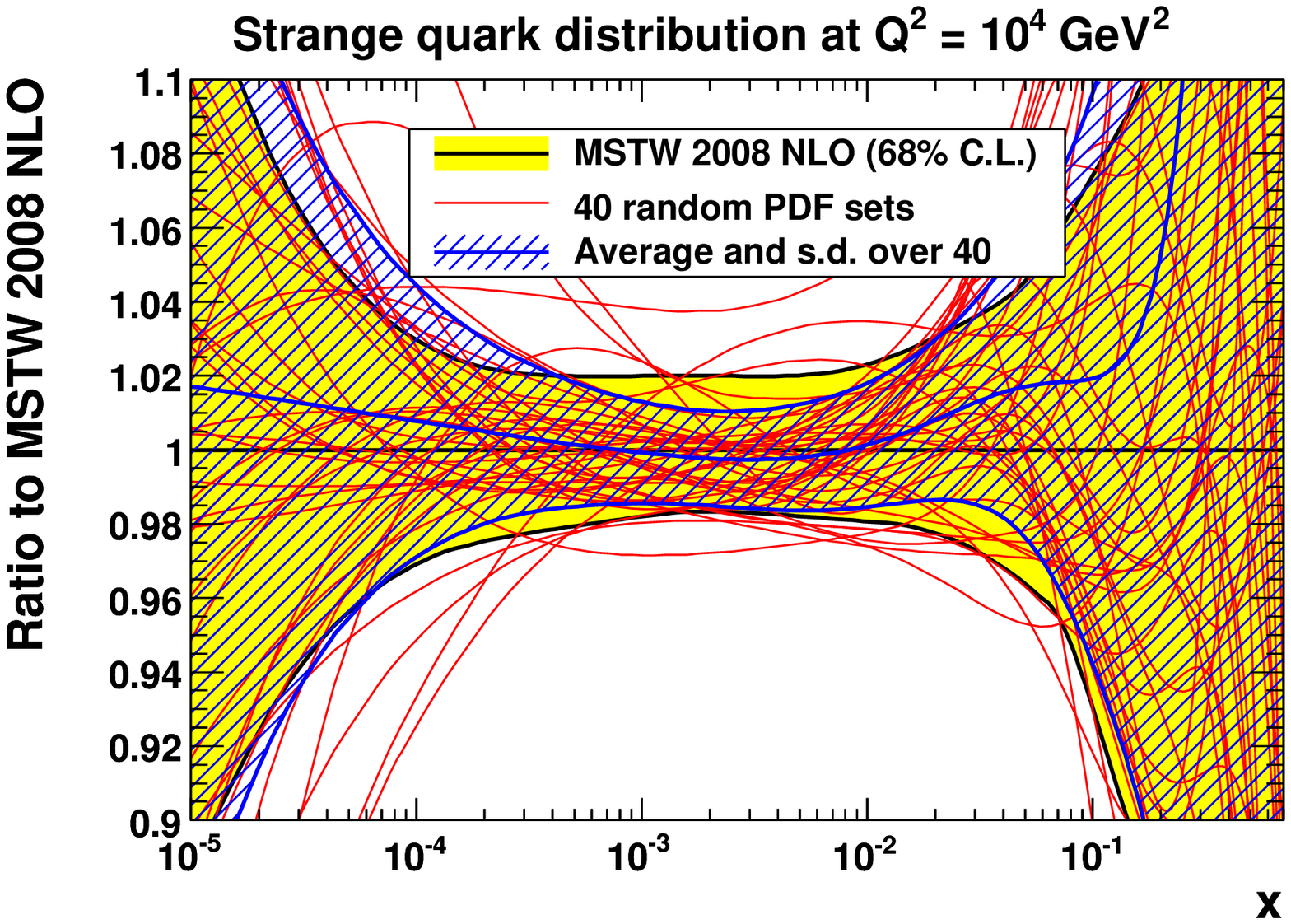}
  \end{minipage}%
  \begin{minipage}{0.5\textwidth}
    (f)\\
    \includegraphics[width=\textwidth]{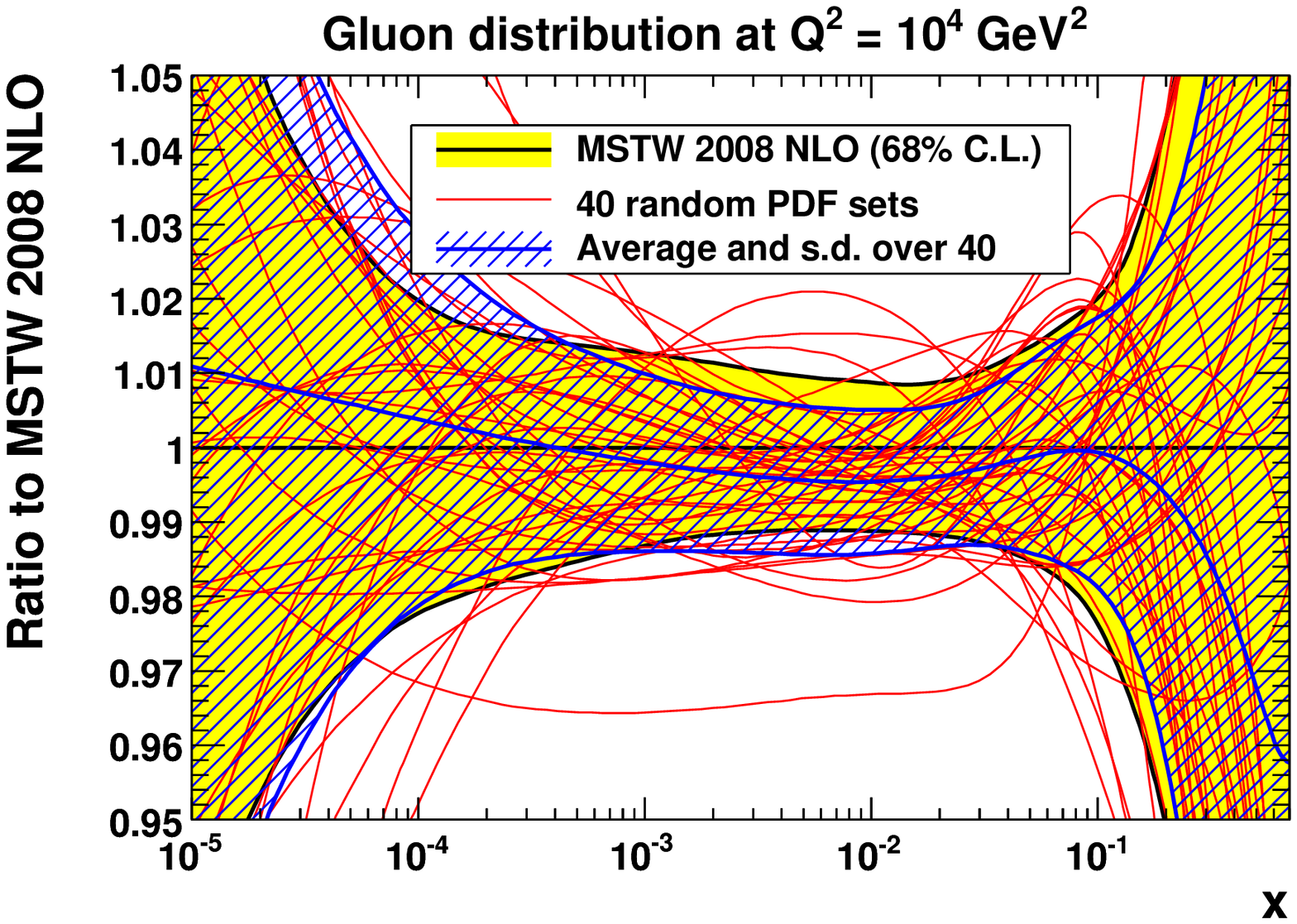}
  \end{minipage}
  \caption{$N_{\rm pdf} = 40$ random sets generated with eq.~\eqref{eq:random} as a ratio to the best-fit PDF set.}
  \label{fig:ratio_random68clAll}
\end{figure}

\begin{figure}
  \centering
  \begin{minipage}{0.5\textwidth}
    (a)\\
    \includegraphics[width=\textwidth]{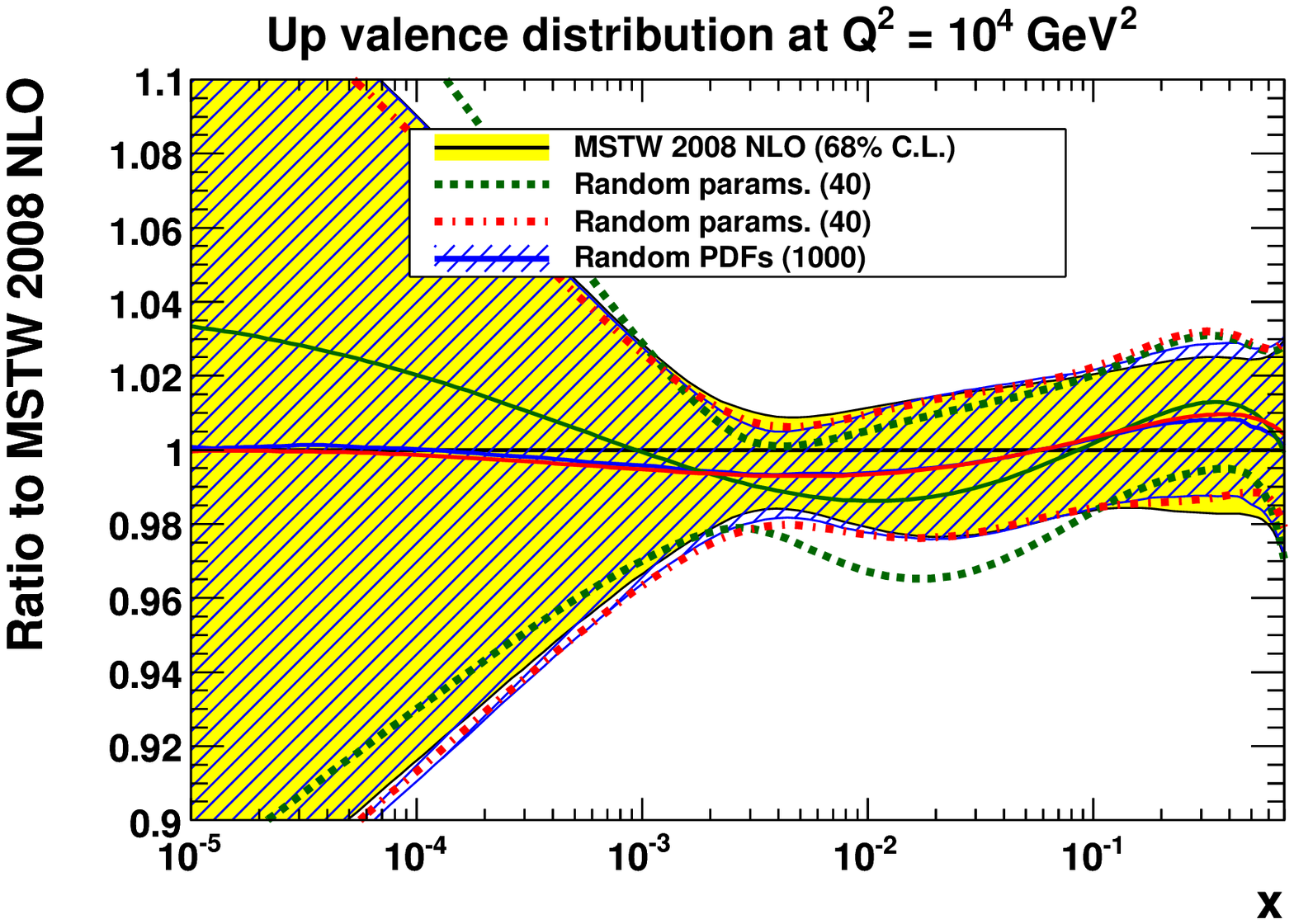}
  \end{minipage}%
  \begin{minipage}{0.5\textwidth}
    (b)\\
    \includegraphics[width=\textwidth]{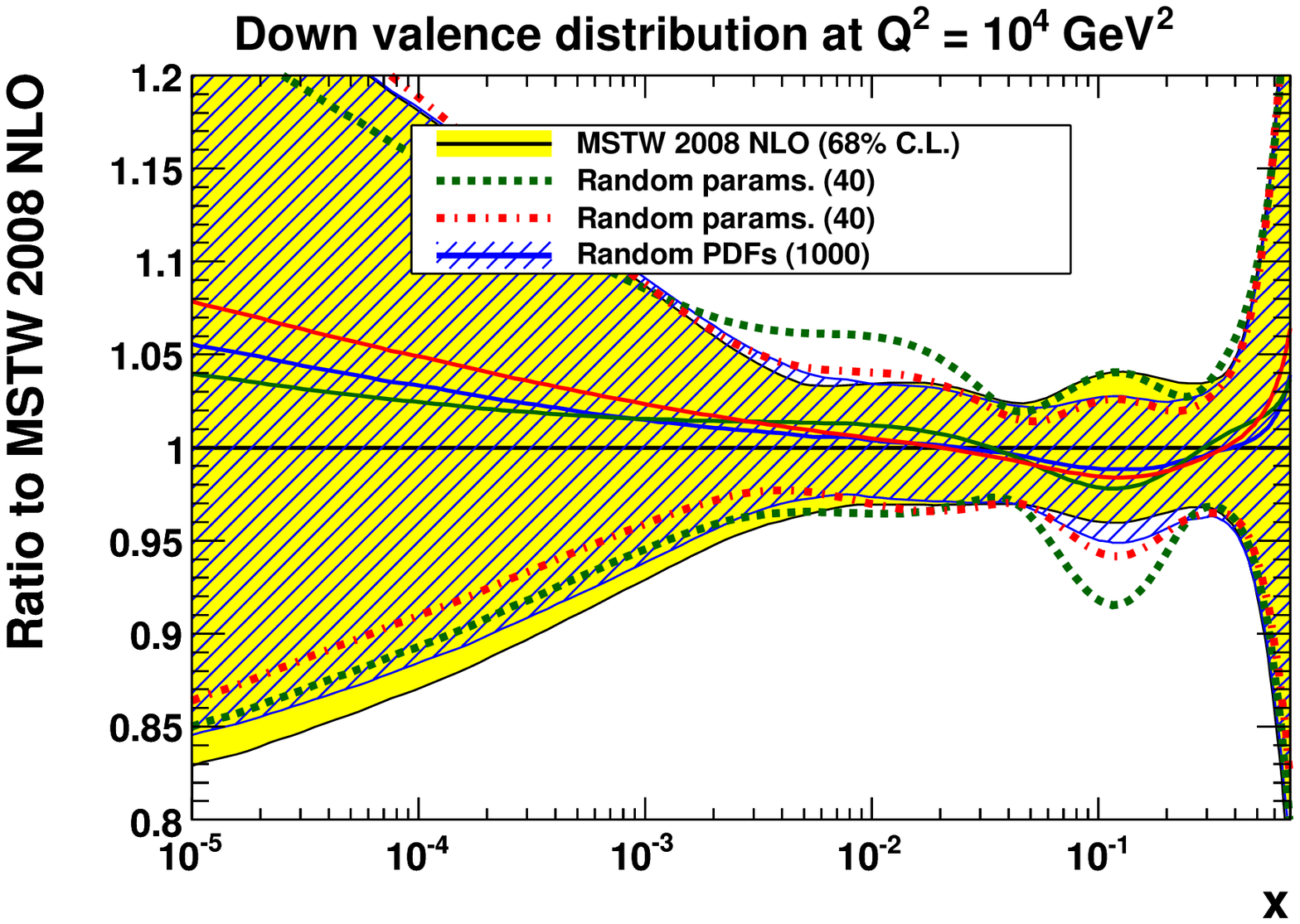}
  \end{minipage}
  \begin{minipage}{0.5\textwidth}
    (c)\\
    \includegraphics[width=\textwidth]{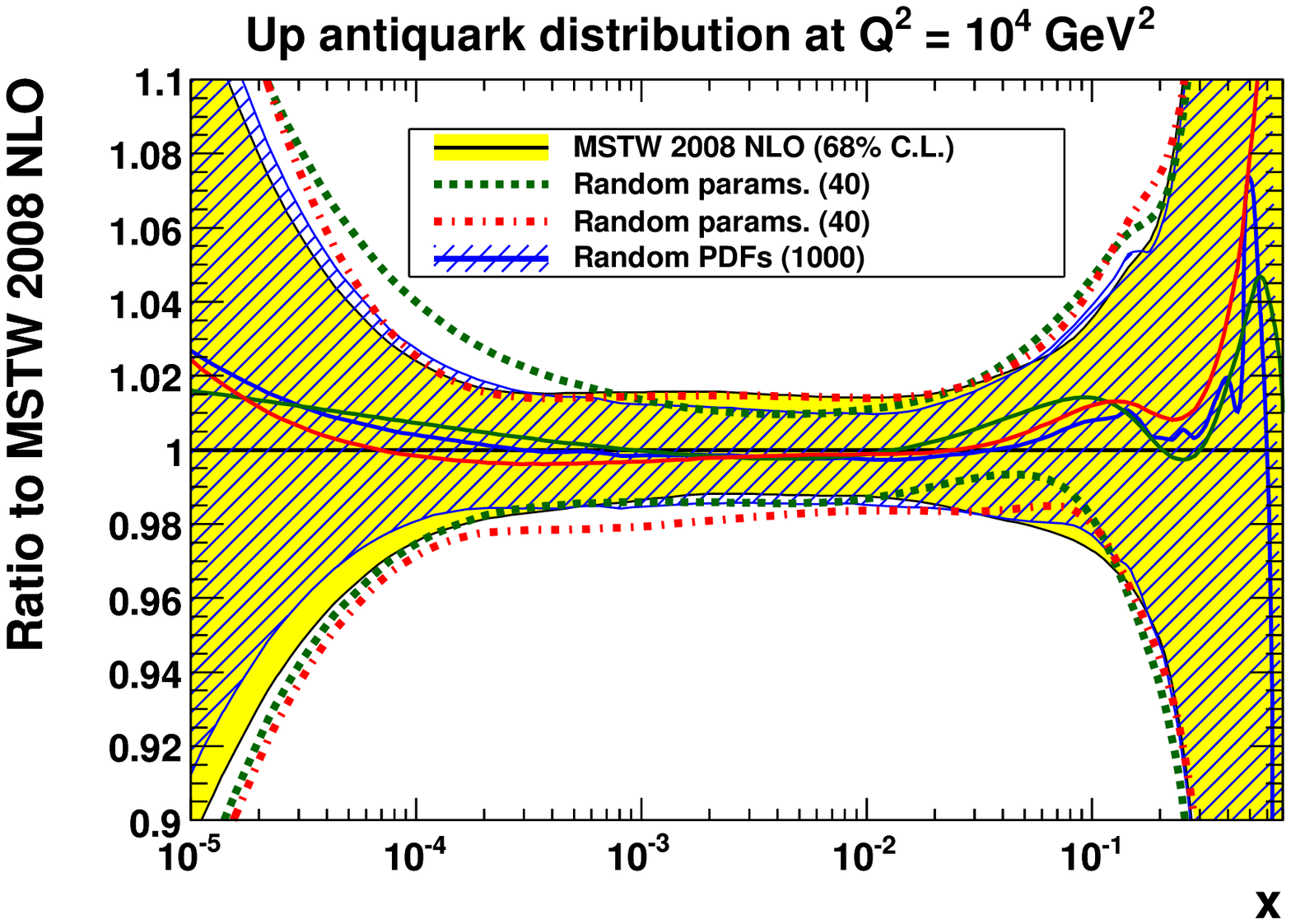}
  \end{minipage}%
  \begin{minipage}{0.5\textwidth}
    (d)\\
    \includegraphics[width=\textwidth]{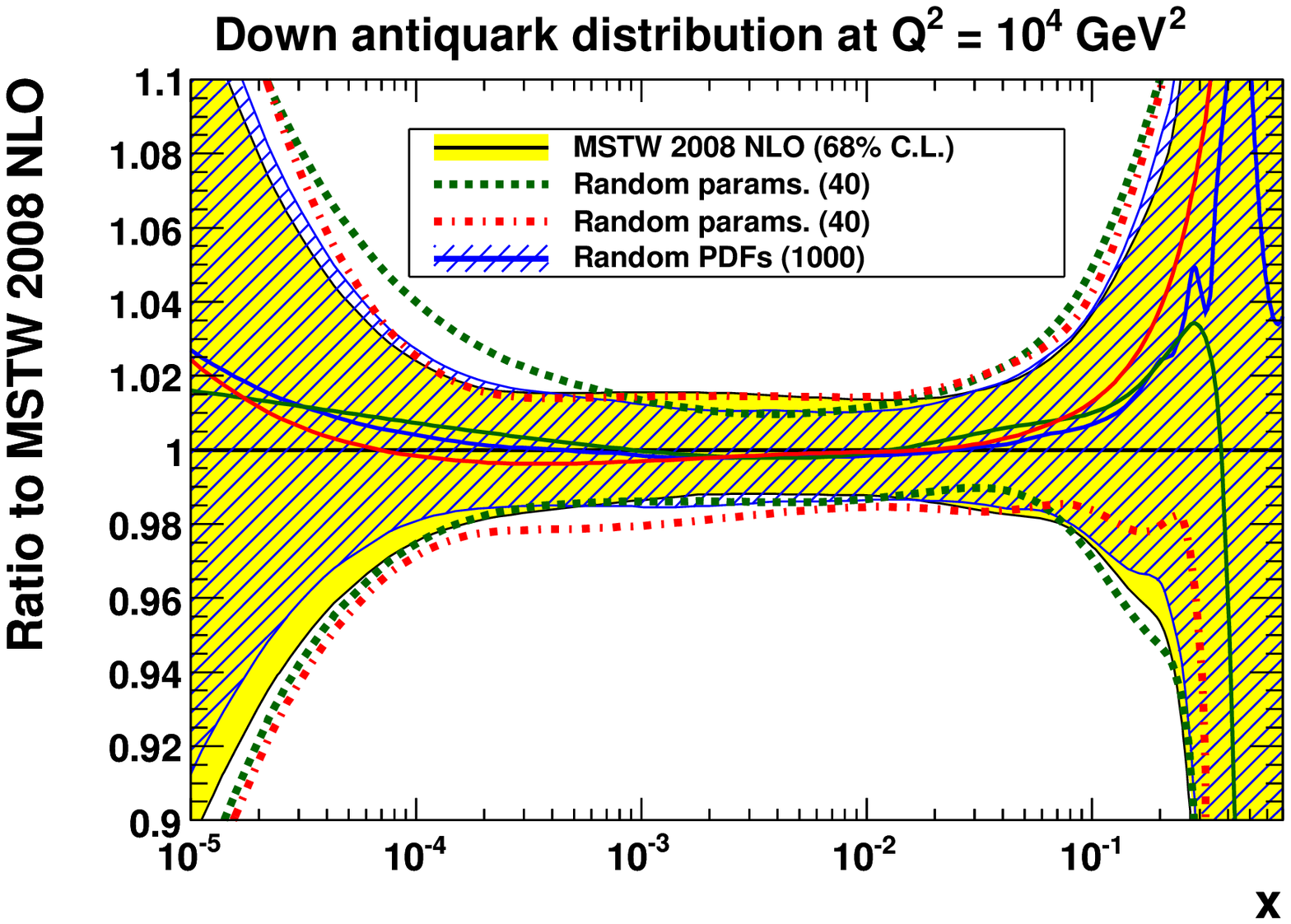}
  \end{minipage}
  \begin{minipage}{0.5\textwidth}
    (e)\\
    \includegraphics[width=\textwidth]{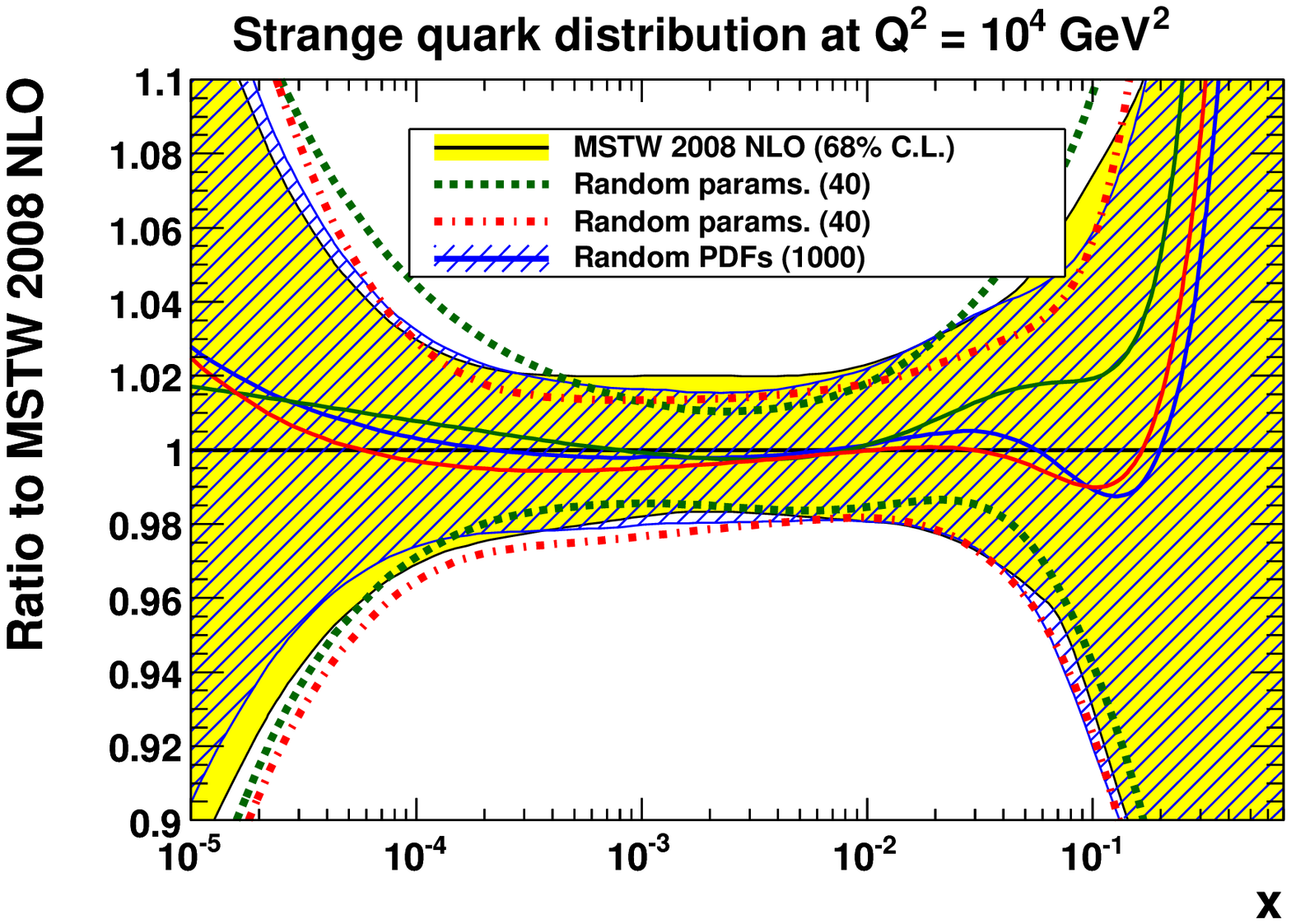}
  \end{minipage}%
  \begin{minipage}{0.5\textwidth}
    (f)\\
    \includegraphics[width=\textwidth]{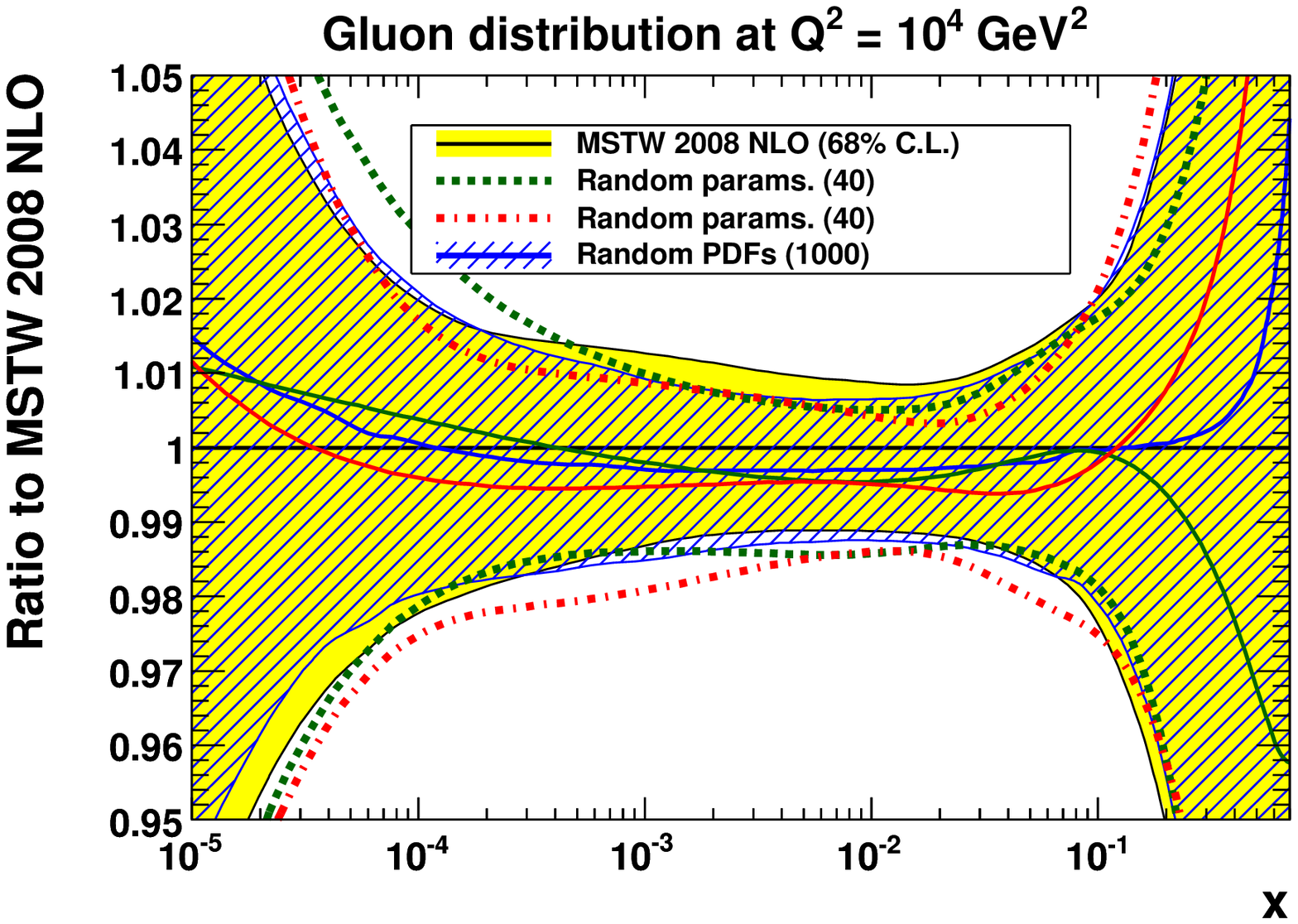}
  \end{minipage}
  \caption{Comparison of best-fit and Hessian uncertainty to the average and standard deviation of two sets of $N_{\rm pdf} = 40$ PDFs generated with different random parameters given by eq.~\eqref{eq:random} and one set of $N_{\rm pdf} = 1000$ random PDFs generated with eq.~\eqref{eq:randomPDF}.}
  \label{fig:ratio_random68cl}
\end{figure}

\begin{figure}
  \centering
  \begin{minipage}{0.5\textwidth}
    (a)\\
    \includegraphics[width=\textwidth]{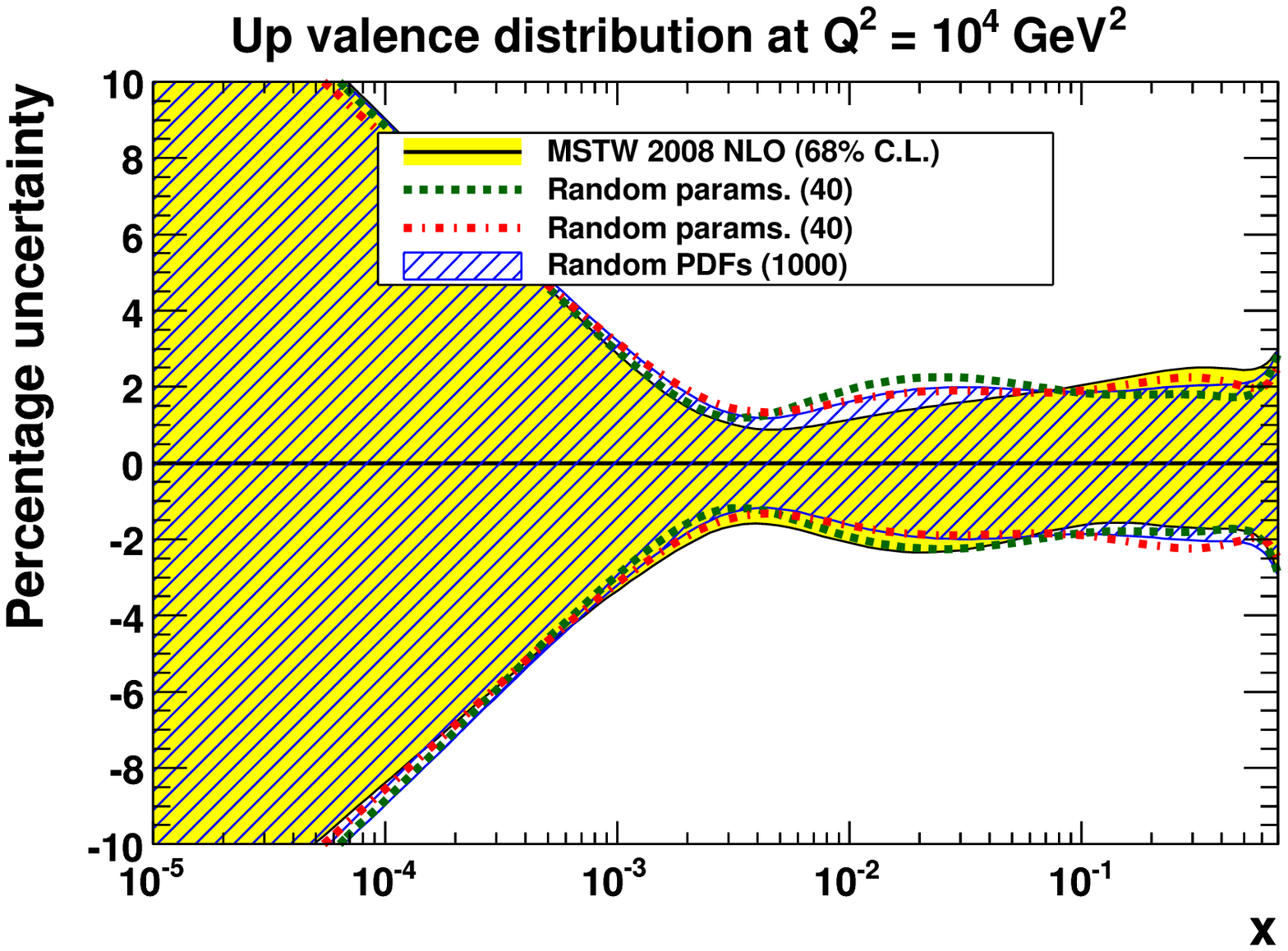}
  \end{minipage}%
  \begin{minipage}{0.5\textwidth}
    (b)\\
    \includegraphics[width=\textwidth]{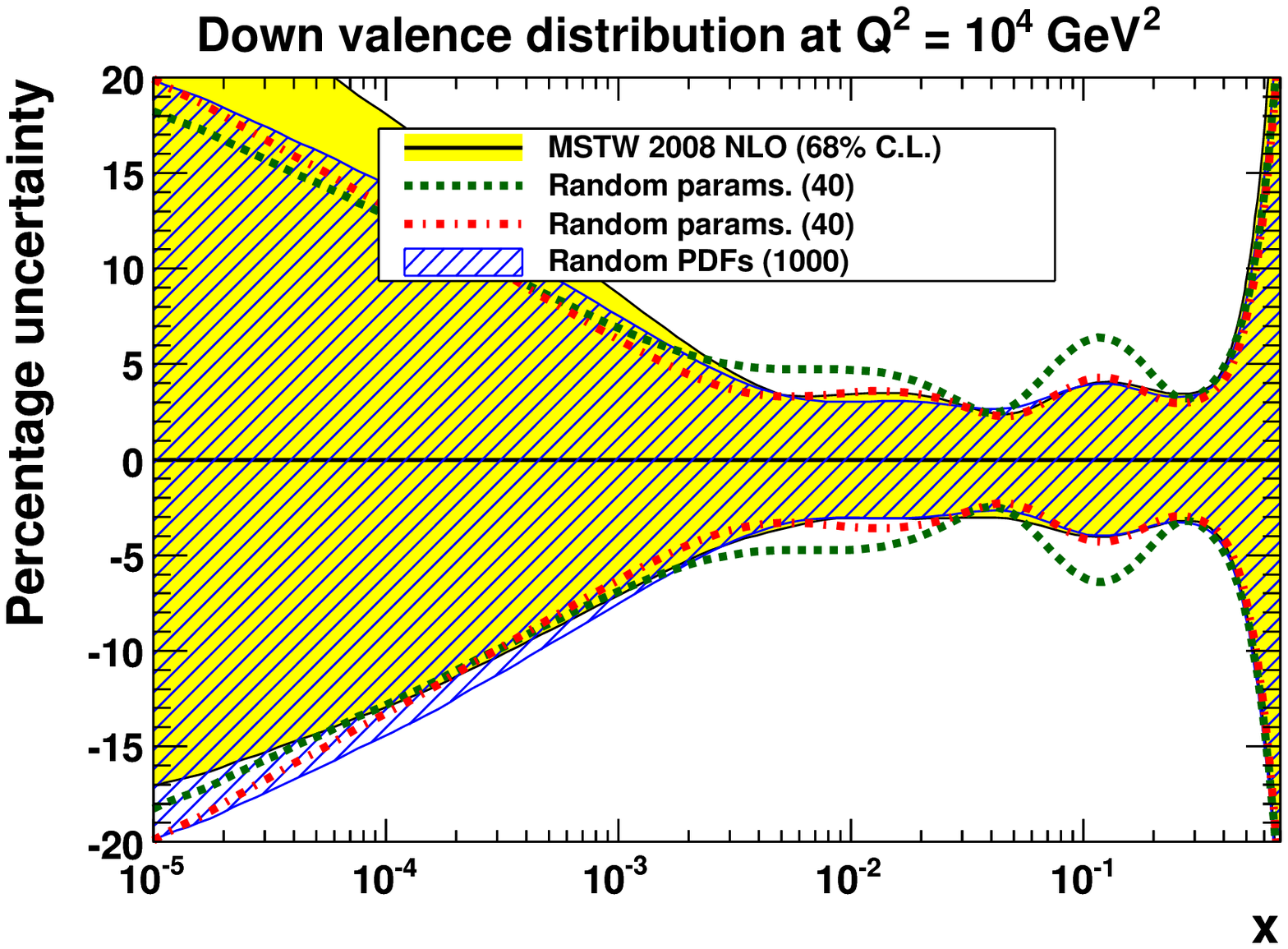}
  \end{minipage}
  \begin{minipage}{0.5\textwidth}
    (c)\\
    \includegraphics[width=\textwidth]{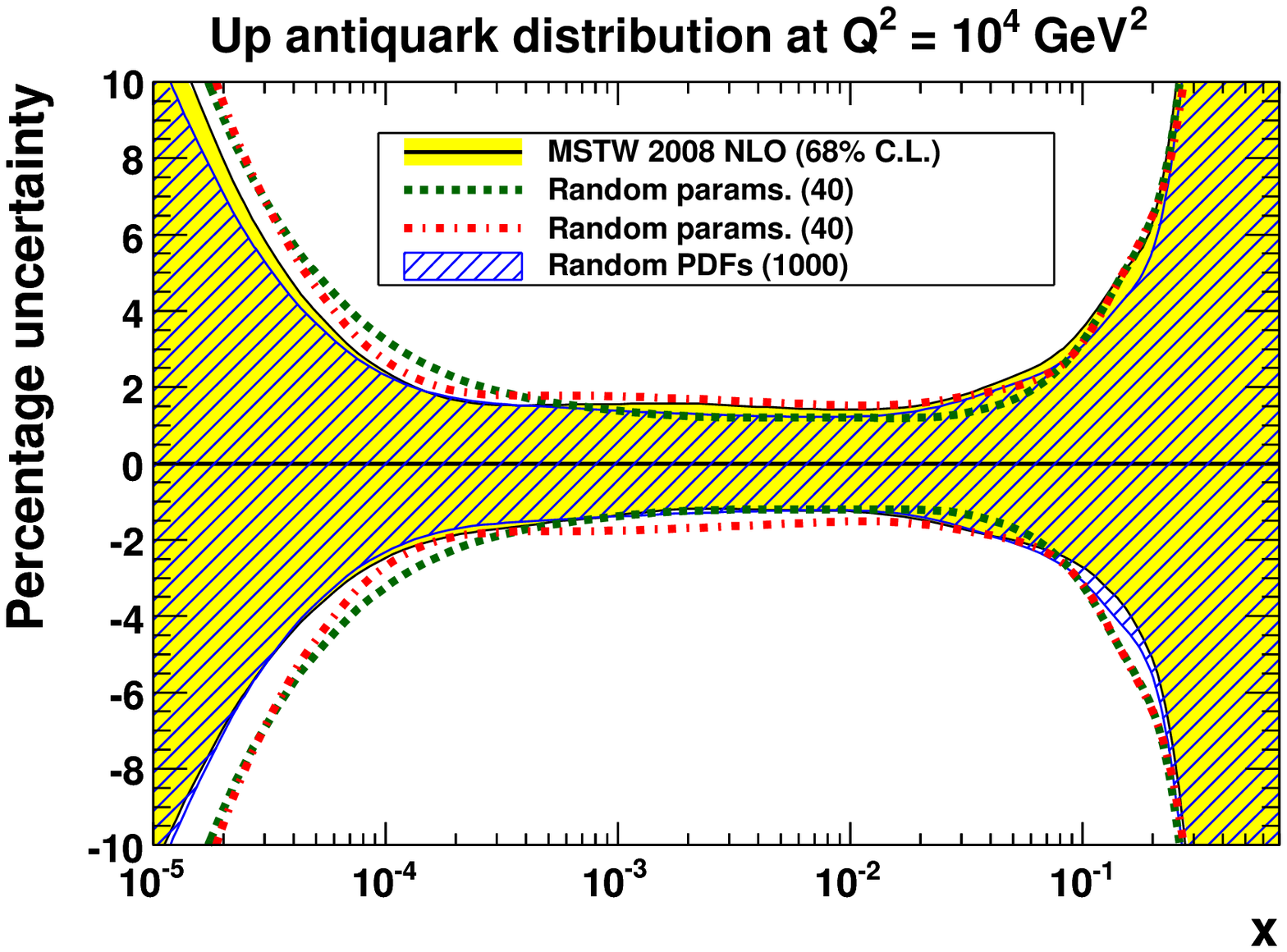}
  \end{minipage}%
  \begin{minipage}{0.5\textwidth}
    (d)\\
    \includegraphics[width=\textwidth]{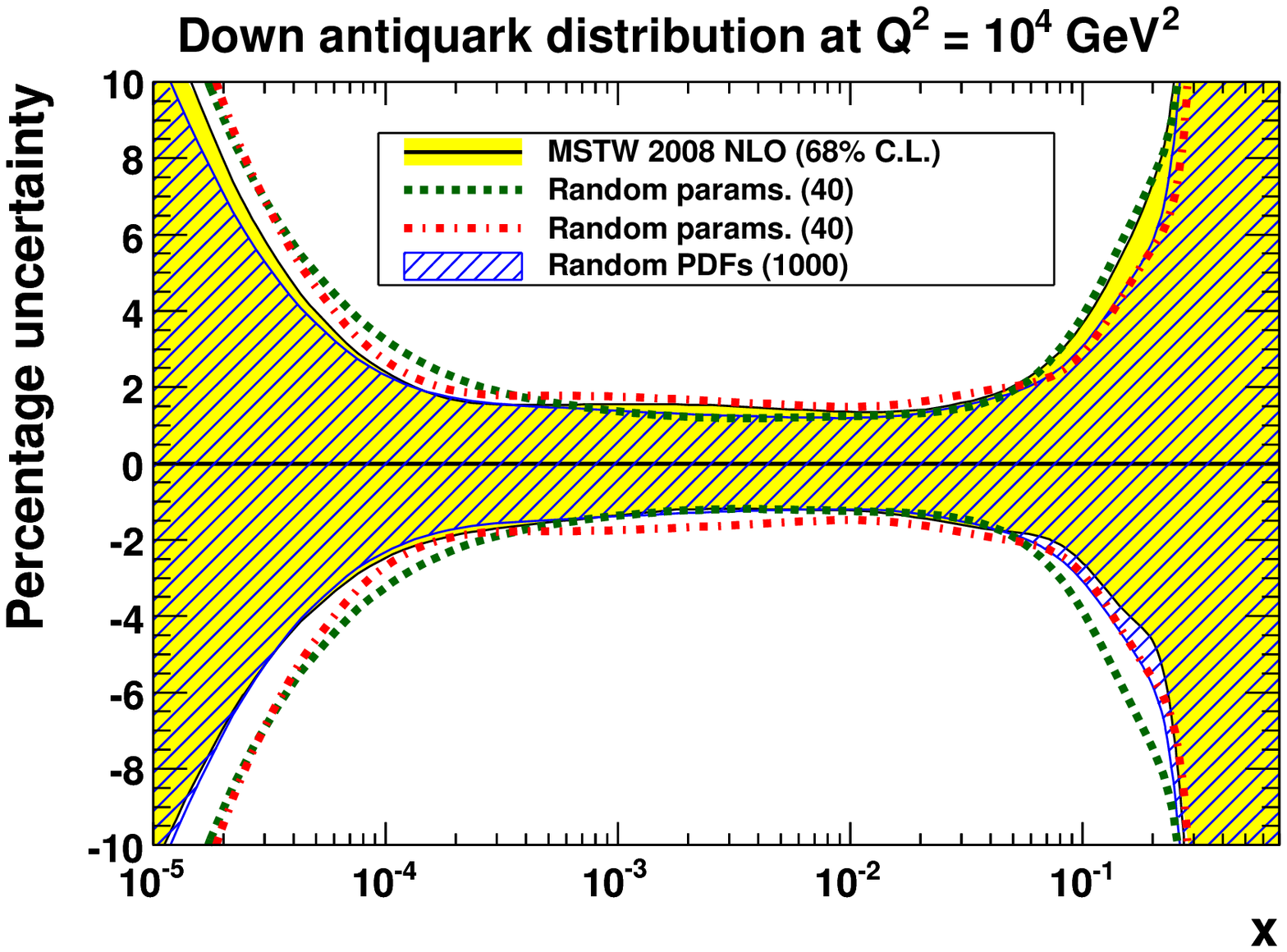}
  \end{minipage}
  \begin{minipage}{0.5\textwidth}
    (e)\\
    \includegraphics[width=\textwidth]{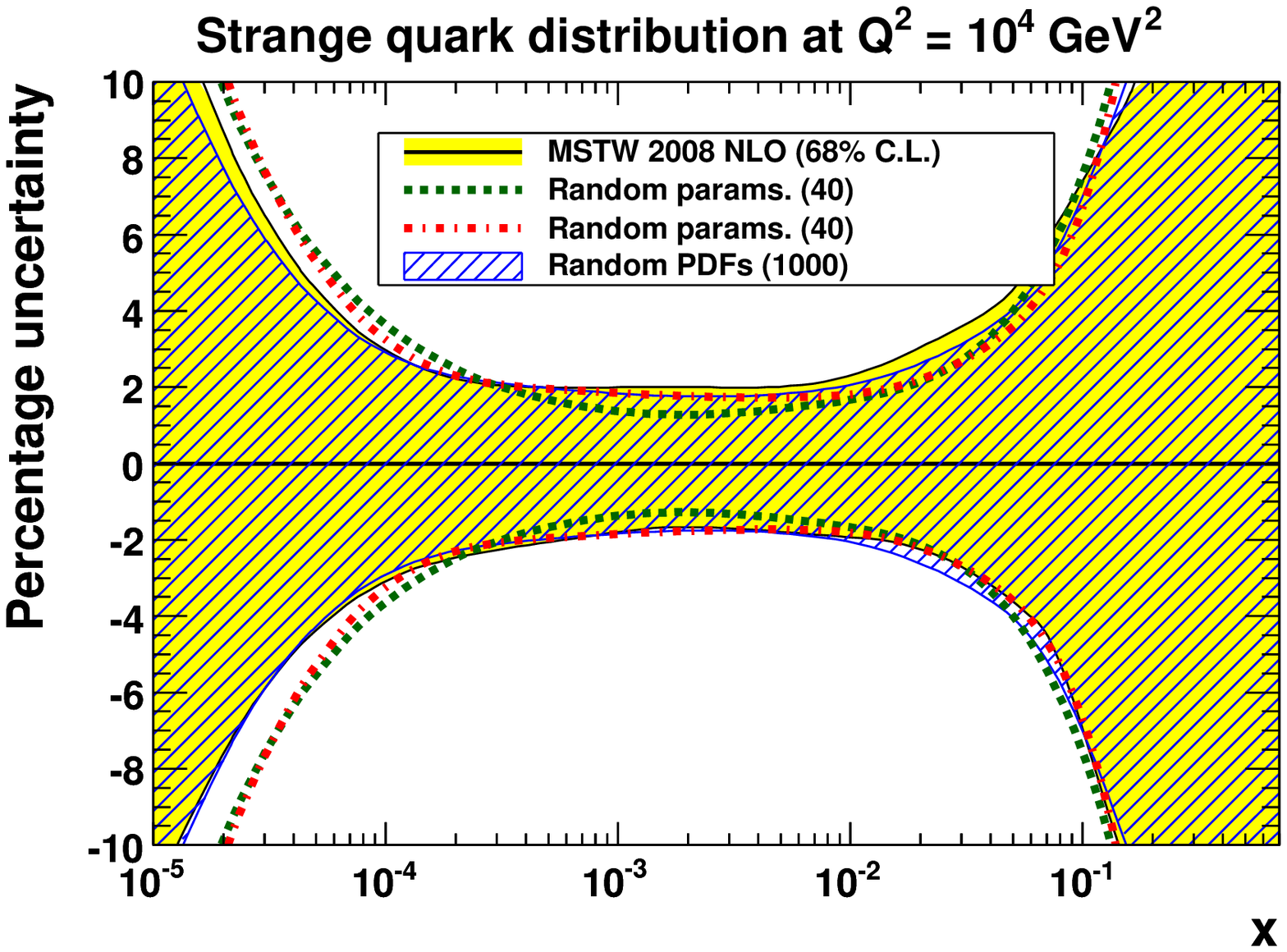}
  \end{minipage}%
  \begin{minipage}{0.5\textwidth}
    (f)\\
    \includegraphics[width=\textwidth]{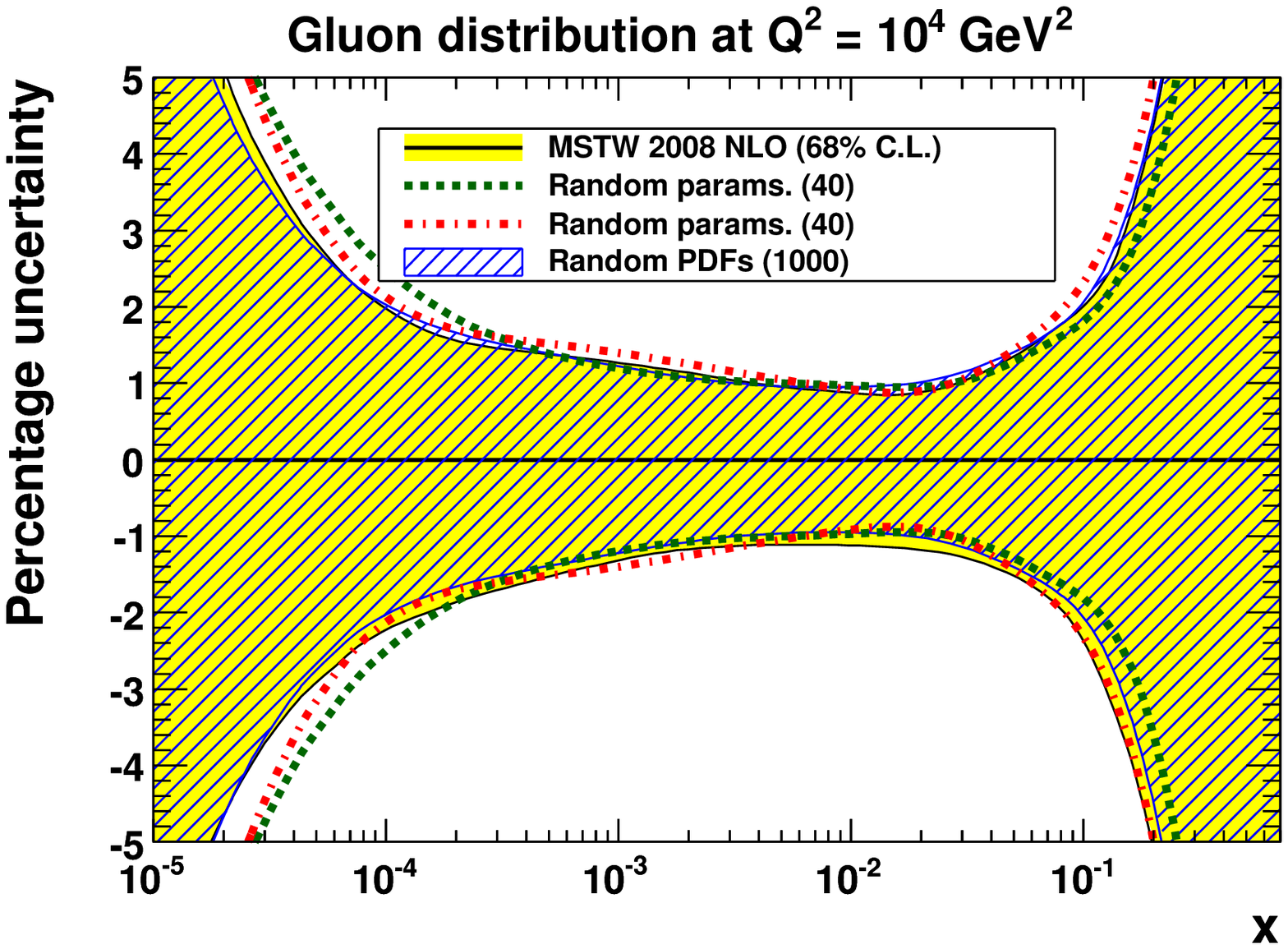}
  \end{minipage}
  \caption{Similar to figure~\ref{fig:ratio_random68cl} but percentage uncertainties rather than the ratio to the best-fit.}
  \label{fig:frac_random68cl}
\end{figure}

Instead of sampling the probability density by working in the space of data, we can produce random PDFs directly in the space of fit parameters.\footnote{We thank H.~Prosper for making this suggestion.}  In fact, this was done in the original work of Giele and Keller~\cite{Giele:1998gw} using the covariance matrix of parameters from Alekhin's fit~\cite{Alekhin:1996za}.  A convenient way to generate the random PDFs is to use the eigenvectors of the covariance matrix.  Recall from eq.~\eqref{eq:eigbasis} that the parameter displacements from the global minimum can be expanded in a basis of the rescaled eigenvectors $e_{ik}\equiv \sqrt{\lambda_k}\,v_{ik}$, that is,
\begin{equation} \label{eq:component}
  a_i - a_i^0 = \sum_{j=1}^n e_{ij}\,z_j,
\end{equation}
with $n=20$ the number of input PDF parameters.  Usually the $\pm k$th eigenvector PDF set is defined by taking $z_j = \left(\pm t_j^\pm\right)\delta_{jk}$ in eq.~(\ref{eq:component}), that is, the usual eigenvector PDF sets are generated with input parameters:
\begin{equation} \label{eq:eigensteptdynamic}
  a_i(S_k^\pm) = a_i^0 \pm t_k^\pm\,e_{ik}\qquad (k=1,\ldots,n),
\end{equation}
with $t_k^\pm$ adjusted to give the desired $T_k^\pm = (\Delta\chi^2_{\rm global})^{1/2}$.  However, we can instead randomly sample the parameter space such that the $k$th random PDF set is generated with input parameters obtained by taking $z_j = \left(\pm t_j^\pm\right)|R_{jk}|$ in eq.~(\ref{eq:component}), that is,
\begin{equation} \label{eq:random}
  a_i(\mathcal{S}_k) = a_i^0 + \sum_{j=1}^n e_{ij}\,\left(\pm t_j^\pm\right)\,|R_{jk}|\qquad (k=1,\ldots,N_{\rm pdf}),
\end{equation}
where $R_{jk}$ is a Gaussian-distributed random number of mean zero and variance one, and either $+t_j^+$ or $-t_j^-$ is chosen depending on the sign of $R_{jk}$.  There are therefore $n=20$ random numbers $R_{jk}$ ($j=1,\ldots,n$) associated with the $k$th random PDF set ($k=1,\ldots,N_{\rm pdf}$).  The number of random PDF sets $N_{\rm pdf}$ is arbitrary, but again we choose $N_{\rm pdf} = 40$ mostly for practical reasons.  Each random PDF set has equal probability defined by the covariance matrix of fit parameters, and therefore statistical quantities such as the mean and standard deviation can easily be calculated using formulae such as eqs.~\eqref{eq:MCav} and \eqref{eq:MCsd} with the obvious replacement $N_{\rm rep}\to N_{\rm pdf}$.  A comparison of the average and standard deviation of $N_{\rm pdf} = 40$ PDFs constructed with eq.~\eqref{eq:random} to the best-fit and Hessian uncertainty is made in figure~\ref{fig:ratio_random68clAll}.  There is generally good agreement, with some shift of the average compared to the best-fit that can be attributed mostly to asymmetric tolerance values ($t_j^+\ne t_j^-$).  We have verified this explanation by repeating the same studies without a tolerance ($T_j^\pm=1$).  Alternative ad hoc treatments of the asymmetric tolerance values are possible.  For example, if $t_j^+>t_j^-$ proportionally more random PDF sets could be produced for a ``$-$'' sign than for a ``$+$'' sign in eq.~\eqref{eq:random} so that the average would be closer to the best-fit, or one could simply symmetrise with the replacement $t_j^\pm\to (t_j^++t_j^-)/2$ in eq.~\eqref{eq:random}.  However, since the degree of asymmetry is generally small, we will not explore these possibilities in practice.
As some measure of the amount of statistical fluctuation, we produce another $N_{\rm pdf} = 40$ PDFs constructed with eq.~\eqref{eq:random} using different random numbers $R_{jk}$.  The results are shown in figures~\ref{fig:ratio_random68cl} and \ref{fig:frac_random68cl} and we conclude that $N_{\rm pdf} = 40$ is enough to avoid significant fluctuations, although there is some moderate variation due to the limited statistics (for example, in $d_v$ at $x\sim0.1$).

In principle, there is some amount of non-linearity in going from the input PDF parameters $a_i$ to the input PDFs $f(x,Q_0^2)$, then to the evolved PDFs $f(x,Q^2)$ and to physical observables $F$ calculated using these evolved PDFs (for example, hadronic cross sections with a quadratic PDF dependence).  However, we find that, in practice, the apparent degree of non-linearity is small, an assumption that is inherent in the Hessian method for propagating uncertainties.  Making this assumption of linearity, an alternative and simpler way to generate random PDFs is to work with the existing eigenvector PDF sets directly at the level of the quantity of interest $F$ such as the evolved PDF or the hadronic cross section.  Then we can build random values of $F$ according to\footnote{cf.~the studies of F.~De Lorenzi: see eq.~(3.1) of ref.~\cite{DeLorenzi:2010zt} or eq.~(6.1) of ref.~\cite{DeLorenzi:2011}.}
\begin{equation} \label{eq:randomPDF}
  F(\mathcal{S}_k) = F(S_0) + \sum_{j=1}^{n}\left[F(S_j^\pm)-F(S_0)\right]\,|R_{jk}|\qquad (k=1,\ldots,N_{\rm pdf}),
\end{equation}
where $S_j^+$ or $S_j^-$ is chosen depending on the sign of $R_{jk}$.  Note that for the case $F=a_i$ in eq.~\eqref{eq:randomPDF}, then $a_i(S_0)\equiv a_i^0$ and inserting $a_i(S_j^\pm)$ from eq.~\eqref{eq:eigensteptdynamic} then we recover eq.~\eqref{eq:random}.  This construction of a random $F(\mathcal{S}_k)$ using eq.~\eqref{eq:randomPDF} can be done ``on the fly'' for an almost arbitrarily large value of $N_{\rm pdf}$, after the initial computation of $F(S_0)$ and $F(S_j^\pm)$ $(j=1,\ldots,n)$ requiring only $2n+1$ ($=41$ for the MSTW 2008 PDFs) evaluations of $F$.  We choose $N_{\rm pdf}=1000$ for the results shown in figures~\ref{fig:ratio_random68cl} and \ref{fig:frac_random68cl}, although the results are similar with a much smaller value.  Here we take ``$F$'' in eq.~\eqref{eq:randomPDF} to be the evolved PDF at $Q=100$~GeV for the particular parton flavour shown in each plot, then we construct $N_{\rm pdf}=1000$ values of $F(\mathcal{S}_k)$ and take the average and standard deviation, finding good agreement with the best-fit and Hessian uncertainty.  Again, the slight shift of the average compared to the best-fit can be attributed mostly to asymmetric tolerance values, which we confirm by repeating the same exercise starting from eigenvector PDF sets generated with $\Delta\chi^2_{\rm global}=1$.  As already mentioned, ad hoc modifications to the procedure could be adopted to better account for asymmetric tolerance values, but we choose not to explore these possibilities in this work given the relatively small size of the effect.  For example, a symmetrised version of eq.~\eqref{eq:randomPDF} could be obtained using
\begin{equation} \label{eq:randomPDFsymm}
  F(\mathcal{S}_k) = F(S_0) + \frac{1}{2}\sum_{j=1}^{n}\left\lvert F(S_j^+)-F(S_j^-)\right\rvert\,R_{jk}\qquad (k=1,\ldots,N_{\rm pdf}),
\end{equation}
analogous to the symmetric formula for PDF uncertainties given in eq.~\eqref{eq:symmunc}.

We note that an unsuccessful attempt to generate random PDFs directly in the space of fit parameters was made in section~6.5 of ref.~\cite{Pumplin:2009nk}.  This attempt was flawed in that all random PDF sets were constructed with the unnecessary constraint of a fixed $\Delta\chi^2=100$, with the $n$ parameters distributed on the surface of an $n$-dimensional hypersphere using the eigenvectors as basis vectors, leading to an envelope of the random PDF sets covering a much smaller range than the usual Hessian uncertainty.  By contrast, if we generate random PDF sets according to eq.~\eqref{eq:random}, then the $\Delta\chi^2$, or equivalently $t_j^\pm$, is only used to define the distance along a particular eigenvector direction.  At a general point in parameter space, given by stepping along all eigenvector directions by a random amount, the $\Delta\chi^2$ is irrelevant and it can be very large.  It is not necessary or desirable that each random PDF set should have $\Delta\chi^2$ below a certain value.  A fixed $\Delta\chi^2$ will only be recovered in the specific (and very unlikely) case that $|R_{jk}| = \delta_{jk}$, then eq.~\eqref{eq:random} reduces to eq.~\eqref{eq:eigensteptdynamic}.

Another argument that a Monte Carlo approach in the space of fit parameters involves exploring a space too wide to be sampled efficiently with a small number of random PDFs was made in section~3.2.1 of ref.~\cite{Forte:2010dt}.  There it was argued that if the probability distribution for each parameter is given as a histogram with three bins, say the one-sigma region around the central value and the two outer regions, then na\"ively one might expect the need to randomly sample $3^n\gtrsim 3\times10^9$ PDF sets for $n=20$ free parameters.  However, the $n$ parameters are certainly not independent, and the complete correlation information is provided by the covariance matrix obtained from the global fit.  Working in the basis of eigenvectors then provides an optimally efficient way to sample the parameter space randomly along each eigenvector direction.

\begin{figure}
  \centering
  \begin{minipage}{0.5\textwidth}
    (a)\\
    \includegraphics[width=\textwidth]{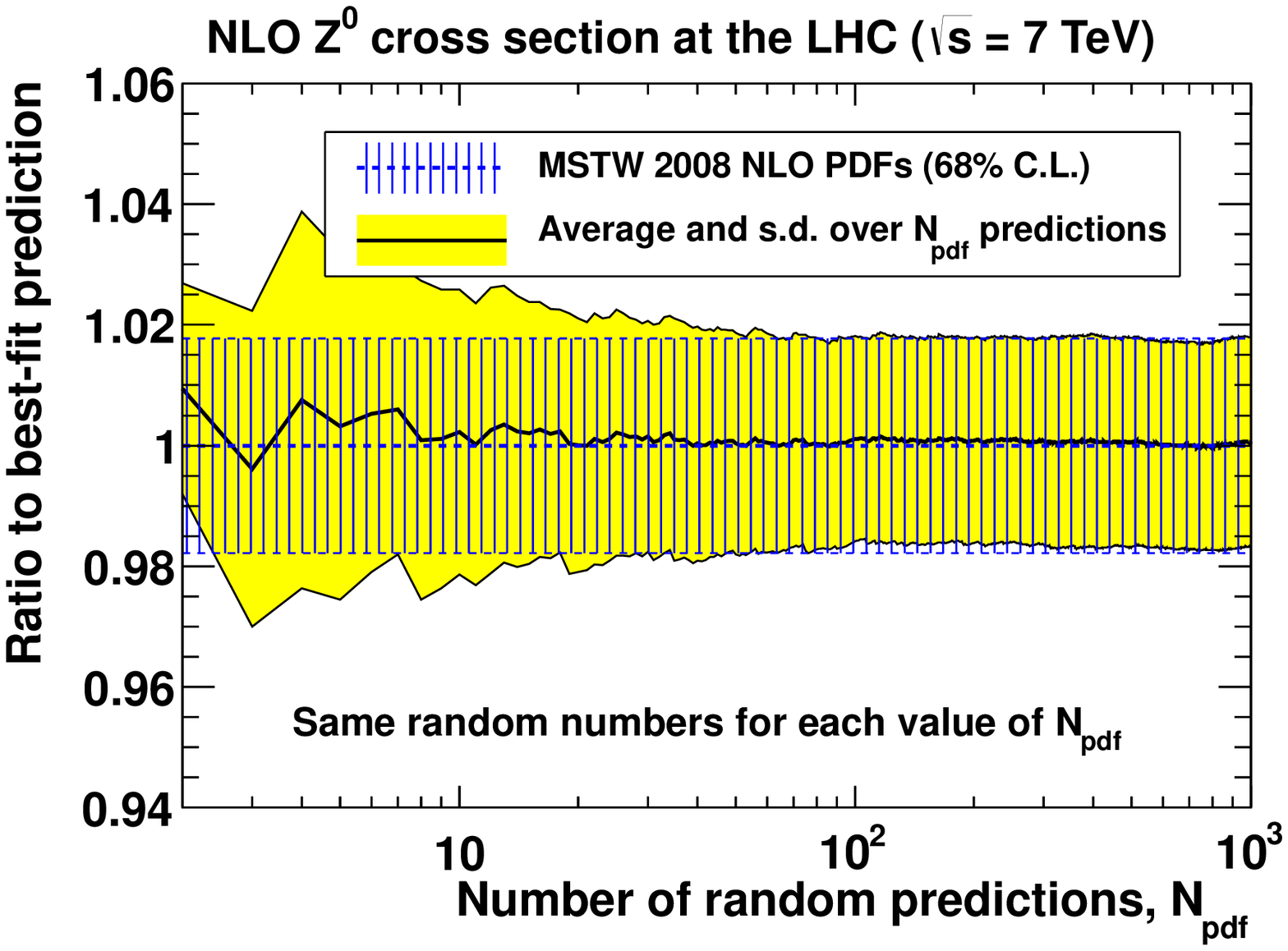}
  \end{minipage}%
  \begin{minipage}{0.5\textwidth}
    (b)\\
    \includegraphics[width=\textwidth]{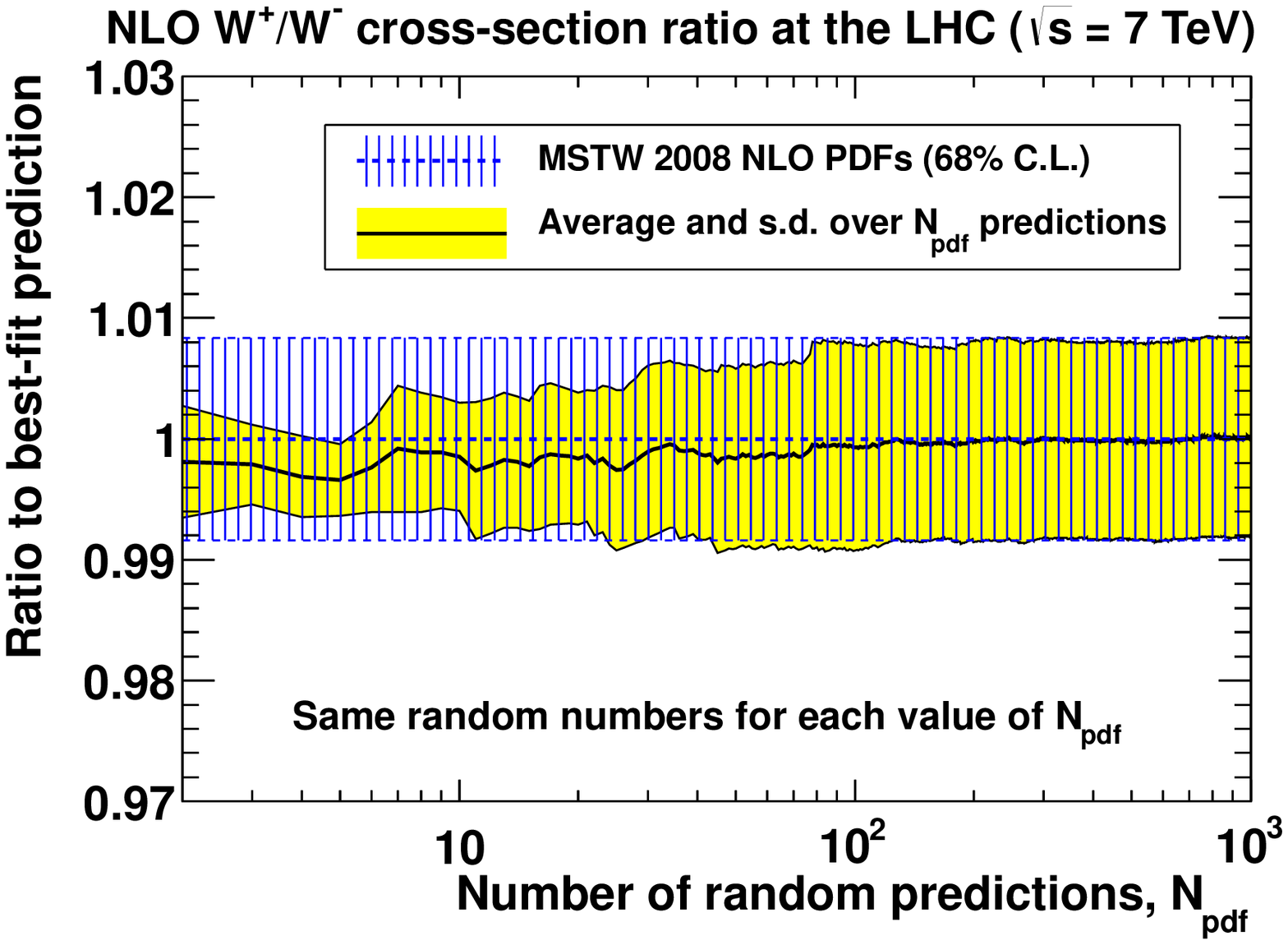}
  \end{minipage}
  \begin{minipage}{0.5\textwidth}
    (c)\\
    \includegraphics[width=\textwidth]{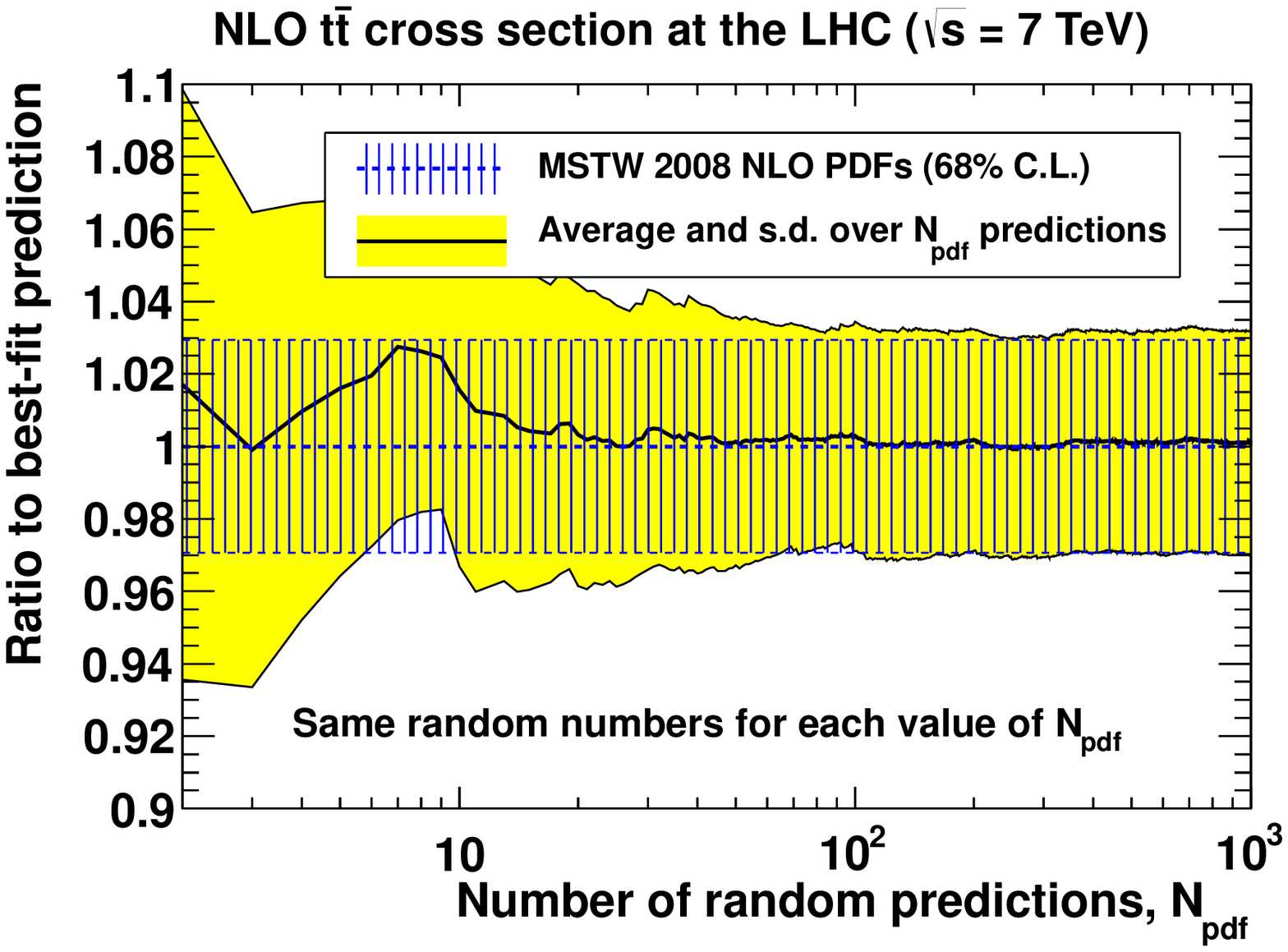}
  \end{minipage}%
  \begin{minipage}{0.5\textwidth}
    (d)\\
    \includegraphics[width=\textwidth]{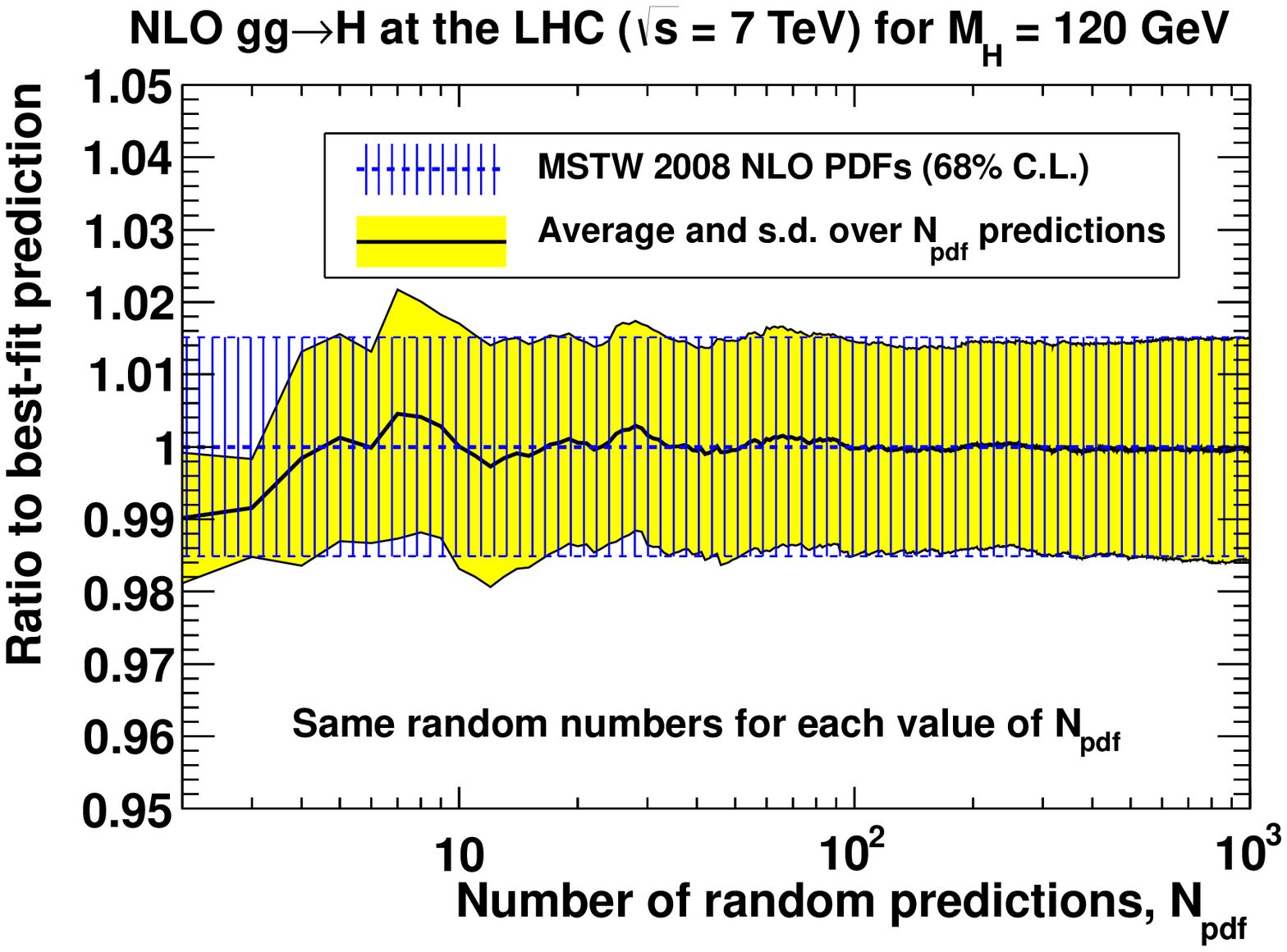}
  \end{minipage}
  \caption{Convergence of average and standard deviation of $N_{\rm pdf}$ random predictions as a function of $N_{\rm pdf}$, each time adding one more random prediction to the $N_{\rm pdf}-1$ previous random predictions, normalised to the best-fit prediction and compared to the Hessian uncertainty.}
  \label{fig:conv_samerand}
\end{figure}
\begin{figure}
  \centering
  \begin{minipage}{0.5\textwidth}
    (a)\\
    \includegraphics[width=\textwidth]{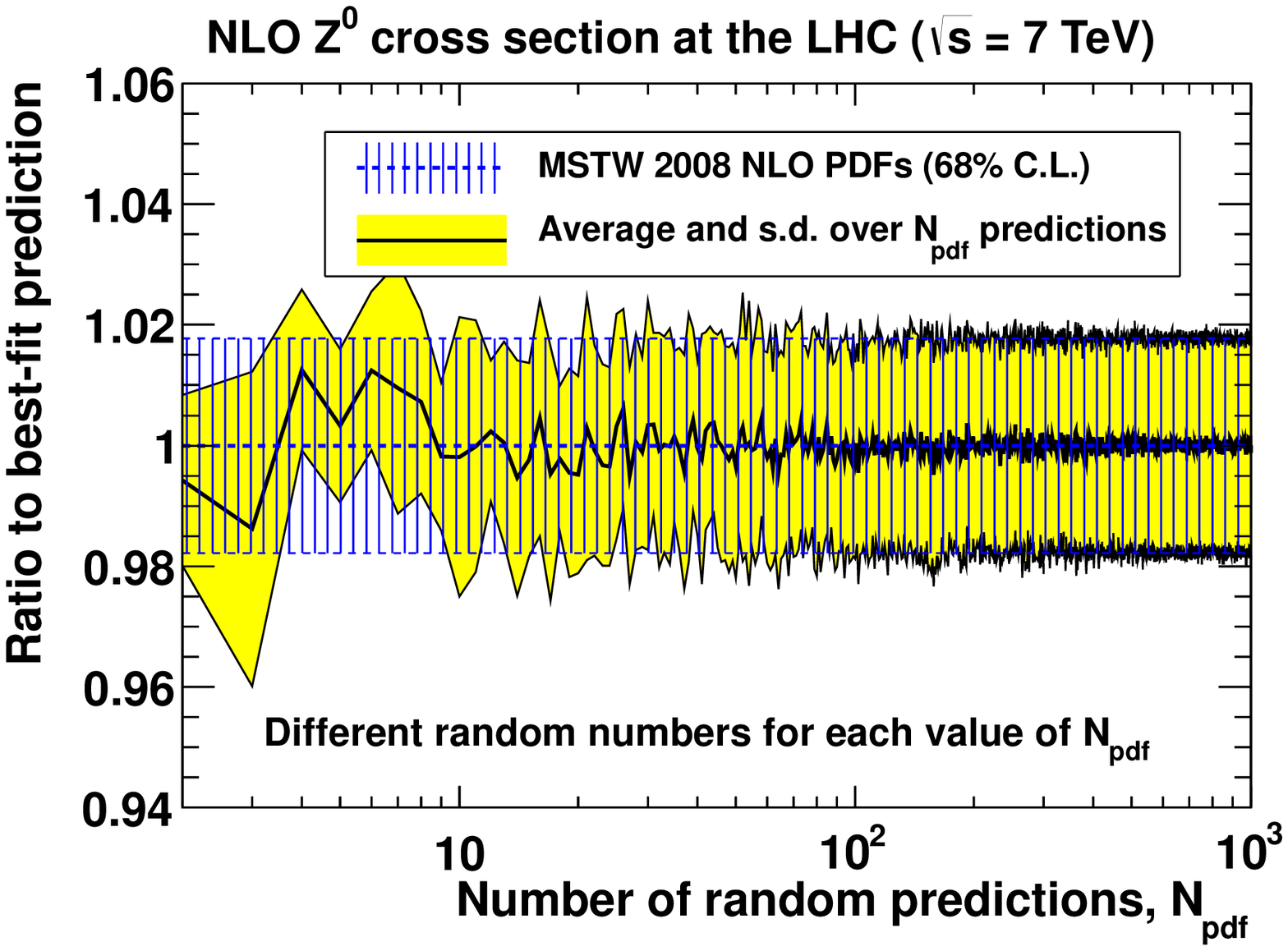}
  \end{minipage}%
  \begin{minipage}{0.5\textwidth}
    (b)\\
    \includegraphics[width=\textwidth]{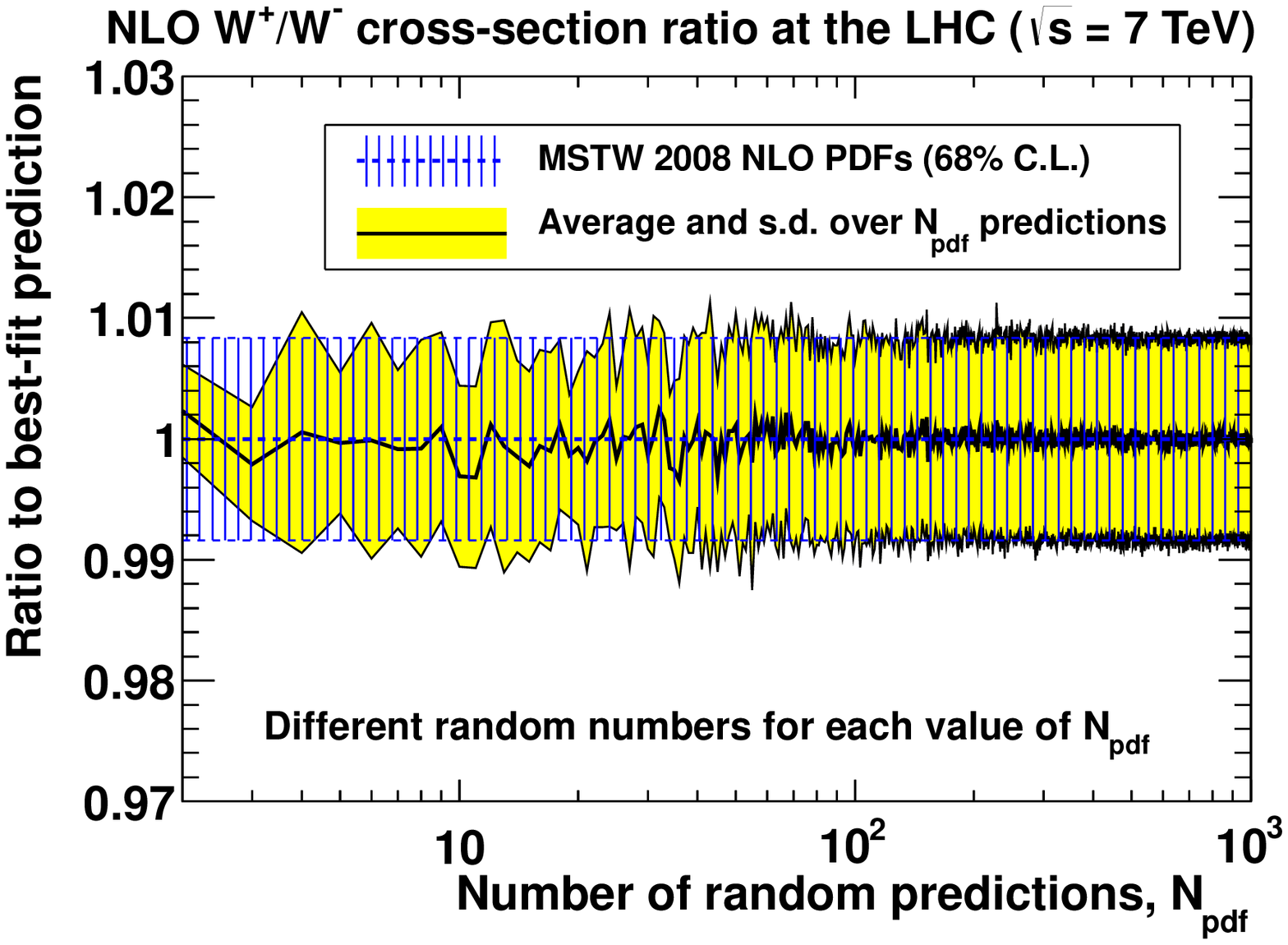}
  \end{minipage}
  \begin{minipage}{0.5\textwidth}
    (c)\\
    \includegraphics[width=\textwidth]{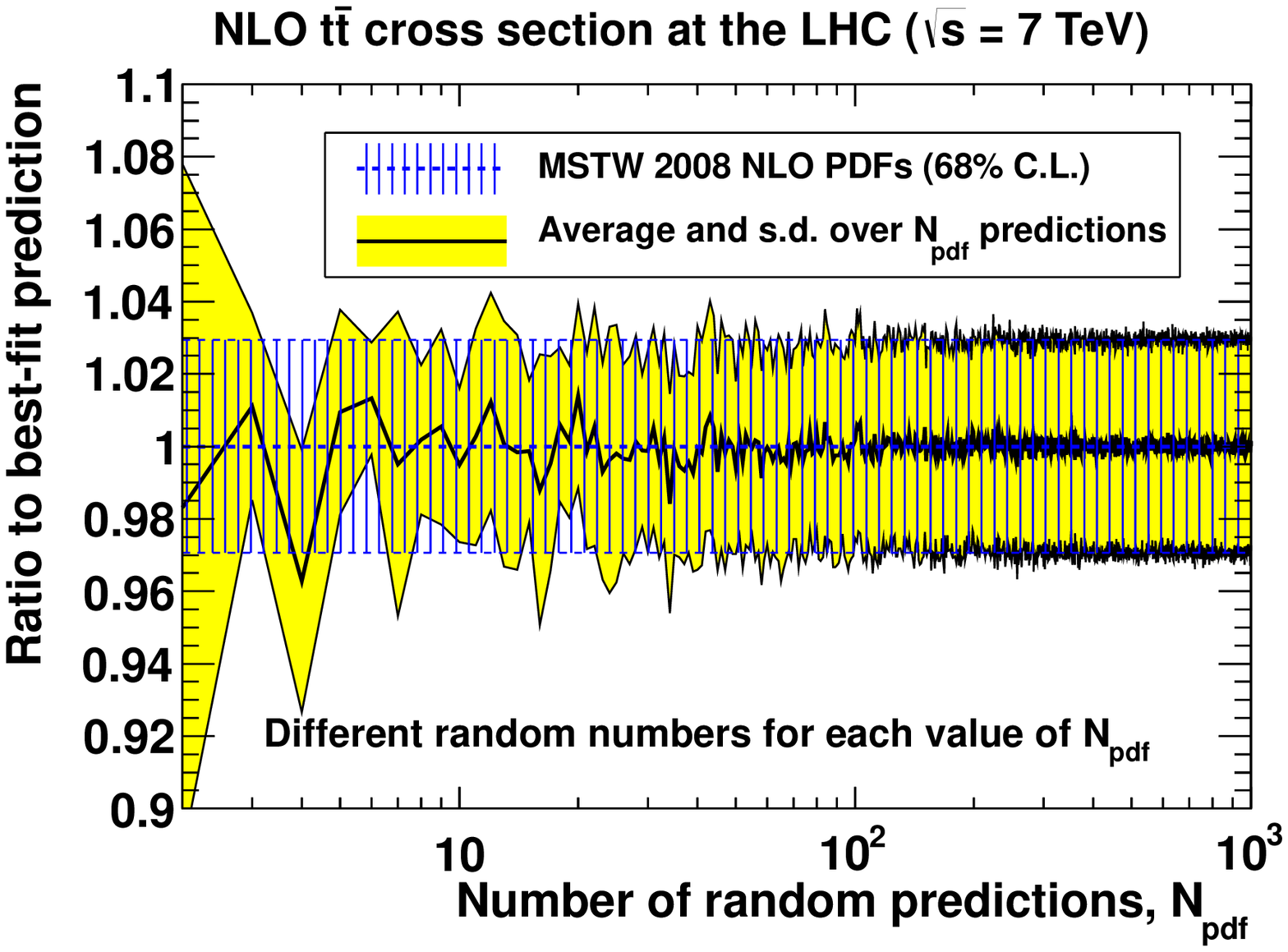}
  \end{minipage}%
  \begin{minipage}{0.5\textwidth}
    (d)\\
    \includegraphics[width=\textwidth]{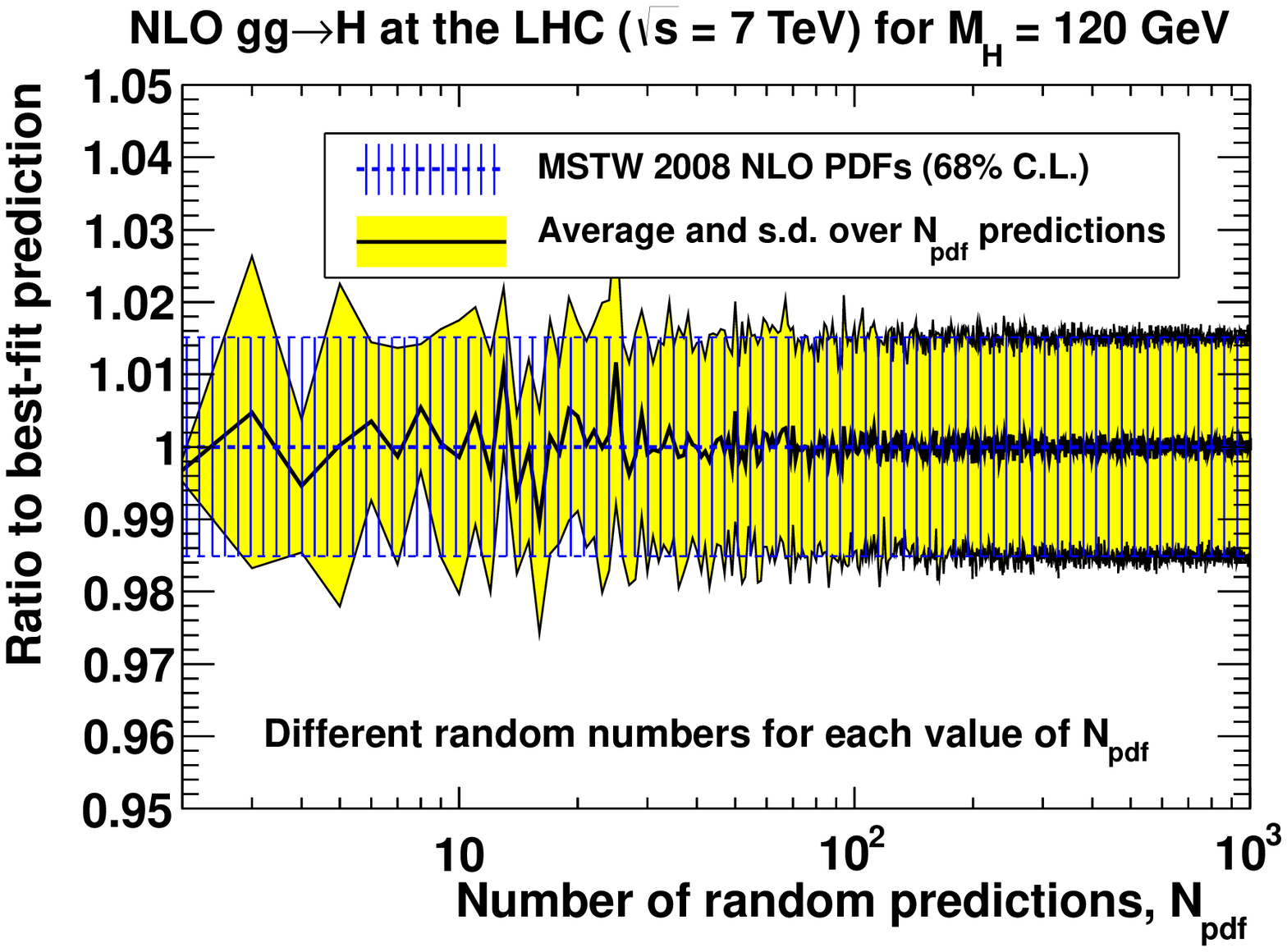}
  \end{minipage}
  \caption{Convergence of average and standard deviation of $N_{\rm pdf}$ random predictions as a function of $N_{\rm pdf}$, each time generating $N_{\rm pdf}$ \emph{independent} random predictions with different random numbers, normalised to the best-fit prediction and compared to the Hessian uncertainty.}
  \label{fig:conv_diffrand}
\end{figure}

Nevertheless, it is instructive to perform a numerical exercise in order to explicitly demonstrate roughly how many random predictions are necessary to adequately sample the parameter space.  We consider the 7~TeV LHC total cross sections for four typical processes corresponding to inclusive production of (a)~$Z^0$ bosons, (b)~$W^+$ relative to $W^-$ bosons, (c)~top-pairs and (d)~Standard Model Higgs bosons  with $M_H=120$~GeV from gluon--gluon fusion.  These four processes are chosen to sample a variety of parton flavours and momentum fractions $x$.  We use the existing NLO calculations from ref.~\cite{Watt:2011kp} with the MSTW 2008 NLO best-fit and Hessian eigenvector PDF sets at 68\% C.L.  For each of the four processes, we generate the minimal $N_{\rm pdf}=2$ random predictions computed using eq.~\eqref{eq:randomPDFsymm} for $F=\{\sigma_{Z^0},\sigma_{W^+}/\sigma_{W^-},\sigma_{t\bar{t}},\sigma_H\}$ and calculate the average and standard deviation.  Then the number of random predictions, $N_{\rm pdf}$, is incremented by one, and the average and standard deviation recomputed, until $N_{\rm pdf}=1000$.  The results are shown in figure~\ref{fig:conv_samerand} normalised to the best-fit prediction and compared with the symmetric Hessian uncertainty of eq.~\eqref{eq:symmunc}.  We use the symmetrised formulae of eqs.~\eqref{eq:symmunc} and \eqref{eq:randomPDFsymm} to allow a direct comparison between the best-fit prediction and the average over the random predictions, without the complications arising from asymmetric tolerance values discussed elsewhere.  We show a similar set of plots in figure~\ref{fig:conv_diffrand} where each value of $N_{\rm pdf}$ now corresponds to the average and standard deviation over $N_{\rm pdf}$ \emph{independent} random predictions.  The results for adjacent $N_{\rm pdf}$ values therefore indicate the size of the statistical fluctuations, which decrease going to larger $N_{\rm pdf}$ values, but are still not completely negligible even for $N_{\rm pdf}\sim1000$.  However, although there is little computational overhead in taking $N_{\rm pdf}$ to be very large when the random predictions are generated ``on the fly'', one would not expect to see noticeable differences when $N_{\rm rep}$ is much larger than around 1000.  In fact, the statistical fluctuations are very small compared to the PDF uncertainty for $N_{\rm pdf}\gtrsim 100$ and even $N_{\rm pdf}=40$ may be sufficiently accurate for many practical purposes.

\section{Reweighting to describe the LHC $W\to\ell\nu$ charge asymmetry data} \label{sec:reweighting}
Updating a PDF set with new data using a Bayesian reweighting method based on statistical inference was originally proposed by Giele and Keller~\cite{Giele:1998gw} and later developed further by the NNPDF Collaboration~\cite{Ball:2010gb,Ball:2011gg}.  Suppose we have a set of $N_{\rm pdf}$ random PDFs $\{\mathcal{S}_k\}$ with equal probability.  It is irrelevant whether they are generated in the space of data (section~\ref{sec:MCgen}) or in the space of parameters (section~\ref{sec:random}).  We can then apply the Bayesian reweighting technique exactly as for the NNPDF sets.  The key formulae are summarised below, but we refer to refs.~\cite{Ball:2010gb,Ball:2011gg} for the derivation and more details of the method.  We first compute the $\chi^2_k$ for the new data set (comprising $N_{\rm pts.}$ data points) using each $\mathcal{S}_k$, then we can calculate the mean value of any PDF-dependent quantity $F(\mathcal{S}_k)$ as:
\begin{equation}
  \langle F\rangle_{\rm old} = \frac{1}{N_{\rm pdf}}\sum_{k=1}^{N_{\rm pdf}}F(\mathcal{S}_k),\quad
  \langle F\rangle_{\rm new} = \frac{1}{N_{\rm pdf}}\sum_{k=1}^{N_{\rm pdf}}w_k(\chi^2_k)\,F(\mathcal{S}_k),
\end{equation}
where the \emph{weights} are given by
\begin{equation} \label{eq:weights}
  w_k(\chi^2_k) = \frac{W_k(\chi^2_k)}{\frac{1}{N_{\rm pdf}}\sum_{j=1}^{N_{\rm pdf}}W_j(\chi^2_j)},\quad W_k(\chi^2_k) \equiv \left(\chi^2_k\right)^{\frac{1}{2}\left(N_{\rm pts.}-1\right)}\,\exp\left(-\frac{1}{2}\chi^2_k\right),
\end{equation}
with the denominator of $w_k(\chi^2_k)$ ensuring the normalisation condition:
\begin{equation}
  \sum_{k=1}^{N_{\rm pdf}}w_k(\chi^2_k) = N_{\rm pdf}.
\end{equation}
Note that the expression for the weights in eq.~\eqref{eq:weights} differs from the original formula in ref.~\cite{Giele:1998gw} due to subtle arguments explained in ref.~\cite{Ball:2010gb}.  The standard deviation $\Delta F$ after reweighting can be calculated using eq.~\eqref{eq:MCsd} with the trivial replacement $N_{\rm rep}\to N_{\rm pdf}$ and using the weighted averages $\langle F^2\rangle_{\rm new}$ and $\langle F\rangle_{\rm new}$.  The effective number of random PDF sets left after reweighting, referred to as the ``Shannon entropy''~\cite{Ball:2010gb}, is given by
\begin{equation} \label{eq:Neff}
  N_{\rm eff} = \exp\left(\frac{1}{N_{\rm pdf}}\sum_{k=1}^{N_{\rm pdf}}w_k\ln\left(\frac{N_{\rm pdf}}{w_k}\right)\right).
\end{equation}

As a simple application of this reweighting technique, we will consider the 7~TeV LHC data from the 2010 running period on the $W\to\ell\nu$ charge asymmetry from CMS~\cite{Chatrchyan:2011jz} and ATLAS~\cite{Aad:2011dm}.  The $W\to\ell\nu$ charge asymmetry is defined differentially as a function of the pseudorapidity $\eta_\ell$ of the charged-lepton from the $W$-boson decay, i.e.
\begin{equation}
  A_\ell(\eta_\ell) = \frac{{\rm d}\sigma(\ell^+)/{\rm d}\eta_\ell-{\rm d}\sigma(\ell^-)/{\rm d}\eta_\ell}{{\rm d}\sigma(\ell^+)/{\rm d}\eta_\ell+{\rm d}\sigma(\ell^-)/{\rm d}\eta_\ell}.
\end{equation}
\begin{figure}
  \centering
  \begin{minipage}{0.5\textwidth}
    (a)\\
    \includegraphics[width=\textwidth]{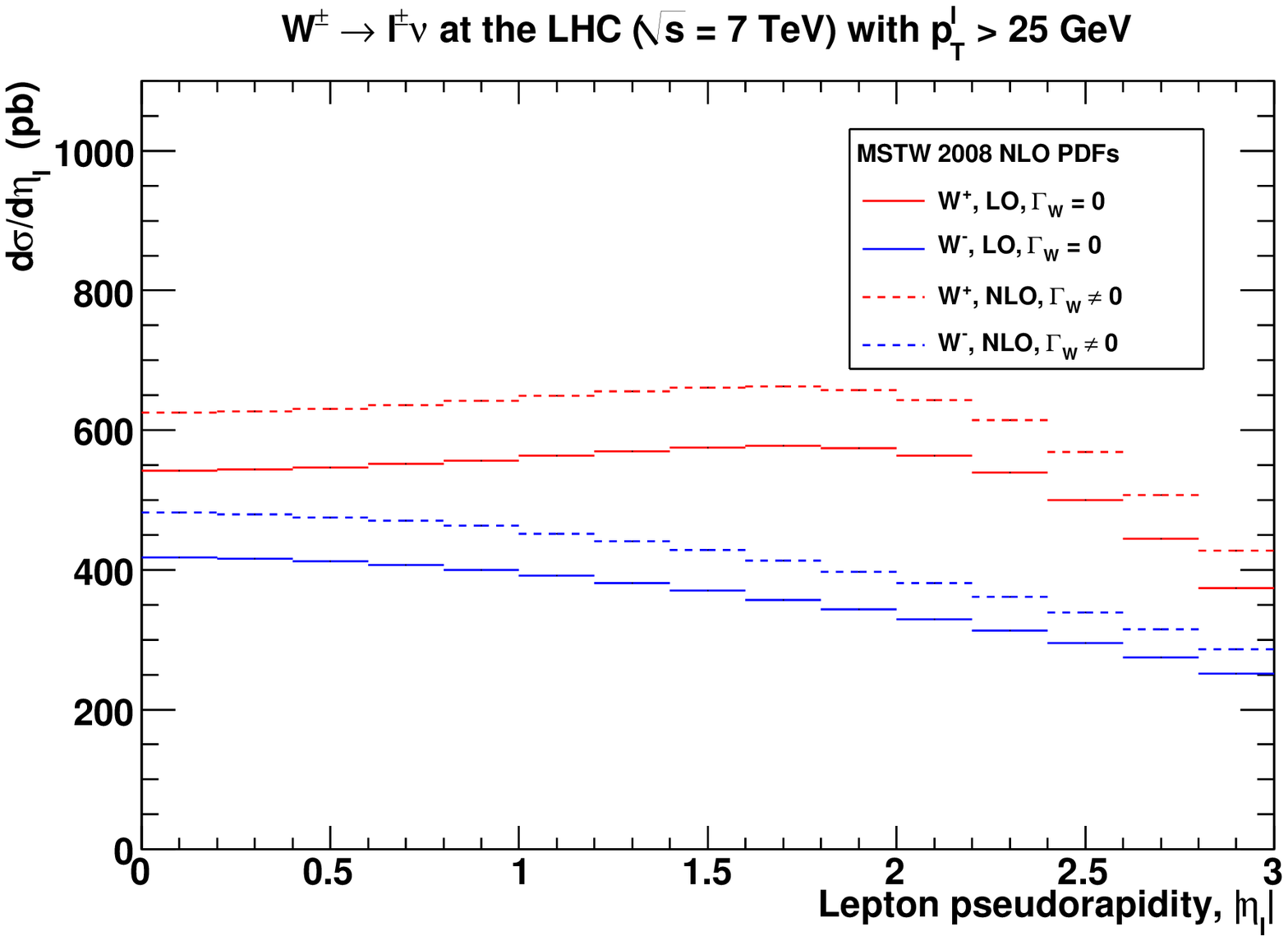}
  \end{minipage}%
  \begin{minipage}{0.5\textwidth}
    (b)\\
    \includegraphics[width=\textwidth]{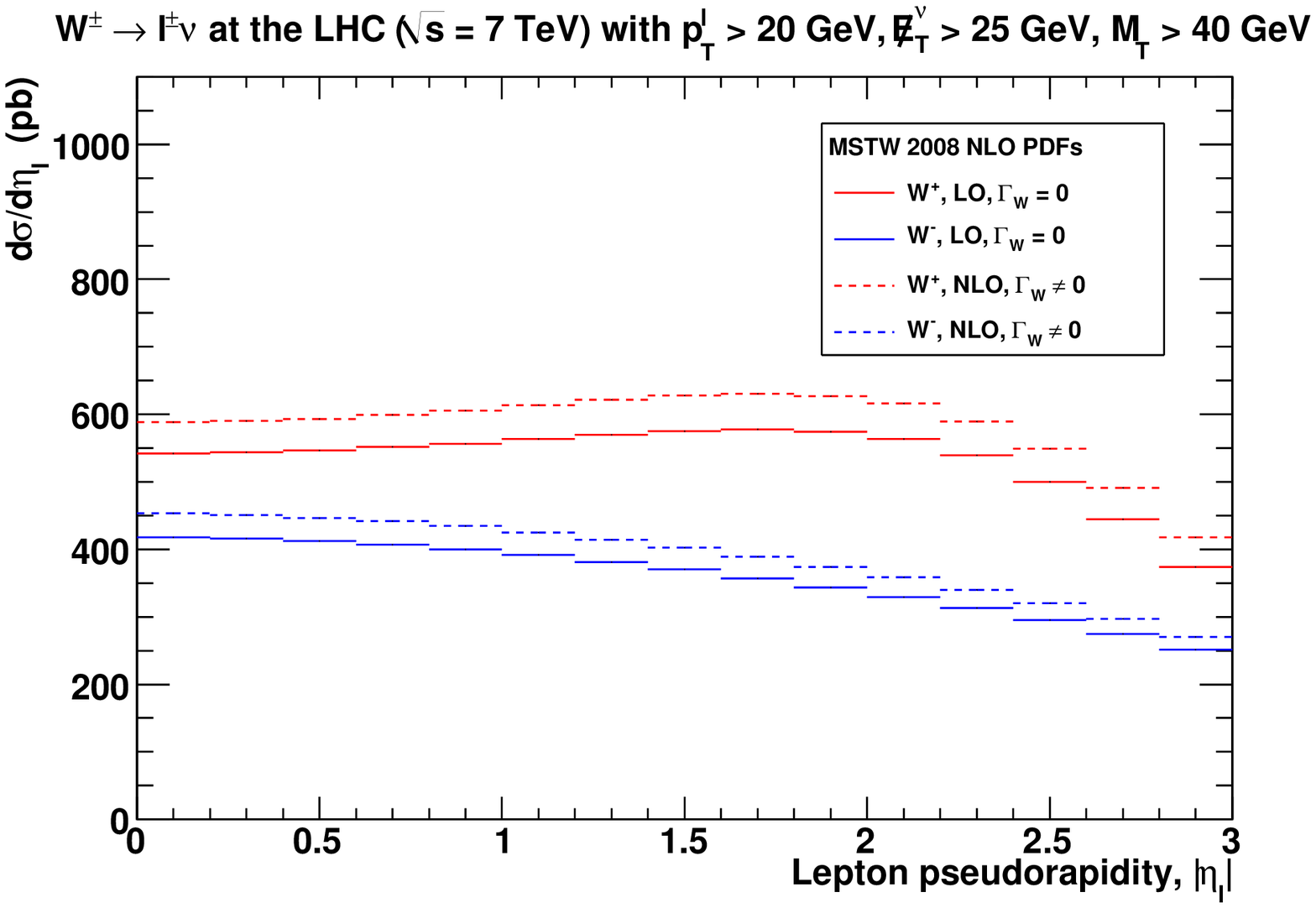}
  \end{minipage}
  \begin{minipage}{0.5\textwidth}
    (c)\\
    \includegraphics[width=\textwidth]{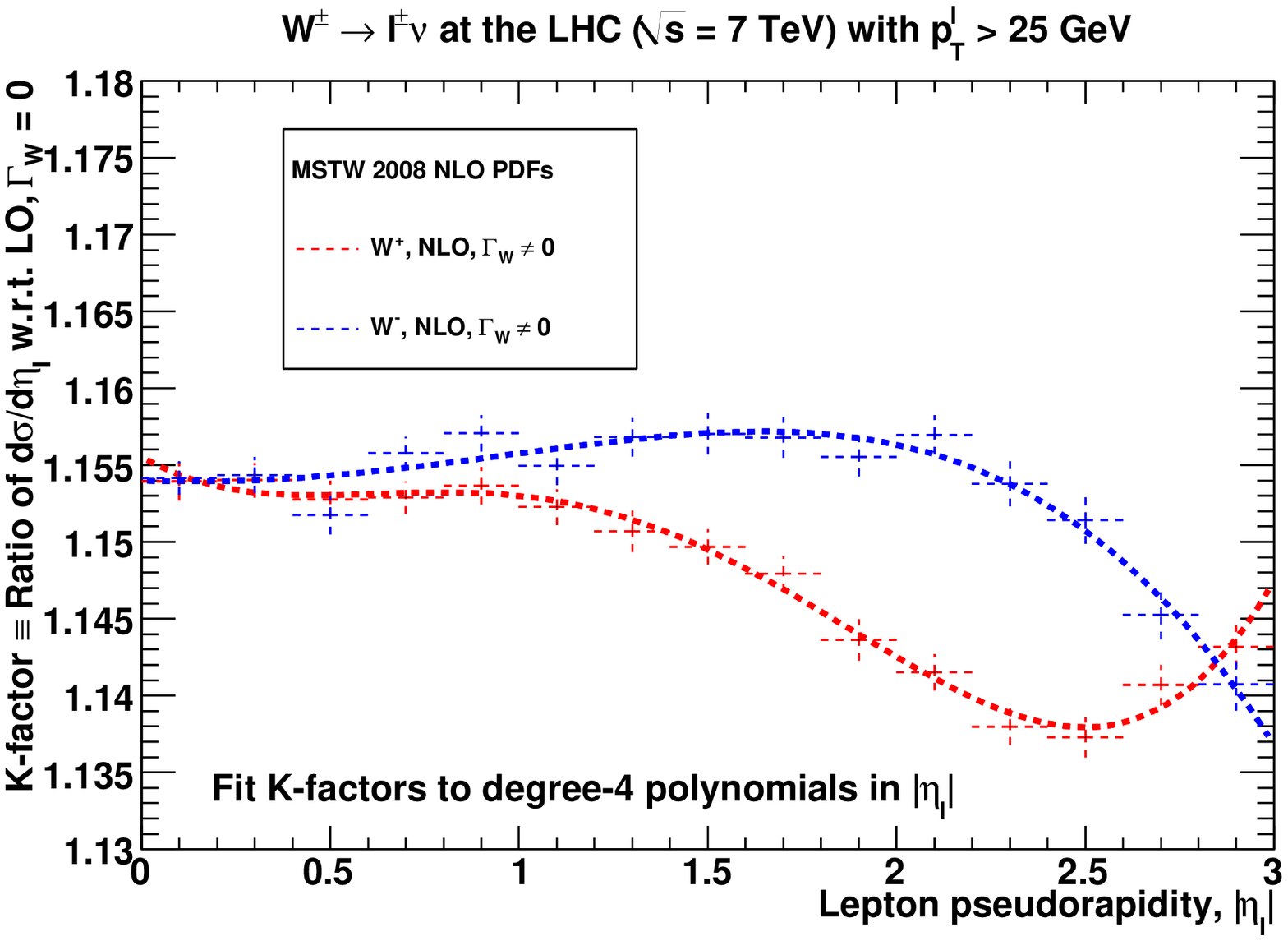}
  \end{minipage}%
  \begin{minipage}{0.5\textwidth}
    (d)\\
    \includegraphics[width=\textwidth]{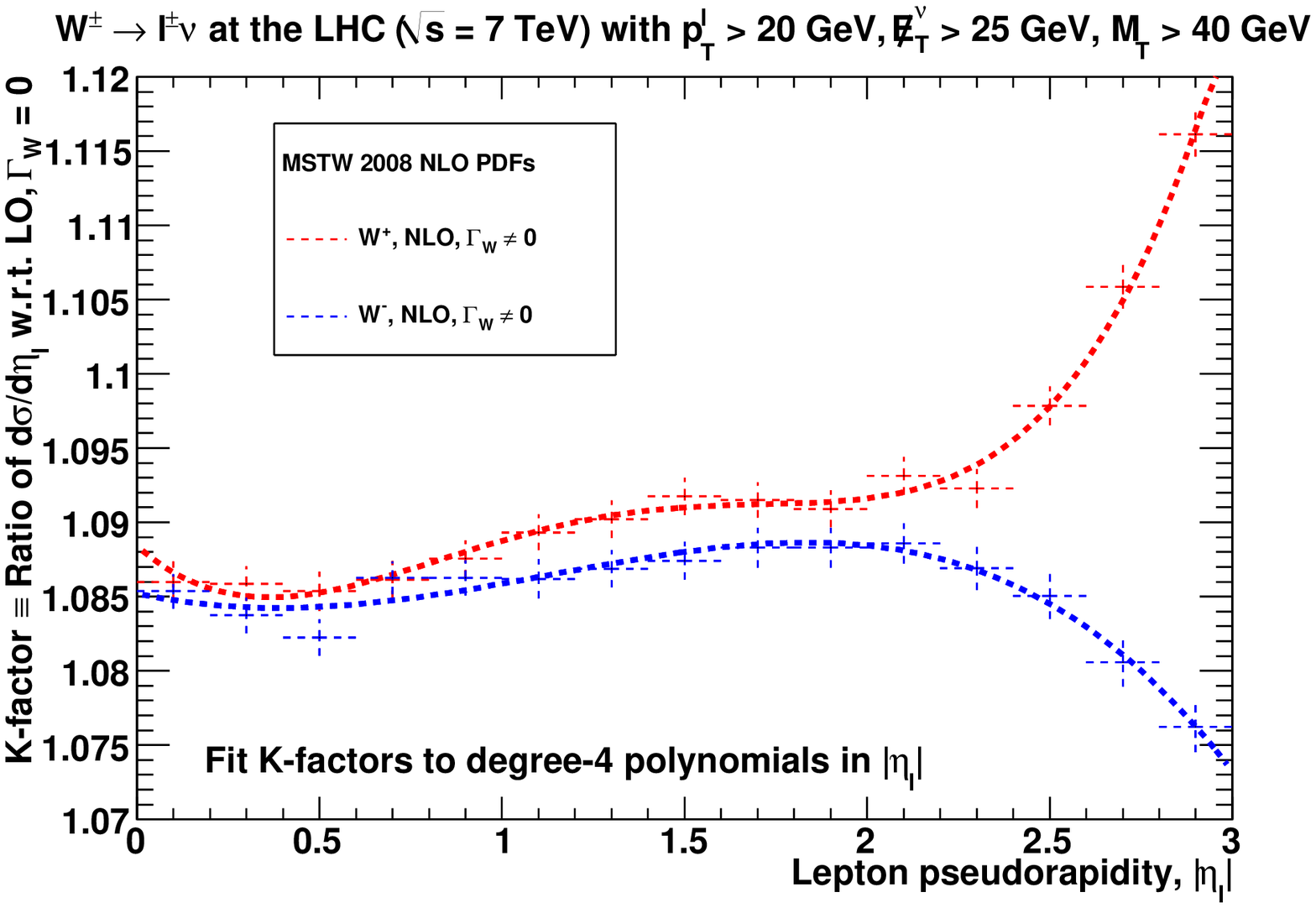}
  \end{minipage}
  \begin{minipage}{0.5\textwidth}
    (e)\\
    \includegraphics[width=\textwidth]{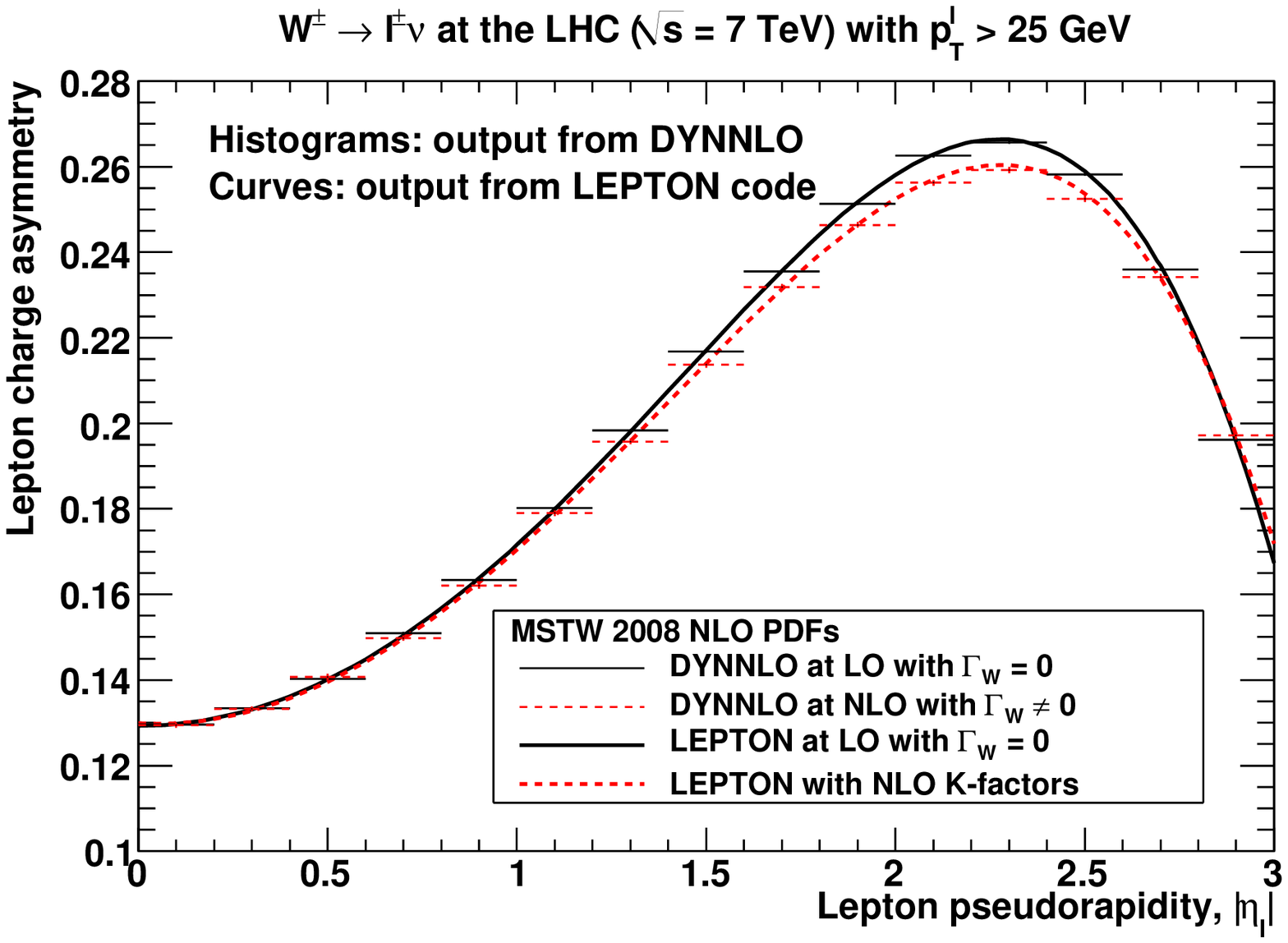}
  \end{minipage}%
  \begin{minipage}{0.5\textwidth}
    (f)\\
    \includegraphics[width=\textwidth]{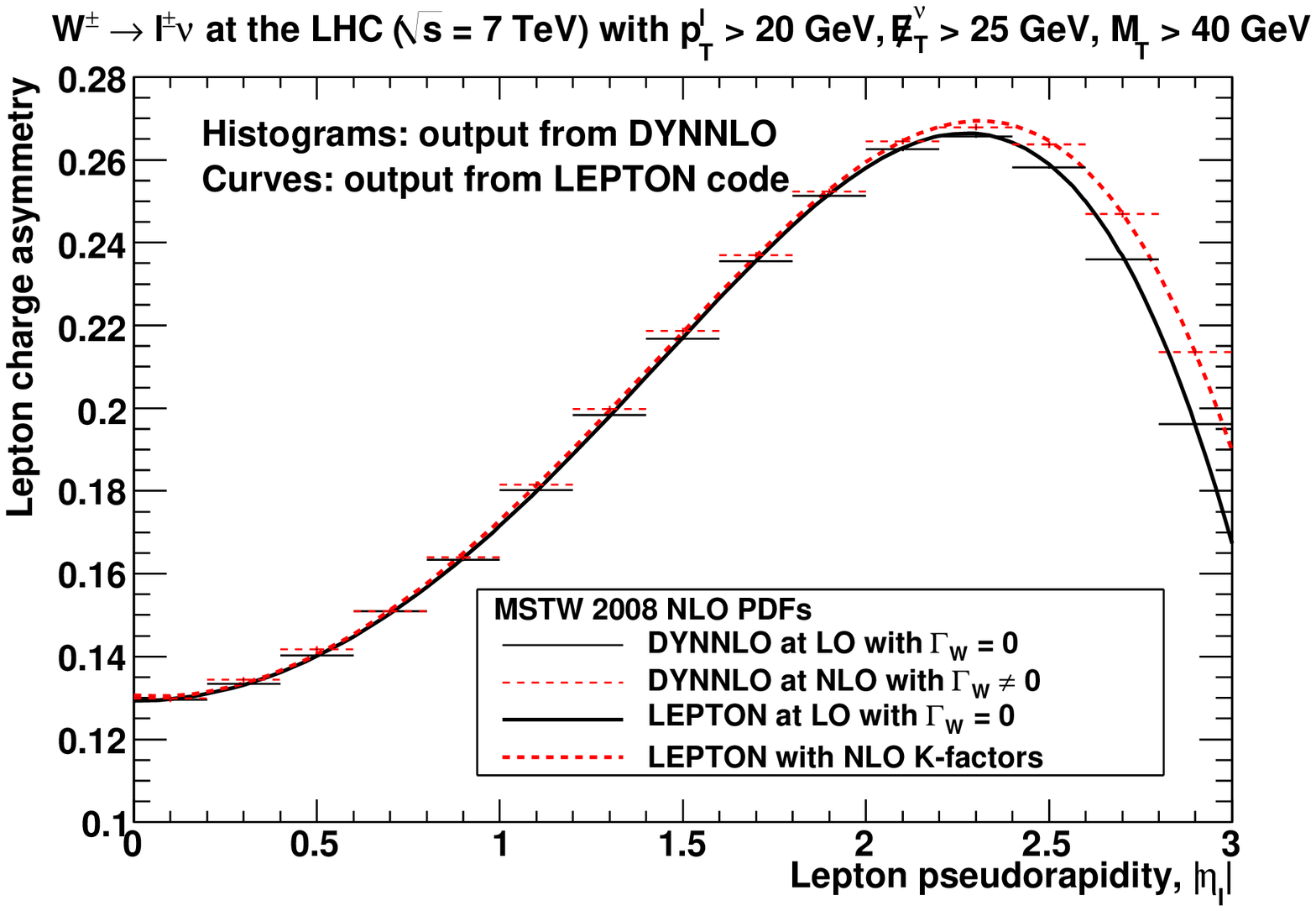}
  \end{minipage}
  \caption{(a,b)~${\rm d}\sigma(\ell^\pm)/{\rm d}\eta_\ell$ distributions, (c,d)~$K$-factors, (e,f)~lepton charge asymmetry, for kinematic cuts corresponding to the (a,c,e)~CMS data~\cite{Chatrchyan:2011jz} and (b,d,f)~ATLAS data~\cite{Aad:2011dm}.}
  \label{fig:Kfactors}
\end{figure}
We will consider the CMS data~\cite{Chatrchyan:2011jz} with charged-lepton transverse momentum cut of $p_T^\ell>25$~GeV in both the electron ($\ell=e$) and muon ($\ell=\mu$) channels.  The ATLAS data~\cite{Aad:2011dm} combine the electron and muon channels with cuts of $p_T^\ell>20$~GeV, missing transverse energy $\not\mathrel{E}_T^\nu>25$~GeV and transverse mass $M_T=\sqrt{2p_T^\ell\not\mathrel{E}_T^\nu(1-\cos\Delta\phi_{\ell\nu})}>40$~GeV, where $\Delta\phi_{\ell\nu}$ is the azimuthal separation between the directions of the charged-lepton and neutrino.  The pseudorapidity distributions, ${\rm d}\sigma(\ell^\pm)/{\rm d}\eta_\ell$, calculated from the public \textsc{dynnlo} code~\cite{Catani:2009sm} using the MSTW 2008 NLO best-fit PDFs with $\mu_R=\mu_F=M_W$, are shown in figure~\ref{fig:Kfactors}(a,b) for (a)~CMS cuts and (b)~ATLAS cuts.  For LO kinematics ($p_T^W=0$) with zero $W$ width ($\Gamma_W=0$), then $p_T^\ell=\not\mathrel{E}_T^\nu$ and $M_T=2p_T^\ell$, and the predictions are identical for the CMS and ATLAS cuts, but not after accounting for NLO and finite $W$ width effects.  In figure~\ref{fig:Kfactors}(c,d) we define a $K$-factor by taking the ratio of the \textsc{dynnlo} histograms, then we fit to quartic polynomials in $|\eta_\ell|$ to provide a convenient parameterisation and to smooth statistical fluctuations from the \textsc{vegas} multidimensional integration.  A fast calculation of the $W\to\ell\nu$ charge asymmetry can then be obtained using a simple LO calculation with zero $W$ width (denoted ``\texttt{LEPTON}''), including the parameterised $K$-factors for ${\rm d}\sigma(\ell^\pm)/{\rm d}\eta_\ell$, making the assumption that the $K$-factors are independent of the PDF choice.  In figure~\ref{fig:Kfactors}(e,f) we compare the \texttt{LEPTON} calculation, without and with the inclusion of $K$-factors, with the \textsc{dynnlo} histograms, finding good agreement (by construction).  It can be seen that the NLO corrections and finite-width effects are very small over most of the $|\eta_\ell|$ range.  The effect on the $W\to\ell\nu$ charge asymmetry of neglecting the PDF dependence of the $K$-factors should then be completely negligible.  We have also computed the NNLO corrections using the \textsc{dynnlo} code and find them to be much smaller than the NLO corrections, but we will consider only NLO QCD in making comparisons to data, as done elsewhere in this paper.

We will focus on demonstrating the reweighting technique rather than aiming to make an exhaustive study of the impact of the LHC data.  With this aim in mind, we will not consider in this work the 2010 CMS data with $p_T^\ell>30$~GeV~\cite{Chatrchyan:2011jz}, the preliminary CMS measurements using 2011 data with $p_T^\mu>25$~GeV~\cite{CMS:muonasymmetry} or $p_T^e>35$~GeV~\cite{CMS:electronasymmetry}, or the recent LHCb measurements using 2010 data with $p_T^\mu>\{20,25,30\}$~GeV~\cite{Aaij:2012vn}.  The ATLAS Collaboration~\cite{Aad:2011dm} provide the differential cross sections, ${\rm d}\sigma(\ell^\pm)/{\rm d}\eta_\ell$, separately for $W^+\to\ell^+\nu$ and $W^-\to\ell^-\bar{\nu}$ with the complete information on correlated systematic uncertainties, which is potentially more useful for PDF fits than simply the asymmetry $A_\ell(\eta_\ell)$.  A future study could perhaps investigate the use of reweighting with the ATLAS $W^\pm$ (and $Z/\gamma^*$) differential cross sections rather than the asymmetry $A_\ell(\eta_\ell)$.  In this study, we simply calculate the $\chi^2_k$ values with all experimental uncertainties added in quadrature.

\begin{figure}
  \centering
  \begin{minipage}{0.5\textwidth}
    (a)\\
    \includegraphics[width=\textwidth]{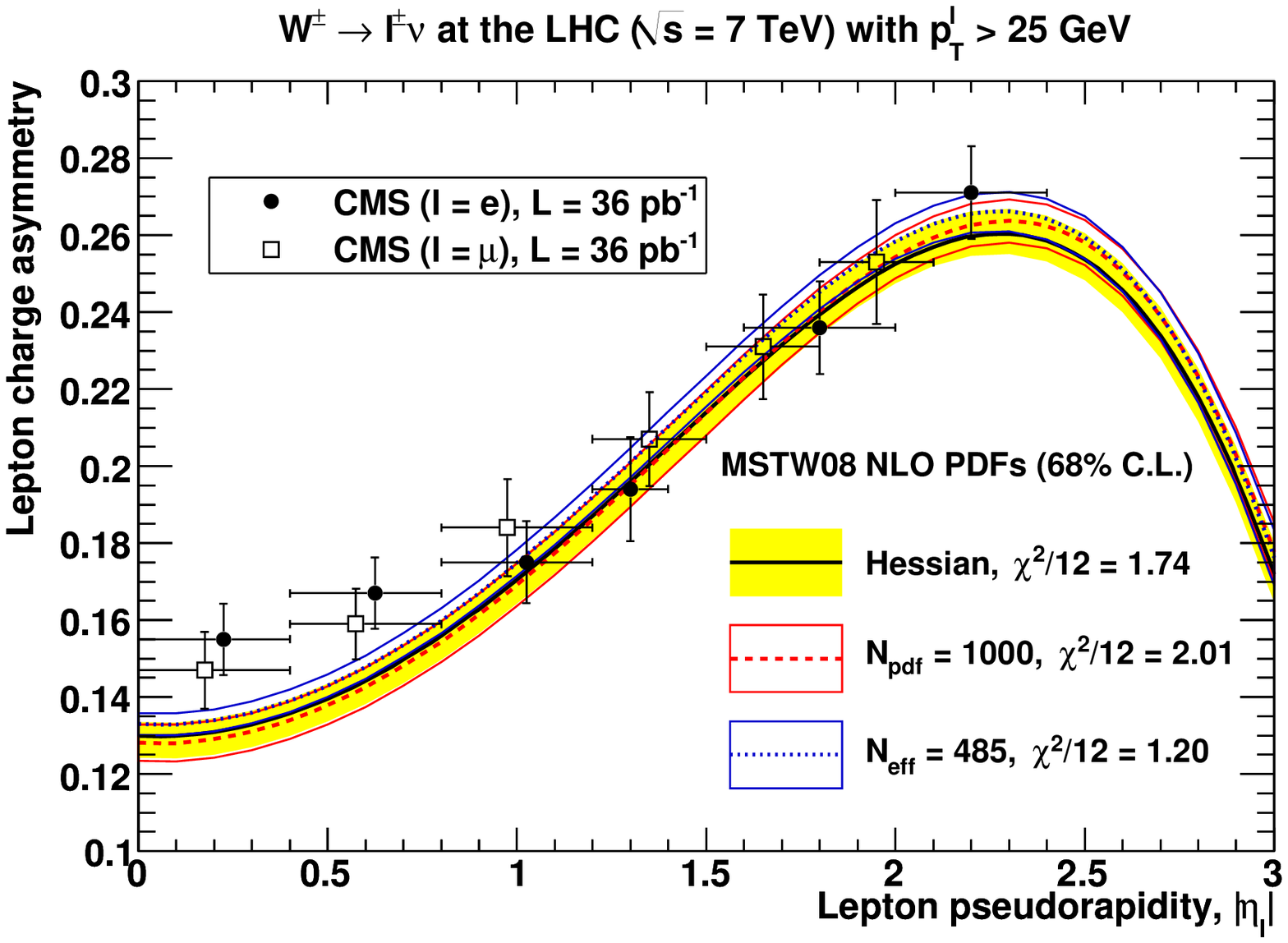}
  \end{minipage}%
  \begin{minipage}{0.5\textwidth}
    (b)\\
    \includegraphics[width=\textwidth]{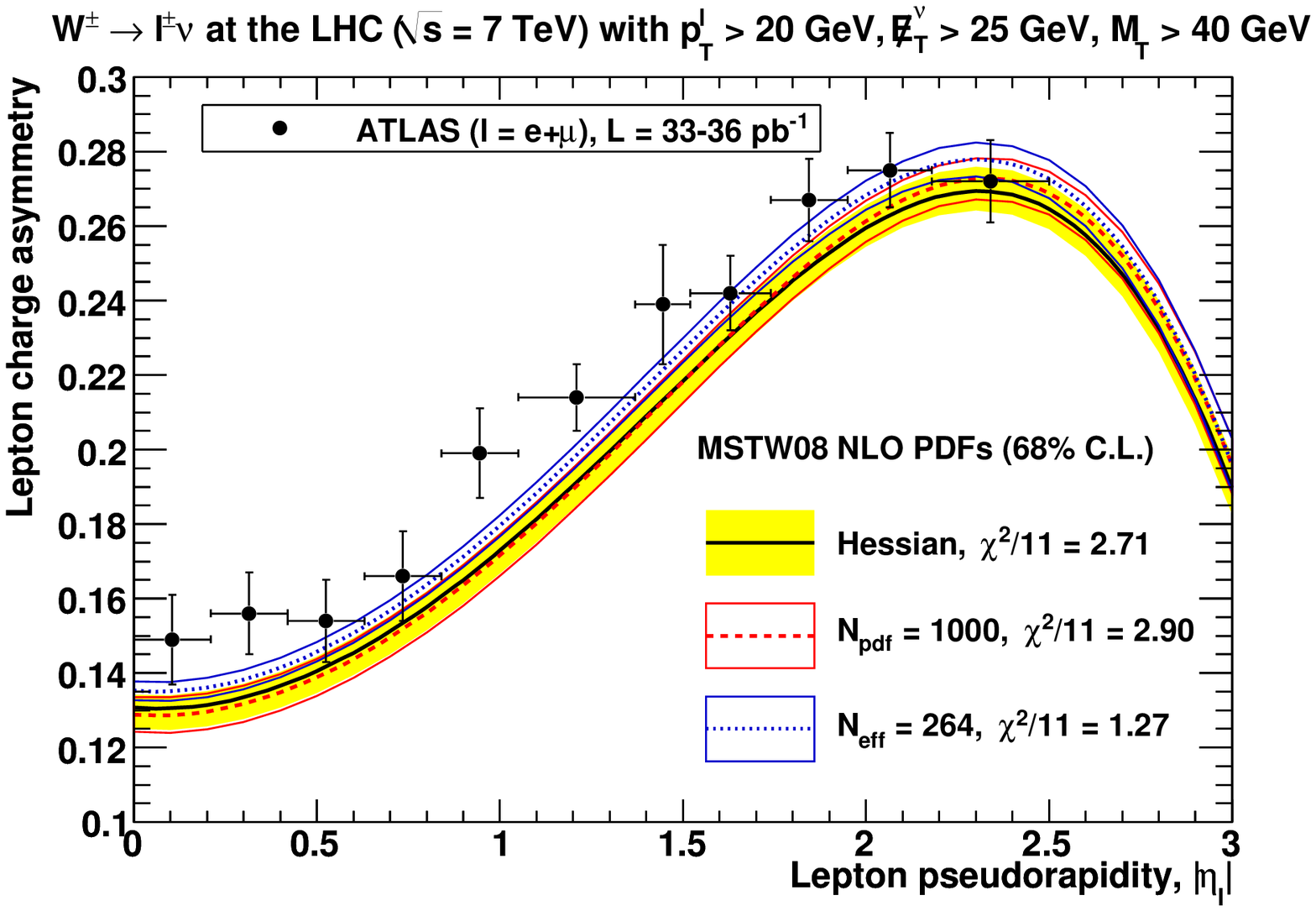}
  \end{minipage}
  \begin{minipage}{0.5\textwidth}
    (c)\\
    \includegraphics[width=\textwidth]{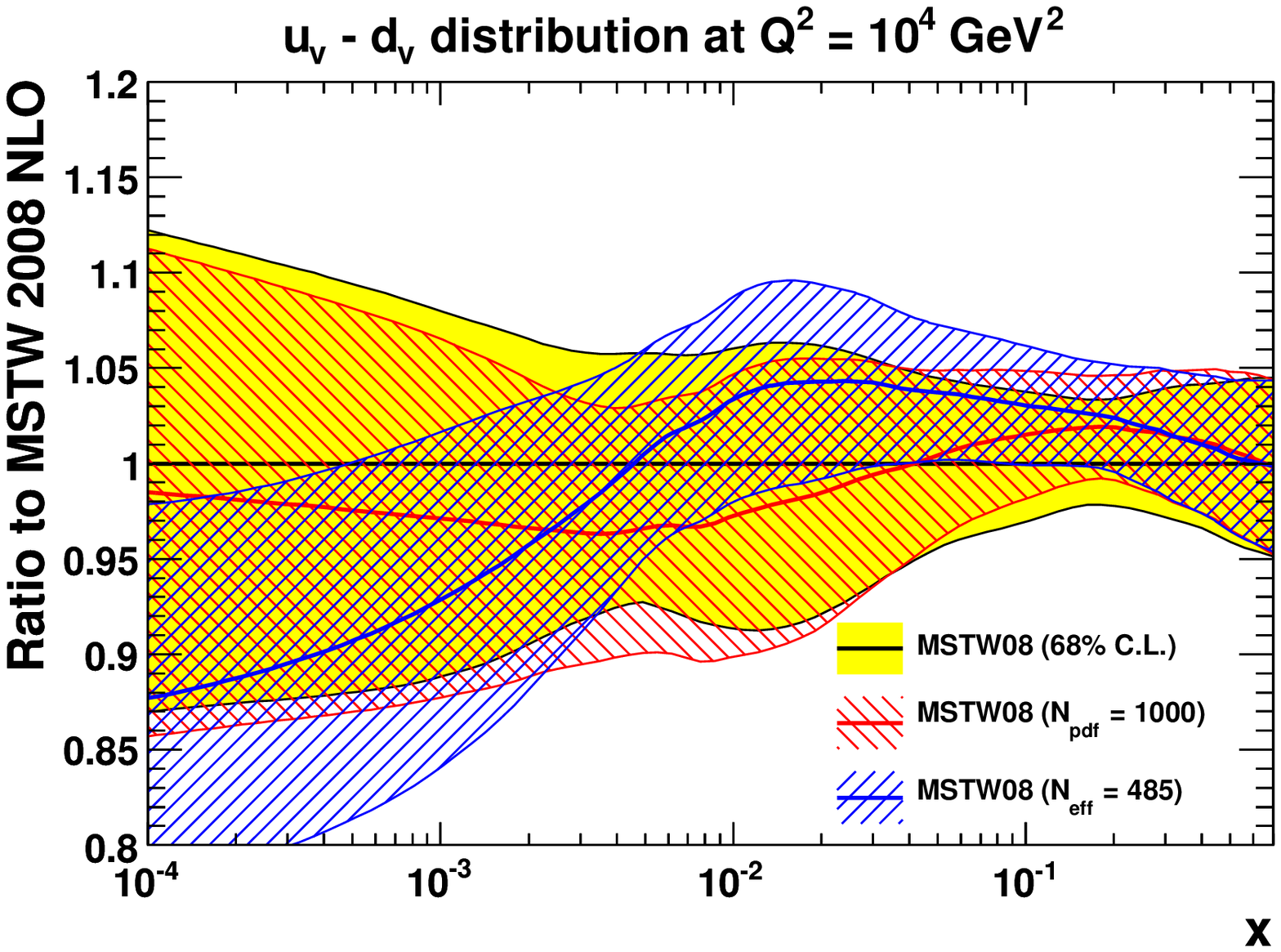}
  \end{minipage}%
  \begin{minipage}{0.5\textwidth}
    (d)\\
    \includegraphics[width=\textwidth]{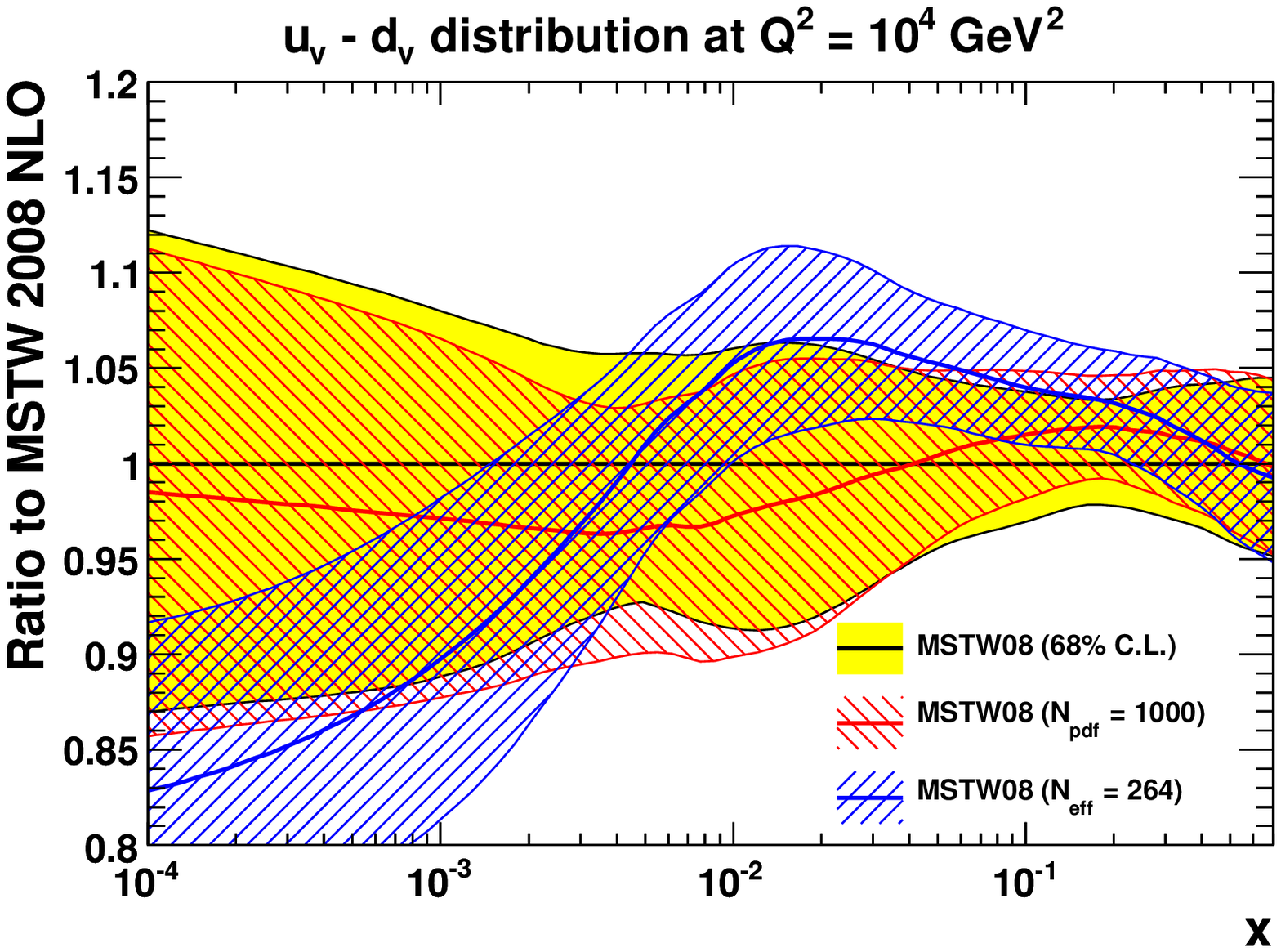}
  \end{minipage}
  \caption{Lepton charge asymmetry $A_\ell(\eta_\ell)$ predictions compared to (a)~CMS~\cite{Chatrchyan:2011jz} and (b)~ATLAS~\cite{Aad:2011dm} data, then change in $u_v-d_v$ after reweighting using (c)~CMS and (d)~ATLAS data.}
  \label{fig:uvdv}
\end{figure}
In figure~\ref{fig:uvdv}(a,b) we compare the (a)~CMS and (b)~ATLAS data on the $W\to\ell\nu$ charge asymmetry to predictions using the MSTW 2008 NLO PDFs, firstly with the usual best-fit and Hessian uncertainty.  We then generate $N_{\rm pdf}=1000$ random predictions for the asymmetry by taking $F=A_\ell(\eta_\ell)$ in eq.~\eqref{eq:randomPDF}, and take the average and standard deviation, giving results slightly different from the best-fit and Hessian uncertainty (mainly due to the asymmetric tolerance values).  The $\chi^2$ values of the average $A_\ell(\eta_\ell)$, displayed in the plot legends, are then slightly larger than the $\chi^2$ of the best-fit predictions.  Next we compute weights for each of the $N_{\rm pdf}$ predictions according to eq.~\eqref{eq:weights}, then finally we plot the weighted average and standard deviation in figure~\ref{fig:uvdv}(a,b).  The $\chi^2$ of the weighted average $A_\ell(\eta_\ell)$ improves significantly compared to the unweighted average.  The effective number of random predictions $N_{\rm eff}$ after reweighting, computed according to eq.~\eqref{eq:Neff}, is about half the original number for CMS and almost a quarter the original number for ATLAS.  The most significant change in the predictions after reweighting is for $\eta_\ell\approx 0$ where $A_\ell(\eta_\ell)$ depends on the combination $u_v-d_v$ at momentum fractions $x$ slightly above $x\sim M_W/\sqrt{s}\sim 0.01$.  In figure~\ref{fig:uvdv}(c,d) we plot this combination for $Q^2=(100~{\rm GeV})^2$ for the same three sets of predictions shown in figure~\ref{fig:uvdv}(a,b).  We compare the best-fit and Hessian uncertainty with the unweighted/weighted average and standard deviation of $N_{\rm pdf}=1000$ random PDFs constructed by taking $F=x(u_v-d_v)(x,Q^2)$ in eq.~\eqref{eq:randomPDF}, with the \emph{same} random numbers $R_{jk}$ and weights $w_k$ used in figure~\ref{fig:uvdv}(a,b).  As expected from figure~\ref{fig:uvdv}(a,b), the shift in $u_v-d_v$ is largest at $x\sim 0.01$, and the average value after reweighting using the ATLAS data even lies outside the original uncertainty band.  There is also a distinct reduction in the size of the uncertainty band after reweighting.

\begin{figure}
  \centering
  \begin{minipage}{0.5\textwidth}
    (a)\\
    \includegraphics[width=\textwidth]{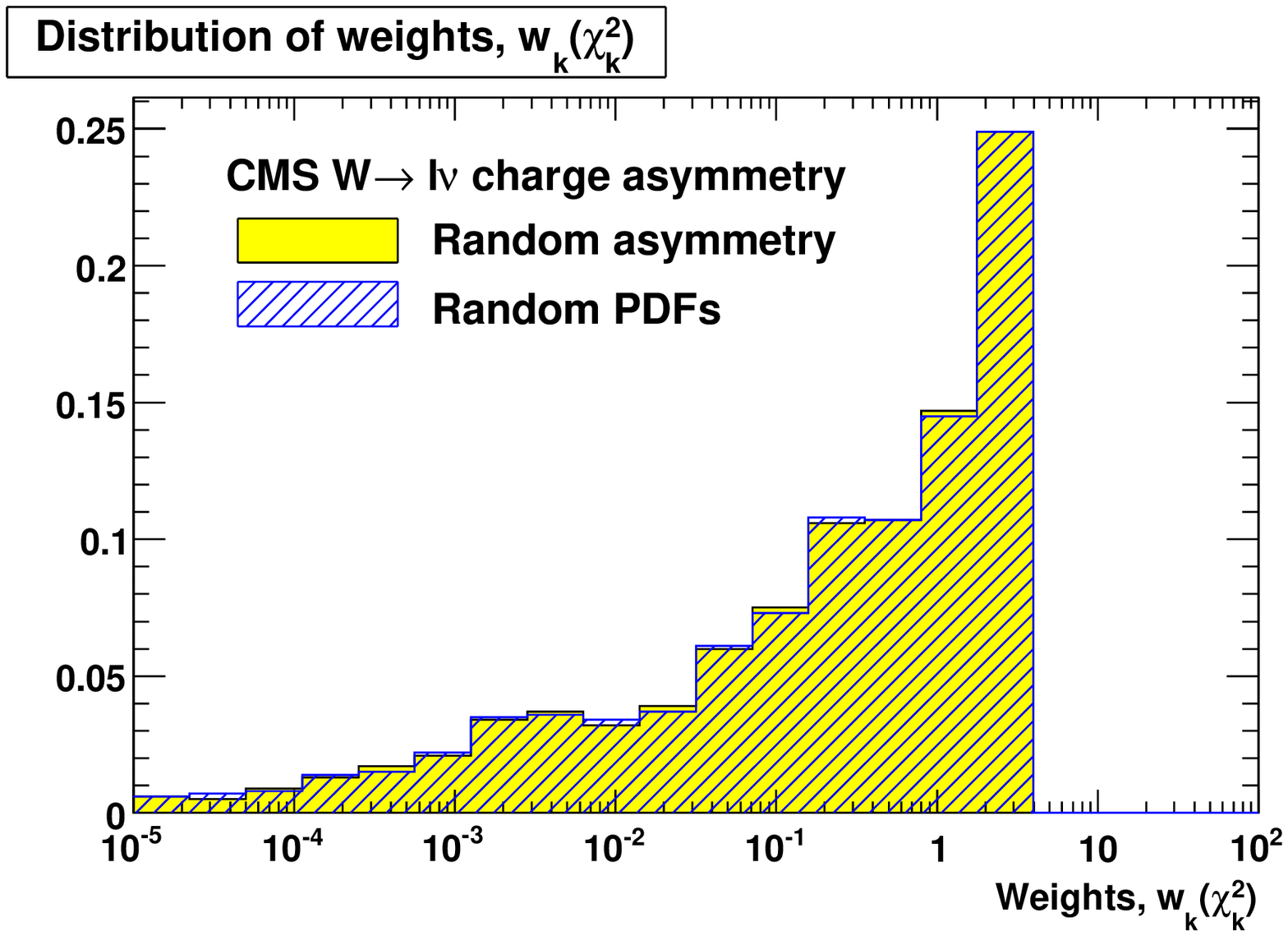}
  \end{minipage}%
  \begin{minipage}{0.5\textwidth}
    (b)\\
    \includegraphics[width=\textwidth]{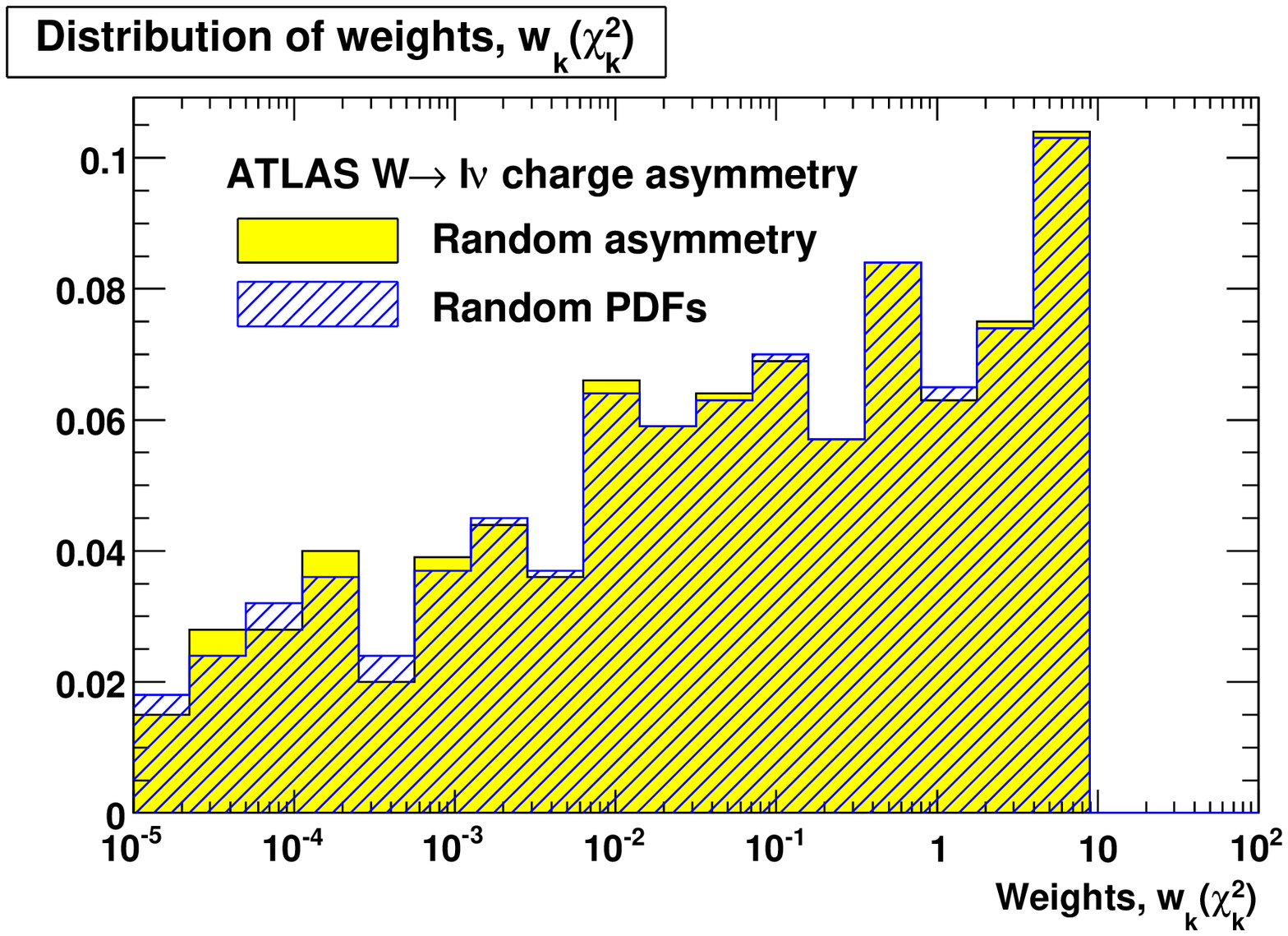}
  \end{minipage}
  \begin{minipage}{0.5\textwidth}
    (c)\\
    \includegraphics[width=\textwidth]{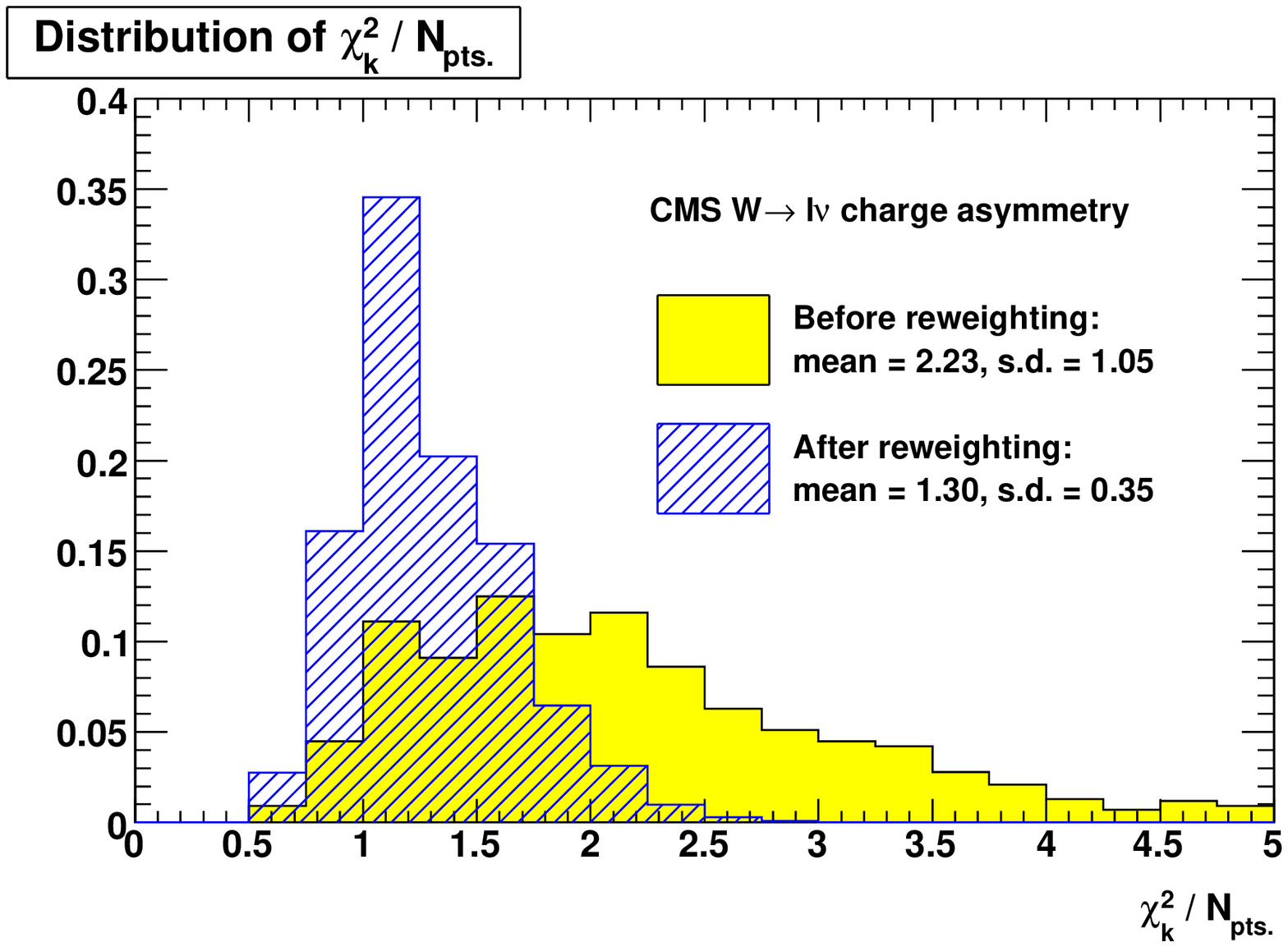}
  \end{minipage}%
  \begin{minipage}{0.5\textwidth}
    (d)\\
    \includegraphics[width=\textwidth]{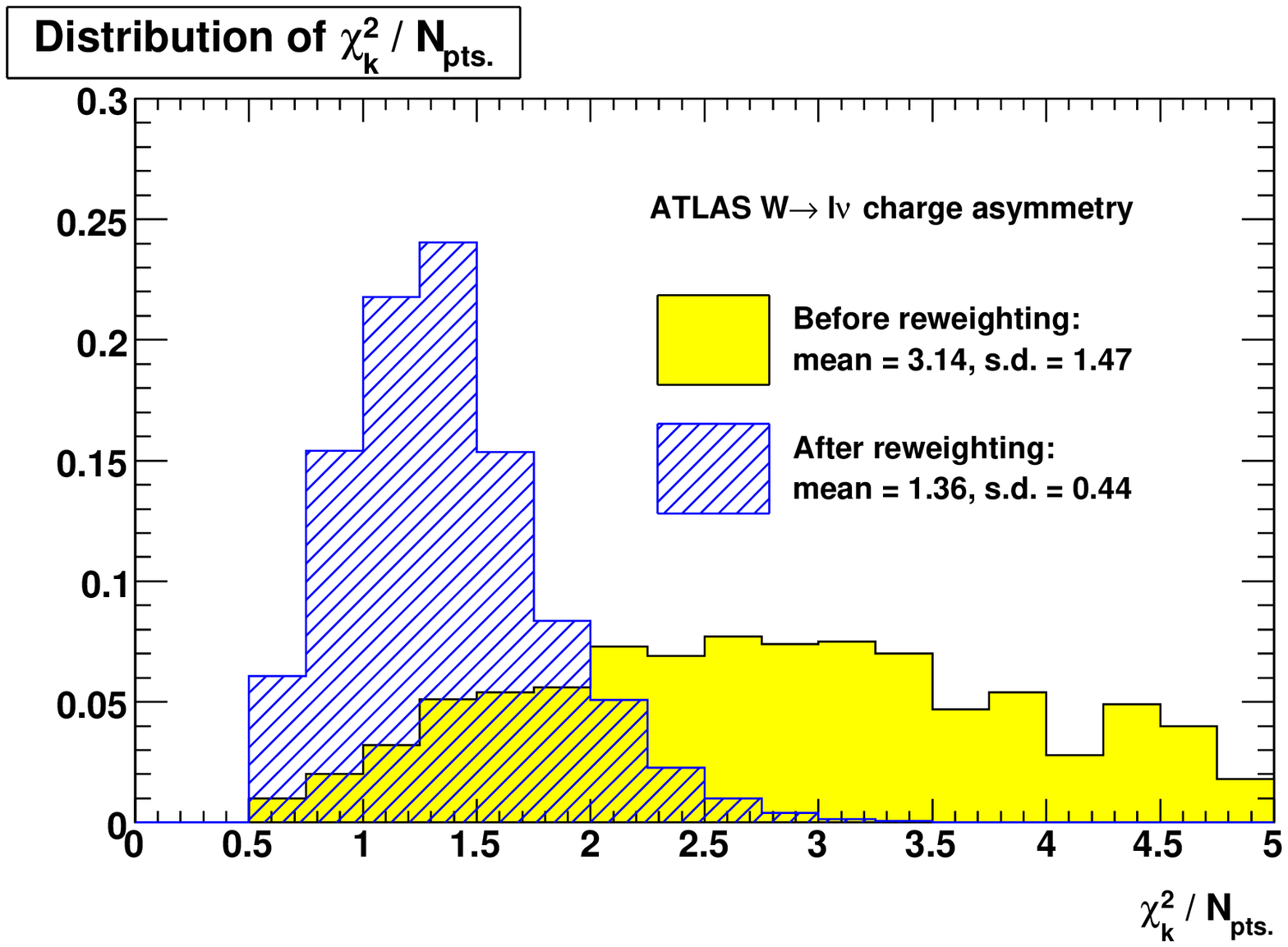}
  \end{minipage}
  \begin{minipage}{0.5\textwidth}
    (e)\\
    \includegraphics[width=\textwidth]{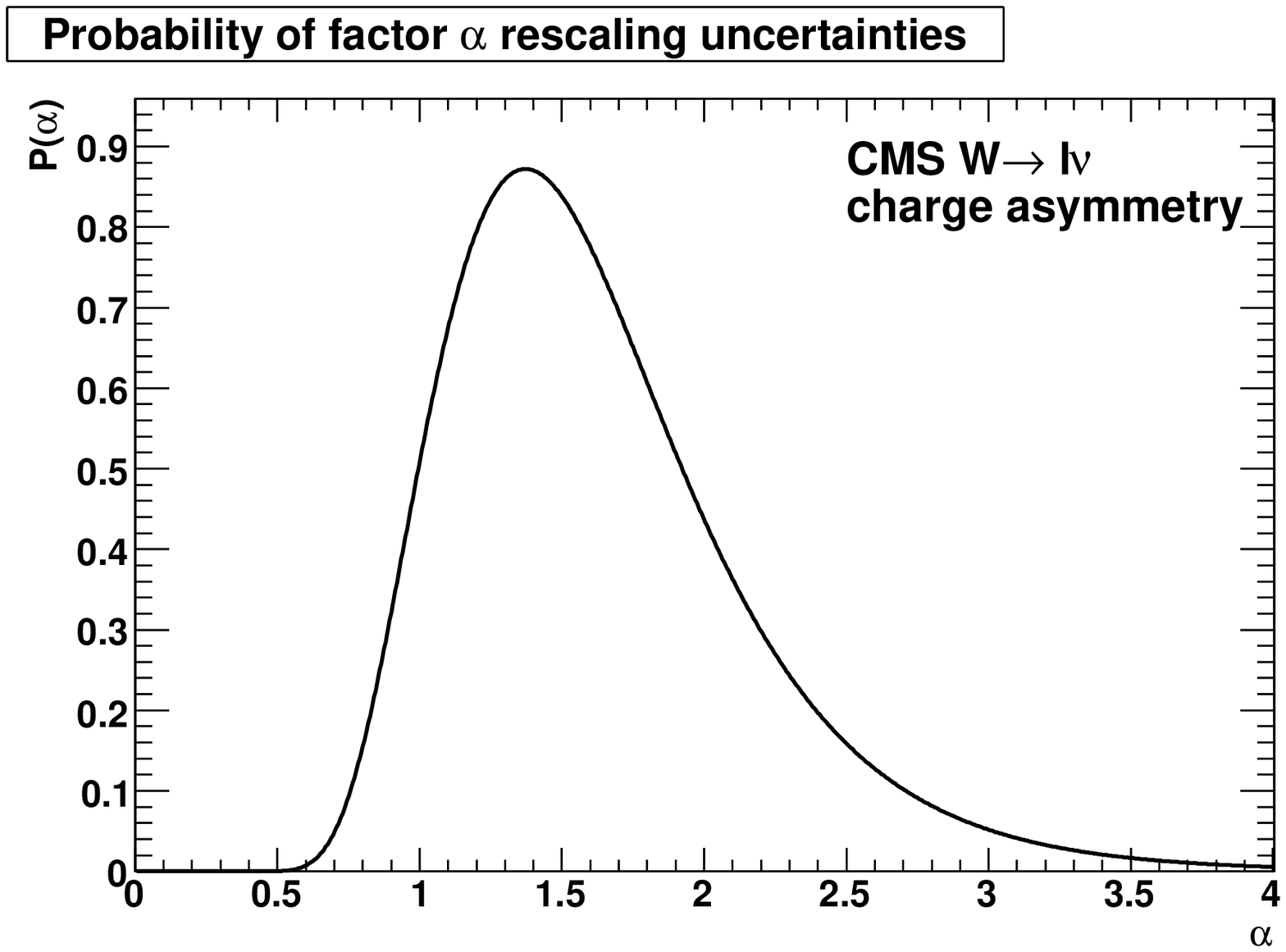}
  \end{minipage}%
  \begin{minipage}{0.5\textwidth}
    (f)\\
    \includegraphics[width=\textwidth]{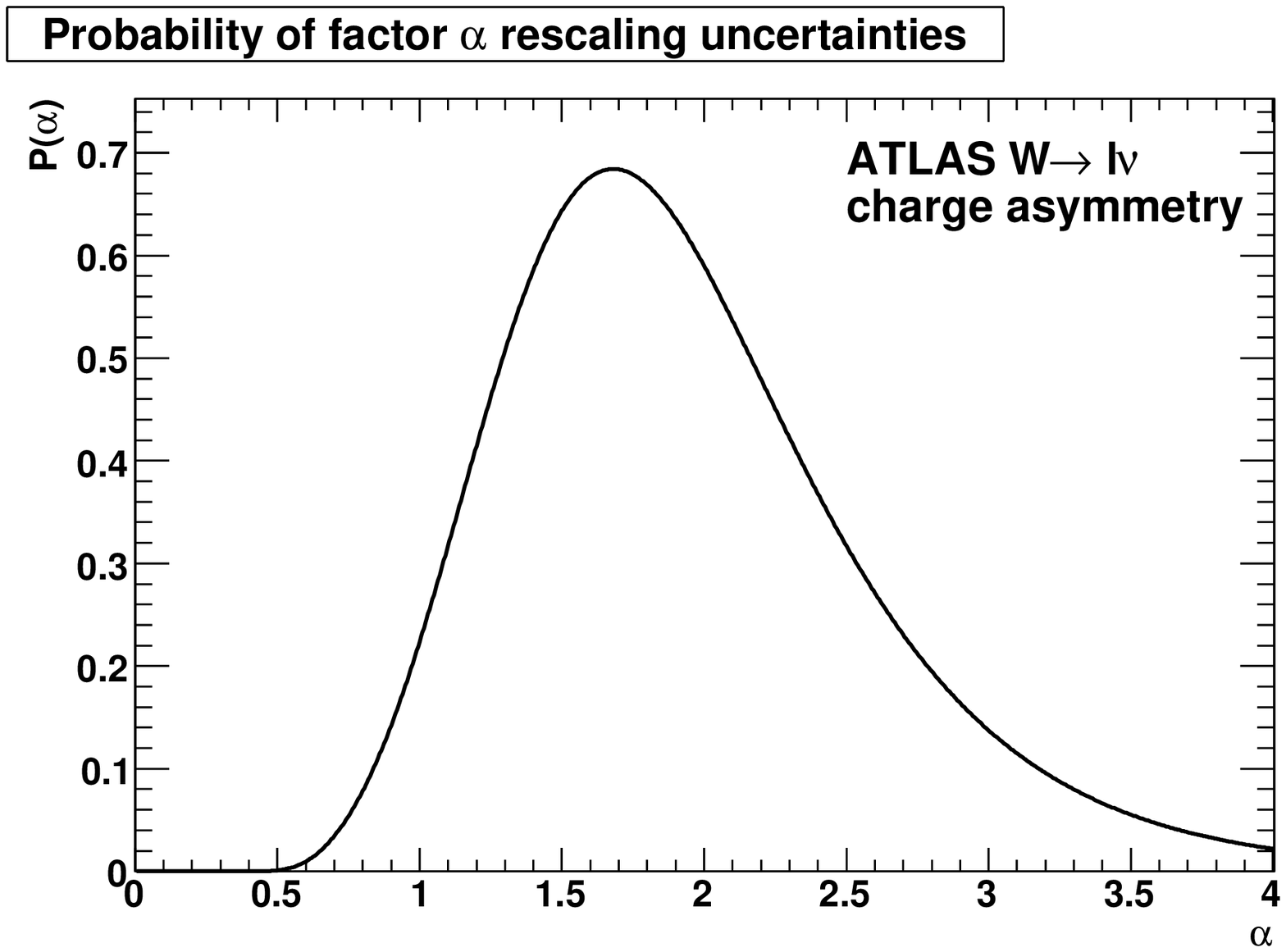}
  \end{minipage}
  \caption{Distributions of (a,b)~$w_k$, (c,d)~$\chi^2_k/N_{\rm pts.}$, (e,f) $\mathcal{P}(\alpha)$, for (a,c,e)~CMS and (b,d,f)~ATLAS.}
  \label{fig:reweight}
\end{figure}
The procedure just described is not completely unambiguous.  Alternative prescriptions could be formulated which are equivalent in a linear approximation, but which might differ due to some degree of non-linearity.  For example, rather than starting by generating random predictions for the asymmetry by taking $F=A_\ell(\eta_\ell)$ in eq.~\eqref{eq:randomPDF}, we could instead generate $N_{\rm pdf}=1000$ random PDF sets by taking $F=xf(x,M_W^2)$ in eq.~\eqref{eq:randomPDF}, where $f=\{g,d,u,s,c,b,\bar{d},\bar{u},\bar{s},\bar{c},\bar{b}\}$, then calculate $A_\ell(\eta_\ell)$ for each of these $N_{\rm pdf}$ random PDF sets, before calculating weights according to eq.~\eqref{eq:weights} as before.  This alternative method will give slightly different results since $A_\ell(\eta_\ell)$ depends on $xf(x,M_W^2)$ in a non-linear manner.  In figure~\ref{fig:reweight}(a,b) we compare the distribution of weights $w_k$ computed using the two different methods, using the same random numbers $R_{jk}$ to allow a direct comparison of individual weights with the same label $k$.  The distribution of weights is very similar using the two methods.  The individual weights typically agree to within a few percent and differ by only a few tens of percent in the worst cases.  The values of $N_{\rm eff}$ agree to the nearest integer and the values of $\chi^2/N_{\rm pts.}$ agree to two decimal places.  The plots of figure~\ref{fig:uvdv} produced using the alternative method are indistinguishable.  We conclude that the degree of non-linearity is small and both techniques may be useful in practice.  For example, it might be useful to first generate $N_{\rm pdf}=1000$ random PDF sets as grid files by taking $F=xf(x,Q^2)$ in eq.~\eqref{eq:randomPDF}, then these grid files can be processed in exactly the same way as the NNPDF grid files.  On the other hand, that method would require substantial disk storage and would require the observable $A_\ell(\eta_\ell)$ to be evaluated $N_{\rm pdf}$ times, which is potentially time-consuming.  With the first method described above, it is unnecessary to store intermediate grid files, and only $2n+1$ ($=41$ for the MSTW 2008 PDFs) evaluations of $A_\ell(\eta_\ell)$ are needed for the best-fit and $2n$ eigenvector PDF sets, exactly as for the usual computation of Hessian uncertainties.  The first method will therefore be used for subsequent results.

The $\chi^2$ distribution of the new data set after reweighting can easily be histogrammed:
\begin{equation}
  \mathcal{P}(\chi^2_a<\chi^2<\chi^2_b) = \frac{1}{N_{\rm pdf}}\sum_{k=1}^{N_{\rm pdf}}w_k(\chi_k^2)\,\Theta(\chi^2_k-\chi_a^2)\Theta(\chi^2_b-\chi_k^2),
\end{equation}
where the $\chi^2$ distribution before reweighting is trivially obtained by setting all weights $w_k$ equal to unity.  Both these distributions are shown in figure~\ref{fig:reweight}(c,d).  The plot legends indicate the mean $\chi^2$ and the standard deviation.  The reweighting procedure shifts the $\chi^2$ distribution so that larger weights are given to the random predictions with $\chi^2_k/N_{\rm pts.}\sim 1$.

If we rescale the data uncertainties by a factor $\alpha$, then the probability density for the rescaling parameter $\alpha$ is given by~\cite{Ball:2010gb}
\begin{equation} \label{eq:palpha}
  \mathcal{P}(\alpha) \propto \frac{1}{\alpha}\sum_{k=1}^{N_{\rm pdf}}W_k\left(\frac{\chi^2_k}{\alpha^2}\right),
\end{equation}
that is, the sum of the unnormalised weights given by eq.~\eqref{eq:weights} with the replacement $\chi^2_k\to \chi^2_k/\alpha^2$.  The overall normalisation of eq.~\eqref{eq:palpha} can be determined from the condition that the integral of $\mathcal{P}(\alpha)$ over $\alpha$ gives unity.  The probability distribution $\mathcal{P}(\alpha)$ is shown in figure~\ref{fig:reweight}(e,f).  These plots suggest that the LHC data on $A_\ell(\eta_\ell)$ are somewhat inconsistent with the data in the MSTW 2008 NLO fit and that the uncertainties on the LHC $A_\ell(\eta_\ell)$ data should be rescaled by a factor 1.37 for CMS and 1.68 for ATLAS to achieve consistency, where these are the most probable values of $\alpha$.  Conversely, a most probable value of $\alpha$ much less than 1 would suggest that the experimental uncertainties are overestimated to some extent.  In that case, it might be desirable to repeat the reweighting procedure with the replacement $\chi^2_k\to \chi^2_k/\alpha^2$ in eq.~\eqref{eq:weights}, where $\alpha$ is the most probable value.

It is clear (see, for example, the discussion in ref.~\cite{Watt:2011kp}) that there is some considerable tension between the LHC $W\to\ell\nu$ charge asymmetry data and some of the data already included in the MSTW 2008 fit, such as the Tevatron $W\to\ell\nu$ asymmetry, the NMC $F_2^d/F_2^p$ ratio, and the E866/NuSea Drell--Yan $\sigma^{pd}/\sigma^{pp}$ ratio.  Other tensions have been observed with the more recent and precise Tevatron data on the $W\to\ell\nu$ charge asymmetry, and partially resolved by more flexible nuclear corrections for deuteron structure functions~\cite{Thorne:2010kj} and extended parameterisation choices for the functional form of the input PDFs.  Indeed, we note that the LHC asymmetry $A_\ell(\eta_\ell)$ depends on valence-quark parameterisations near $x\sim 0.01$, and the studies in section~\ref{sec:parambias} suggested that this is the single place where the MSTW 2008 parameterisation is likely to be inadequate.  Further attempts to resolve these tensions will be necessary for any future update of the MSTW 2008 fit.  Therefore, the reweighting technique is instructive, but does not indicate the ultimate impact of including the new data in a global PDF fit after closer scrutiny of potential sources of tension.  Nevertheless, we hope that the new method presented in this section of generating random predictions on-the-fly from the existing eigenvector PDF sets, followed by subsequent Bayesian reweighting, will be useful for a wide range of potential studies by third parties from both the experimental and theoretical communities.

\section{Conclusions} \label{sec:conclusions}
We have made a first study of the Monte Carlo approach to experimental uncertainty propagation in the context of the MSTW 2008 NLO PDF fit~\cite{Martin:2009iq}, either using data replicas or alternatively working directly in parameter space.  The main findings of this study are as follows:
\begin{itemize}
\item The Hessian method and the Monte Carlo method using data replicas are approximately equivalent in a global fit when using the same parameterisation and (lack of) tolerance, i.e.~$\Delta\chi^2=1$.  Similar findings have previously been observed in a fit only to H1 data~\cite{Dittmar:2009ii}.
\item The Monte Carlo approach using data replicas is better suited to exploring parameterisation bias due to the potentially restrictive input functional form.  Increasing the number of parameters from $20\to28$ has only a small effect on PDF uncertainties, with the exception of the valence-quark distributions at low $x$ values where there is a moderate increase in PDF uncertainties.  This gives some confidence that, in general, PDF uncertainties in the MSTW 2008 fit are not significantly underestimated due to parameterisation bias, with the possible exception of the strange-quark and -antiquark distributions where the imposed parameterisation constraint is more severe due to the lack of available data constraints.
\item The previous findings raise the question why the MSTW/CTEQ uncertainties (\emph{with} a tolerance) are similar to the NNPDF uncertainties (\emph{without} a tolerance)~\cite{Watt:2011kp}, if the tolerance in the former is not compensating for the more restricted functional-form parameterisation rather than the more flexible neural-network parameterisation.  One possibility is that the procedural uncertainties for NNPDF associated with splitting data into training and validation sets mimic the effect of a tolerance for MSTW/CTEQ (see discussion in section 3.2 of ref.~\cite{DeRoeck:2011na}).  Further investigation would be needed by the NNPDF Collaboration to clarify this possible explanation.
\item The Monte Carlo approach using data replicas is also better suited when making fits to limited data sets without the need to restrict the input parameterisation.  We compared the global-fit PDFs to those extracted using a similar flexible parameterisation from more limited data sets either excluding HERA data, including only HERA data, or including only collider (HERA and Tevatron) data.  The fits to limited data sets gave much larger PDF uncertainties for some parton combinations, implying that we need data from HERA, the Tevatron \emph{and} the fixed-target experiments to get a meaningful result.  The PDF uncertainty bands from the fits to the limited data sets are not close to overlapping in many cases, implying that some kind of \emph{tolerance} is needed to accommodate inconsistencies between the various data subsets.
\item As a further exercise to examine the effect of data set inconsistency, we generated idealised pseudodata from the best-fit theory predictions, then we introduced deliberate inconsistencies.  The fractional PDF uncertainties were very similar whether fitting the real data, the consistent pseudodata or the inconsistent pseudodata.  On the other hand, the central values obtained when fitting the inconsistent pseudodata were incompatible accounting for the uncertainty bands, even though the $\chi^2_{\rm global}$ only increased by around 10\% and the $\chi^2/N_{\rm pts.}$ for individual data sets did not deviate far above unity.  Given that a good fit should have $\chi^2/N_{\rm pts.}$ approximately in the range $1\pm \sqrt{2/N_{\rm pts.}}$~\cite{Collins:2001es}, giving $1.0\pm0.1$ for $N_{\rm pts.}\sim 200$, it is far from obvious to spot genuine inconsistencies in the real data of the size we introduced into the idealised pseudodata.  It is definitely not the case that the PDF uncertainties will automatically expand to accommodate any inconsistencies.  Again, this suggests the need for a tolerance to accommodate potential data set inconsistencies in the real data.
\item Having established the need for an appropriate tolerance, we pointed out that it could be introduced by rescaling all experimental uncertainties by a common factor (say, 3) in the generation of data replicas.  However, the introduction of a ``dynamic'' tolerance for each eigenvector direction is not possible, since no use is made of the covariance matrix of fit parameters in the Monte Carlo error propagation.
\item Instead, we proposed sampling the covariance matrix of fit parameters by stepping along each eigenvector direction by a random amount, including the appropriate tolerance values.  This method of generating random PDF sets is close to the usual generation of eigenvector PDF sets in the Hessian method where one steps along each eigenvector direction in turn by a fixed amount.
\item In fact, assuming linearity between the input PDF parameters and derived quantities such as evolved PDFs or hadronic cross sections, an assumption that is inherent in the Hessian method, then it is more convenient to generate random predictions on-the-fly from the existing eigenvector PDF sets.
\item As a simple example application to demonstrate the benefits of having randomly-distributed theory predictions, we used Bayesian reweighting to investigate the effect on the PDFs of recent LHC data on the $W\to\ell\nu$ charge asymmetry.  Similar studies can now easily be performed by third parties using any PDF determination where eigenvector PDF sets are provided.  The reweighting technique is therefore no longer limited only to using the PDF sets provided by the NNPDF Collaboration.
\end{itemize}
We conclude that the Monte Carlo method using data replicas is, on balance, not superior to the Hessian method in a global fit when using a conventional functional-form parameterisation of the input PDFs.  In particular, one of the key benefits of the Monte Carlo approach, namely the use of Bayesian reweighting, can even be accomplished more efficiently using the existing eigenvector PDF sets.  Therefore, any future update of the ``MSTW 2008'' analysis will continue to use the Hessian method with a ``dynamic'' tolerance.

\acknowledgments
We thank J.~Andersen, R.~Cousins, G.~Cowan, S.~Forte, S.~Lauritzen, L.~Lyons, A.~D.~Martin, R.~McNulty, H.~Prosper, J.~Pumplin, J.~Rojo, G.~Salam and W.~J.~Stirling for useful discussions.  The work of R.S.T.~is supported partly by the London Centre for Terauniverse Studies (LCTS), using funding from the European Research Council via the Advanced Investigator Grant 267352.

\bibliographystyle{JHEP}
\bibliography{mcrep}

\providecommand{\href}[2]{#2}\begingroup\raggedright\begin{thebibliography}{10}

\bibitem{Watt:2011kp}
G.~Watt, {\it {Parton distribution function dependence of benchmark Standard
  Model total cross sections at the 7 TeV LHC}},  {\em JHEP} {\bf 1109} (2011)
  069, [\href{http://arxiv.org/abs/1106.5788}{{\tt arXiv:1106.5788}}].

\bibitem{Thorne:2011kq}
R.~S. Thorne and G.~Watt, {\it {PDF dependence of Higgs cross sections at the
  Tevatron and LHC: Response to recent criticism}},  {\em JHEP} {\bf 1108}
  (2011) 100, [\href{http://arxiv.org/abs/1106.5789}{{\tt arXiv:1106.5789}}].

\bibitem{Giele:1998gw}
W.~T. Giele and S.~Keller, {\it {Implications of hadron collider observables on
  parton distribution function uncertainties}},  {\em Phys.Rev.} {\bf D58}
  (1998) 094023, [\href{http://arxiv.org/abs/hep-ph/9803393}{{\tt
  hep-ph/9803393}}].

\bibitem{Giele:2001mr}
W.~T. Giele, S.~A. Keller, and D.~A. Kosower, {\it {Parton distribution
  function uncertainties}},  \href{http://arxiv.org/abs/hep-ph/0104052}{{\tt
  hep-ph/0104052}}.

\bibitem{Ball:2011eq}
{\bf NNPDF} Collaboration, R.~D. Ball {\em et.~al.}, {\it {Parton
  distributions: determining probabilities in a space of functions}},
  \href{http://arxiv.org/abs/1110.1863}{{\tt arXiv:1110.1863}}.

\bibitem{Martin:2009iq}
A.~D. Martin, W.~J. Stirling, R.~S. Thorne, and G.~Watt, {\it {Parton
  distributions for the LHC}},  {\em Eur.Phys.J.} {\bf C63} (2009) 189--285,
  [\href{http://arxiv.org/abs/0901.0002}{{\tt arXiv:0901.0002}}].

\bibitem{Chatrchyan:2011jz}
{\bf CMS} Collaboration, S.~Chatrchyan {\em et.~al.}, {\it {Measurement of the
  lepton charge asymmetry in inclusive $W$ production in pp collisions at
  $\sqrt{s} = 7$ TeV}},  {\em JHEP} {\bf 1104} (2011) 050,
  [\href{http://arxiv.org/abs/1103.3470}{{\tt arXiv:1103.3470}}].

\bibitem{Aad:2011dm}
{\bf ATLAS} Collaboration, G.~Aad {\em et.~al.}, {\it {Measurement of the
  inclusive $W^\pm$ and $Z/\gamma^*$ cross sections in the electron and muon
  decay channels in $pp$ collisions at $\sqrt{s} = 7$~TeV with the ATLAS
  detector}},  {\em Phys. Rev. D85,} {\bf 072004} (2012)
  [\href{http://arxiv.org/abs/1109.5141}{{\tt arXiv:1109.5141}}].

\bibitem{Pumplin:2001ct}
J.~Pumplin, D.~Stump, R.~Brock, D.~Casey, J.~Huston, {\em et.~al.}, {\it
  {Uncertainties of predictions from parton distribution functions. 2. The
  Hessian method}},  {\em Phys.Rev.} {\bf D65} (2001) 014013,
  [\href{http://arxiv.org/abs/hep-ph/0101032}{{\tt hep-ph/0101032}}].

\bibitem{Pumplin:2002vw}
J.~Pumplin, D.~R. Stump, J.~Huston, H.~L. Lai, P.~M. Nadolsky, {\em et.~al.},
  {\it {New generation of parton distributions with uncertainties from global
  QCD analysis}},  {\em JHEP} {\bf 0207} (2002) 012,
  [\href{http://arxiv.org/abs/hep-ph/0201195}{{\tt hep-ph/0201195}}].

\bibitem{Martin:2002aw}
A.~D. Martin, R.~G. Roberts, W.~J. Stirling, and R.~S. Thorne, {\it
  {Uncertainties of predictions from parton distributions. 1: Experimental
  errors}},  {\em Eur.Phys.J.} {\bf C28} (2003) 455--473,
  [\href{http://arxiv.org/abs/hep-ph/0211080}{{\tt hep-ph/0211080}}].

\bibitem{Collins:2001es}
J.~C. Collins and J.~Pumplin, {\it {Tests of goodness of fit to multiple data
  sets}},  \href{http://arxiv.org/abs/hep-ph/0105207}{{\tt hep-ph/0105207}}.

\bibitem{Nakamura:2010zzi}
{\bf Particle Data Group} Collaboration, K.~Nakamura {\em et.~al.}, {\it
  {Review of particle physics}},  {\em J.Phys.G} {\bf G37} (2010) 075021.

\bibitem{Devenish:2004pb}
R.~Devenish and A.~Cooper-Sarkar, {\em {Deep inelastic scattering}}.
\newblock {Oxford University Press}, 2004.

\bibitem{Dittmar:2009ii}
J.~Feltesse, A.~Glazov, and V.~Radescu, {\it {3.2 Experimental Error
  Propagation}},  in {\em {Parton Distributions}} (M.~Dittmar, S.~Forte,
  A.~Glazov, and S.~Moch, eds.), 2009.
\newblock \href{http://arxiv.org/abs/0901.2504}{{\tt arXiv:0901.2504}}.

\bibitem{Tung:2006tb}
W.~K. Tung, H.~L. Lai, A.~Belyaev, J.~Pumplin, D.~Stump, {\em et.~al.}, {\it
  {Heavy Quark Mass Effects in Deep Inelastic Scattering and Global QCD
  Analysis}},  {\em JHEP} {\bf 0702} (2007) 053,
  [\href{http://arxiv.org/abs/hep-ph/0611254}{{\tt hep-ph/0611254}}].

\bibitem{Pumplin:2009bb}
J.~Pumplin, {\it {Parametrization dependence and $\Delta\chi^2$ in parton
  distribution fitting}},  {\em Phys.Rev.} {\bf D82} (2010) 114020,
  [\href{http://arxiv.org/abs/0909.5176}{{\tt arXiv:0909.5176}}].

\bibitem{Glazov:2010bw}
A.~Glazov, S.~Moch, and V.~Radescu, {\it {Parton Distribution Uncertainties
  using Smoothness Prior}},  {\em Phys.Lett.} {\bf B695} (2011) 238--241,
  [\href{http://arxiv.org/abs/1009.6170}{{\tt arXiv:1009.6170}}].

\bibitem{HERA:2009wt}
{\bf H1 and ZEUS} Collaboration, F.~D. Aaron {\em et.~al.}, {\it {Combined
  Measurement and QCD Analysis of the Inclusive $e^\pm p$ Scattering Cross
  Sections at HERA}},  {\em JHEP} {\bf 1001} (2010) 109,
  [\href{http://arxiv.org/abs/0911.0884}{{\tt arXiv:0911.0884}}].

\bibitem{HERA:2010}
{\bf H1 and ZEUS} Collaboration, {\it {PDF fits including HERA-II high $Q^2$
  data}},  2010.
\newblock \texttt{H1prelim-10-142}, \texttt{ZEUS-prel-10-018}.

\bibitem{HERA:2011}
{\bf H1 and ZEUS} Collaboration, {\it {HERAPDF1.5 NNLO}},  2011.
\newblock \texttt{H1prelim-11-042}, \texttt{ZEUS-prel-11-002}.

\bibitem{Watt:2012fj}
G.~Watt, {\it {MSTW PDFs and impact of PDFs on cross sections at Tevatron and
  LHC}},  {\em Nucl.Phys.Proc.Suppl.} {\bf 222-224} (2012) 61--80,
  [\href{http://arxiv.org/abs/1201.1295}{{\tt arXiv:1201.1295}}].

\bibitem{Ball:2011uy}
{\bf NNPDF} Collaboration, R.~D. Ball {\em et.~al.}, {\it {Unbiased global
  determination of parton distributions and their uncertainties at NNLO and at
  LO}},  {\em Nucl.Phys.} {\bf B855} (2012) 153--221,
  [\href{http://arxiv.org/abs/1107.2652}{{\tt arXiv:1107.2652}}].

\bibitem{Pumplin:2009sc}
J.~Pumplin, {\it {Experimental consistency in parton distribution fitting}},
  {\em Phys.Rev.} {\bf D81} (2010) 074010,
  [\href{http://arxiv.org/abs/0909.0268}{{\tt arXiv:0909.0268}}].

\bibitem{Aad:2012sb}
{\bf ATLAS} Collaboration, G.~Aad {\em et.~al.}, {\it {Determination of the
  strange quark density of the proton from ATLAS measurements of the
  $W\to\ell\nu$ and $Z\to\ell\ell$ cross sections}},  {\em Phys.Rev.Lett.} {\bf
  109} (2012) 012001, [\href{http://arxiv.org/abs/1203.4051}{{\tt
  arXiv:1203.4051}}].

\bibitem{Hartland:2012}
N.~Hartland, {\it {LHC data and the proton strangeness}},
  \href{http://arxiv.org/abs/1205.3508}{{\tt arXiv:1205.3508}}.

\bibitem{Alekhin:1996za}
S.~Alekhin, {\it {Extraction of parton distributions and $\alpha_S$ from DIS
  data within the Bayesian treatment of systematic errors}},  {\em Eur.Phys.J.}
  {\bf C10} (1999) 395--403, [\href{http://arxiv.org/abs/hep-ph/9611213}{{\tt
  hep-ph/9611213}}].

\bibitem{DeLorenzi:2010zt}
F.~De~Lorenzi, {\it {Parton Distribution Function sensitivity studies using
  electroweak processes at LHCb}},  \href{http://arxiv.org/abs/1011.4260}{{\tt
  arXiv:1011.4260}}.

\bibitem{DeLorenzi:2011}
F.~De~Lorenzi, {\em {Parton Distribution Function Studies and a Measurement of
  Drell--Yan Produced Muon Pairs at LHCb}}.
\newblock PhD thesis, University College Dublin, 2011.
\newblock {\href{http://cdsweb.cern.ch/record/1449128}{CERN-THESIS-2011-237}}.

\bibitem{Pumplin:2009nk}
J.~Pumplin, J.~Huston, H.~Lai, P.~Nadolsky, W.-K. Tung, {\em et.~al.}, {\it
  {Collider Inclusive Jet Data and the Gluon Distribution}},  {\em Phys.Rev.}
  {\bf D80} (2009) 014019, [\href{http://arxiv.org/abs/0904.2424}{{\tt
  arXiv:0904.2424}}].

\bibitem{Forte:2010dt}
S.~Forte, {\it {Parton distributions at the dawn of the LHC}},  {\em Acta
  Phys.Polon.} {\bf B41} (2010) 2859--2920,
  [\href{http://arxiv.org/abs/1011.5247}{{\tt arXiv:1011.5247}}].

\bibitem{Ball:2010gb}
{\bf NNPDF} Collaboration, R.~D. Ball {\em et.~al.}, {\it {Reweighting NNPDFs:
  the $W$ lepton asymmetry}},  {\em Nucl.Phys.} {\bf B849} (2011) 112--143,
  [\href{http://arxiv.org/abs/1012.0836}{{\tt arXiv:1012.0836}}].

\bibitem{Ball:2011gg}
{\bf NNPDF} Collaboration, R.~D. Ball {\em et.~al.}, {\it {Reweighting and
  Unweighting of Parton Distributions and the LHC $W$ lepton asymmetry data}},
  {\em Nucl.Phys.} {\bf B855} (2012) 608--638,
  [\href{http://arxiv.org/abs/1108.1758}{{\tt arXiv:1108.1758}}].

\bibitem{Catani:2009sm}
S.~Catani, L.~Cieri, G.~Ferrera, D.~de~Florian, and M.~Grazzini, {\it {Vector
  boson production at hadron colliders: A Fully exclusive QCD calculation at
  NNLO}},  {\em Phys.Rev.Lett.} {\bf 103} (2009) 082001,
  [\href{http://arxiv.org/abs/0903.2120}{{\tt arXiv:0903.2120}}].

\bibitem{CMS:muonasymmetry}
{\bf CMS} Collaboration, {\it {Measurement of the Muon Charge Asymmetry in
  Inclusive $W$ Production in $pp$ Collisions at $\sqrt{s}=7$~TeV}},  25th
  August 2011.
\newblock {\href{http://cdsweb.cern.ch/record/1377410}{CMS PAS EWK-11-005}}.

\bibitem{CMS:electronasymmetry}
{\bf CMS} Collaboration, {\it {Measurement of the Electron Charge Asymmetry in
  Inclusive $W$ Production in $pp$ Collisions at $\sqrt{s} = 7$~TeV}},  7th
  March 2012.
\newblock {\href{http://cdsweb.cern.ch/record/1429932}{CMS PAS SMP-12-001}}.

\bibitem{Aaij:2012vn}
{\bf LHCb} Collaboration, R.~Aaij {\em et.~al.}, {\it {Inclusive $W$ and $Z$
  production in the forward region at $\sqrt{s} = 7$~TeV}},  {\em JHEP} {\bf
  1206} (2012) 058, [\href{http://arxiv.org/abs/1204.1620}{{\tt
  arXiv:1204.1620}}].

\bibitem{Thorne:2010kj}
R.~S. Thorne, A.~D. Martin, W.~J. Stirling, and G.~Watt, {\it {The effects of
  combined HERA and recent Tevatron $W\to\ell\nu$ charge asymmetry data on the
  MSTW PDFs}},  {\em PoS} {\bf DIS2010} (2010) 052,
  [\href{http://arxiv.org/abs/1006.2753}{{\tt arXiv:1006.2753}}].

\bibitem{DeRoeck:2011na}
A.~De~Roeck and R.~S. Thorne, {\it {Structure Functions}},  {\em
  Prog.Part.Nucl.Phys.} {\bf 66} (2011) 727--781,
  [\href{http://arxiv.org/abs/1103.0555}{{\tt arXiv:1103.0555}}].

\end{thebibliography}\endgroup

\end{document}